\documentclass[12pt,a4paper,twoside,openright]{report}

\usepackage{ngerman} 
\usepackage{calc} 
\usepackage{fancyhdr} 
\usepackage{exscale,amssymb} 
\usepackage[intlimits]{amsmath} 
\usepackage{mathdefs} 
\usepackage{graphicx} 
\usepackage[sort]{cite} 
\usepackage[hang,bf]{caption2} 

\graphicspath{{fig/}{axodraw/}}

\allowdisplaybreaks

\fancypagestyle{plain}{%
  \fancyhf{}
  \fancyfoot[C]{\thepage}

}

\renewcommand{\chaptermark}[1]{%
  \markboth{\chaptername\ \thechapter.\hspace{1em}#1}{}}

\newcommand{\clearemptypage}{\clearpage{\pagestyle{empty}\cleardoublepage}}

\newcommand{\valignbox}[2][]{\begin{tabular}[#1]{@{}c@{}} #2 \end{tabular}}

\newcommand{\vcentergraphics}[2][]{%
  \valignbox{\raisebox{-2ex}{\includegraphics[#1]{#2}}}}

\begin{document}



\begin{titlepage}
\centering
\large

\textbf{\Huge
  Sudakov-Logarithmen\\
  in der\\[1ex]
  elektroschwachen Wechselwirkung}

\vfill

Zur Erlangung des akademischen Grades eines\\
\textsc{\Large Doktors der Naturwissenschaften}\\
von der Fakult"at f"ur Physik der\\
Universit"at Karlsruhe (TH)\\[3ex]
genehmigte\\[1.5cm]
\textsc{\Large Dissertation}\\[1.5cm]
von\\[1.5cm]
\textbf{\Large Dipl.-Phys. Bernd Joachim Jantzen}\\[1ex]
aus Backnang

\vfill

\begin{tabular}{l@{\hspace{1em}}l}
  Tag der m"undlichen Pr"ufung: & 17. Juni 2005 \\
  Referent: & Prof. Dr. Johann H. K"uhn \\
  Korreferent: & Prof. Dr. Dieter Zeppenfeld \\
\end{tabular}

\end{titlepage}

\clearemptypage


\tableofcontents
\addtocontents{toc}{\protect\markboth{\contentsname}{\contentsname}}


\clearemptypage

\chapter*{Einleitung}
\addcontentsline{toc}{chapter}{Einleitung}
\markboth{Einleitung}{Einleitung}

In der Teilchenphysik werden die fundamentalen Wechselwirkungen zwischen den
Elementarteilchen, mit Ausnahme der Gravitation, im Rahmen des
Standardmodells betrachtet.
Dieses liefert eine quantenfeldtheoretische Beschreibung der
elektromagnetischen, der schwachen und der starken Wechselwirkung.
In den vergangenen Jahrzehnten wurde das Standardmodell in vielen Experimenten
getestet und hat sich
als ein "au"serst erfolgreiches theoretisches Modell erwiesen.
Pr"azisionsmessungen konnten sowohl die theoretischen Vorhersagen
in den bisher experimentell zug"anglichen Energiebereichen best"atigen
als auch die Parameter des Modells mit gro"ser Genauigkeit bestimmen.
Die Genauigkeit der Experimente verlangt auch auf theoretischer Seite immer
bessere Vorhersagen.

Von besonderem Interesse sind heute
neben der Suche nach dem noch nicht entdeckten Higgs-Boson
m"ogliche Erweiterungen des Standardmodells
bzw. die Betrachtung des Standardmodells als Niederenergielimes
einer zugrunde liegenden Theorie, die ggf. auch die Gravitation mit
einschlie"st.
Viele neue Effekte dieser Erweiterungsmodelle
werden erst bei Energien, die weit oberhalb des experimentell zug"anglichen
Bereichs liegen, relevant.
Aber sie "au"sern sich bereits im Niederenergiebereich durch kleine Korrekturen
zu den Standardmodellvorhersagen.
Bei den Pr"azisionsmessungen an k"unftigen Teilchenbeschleunigern
hofft man insbesondere, solche Abweichungen sehen zu k"onnen.

Die elektroschwache Wechselwirkung wird durch
das masselose Photon und die massiven
$W$- und $Z$-Bosonen ($M_W \approx 80\,\GeV$, $M_Z \approx 91\,\GeV$)
vermittelt.
Diese Wechselwirkung wurde experimentell von sehr niedrigen
Energieskalen
bis hin zu Energieskalen im Bereich der $W$- und $Z$-Masse untersucht.
Der LEP-Beschleuniger am CERN erreichte vor seiner Abschaltung eine maximale
Schwerpunktsenergie der Elektron-Positron-Kollisionen von etwa 210~GeV.
Das Tevatron am Fermilab beschleunigt Protonen und Antiprotonen bis etwa 2~TeV.
Jedoch geht bei einem Hadron-Beschleuniger nur ein Teil dieser Energie in
die jeweilige harte Streureaktion ein.
Bei diesen Experimenten liegen die Impuls"ubertr"age der elektroschwachen
Prozesse noch in der Gr"o"senordnung von $M_W$ und~$M_Z$.

Zuk"unftige bzw. geplante Elektron-Positron-Linearbeschleuniger wie der
\emph{Inter\-national Linear Collider} (ILC) mit Energien weit "uber den
Eichboson-Massen und Hadron-Beschleuniger wie der LHC mit einer
Schwerpunktsenergie von 14~TeV er"offnen den Zugang zu einem neuen
Energiebereich.
In exklusiven Prozessen, d.h. mit Endzust"anden ohne zus"atzliche Abstrahlung
von reellen Eichbosonen, werden bei hohen Impuls"ubertr"agen
doppelt-logarithmische Korrekturen, sogenannte \emph{Sudakov-Logarithmen}
relevant.
Diese r"uhren vom Austausch virtueller Eichbosonen her
und wurden zuerst von Sudakov~\cite{Sudakov:1954sw} in der
Quantenelektrodynamik untersucht.

In den Strahlungskorrekturen zu exklusiven Prozessen
mit Eichbosonen der Masse~$M$ bei hohen Impuls"ubertr"agen~$\sqrt s$
erh"alt man Potenzen des Logarithmus $\ln(s/M^2)$.
Terme der Ordnung~$M^2/s$ werden im Hochenergielimes vernachl"assigt.
Die N"aherung der f"uhrenden Logarithmen (LL = leading logarithmic)
umfasst in jeder Ordnung~$\alpha^n$ der St"orungstheorie die Terme
mit den h"ochsten Logarithmen"=Potenzen, $\alpha^n \ln^{2n}(s/M^2)$.
Weitere N"aherungen schlie"sen die n"achstf"uhrenden Logarithmen
$\alpha^n \ln^{2n-1}(s/M^2)$ (NLL = next-to-leading logarithmic),
die n"achstn"achstf"uhrenden Logarithmen
$\alpha^n \ln^{2n-2}(s/M^2)$ (NNLL = next-to-next-to-leading logarithmic)
usw. mit ein.
Im Limes $s \to \infty$ stellen die f"uhrenden Logarithmen den gr"o"sten
Beitrag
zu den Strahlungskorrekturen dar.
Die ph"anomenologische Untersuchung von Sudakov-Logarithmen in Bezug auf
elektroschwache Prozesse bei hohen Energien hat in den vergangenen Jahren
ein deutliches Interesse erfahren\cite{Beccaria:1998qe,Ciafaloni:1998xg,
  Kuhn:1999de,Fadin:2000bq,Kuhn:2000nn,Kuhn:2000hx,
  Beccaria:1999xd,Beccaria:1999fk,Beccaria:2000jz,Beccaria:2001yf,
  Denner:2001jv,Denner:2001gw,Melles:2000ia,
  Hori:2000tm,Beenakker:2001kf,Denner:2003wi,Pozzorini:2004rm,
  Ciafaloni:2000df,Ciafaloni:2001vt,Ciafaloni:2000rp,
  Kuhn:2001hz,Feucht:2003yx,Feucht:2004rp,
  Kuhn:2004em,Accomando:2004de,Denner:2004iz,
  Jantzen:2005xi}.

In bisherigen Arbeiten wurden der Formfaktor eines abelschen Vektorstroms,
die Vierfermionamplitude sowie Wirkungsquerschnitte f"ur massive abelsche
und nichtabelsche Theorien in LL-, NLL- und NNLL-N"aherung
berechnet~\cite{Kuhn:1999de,Fadin:2000bq,Kuhn:2000nn,Kuhn:2000hx,Kuhn:2001hz}.
Bei den experimentell relevanten Energien im TeV-Bereich
spielen jedoch auch die darauffolgenden Terme der Entwicklung in
Logarithmenpotenzen eine wichtige Rolle.
F"ur Pr"azisionsergebnisse ist die Kenntnis
aller logarithmischen Terme im Hochenergielimes erforderlich.

Die vorliegende Arbeit befasst sich mit abelschen und nichtabelschen
Zweischleifenkorrekturen zum Formfaktor eines abelschen Vektorstroms.
Diese Ergebnisse werden mit Hilfe von Evolutionsgleichungen zur
Vierfermionamplitude kombiniert und zur Berechnung von Korrekturen zu
elektroschwachen Wirkungsquerschnitten von Vierfermionprozessen
$f \bar f \to f' \bar f'$ verwendet.
Die Ergebnisse der Arbeit wurden
in Zusammenarbeit mit Johann H. K"uhn, Sven Moch, Alexander A. Penin und
Vladimir A. Smirnov erzielt und teilweise
in\cite{Feucht:2003yx,Feucht:2004rp,Jantzen:2005xi}
ver"offentlicht{\footnotemark}
bzw. zur Ver"offentlichung eingereicht,
weitere Ver"offentlichungen sind in Vorbereitung.
\footnotetext{Der Name des Autors hat sich durch Heirat von B.~Feucht
  nach B.~Jantzen ge"andert, so dass nun Ver"offentlichungen unter
  beiden Namen vorliegen.}

Die Arbeit ist wie folgt gegliedert:
Zun"achst werden in Kapitel~\ref{chap:grundlagen} die Grundlagen der
Rechnungen vorgestellt: das Standardmodell und das in dieser Arbeit
verwendete $SU(2)\times U(1)$-Modell, die Behandlung elektroschwacher
Strahlungskorrekturen, die Sudakov-Logarithmen in der Vierfermionamplitude,
der Formfaktor des abelschen Vektorstroms und die sp"ater ben"otigten
Einschleifenergebnisse.

Danach folgen drei Kapitel, in denen verschiedene Beitr"age zum
$SU(2)$-Formfaktor vorgestellt und berechnet werden:
in Kapitel~\ref{chap:nfns} die Beitr"age aus Fermionschleifen und Schleifen
skalarer Teilchen,
in Kapitel~\ref{chap:abelsch} die abelschen Beitr"age, die auch in einem
reinen $U(1)$-Modell vorkommen,
und in Kapitel~\ref{chap:nichtabelsch} Beitr"age aus der nichtabelschen
$SU(2)$-Gruppenstruktur sowie die Higgs-Beitr"age.

Kapitel~\ref{chap:SU2U1} kombiniert die schwache
$SU(2)$"=Wechselwirkung mit der $U(1)$-Wechsel\-wirkung der Hyperladung,
zeigt die Faktorisierung der infraroten Singularit"aten
und erl"autert eine M"oglichkeit zur Berechnung des Formfaktors mit
massiven und masselosen Eichbosonen durch die Auswertung von Korrekturen
mit rein massiven Bosonen.

Abschlie"send "ubertr"agt Kapitel~\ref{chap:EW2loop} die Ergebnisse f"ur
den Formfaktor auf die Vierfermionamplitude im elektroschwachen
Standardmodell, diskutiert die Ergebnisse f"ur Wirkungsquerschnitte von
Vierfermionprozessen und fasst die Resultate dieser Arbeit zusammen.

Im Anhang~\ref{chap:feynman} sind die Feynman-Regeln des $SU(2) \times
U(1)$-Modells mit spontaner Symmetriebrechung angegeben.
Anhang~\ref{chap:math} erl"autert die in dieser Arbeit verwendeten
mathematischen Funktionen und Konstanten.
Anhang~\ref{chap:methoden} geht auf Methoden ein, die f"ur die
Auswertung der Schleifenintegrale zur Anwendung kommen.
Die Ergebnisse der einzelnen skalaren Integrale aus den Kapiteln
\ref{chap:abelsch} und \ref{chap:nichtabelsch} sind im
Anhang~\ref{chap:Skalar} aufgelistet.
Das Literaturverzeichnis und die Danksagung finden sich am Ende der Arbeit
nach den Anh"angen.


\clearemptypage

\chapter{Elektroschwache Strahlungskorrekturen bei hohen Energien}
\label{chap:grundlagen}

In diesem Kapitel werden die theoretischen Grundlagen der vorliegenden
Arbeit erl"autert.
Am Beginn steht eine Beschreibung der f"ur diese Arbeit relevanten Aspekte
des Standardmodells.
Der nachfolgende Abschnitt behandelt das f"ur die Rechnungen verwendete
$SU(2) \times U(1)$-Modell.
Abschnitt~\ref{sec:Schleifen} geht auf die Behandlung von
elektroschwachen Strahlungskorrekturen ein.
Abschnitt~\ref{sec:Renormierung} erl"autert das verwendete Renormierungsschema.
In Abschnitt~\ref{sec:Sudakov} werden Sudakov-Logarithmen in der
Vierfermionstreuung, die Vierfermionamplitude und der Formfaktor
betrachtet.
Das Kapitel schlie"st mit einer Zusammenstellung der f"ur die weiteren
Rechnungen relevanten Einschleifenergebnisse.

%
%
\section{Das Standardmodell und die elektroschwache Wechselwirkung}
\label{sec:Standardmodell}

Das Standardmodell der Teilchenphysik beschreibt die fundamentalen
Wechselwirkungen zwischen den Elementarteilchen (siehe
z.B. \cite{Peskin:1995ev,Nachtmann:1986ib,Cheng:1984bj,Bohm:2001yx}).
Dazu geh"oren die elektromagnetische, die schwache und die starke
Wechselwirkung. Die Gravitation ist im Standardmodell nicht enthalten.
Auf die Theorie der Quantenchromodynamik (QCD) f"ur die starke
Wechselwirkung soll hier nicht eingegangen werden,
da bei der vorliegenden Arbeit nur elektroschwache Effekte
ber"ucksichtigt werden.

Das Modell von Glashow~\cite{Glashow:1961tr}, Salam~\cite{Salam:ed.rm}
und Weinberg~\cite{Weinberg:1967tq}
vereinigt elektromagnetische und schwache Wechselwirkung zur
elektroschwachen Wechselwirkung.
Die Fermionen dieses Modells sind in 3~Familien von Leptonen und
3~Familien von Quarks angeordnet.
Sie sind Spin"~$\frac12$-Teilchen, ihre Helizit"at
(Projektion des Spins auf die
Impulsrichtung) kann die Werte $+\frac12$ ("`rechtsh"andig"') und
$-\frac12$ ("`linksh"andig"') annehmen.
Bez"uglich der schwachen Wechselwirkung werden die linksh"andigen Leptonen
jeder Familie~$i$ ($l_i^L$ und zugeh"origes Neutrino $\nu_i^L$)
und die linksh"andigen Quarks ($u_i^L$ und $d_i^L$) in SU(2)-Dubletts
\[
  L_i^L = \binom{\nu_i^L}{l_i^L}
  \quad \text{und} \quad
  Q_i^L = \binom{u_i^L}{d_i^L}
\]
mit Isospin~$\frac12$ zusammengefasst,
w"ahrend sich die rechtsh"andigen Fermionen in Singletts
\[
  l_i^R \,, \quad u_i^R \quad \text{und} \quad d_i^R
\]
mit Isospin~0 befinden.
Dies erm"oglicht die unterschiedliche Kopplung der schwachen Wechselwirkung an
links- und rechtsh"andige Fermionen.
Rechtsh"andige Neutrinos sind urspr"unglich im Standardmodell nicht enthalten,
k"onnen aber beispielsweise
als zus"atzliche Isospin-Singletts eingef"uhrt werden.
Insgesamt beinhaltet das Modell folgende Fermion-Dubletts und
Fermion-Singletts:
\begin{gather*}
  L_i^L = \binom{\nu_e}{e}^L, \binom{\nu_\mu}{\mu}^L,
    \binom{\nu_\tau}{\tau}^L \,;\quad
  Q_i^L = \binom{u}{d}^L, \binom{c}{s}^L, \binom{t}{b}^L
  \\
  l_i^R = e^R, \mu^R, \tau^R \,;\quad
  u_i^R = u^R, c^R, t^R \,;\quad d_i^R = d^R, s^R, b^R
  \,.
\end{gather*}

Die Symmetriegruppe der elektroschwachen Wechselwirkung ist
$SU(2)_W \times U(1)_Y$.
Generatoren der $SU(2)_W$-Symmetrie sind die Komponenten~$I^a_W$ ($a=1,2,3$)
des schwachen Isospins.
F"ur die Dubletts gilt $I_W^a = \frac12 \sigma^a$
mit den Pauli"=Matrizen~$\sigma^a$.
F"ur die Singletts ist $I_W^a = 0$.

Zur $U(1)_Y$-Gruppe geh"ort der Generator $Y$ mit der
Hyperladung als Quantenzahl.
Die Hyperladung h"angt mit der Komponente $I_W^3$ des Isospins und der
elektrischen Ladung~$Q$ (in Einheiten der Elementarladung)
"uber die Gell-Mann-Nishijima-Relation
\begin{equation}
  Q = I_W^3 + \frac{Y}{2}
\end{equation}
zusammen.
Die Quantenzahlen der elektroschwachen Wechselwirkung sind in
Tabelle~\ref{tab:SMfermionen} dargestellt.
\begin{table}[ht]
  \[
    \renewcommand{\arraystretch}{1.3}
    \renewcommand{\box}[1]{\makebox[1.5cm]{$#1$}}
    \begin{array}{|c|c|c||c|c|c|}
    \hline
      \multicolumn{3}{|c||}{\text{Familie}} &
        \multicolumn{3}{c|}{\text{Quantenzahlen}} \\
      \box{1} & \box{2} & \box{3} &
        \box{I_W^3} & \box{\frac{Y}{2}} & \box{Q} \\ \hline
    \hline
      \nu_e^L & \nu_\mu^L & \nu_\tau^L & \frac12 & -\frac12 & 0 \\ \hline
      e^L & \mu^L & \tau^L & -\frac12 & -\frac12 & -1 \\ \hline
      e^R & \mu^R & \tau^R & 0 & -1 & -1 \\ \hline
    \hline
      u^L & c^L & t^L & \frac12 & \frac16 & \frac23 \\ \hline
      d^L & s^L & b^L & -\frac12 & \frac16 & -\frac13 \\ \hline
      u^R & c^R & t^R & 0 & \frac23 & \frac23 \\ \hline
      d^R & s^R & b^R & 0 & -\frac13 & -\frac13 \\ \hline
    \end{array}
  \]
  \caption{\label{tab:SMfermionen}
    Standardmodell-Fermionen mit Quantenzahlen}
\end{table}
Zudem gibt es jedes Quark in 3~Farben, zwischen denen die starke Wechselwirkung
transformiert. Die Generatoren der elektroschwachen Wechselwirkung sind
diagonal im Farbraum, so dass die Farben lediglich bei der Anzahl der
Fermionen ber"ucksichtigt werden m"ussen.

Die kinetische Lagrange-Dichte $\Lc = \bar\psi \, i\dslash\partial \, \psi$
eines Fermion-Dubletts oder -Singletts~$\psi$ ist unter einer globalen
$SU(2)_W \times U(1)_Y$-Transformation invariant.
F"ur die Invarianz unter einer lokalen, d.h. ortsabh"angigen
Transformation
wird die gew"ohnliche Ableitung~$\partial_\mu$ durch die kovariante
Ableitung~$D_\mu$ ersetzt:
\begin{equation}
\label{eq:Dkov}
  D_\mu = \partial_\mu - ig\,I_W^a\,W^a_\mu - ig'\,\frac{Y}{2}\,B_\mu
  \,.
\end{equation}
Die Transformation der Eichfelder $W^a_\mu$ und~$B_\mu$ gew"ahrleistet die
lokale $SU(2)_W \times U(1)_Y$-Eichsymmetrie der Lagrange-Dichte.
Die kinetische Lagrange-Dichte der Fermionen im Standardmodell lautet:
\begin{equation}
\label{eq:LSMFermion}
  \Lc_f = \bar L_i^L \, i\dslash D \, L_i^L
    + \bar Q_i^L \, i\dslash D \, Q_i^L
    + \bar l_i^R \, i\dslash D \, l_i^R
    + \bar u_i^R \, i\dslash D \, u_i^R + \bar d_i^R \, i\dslash D \, d_i^R
  \,.
\end{equation}

Die Eichbosonen $W^a$ und $B$ vermitteln die elektroschwache
Wechselwirkung. Sie erhalten in der Lagrange-Dichte den kinetischen Term
\begin{equation}
\label{eq:LEich}
  \Lc_{\text{Eich}} = -\frac{1}{4} W^a_{\mu\nu} W^{a\mu\nu}
    - \frac{1}{4} B_{\mu\nu} B^{\mu\nu}
  \,,
\end{equation}
mit
\begin{equation}
  W^a_{\mu\nu} = \partial_\mu W^a_\nu - \partial_\nu W^a_\mu
    + g\,f^{abc} \, W^b_\mu W^c_\nu
  \,,\quad
  B_{\mu\nu} = \partial_\mu B_\nu - \partial_\nu B_\mu
  \,,
\end{equation}
wobei $f^{abc} = \eps^{abc}$ die $SU(2)$-Strukturkonstanten sind.
Die Kopplung der Eichbosonen an die Fermionfelder geschieht
"uber die kovariante
Ableitung~(\ref{eq:Dkov}) in der Lagrange-Dichte~(\ref{eq:LSMFermion})
und wird durch die Kopplungskonstanten $g$ und $g'$ bestimmt.

Die Massen der Eichbosonen und Fermionen werden im Standardmodell "uber
den Higgs-Mechanismus eingef"uhrt.
Das Higgs-Feld~$\Phi$ ist ein $SU(2)_W$-Dublett komplexer Skalarfelder
mit Isospin~$\frac12$ und Hyperladung $Y_\Phi = 1$.
Durch ein Higgs-Potential, dessen Minimum bei $\Phi\ne0$ liegt,
erh"alt eine Komponente des Higgs-Dubletts einen endlichen
Vakuumerwartungswert.
Dies f"uhrt zur spontanen Brechung der \mbox{$SU(2)_W \times
U(1)_Y$}-Symmetrie, da der Vakuumerwartungswert nur unter der
$U(1)_{\text{em}}$-Symmetrie invariant bleibt, deren Generator die
elektrische Ladung~$Q$ ist.

Die kinetische Energie des Higgs-Dubletts lautet
$(D_\mu\Phi)^\dag (D^\mu\Phi)$ mit der kovarianten Ableitung~(\ref{eq:Dkov}).
Dadurch koppelt das Higgs-Dublett an die Eichbosonen, so dass
der Vakuumerwartungswert zu Massentermen f"ur die Eichbosonen in der
Lagrange-Dichte f"uhrt.
Weil das Higgs-Dublett sowohl an die $SU(2)_W$-Eichbosonen als auch an das
$U(1)_Y$-Eichboson koppelt, mischen diese.
Masseneigenzust"ande sind das masselose Photon sowie
die massiven Eichbosonen $W^\pm$ und $Z$
mit
\begin{equation}
  M_W = M_Z \, \cos\theta_W
  \,,
\end{equation}
wobei $\theta_W$ der schwache Mischungswinkel ist, der auch die $SU(2)_W$-
und $U(1)_Y$-Kopplungskonstanten mit der elektrischen Elementarladung~$e$
in Beziehung setzt:
\begin{equation}
\label{eq:eggprel}
  e = \frac{g g'}{\sqrt{g^2+{g'}^2}} = g\,\sin\theta_W = g'\,\cos\theta_W
  \,.
\end{equation}

Die Fermionen erhalten ihre Massen im Standardmodell durch
$SU(2)_W \times U(1)_Y$-invariante Yukawa-Kopplungen des Higgs-Dubletts an
die Fermionfelder.

%
%
\section[Ein $SU(2) \times U(1)$-Modell ohne Mischung]
  {\boldmath Ein $SU(2) \times U(1)$-Modell ohne Mischung}
\label{sec:SU2U1}

Die Berechnung von Strahlungskorrekturen im Rahmen der elektroschwachen
Wechselwirkung wird insbesondere durch die Vielzahl von verschiedenen
Massen erschwert. Allein bei den Eichbosonen gibt es das masselose Photon
sowie die schweren Eichbosonen $W$ und $Z$ mit "ahnlicher, aber
verschiedener Masse.

Um Zweischleifenvorhersagen zu erm"oglichen, arbeiten wir in einem
vereinfachten Modell, in dem alle schweren Eichbosonen die gleiche Masse
$M_W = M_Z \equiv M$ besitzen.
Dies kann im Rahmen des Higgs-Mechanismus erreicht werden, wenn die Eichbosonen
der verschiedenen Symmetriegruppen nicht mischen. Deshalb betrachten wir
ein Higgs-Dublett mit Hyperladung $Y_\Phi = 0$, das nur an die schwachen
Eichbosonen~$W^a$, nicht aber an das $U(1)$-Eichboson~$B$ koppelt.

Die folgende Beschreibung des $SU(2)\times U(1)$-Modells entspricht in weiten
Teilen dem Standardmodell. Der Unterschied zum Standardmodell
liegt vor allem in den Masseneigenzust"anden der Eichbosonen. Dies wird
am Ende dieses Abschnitts nochmals Thema sein.
Zun"achst wird die Lagrangedichte des Modells detailliert beschrieben.

Der kinetische Term der Eichbosonen in der Lagrange-Dichte ist weiterhin
durch Gl.~(\ref{eq:LEich}) gegeben.
Dieser enth"alt auch die Selbstkopplungen der Eichbosonen.
Der kinetische Term von allgemein $n_f$~linksh"andigen Fermion-Dubletts in der
fundamentalen Darstellung der $SU(2)$ und der $U(1)$,
\begin{equation}
  \Lc_f = \sum_{i=1}^{n_f} \, \bar\Psi_i^L \, i\dslash D \, \Psi_i^L
  \,,
\end{equation}
mit der kovarianten Ableitung
\begin{equation}
\label{eq:DkovSU2U1}
  D_\mu = \partial_\mu - ig\,t^a\,W^a_\mu - ig'\,\frac{Y}{2}\,B_\mu
  \,,
\end{equation}
liefert die Kopplungen zwischen Eichbosonen und Fermionen.
Dabei sind die $t^a$ ($a=1,2,3$) die $SU(2)$-Generatoren
in der fundamentalen Darstellung
mit der Lie-Algebra $[ t^a, t^b ] = i\,f^{abc}\,t^c$,
wobei die $f^{abc}$ die Strukturkonstanten der Gruppe und zugleich
(bis auf einen Faktor $i$) die
Generatoren der adjungierten Darstellung sind.
Allgemein gilt
\begin{equation}
\label{eq:Casimir}
  t^a t^a = C_F \, \unity \,,\quad
  f^{acd} f^{bcd} = C_A \, \delta^{ab} \,,\quad
  \Tr(t^a t^b) = T_F \, \delta^{ab}
  \,,
\end{equation}
mit dem Einheitsoperator~$\unity$ und den Casimir-Operatoren $C_F$ und
$C_A$ der fundamentalen und der adjungierten Darstellung.
F"ur eine $SU(N)$-Gruppe ist
\begin{equation}
  C_F = \frac{N^2-1}{2N} \,,\quad
  C_A = N \,,\quad
  T_F = \frac{1}{2}
  \,,
\end{equation}
oder speziell f"ur die $SU(2)$-Eichgruppe der schwachen Wechselwirkung:
\begin{equation}
\label{eq:CasimirSU2}
  C_F = \frac{3}{4} \,,\quad C_A = 2 \,,\quad T_F = \frac{1}{2}
  \,.
\end{equation}
Indem man anstelle der speziellen
$SU(2)$-Generatoren~$\frac12 \sigma^a$ die allgemeineren
$SU(N)$-Generatoren~$t^a$ und die daraus resultierenden Farbfaktoren mit
$C_F$, $C_A$ und $T_F$ benutzt,
kann man die Beitr"age zu den Strahlungskorrekturen nach verschiedenen
Farbfaktoren trennen sowie die
Rechnungen auf eine abelsche $U(1)$-Gruppe mit
\begin{equation}
\label{eq:CasimirU1}
  C_F = 1 \,,\quad C_A = 0 \,,\quad T_F = 1
\end{equation}
"ubertragen.

In der kovarianten Ableitung~(\ref{eq:DkovSU2U1}) muss f"ur $Y$ die
$U(1)$-Hyperladung des jeweiligen Fermion-Dubletts entsprechend der
Tabelle~\ref{tab:SMfermionen} eingesetzt werden.
Zus"atzlich werden kinetische Terme f"ur die rechtsh"andigen Fermion-Singletts
entsprechend~(\ref{eq:LSMFermion}) eingef"uhrt, die nur Kopplungen an das
$B$-Feld enthalten.

Das Higgs-Feld ist ein komplexes skalares Dublett in der fundamentalen
Darstellung der $SU(2)$ und wird durch die Lagrange-Dichte
\begin{equation}
\label{eq:LHiggs}
  \Lc_H = (D_\mu\Phi)^\dag (D^\mu\Phi) - V(\Phi)
  \,,
\end{equation}
mit der kovarianten Ableitung~(\ref{eq:DkovSU2U1}) f"ur $Y_\Phi = 0$
und dem Higgs-Potential
\begin{equation}
  V(\Phi) =
  \frac{\lambda^2}{2} \left( \Phi^\dag\Phi - \frac{v^2}{2} \right)^2
  \,,\quad
  \lambda,v > 0 \,,
\end{equation}
beschrieben.
Das Minimum des Higgs-Potentials liegt bei $\Phi^\dag \Phi = v^2/2$,
so dass das Higgs-Dublett einen Vakuumerwartungswert bekommt, der
per Konvention als
\begin{equation}
  \Phi_0 = \frac{1}{\sqrt 2} \binom{0}{v}
\end{equation}
gew"ahlt wird.
Die Entwicklung des Higgs-Dubletts um den Vakuumerwartungswert kann durch die
Anwendung  einer $SU(2)$-Transformation auf $\Phi_0$ gewonnen werden:
\begin{equation}
  \Phi(x) =
  \left( 1 + \frac{1}{v} H(x) + \frac{2i}{v} t^a \phi^a(x) \right) \Phi_0
  = \frac{1}{\sqrt 2} \binom{i\phi^1(x) + \phi^2(x)}{v + H(x) - i\phi^3(x)}
  \,,
\end{equation}
mit dem reellen, physikalischen Higgs-Feld~$H$ und den reellen
Goldstone-Bosonen~$\phi^a$.
Dies ist auch im Fall eines $U(1)$-Higgs-Modells m"oglich, kann aber nicht
allgemein auf $SU(N)$-Modelle "ubertragen werden.
Die Felder~$\phi^a$ sind unphysikalische Freiheitsgrade, die unter
Ausnutzung der $SU(2)$-Symmetrie wegtransformiert werden k"onnen.
Sie m"ussen aber je nach der gew"ahlten Eichfixierung (siehe unten)
ber"ucksichtigt werden.

Neben den Kopplungen zwischen den Feldern enth"alt die
Lagrange-Dichte~(\ref{eq:LHiggs}) auch Massenterme f"ur das physikalische
Higgs-Feld und f"ur die $SU(2)$-Eichbosonen:
\begin{equation}
  M_H = \lambda v \,,\quad
  M \equiv M_W = \frac{g v}{2}
  \,.
\end{equation}
Das $B$-Feld bleibt masselos.
Dies ist der Unterschied zum Standardmodell, in dem die Felder $W^3$ und
$B$ die Masseneigenzust"ande $Z$ und Photon bilden.

Die Eichfixierung wird entsprechend der Klasse der $R_\xi$-Eichungen
durch den Zusatzterm
\begin{equation}
\label{eq:Lfix}
  \Lc_{\text{fix}} =
    -\frac{1}{2\xi} G^a G^a - \frac{1}{2\xi} (\partial^\mu B_\mu)^2
  \,,\quad
  G^a = \partial^\mu W^a_\mu + \xi \frac{gv}{2} \phi^a
  \,,
\end{equation}
in der Lagrange-Dichte erreicht.
Mit der BRST-Symmetrie\cite{Becchi:1975nq,Tyutin:1975,Iofa:1976je}
kann gezeigt werden, dass physikalische Ergebnisse
wie S-Matrixelemente nicht von der Wahl des Parameters~$\xi$ abh"angen.
Der Eichfixierungsterm~(\ref{eq:Lfix}) gibt den unphysikalischen
Goldstone-Bosonen~$\phi^a$ die Masse~$\sqrt\xi M$.
Die Eichabh"angigkeit von $\Lc_{\text{fix}}$ erzwingt die Einf"uhrung
antivertauschender Geistfelder~$c^a$ in der adjungierten Darstellung der
$SU(2)$-Gruppe:
\begin{equation}
  \Lc_{\text{Geist}} = \bar c^a \left[
    \left( -\partial^2 - \xi M^2 - \xi M \frac{g}{2} H \right) \delta^{ab}
    + \left( g \partial^\mu W^c_\mu + \xi M \frac{g}{2} \phi^c \right)
      f^{abc}
    \right] c^b
  \,.
\end{equation}
Wie die Goldstone-Bosonen haben die Geistfelder die Masse~$\sqrt\xi M$.
Im Hochenergielimes (siehe Abschnitt~\ref{sec:Sudakov}) werden die Massen
der Fermionen vernachl"assigt, so dass auf die Einf"uhrung der
Yukawa-Kopplungen hier verzichtet wird.

Aus der Lagrange-Dichte erh"alt man die Feynman-Regeln, die
im Anhang~\ref{chap:feynman} zusammengestellt sind.
In dieser Arbeit wird ausschlie"slich in der Feynman-'t~Hooft-Eichung mit
$\xi=1$ gearbeitet, in der die Eichbosonpropagatoren eine besonders
einfache Form haben.

Dieses $SU(2) \times U(1)$-Modell ist dem Standardmodell sehr "ahnlich.
Ungew"ohnlich ist, dass hier mit den $SU(2)$-Eichbosonen $W^1$, $W^2$ und
$W^3$ sowie mit dem $U(1)$-Eichboson~$B$  gerechnet wird anstatt mit
$W^\pm$, $Z$ und dem Photon.
Dies vereinfacht die Rechnungen erheblich, denn anstelle von einzelnen,
verschiedenen Eichbosonen braucht neben dem $U(1)$-Eichboson~$B$ nur das
$SU(2)$-Triplett~$W^a$ in der adjungierten Darstellung betrachtet zu werden.

Durch die Vernachl"assigung der Mischung zwischen den beiden Eichgruppen
wird ein relativer Fehler in der Gr"o"senordnung von
$\sin^2\theta_W \approx 0{,}2$ gemacht.
Dieser wird bei der Zusammenfassung der Ergebnisse in
Kapitel~\ref{chap:EW2loop} diskutiert.

%
%
\section{Strahlungskorrekturen und IR-Divergenzen}
\label{sec:Schleifen}

\begin{figure}[ht]
  \centering
  \includegraphics{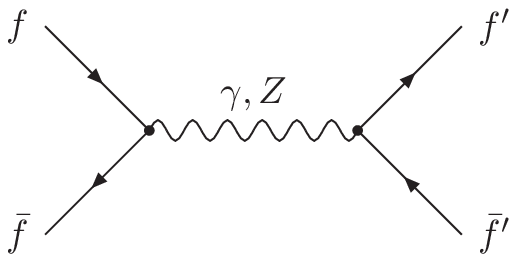}
  \caption{Born-N"aherung des Vierfermionprozesses $f \bar f \to f' \bar f'$}
  \label{fig:4fBorn} 
\end{figure}

Das Standardmodell erlaubt die Beschreibung von Prozessen, die in der
Teilchenphysik eine Rolle spielen.  In dieser Arbeit wird die
Vernichtung eines Fermion-Antifermion-Paares unter Erzeugung eines anderen
Fermion-Antifermion-Paares betrachtet.  In der Born-N"aherung l"asst sich
dieser Vierfermionprozess mit dem Feynman-Diagramm in Abb.~\ref{fig:4fBorn}
darstellen, bei dem die elektroschwache Wechselwirkung von einem virtuellen
Photon ($\gamma$) oder einem $Z$-Boson vermittelt wird.

Die Feynman-Regeln f"ur dieses Diagramm liest man aus der Lagrange-Dichte des
Standardmodells bzw. f"ur diese Arbeit des $SU(2) \times U(1)$-Modells
(siehe Anhang~\ref{chap:feynman}) ab.
Aus der Born-N"aherung erh"alt man einen Wert f"ur den Wirkungsquerschnitt
dieses Prozesses, der f"ur viele Zwecke bereits ausreicht. Dazu geh"oren:
die Absch"atzung der Gr"o"senordnung, mit denen Teilprozesse zum
Gesamtergebnis beitragen;
die grobe experimentelle "Uberpr"ufung der theoretischen Vorhersagen und
der Relationen zwischen den Parametern des Modells;
die Absch"atzung des Hintergrunds einer Messung.

Um aber Parameter des Modells mit m"oglichst gro"ser Genauigkeit
bestimmen oder die Vorhersagen von Modellen ausschlie"sen zu k"onnen,
muss man Pr"azisionsmessungen durchf"uhren.
Dazu ist sowohl auf experimenteller wie auf theoretischer Seite eine hohe
Genauigkeit der Ergebnisse n"otig.
Die Pr"azisionstests der vergangenen Jahrzehnte haben das Standardmodell
im experimentell zug"anglichen Energiebereich eindrucksvoll best"atigt.
\mbox{Eine} der
heutigen Herausforderungen der Hochenergiephysik besteht in der Suche nach
kleinsten Abweichungen der experimentellen Ergebnisse von Vorhersagen des
Standardmodells, um daraus z.B. etwas "uber m"ogliche
Standardmodell-Erweiterungen zu lernen.

Es gibt verschiedene Ans"atze, die auf der Basis des Standardmodells "uber die
Genauigkeit der Born-N"aherung hinausgehen.
Dazu geh"oren neben der St"orungstheorie, die im Folgenden erl"autert wird,
auch nichtperturbative Methoden und Gitterrechnungen.
In st"orungstheoretischen Rechnungen betrachtet man die Wechselwirkungsterme
in der Lagrange-Dichte als kleine St"orungen und entwickelt die Ergebnisse
in Potenzen der Kopplungskonstanten.
Dies f"uhrt zu einer Serie von Feynman-Diagrammen, zu denen neben dem
Baum-Diagramm ("`tree-level"') der Born-N"aherung
Feynman-Diagramme mit "`Schleifen"' geh"oren.
In dieser Arbeit wird die St"orungsreihe in Potenzen von
$\frac{\alpha}{4\pi}$ bzw. $\frac{\alpha'}{4\pi}$ definiert mit
\begin{equation}
  \alpha = \frac{g^2}{4\pi} \,,\qquad
  \alpha' = \frac{{g'}^2}{4\pi} \,.
\end{equation}
In h"oheren Ordnungen der St"orungstheorie kommt mit jeder Schleife durch
zus"atzliche Vertizes ein Faktor~$\alpha$ oder $\alpha'$ hinzu.  Diese
h"oheren Ordnungen der St"orungstheorie stellen Korrekturen zum Born-Term
dar.
Elektroschwache Einschleifenkorrekturen wurden ausgiebig untersucht;
f"ur Ergebnisse dazu wird z.B. auf~\cite{Bardin:1999ak}
verwiesen.

Die Eichbosonen des elektroschwachen Standardmodells sind
das masselose Photon und die massiven Bosonen $W^\pm$ und~$Z$.
Dem entsprechen im $SU(2) \times U(1)$-Modell dieser Arbeit das masselose
Hyperladungs-Eichboson~$B$ und die massiven $SU(2)$-Bosonen~$W^a$.
Alle Eichbosonen k"onnen als reelle Teilchen im Endzustand der Reaktion
abgestrahlt werden oder in virtuellen Korrekturen auftreten.

Bei einem masselosen Eichboson wie dem Photon ist die reelle Abstrahlung
f"ur sich allein betrachtet divergent. Infrarote (IR) Divergenzen k"onnen
aus Phasenraumbereichen kommen, in denen das Photon weich ("`soft"') ist,
also sein
Impuls verschwindet.  Oder es treten kollineare Divergenzen
in Phasenraumbereichen auf, in denen
das Photon mit einem der anderen Teilchen im Anfangs- oder Endzustand
kollinear ist. Da
ein Teilchendetektor nur eine endliche Energieaufl"osung hat, kann er
weiche Photonen mit einer Energie unterhalb einer bestimmten Schwelle nicht
wahrnehmen. Die IR-Singularit"aten kommen also von unbeobachtbaren
Photonen. Auch der exklusive Prozess ohne zus"atzliche Photonen im
Endzustand, bei dem nur virtuelle Korrekturen durch Photonen
ber"ucksichtigt werden, besitzt infrarote und kollineare Singularit"aten.
Nach dem Theorem von Kino\-shita, Lee und
Nauenberg~\cite{Kinoshita:1962ur,Lee:1964is} sind jedoch inklusive
Observablen IR-endlich.
Diese ber"ucksichtigen sowohl reelle als
auch virtuelle Korrekturen. Die Divergenzen aus reellen und virtuellen
Beitr"agen kompensieren sich.  Nur die Summe beider Beitr"age stellt eine
physikalische Observable dar.

Abgestrahlte massive Eichbosonen wie $W$ und $Z$ (bzw. ihre
Zerfallsprodukte) k"onnen jedoch aufgrund ihrer endlichen Masse auch dann
im Detektor beobachtet werden, wenn ihre kinetische Energie verschwindet.
Die reelle Abstrahlung und die virtuellen Korrekturen besitzen jeweils
separat keine IR-Singularit"aten.
Bei massiven Eichbosonen k"onnen also exklusive Prozesse mit rein
virtuellen Korrekturen getrennt betrachtet werden. Diese exklusiven
Observablen enthalten im Grenzfall gro"ser Impuls"ubertr"age $\sqrt s \gg M$
Potenzen des Logarithmus $\ln(s/M^2)$,
in den das Verh"altnis zwischen
$s$ und dem Quadrat der Eichboson-Masse~$M$
eingeht.

Die Massen der $W$- und $Z$-Bosonen ($M_W \approx 80\,\GeV$,
$M_Z \approx 91\,\GeV$)
liegen in der Gr"o"senordnung heute erreichbarer Energien bei
Elektron-Positron-Beschleunigern.
Solange sich die Schwerpunktsenergie eines Streuprozesses
unterhalb der Eichboson-Massen befindet,
tragen Korrekturen mit virtuellen $W$"~~und $Z$"~Bosonen nur wenig
zur Streuamplitude bei.
Wenn aber die Impuls"ubertr"age weit "uber den Massen der Eichbosonen
liegen, also im TeV-Bereich wie beim \emph{Large Hadron Collider} (LHC)
oder beim geplanten internationalen $e^+ e^-$-Linearbeschleuniger (ILC),
dann treten neue Effekte auf, die mit zunehmender Energie dominant werden.
Im Limes $s \gg M_{W,Z}^2$ bestimmen die Potenzen von
$\ln(s/M_{W,Z}^2)$ das Verhalten der Observablen.

%
%
\section{Renormierung}
\label{sec:Renormierung}

Neben IR-Singularit"aten k"onnen Strahlungskorrekturen auch ultraviolette
(UV) Divergenzen produzieren.
Diese kommen aus den Integrationsbereichen gro"ser Schleifenimpulse.
UV-Divergenzen k"onnen systematisch durch die Renormierung in die Parameter
des Modells absorbiert werden.
Die Renormierung stellt die Zusammenh"ange her zwischen den "`nackten"'
Parametern aus der Lagrange-Dichte und den physikalischen Gr"o"sen,
die im Experiment gemessen werden k"onnen und gegen"uber den nackten
Parametern UV-divergente Strahlungskorrekturen erfahren.

Grunds"atzlich k"onnen die Strahlungskorrekturen entweder mit der nackten
Lagrange-Dichte berechnet und die nackten Parameter anschlie"send auf die
physikalischen Gr"o"sen transformiert werden.
Oder man betrachtet eine renormierte Lagrange-Dichte, die neben den
physikalischen Gr"o"sen sogenannte Counterterme enth"alt, welche die
UV-Divergenzen kompensieren.
In dieser Arbeit wird der erste dieser beiden Ans"atze gew"ahlt.

Die UV-Divergenzen werden mit der Methode der \emph{dimensionalen
Regularisierung}~\cite{'tHooft:1972fi} behandelt.
Dabei werden die Schleifenintegrale statt in 4~Raum-Zeit-Dimensionen
in $d = 4-2\eps$ Dimensionen ausgef"uhrt, wobei $\eps$ ein
infinitesimaler komplexer Parameter ist.
UV-Singularit"aten von Einschleifenintegralen "au"sern sich dann in
$1/\eps$-Polen im Ergebnis,
eine h"ohere Anzahl von Schleifen produziert $\eps$-Pole h"oherer Ordnung.
Auch IR-Singularit"aten k"onnen mit dieser Methode regularisiert werden und
f"uhren zu Polen in~$\eps$.

Im Gegensatz zur Regularisierung mit einem Cutoff-Parameter, der die
Impulsintegration zu gro"sen Impulsen hin abschneidet,
bleiben bei der dimensionalen Regularisierung die
Ward-Identit"aten erhalten.
Au"serdem k"onnen im Rahmen der dimensionalen Regularisierung alle skalenlosen
Integrale
(d.h. Integrale, deren Ergebnis keinen Parameter mit Massendimension
mehr enth"alt) zu null gesetzt werden~\cite{Collins:1984xc,Leibbrandt:1975dj}.

Durch die Verallgemeinerung der Raum-Zeit auf $d$~Dimensionen "andern sich auch
die Dimen\-sionen der Lagrange-Dichte, der Felder und der Kopplungskonstanten.
Aus dem Wirkungsintegral $S = \int\!d^dx \, \Lc$, das dimensionslos
bleiben muss, kann man die Massen\-dimen\-sion~$d$ der Lagrange-Dichte~$\Lc$
ablesen.
Die Dimension der Felder erkennt man dann an den kinetischen Termen;
sie ist $\frac{d-1}{2}$ f"ur Fermionfelder und $\frac{d}{2}-1$ f"ur
skalare Felder und Vektorfelder.
Die $SU(2)$- und $U(1)$-Kopplungskonstanten schlie"slich sind nicht mehr
dimensionslos, sondern besitzen die Dimension $2-\frac{d}{2} = \eps$.
Entsprechend tr"agt $\alpha$ (und genauso $\alpha'$) die Dimension~$2\eps$.
Um ein dimensionsloses $\alpha$ zu behalten,
ersetzt man konventionsgem"a"s
\begin{equation}
\label{eq:alpha-mualpha}
  \alpha \to \mu^{2\eps} \, \alpha \,,
\end{equation}
wobei $\mu$ eine Konstante mit der Dimension einer Masse ist.
Dadurch erh"alt man pro Schleifenintegral mit~$\alpha$ auch den
Faktor~$\mu^{2\eps}$.
Dieser Faktor wird gem"a"s Gl.~(\ref{eq:muloopint}) mit dem
Schleifenintegral kombiniert.

Dem Parameter~$\mu$ kommt die Bedeutung einer Renormierungsskala zu.
Er wird so gew"ahlt, dass die Effekte h"oherer Ordnungen der
St"orungstheorie m"oglichst klein werden. Dazu w"ahlt man f"ur $\mu$ in der
Regel einen Wert in der Gr"o"senordnung der im konkreten Prozess
auftretenden Energieskalen.
Physikalische Observablen d"urfen nicht vom Renormierungsparameter~$\mu$
abh"angen. Diese Tatsache wird von Renormierungsgruppengleichungen benutzt,
die Ableitungen bestimmter Gr"o"sen nach~$\mu$ oder anderen Parametern
untersuchen.

In der vorliegenden Arbeit werden die Kopplungskonstanten nach dem
\MSbar-Schema renormiert.
Das diesem zugrunde liegende MS-Schema (minimal subtraction) w"ahlt den
Zusammenhang zwischen nackten und renormierten Gr"o"sen so, dass nur die
$1/\eps$-Pole aus den UV-divergenten Ausdr"ucken subtrahiert werden und die
endlichen Anteile "ubrig bleiben.
Im davon abgewandelten \MSbar-Schema (modified minimal subtraction) wird
zus"atzlich pro Schleifenintegration einmal der Faktor
$S_\eps = (4\pi)^\eps \, e^{-\eps\gamma_E}$ durch die Reskalierung
\begin{equation}
\label{eq:muMSbar}
  \mu^2 \to \mu^2 \, \frac{e^{\gamma_E}}{4\pi}
\end{equation}
in die Renormierungsskala~$\mu$ absorbiert.
Dann h"angen die Ergebnisse insbesondere nicht mehr von der Eulerschen
Konstanten $\gamma_E \approx 0{,}577216$ (\ref{eq:Gammadev}) ab.

Der Zusammenhang zwischen der nackten Kopplung~$\alpha_0$ und der
renormierten Kopplung~$\alpha$ kann multiplikativ durch
\begin{equation}
\label{eq:alpharen}
  \alpha_0 = Z_\alpha \, \alpha
\end{equation}
angegeben werden.
Der Faktor~$Z_\alpha$ h"angt mit der $\beta$-Funktion zusammen, welche die
Abh"angigkeit der renormierten Kopplung von~$\mu$ beschreibt.
F"ur das MS- oder \MSbar-Schema in $d=4-2\eps$ Dimensionen gilt:
\begin{equation}
\label{eq:betaalpha}
  \beta(\alpha)
  \equiv \frac{1}{4\pi} \, \frac{\partial\alpha}{\partial\ln\mu^2}
  = -\frac{1}{4\pi} \,
    \frac{\eps \, Z_\alpha \, \alpha}
      {Z_a + \alpha \, \frac{\partial Z_\alpha}{\partial\alpha}}
  = -\eps \, \frac{\alpha}{4\pi}
    - \beta_0 \left(\frac{\alpha}{4\pi}\right)^2
    + \Oc(\alpha^3)
  \,,
\end{equation}
woraus folgt:
\begin{equation}
\label{eq:Zalpha}
  Z_\alpha = 1 - \frac{\beta_0}{\eps} \, \frac{\alpha}{4\pi}
  + \Oc(\alpha^2)
  \,.
\end{equation}
Die $\beta$-Funktion ist f"ur abelsche und nichtabelsche Eichtheorien aus
der QCD bis zur Vierschleifenordnung bekannt%
\cite{Gross:1973id,Politzer:1973fx,'tHooft:1972??, 
  Caswell:1974gg,Jones:1974mm,Egorian:1979zx, 
  Tarasov:1980au,Larin:1993tp, 
  vanRitbergen:1997va}. 
In dieser Arbeit wird nur der erste Koeffizient ben"otigt:
\begin{equation}
\label{eq:beta0}
  \beta_0 = \frac{11}{3} C_A - \frac{4}{3} T_F n_f - \frac{1}{3} T_F n_s
\end{equation}
f"ur ein Modell mit $n_f$~Dirac-Fermionen und $n_s$~skalaren Multipletts in
der fundamentalen Darstellung der Eichgruppe.
Da f"ur die UV-Renormierung Effekte aus dem Niederenergiebereich wie die
spontane Symmetriebrechung keine Rolle spielen, gilt (\ref{eq:beta0})
mit $n_s = 1$ auch f"ur ein Higgs-Modell.

Au"serdem m"ussen die Feldst"arken der beteiligten Teilchen renormiert
werden. So gilt z.B. f"ur den Zusammenhang zwischen den nackten und
renormierten Fermionfeldern:
\begin{equation}
  \psi_0 = \sqrt{Z_f} \, \psi \,.
\end{equation}
Der Faktor~$Z_f$ kann anhand des vollst"andigen Fermionpropagators
inklusive aller Schleifenkorrekturen ermittelt werden.
Dieser ist in der unrenormierten Theorie f"ur masselose Fermionen durch
\begin{equation}
\label{eq:fproptotZ}
  \vcentergraphics{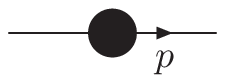} \stackrel{\dslash p \to 0}{\sim}
  Z_f \, \frac{i}{\dslash p} + \Oc(\dslash p^0)
\end{equation}
gegeben.
Andererseits erh"alt man den gleichen Propagator aus den
Selbstenergiekorrekturen, die f"ur masselose Fermionen die Form
\begin{equation}
  \tilde\Sigma(p) = -i \dslash p \, \Sigma(p^2)
\end{equation}
haben. Es gilt:
\begin{equation}
\label{eq:fproptotSigma}
  \vcentergraphics{fprop-tot} =
  \frac{i}{\dslash p}
  + \frac{i}{\dslash p} \, \tilde\Sigma(p) \, \frac{i}{\dslash p}
  = \frac{i}{\dslash p} \, \bigl[ 1 + \Sigma(p^2) \bigr]
  \,.
\end{equation}
Die Identifizierung der Ausdr"ucke (\ref{eq:fproptotZ}) und
(\ref{eq:fproptotSigma}) ergibt:
\begin{equation}
\label{eq:ZfSigma}
  Z_f = 1 + \Sigma(p^2=0) \,.
\end{equation}
F"ur die Fermion-Feldst"arkerenormierung ist also die
Selbstenergiekorrektur beim Impuls $p^2=0$ n"otig.

"Uber die LSZ-Reduktionsformel\cite{Lehmann:1955rq}
geht die Feldst"arkerenormierung auch in die Berechnung von S"~Matrixelementen
ein.
In der unrenormierten Theorie berechnet man zun"achst die Amplituden aller
beitragenden Feynman-Diagramme ohne Selbst\-energie\-korrekturen der externen
Teilchen. Es werden also nur solche Diagramme betrachtet, von denen man an
den "au"seren Linien keine Korrekturen nur durch Entfernen einer Linie
abtrennen kann.
Dieses Ergebnis muss dann f"ur jedes externe Teilchen mit einem
Faktor~$\sqrt Z$ der jeweiligen Feldst"arkerenormierung multipliziert
werden, um das S"~Matrixelement zu erhalten.

Der letzte Beitrag zur Renormierung in der vorliegenden Arbeit betrifft die
Eichbosonmasse.
Der Zusammenhang zwischen der nackten
Eichbosonmasse~$M_0$ und der physikalischen Masse~$M$ ergibt sich wieder
aus dem Propagator,
wobei hier Einschleifenkorrekturen gen"ugen.
Schreibt man die Eichboson-Selbstenergiekorrekturen in der Form
\begin{equation}
\label{eq:Pimunu}
  \tilde\Pi^{\mu\nu,ab}(k) =
  i \, \delta^{ab} \, g^{\mu\nu} \, k^2 \, \Pi_1(k^2)
  + (\text{Terme} \propto k^\mu k^\nu) + \Oc(\alpha^2)
  \,,
\end{equation}
dann gilt (ohne Terme $\propto k^\mu k^\nu$):
\begin{align}
  \vcentergraphics{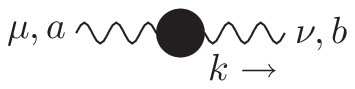}
  &= \frac{-i \, \delta^{ab} \, g_{\mu\nu}}{k^2-M_0^2}
    + \frac{-i \, \delta^{ac} \, g_{\mu\alpha}}{k^2-M_0^2} \,
      \tilde\Pi^{\alpha\beta,cd}(k) \,
      \frac{-i \, \delta^{db} \, g_{\beta\nu}}{k^2-M_0^2}
  \nonumber \\*
  &= \frac{-i \, \delta^{ab} \, g_{\mu\nu}}{k^2-M_0^2}
    \left(1 + \frac{k^2 \, \Pi_1(k^2)}{k^2-M_0^2}\right) + \Oc(\alpha^2)
  \nonumber \\*
  &= \frac{-i \, \delta^{ab} \, g_{\mu\nu}}
       {k^2 - M_0^2 - k^2\,\Pi_1(k^2) + \Oc(\alpha^2)}
  \,.
\end{align}
Der Propagator muss einen Pol bei $k^2 = M^2 - iM\Gamma$ haben,
wobei $M$ die physikalische Eichbosonmasse und $\Gamma$ die totale
Zerfallsbreite ist.
Da $\Gamma$ bereits von der Ordnung~$\alpha$ ist, ergibt sich in
Einschleifenn"aherung:
\begin{align}
\label{eq:Mren}
  M_0^2 &= M^2 \, \bigl[ 1 - \Rep \Pi_1(M^2) \bigr] + \Oc(\alpha^2) \,,
  \\
  \Gamma &= -M \, \Imp \Pi_1(M^2) + \Oc(\alpha^2) \,.
\end{align}

In dieser Arbeit werden die Schleifenrechnungen in der unrenormierten
Theorie mit den nackten Parametern $\alpha_0$ (bzw. $\alpha'_0$) und $M_0$
durchgef"uhrt, auch wenn die Parameter nicht ausdr"ucklich mit dem Index~0
bezeichnet sind.
Die Transformation der nackten auf die physikalischen Parameter in den
Einschleifenergebnissen gem"a"s (\ref{eq:alpharen}) und (\ref{eq:Mren})
ergibt Beitr"age in Zweischleifenordnung (siehe Abschnitt~\ref{sec:1loop}).

%
%
\section[Sudakov-Logarithmen in der Vierfermionstreuung]
  {Sudakov-Logarithmen in der\\ Vierfermionstreuung}
\label{sec:Sudakov}

Die vorliegende Arbeit besch"aftigt sich mit Vorhersagen f"ur Prozesse, die
zusammenfassend als \emph{Vierfermionstreuung} bezeichnet werden.
\begin{figure}[ht]
  \centering
  \includegraphics{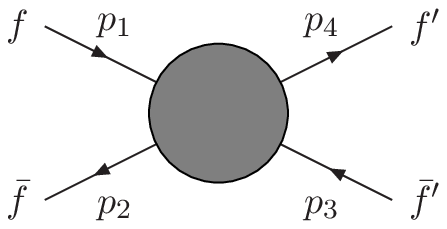}
  \caption{Vierfermionstreuung $f\bar f \to f'\bar f'$}
  \label{fig:4ftot}
\end{figure}
Wie in Abb.~\ref{fig:4ftot} dargestellt, geht dabei ein
Fermion-Antifermion-Paar~$f\bar f$ im Anfangszustand in ein anderes
Fermion-Antifermion-Paar~$f'\bar f'$ im Endzustand "uber.
Es wird nur der Fall $f \ne f'$ betrachtet. Man beschr"ankt sich damit auf
Prozesse, die in Born-N"aherung nur einen Beitrag
aus dem $s$-Kanal mit dem neutralen Strom haben.

In diesem Abschnitt und in den n"achsten Kapiteln wird zun"achst nur die
schwache $SU(2)$-Wechselwirkung mit massiven Eichbosonen untersucht.
Die $U(1)$-Eichgruppe der Hyperladung wird vorerst weggelassen.
Die Hinzunahme der Effekte durch die $U(1)$-Wechselwirkung
mit einem masselosen Eichboson wird in Kapitel~\ref{chap:SU2U1}
beschrieben.
Au"serdem wird zun"achst eine $SU(2)$-Eichgruppe verwendet, die an links- und
rechtsh"andige Fermionen gleicherma"sen koppelt. Es wird also eine
parit"atserhaltende Theorie angenommen, keine chirale Theorie.
Die Unterscheidung nach Chiralit"at wird sp"ater in
Kapitel~\ref{chap:EW2loop} hinzukommen.

Die nachfolgend beschriebene Untersuchung der Vierfermionamplitude und
des Formfaktors sowie die Notation beruhen auf einer Reihe von "alteren
Ver"offentlichungen~\cite{Kuhn:1999de,Kuhn:2000nn,Kuhn:2000hx,Kuhn:2001hz}.
Die in den Gleichungen~(\ref{eq:chi2erg1}),
(\ref{eq:chi2ergbeta0}) und (\ref{eq:zeta2}) zusammengefassten Resultate
vervollst"andigen diese "alteren Arbeiten. Eine Ver"offentlichung
ist in Vorbereitung.

\subsection{Die Vierfermionamplitude}
\label{sec:4famplitude}

Die Vierfermionamplitude, also das S"~Matrixelement des Prozesses,
ist aus zwei verschiedenen $SU(2)$-Amplituden zusammengesetzt.
Die Basis im Isospinraum wird folgenderma"sen gew"ahlt:
\begin{equation}
\label{eq:4fbasisfarbe}
  \begin{aligned}
    \Ac^\lambda &=
      \bar v(-p_2) \, \gamma^\mu t^a \, u(p_1) \cdot
      \bar u(p_4) \, \gamma_\mu t^a \, v(-p_3) \,,
    \\
    \Ac^d &=
      \bar v(-p_2) \, \gamma^\mu \, u(p_1) \cdot
      \bar u(p_4) \, \gamma_\mu \, v(-p_3) \,.
  \end{aligned}
\end{equation}
Die Impulse der 4~externen Fermionen sind in Abb.~\ref{fig:4ftot}
angegeben. $u(p)$ und $v(p)$ sind die Spinoren respektive eines Fermion-
oder Antifermion-Dubletts mit Impuls~$p$.
Zus"atzlich muss f"ur die Teilchen im Anfangs- wie im Endzustand zwischen
links- und rechtsh"andigen Fermionen unterschieden werden, z.B.:
\begin{equation}
\label{eq:4fbasischiral}
  \begin{aligned}
    \Ac^\lambda_{LL} &=
      \bar v^L(-p_2) \, \gamma^\mu t^a \, u^L(p_1) \cdot
      \bar u^L(p_4) \, \gamma_\mu t^a \, v^L(-p_3) \,,
    \\
    \Ac^d_{LR} &=
      \bar v^L(-p_2) \, \gamma^\mu \, u^L(p_1) \cdot
      \bar u^R(p_4) \, \gamma_\mu \, v^R(-p_3) \,.
  \end{aligned}
\end{equation}
Weil eine parit"atserhaltende Theorie betrachtet wird, gen"ugen diese
beiden Chiralit"atsstrukturen, da sich $\Ac_{RL}$ und $\Ac_{RR}$ darauf
zur"uckf"uhren lassen.
Die gesamte Amplitude ist ein Vektor im Raum der Isospin- und
Chiralit"atsstruktur.

Die kinematischen Variablen f"ur die Beschreibung des Vierfermionprozesses
sind $s$, $t$ und $u$.
F"ur Fermionen, die im Hochenergielimes als masselos betrachtet werden,
gilt $p_i^2 = 0$ ($i=1,2,3,4$). Daraus folgt:
\begin{equation}
\begin{aligned}
  s &= (p_1-p_2)^2 \,, \\
  t &= (p_1-p_4)^2 = -s \, x_- \,, \\
  u &= (p_1+p_3)^2 = -s \, x_+ \,, \\
\end{aligned}
\end{equation}
mit
\begin{equation}
\label{eq:xpm}
  x_\pm = \frac{1 \pm \cos\theta}{2} \,,
\end{equation}
wobei $\theta$ der Winkel zwischen den r"aumlichen Komponenten von $p_1$
und $p_4$ im Schwerpunktsystem ist.
Die Born-Amplitude ist f"ur $s \gg M^2$ durch den Ausdruck
\begin{equation}
\label{eq:ABorn}
  \Ac_B = \frac{ig^2}{s} \, \Ac^\lambda
\end{equation}
gegeben und entspricht dem Feyman-Diagramm in Abb.~\ref{fig:4fBorn} (mit
virtuellem~$W^3$).

Das Interesse dieser Arbeit liegt im Hochenergielimes der
Vierfermionprozesse, in dem alle kinematischen Invarianten sehr viel
gr"o"ser als die Eichbosonmasse sind:
\begin{equation}
  |s| \sim |t| \sim |u| \gg M^2 \,.
\end{equation}
Im Fall verschwindender Eichbosonmasse, $M=0$, besitzt die
Vierfermionamplitude infrarote und kollineare Divergenzen.
Die kollinearen Divergenzen der virtuellen Korrekturen faktorisieren.
Aufgrund des engen Zusammenhangs zwischen kollinearen Divergenzen im
masselosen Fall und den entsprechenden kollinearen Logarithmen im Fall
endlicher Eichbosonmasse~$M$, faktorisieren auch diese Logarithmen.
Die mit den kollinearen Divergenzen zusammenh"angende doppelt-logarithmische
Struktur (mit der h"ochsten Logarithmenpotenz~$2n$ in $n$-Schleifenordnung)
h"angt nur von den Eigenschaften der externen Teilchen ab, nicht aber vom
speziellen Prozess
\cite{Cornwall:1975aq,Cornwall:1975ty,Frenkel:1976bj,Amati:1978by,
  Mueller:1979ih,Collins:1980ih,Collins:1989bt,Sen:1981sd,Sen:1983bt}.
Am deutlichsten erkennbar ist dies in einer physikalischen Eichung
(Coulomb-Eichung oder axiale Eichung), in der kollineare Divergenzen nur in
den Selbstenergiekorrekturen der externen Teilchen auftreten
\cite{Frenkel:1976bj,Sen:1981sd,Sen:1983bt}.
Deshalb sind die kollinearen Logarithmen jedes Fermion-Antifermion-Paares
in der Vierfermionamplitude die gleichen wie diejenigen im Formfaktor,
der ein Eichboson an das Fermion-Antifermion-Paar koppelt.

\begin{figure}[ht]
  \centering
  \includegraphics{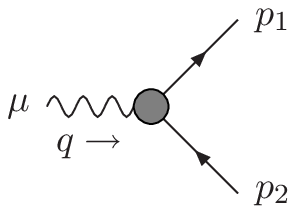}
  \caption{Vertexformfaktor}
  \label{fig:formfaktor}
\end{figure}

Dieser Formfaktor ist durch das Vertexdiagramm in Abb.~\ref{fig:formfaktor}
dargestellt.
Es gen"ugt hierbei, den Formfaktor des Vektorstroms zu betrachten, der an
ein externes abelsches Vektorfeld mit Impuls~$q$ koppelt.
Seine Born-Amplitude ist durch
\begin{equation}
  \Fc_B^\mu = \bar u(p_1) \, \gamma^\mu \, v(-p_2)
\end{equation}
gegeben und enth"alt keinen $SU(2)$-Generator.
Obwohl die Vertizes der Vierfermionamplitude in der Born-N"aherung~$\Ac_B$
durch die $SU(2)$-Generatoren~$t^a$ eine nichtabelsche Natur haben, reicht
f"ur die Faktorisierung der Amplitude der oben angegebene Formfaktor
des \emph{abelschen}
Vektorstroms aus. Denn dieser stellt das einfachste Objekt dar, das bereits
alle universalen kollinearen Logarithmen enth"alt.
Au"serdem ist der Formfaktor des abelschen Vektorstroms separat
eichinvariant, auch wenn nichtabelsche Korrekturen dazu betrachtet werden.
Die Kopplungskonstante zwischen dem externen abelschen Feld und dem
Fermionpaar wird nicht als Teil des Formfaktors, sondern als zur restlichen
Amplitude geh"orig betrachtet. Sie geht deshalb auch nicht in die
Renormierung des Formfaktors ein.

Die Amplitude der Vertexkorrekturen in Abb.~\ref{fig:formfaktor} ist f"ur
masselose Fermionen proportional zur Born-Amplitude,
\begin{equation}
\label{eq:vertexformfaktor}
  \Fc^\mu = F(Q^2) \cdot \bar u(p_1) \, \gamma^\mu \, v(-p_2)
  = F(Q^2) \cdot \Fc_B^\mu
  \,,
\end{equation}
mit dem Formfaktor~$F(Q^2)$. Wegen $p_1^2 = p_2^2 = 0$ h"angt dieser (neben der
Eichbosonmasse~$M$) nur von
\begin{equation}
  Q^2 = -q^2 = -(p_1-p_2)^2 = 2p_1 \cdot p_2
\end{equation}
ab, wobei $Q^2$ als euklidisches Impulsquadrat so definiert wurde, dass f"ur
raumartige Impulse~$q$, bei denen der Formfaktor reell ist, $Q^2 > 0$ gilt.

Alle kollinearen Logarithmen der Vierfermionamplitude
sind Teil der beiden Formfaktoren der
Fermionpaare im Anfangs- und
Endzustand.
%
Die Amplitude kann folgenderma"sen
geschrieben werden:
\begin{equation}
\label{eq:Afaktor}
  \Ac(s) = \frac{ig(s)^2}{s} \, F(Q^2)^2 \, \tilde\Ac(Q^2)
  \,,
\end{equation}
mit $Q^2 = -s$.
Die kollinearen Logarithmen sind mit dem Formfaktorquadrat~$F^2$
abfaktorisiert worden. Die reduzierte Amplitude~$\tilde\Ac$ enth"alt nur
noch Logarithmen, die mit den soften Divergenzen im Fall $M=0$
zusammenh"angen, und Logarithmen die vom Laufen der
Kopplung in $\tilde\Ac$ mit der Renormierungsgruppe herr"uhren.
Das Laufen der Kopplung in der Born-Amplitude ist bereits durch
die Wahl der Renormierungsskala~$s$ in~$g^2$ ber"ucksichtigt.
Alle Effekte von der nichtabelschen Natur der Vertizes in der
Born-Amplitude~$\Ac_B$ sind Teil der reduzierten Amplitude~$\tilde\Ac$,
nicht des Formfaktors.

Die reduzierte Amplitude~$\tilde\Ac$ ist wie $\Ac$ ein Vektor im Raum der
Isospin- und Chiralit"ats\-struktur. Sie erf"ullt f"ur $Q^2 \gg M^2$ die
folgende Evolutionsgleichung\cite{Sen:1983bt,Sterman:1987aj,Botts:1989kf}:
\begin{equation}
\label{eq:Aredevol}
  \frac{\partial\tilde\Ac(Q^2)}{\partial\ln Q^2} =
  \chi\bigl(\alpha(Q^2)\bigr) \, \tilde\Ac(Q^2)
  \,,
\end{equation}
wobei $\chi(\alpha)$ eine Matrix im Raum der Isospin- und Chiralit"atsstruktur
ist und als Matrix der soften anomalen Dimensionen
bezeichnet wird.
Mit $\alpha(Q^2)$ ist die laufende Kopplung an der Skala~$Q^2$ gemeint.
Die L"osung dieser Differentialgleichung lautet:
\begin{equation}
\label{eq:Arederg}
  \tilde\Ac(Q^2) =
  P \, \exp\!\left[\, \int_{M^2}^{Q^2}\!\frac{\dd x}{x} \,
         \chi\bigl(\alpha(x)\bigr) \right]
  \tilde\Ac_0\bigl(\alpha(M^2)\bigr)
  \,,
\end{equation}
wobei $\tilde\Ac_0(\alpha(M^2)) = \tilde\Ac(M^2)$ durch die
Anfangsbedingungen bestimmt werden muss.
Das Symbol~$P$ sorgt f"ur pfadgeordnete Matrizenprodukte in der
Exponentialfunktion:
\begin{equation}
  P \, \chi\bigl(\alpha(x)\bigr) \, \chi\bigl(\alpha(x')\bigr) =
  \begin{cases}
    \chi\bigl(\alpha(x)\bigr) \, \chi\bigl(\alpha(x')\bigr) \,,\;
      x > x' \,,\\
    \chi\bigl(\alpha(x')\bigr) \, \chi\bigl(\alpha(x)\bigr) \,,\;
      x' > x \,.
  \end{cases}
\end{equation}
Die Matrix~$\chi(\alpha)$ wird in der Kopplungskonstanten entwickelt:
\begin{equation}
\label{eq:expchi}
  \chi(\alpha) = \frac{\alpha}{4\pi} \, \chi^{(1)}
  + \left(\frac{\alpha}{4\pi}\right)^2 \chi^{(2)}
  + \Oc(\alpha^3)
  \,.
\end{equation}
In $d=4$ Dimensionen folgt aus der Definition der $\beta$-Funktion in
Gl.~(\ref{eq:betaalpha}):
\begin{equation}
\label{eq:alpharun}
  \frac{\alpha(x)}{4\pi} = \frac{\alpha(M^2)}{4\pi}
    - \beta_0 \, \ln\left(\frac{x}{M^2}\right)
      \left(\frac{\alpha(M^2)}{4\pi}\right)^2
    + \Oc(\alpha^3)
  \,.
\end{equation}
Damit k"onnen die Integrale in~(\ref{eq:Arederg}) ausgef"uhrt werden,
und man erh"alt die reduzierte Amplitude in Zweischleifenn"aherung:
\begin{align}
\label{eq:Atildechis}
  \tilde\Ac(Q^2) &=
  \biggl\{ 1 + \frac{\alpha}{4\pi} \, \chi^{(1)} \, \lqm(Q^2)
\nonumber \\* & \qquad
    + \left(\frac{\alpha}{4\pi}\right)^2 \left[
      \frac{1}{2} \left(\bigl(\chi^{(1)}\bigr)^2
        - \beta_0 \, \chi^{(1)}\right) \lqm^2(Q^2)
      + \chi^{(2)} \, \lqm(Q^2) \right]
  \biggr\} \,
  \tilde\Ac_0\bigl(\alpha(M^2)\bigr)
\nonumber \\* & \quad
  + \Oc(\alpha^3)
  \,,
\end{align}
mit $\lqm(Q^2) = \ln(Q^2/M^2)$
und $\alpha \equiv \alpha(M^2)$ renormiert an der Skala der Eichbosonmasse.

Der Vektor~$\tilde\Ac_0(\alpha)$ wird in der Kopplungskonstanten entwickelt:
\begin{equation}
\label{eq:A0exp}
  \tilde\Ac_0(\alpha) =
  \Ac^\lambda + \frac{\alpha}{4\pi} \, \tilde\Ac_0^{(1)}
  + \left(\frac{\alpha}{4\pi}\right)^2 \tilde\Ac_0^{(2)}
  + \Oc(\alpha^3)
  \,,
\end{equation}
und man definiert die Aufspaltung
\begin{equation}
  \tilde\Ac_0^{(1)} =
  \tilde A_0^{(1)}{}^\lambda_{LL} \, \Ac^\lambda_{LL}
  + \tilde A_0^{(1)}{}^\lambda_{LR} \, \Ac^\lambda_{LR}
  + \tilde A_0^{(1)}{}^d_{LL} \, \Ac^d_{LL}
  + \tilde A_0^{(1)}{}^d_{LR} \, \Ac^d_{LR}
  \,.
\end{equation}
F"ur die vorliegende Analyse wird $\tilde\Ac_0^{(2)}$ nicht ben"otigt.
Au"serdem gen"ugt der Realteil von $\tilde\Ac_0^{(1)}$,
der in \cite{Kuhn:2001hz} f"ur eine $SU(N)$-Theorie
berechnet wurde:
\begin{align}
\label{eq:A01erg}
  \Rep \tilde A_0^{(1)}{}^\lambda_{LL} &=
    \left(C_F - \frac{T_F}{N}\right) f(x_+,x_-)
    + C_A \left(\pi^2 + \frac{85}{9}\right)
    - \frac{20}{9} T_F n_f - \frac{8}{9} T_F n_s
    \,,
\nonumber \\
  \Rep \tilde A_0^{(1)}{}^\lambda_{LR} &=
    - \left(C_F - \frac{T_F}{N} - \frac{C_A}{2}\right) f(x_-,x_+)
    + C_A \left(\pi^2 + \frac{85}{9}\right)
    - \frac{20}{9} T_F n_f - \frac{8}{9} T_F n_s
    \,,
\nonumber \\
  \Rep \tilde A_0^{(1)}{}^d_{LL} &=
    \frac{C_F T_F}{N} f(x_+,x_-)
    \,,
\nonumber \\
  \Rep \tilde A_0^{(1)}{}^d_{LR} &=
    -\frac{C_F T_F}{N} f(x_-,x_+)
    \,,
\end{align}
mit
\begin{equation}
  f(x_+,x_-) =
  \frac{2}{x_+} \ln(x_-)
  + \frac{x_- - x_+}{x_+^2} \ln^2(x_-)
\end{equation}
und $x_\pm = \frac12(1 \pm \cos\theta)$ aus~(\ref{eq:xpm}).
Im $SU(2)$-Higgs-Modell sind $N=2$, $n_s=1$ sowie $C_F$, $C_A$ und $T_F$ aus 
Gl.~(\ref{eq:CasimirSU2}) einzusetzen; $n_f$ steht f"ur die Zahl der
Fermionen.

Die Matrix~$\chi(\alpha)$ ist universal und nicht davon abh"angig,
ob die Eichbosonen massiv oder masselos sind.
Insbesondere stimmen die Koeffizienten vor den Logarithmen in der
Exponentialdarstellung von $\tilde\Ac$,
\begin{multline*}
  \tilde\Ac(Q^2) =
  \exp\biggl\{
    \frac{\alpha}{4\pi} \, \chi^{(1)} \ln\left(\frac{Q^2}{M^2}\right)
\\*
    + \left(\frac{\alpha}{4\pi}\right)^2 \left[
      - \frac{\beta_0}{2} \chi^{(1)} \ln^2\left(\frac{Q^2}{M^2}\right)
      + \chi^{(2)} \ln\left(\frac{Q^2}{M^2}\right) \right]
    \biggr\} \,
    \tilde\Ac(M^2)
  + \Oc(\alpha^3)
  \,,
\end{multline*}
mit den entsprechenden Koeffizienten vor den Logarithmen~$\ln(Q^2/\mu^2)$
im masselosen Fall "uberein.
Aufgrund des einfachen Zusammenhangs zwischen Logarithmen und $\eps$"~Polen
in der masselosen Amplitude k"onnen die Matrixkoeffizienten
$\chi^{(1)}$ und $\chi^{(2)}$ aus den $1/\eps$-Polen des masselosen
Ein- und Zweischleifenergebnisses extrahiert werden.

Der Koeffizient~$\chi^{(1)}$ wurde in \cite{Kuhn:2000nn,Kuhn:2000hx}
berechnet. Er h"angt nicht von der Chiralit"at ab.
Seine Matrixelemente im Isospinraum lauten:
\begin{align}
\label{eq:chi1erg}
  \chi^{(1)}_{\lambda\lambda} &=
    -2 C_A \, \bigl(\ln(x_+) + i\pi\bigr)
    + 4 \left(C_F - \frac{T_F}{N}\right)
      \ln\left(\frac{x_+}{x_-}\right)
    \,,
\nonumber \\
  \chi^{(1)}_{\lambda d} &=
    4 \frac{C_F T_F}{N} \ln\left(\frac{x_+}{x_-}\right)
    \,,
\nonumber \\
  \chi^{(1)}_{d\lambda} &=
    4 \ln\left(\frac{x_+}{x_-}\right)
    \,,
\nonumber \\
  \chi^{(1)}_{dd} &= 0
  \,.
\end{align}
Die Indizes der Matrixelemente wurden so definiert, dass gilt:
\[
  \Ac = A^\lambda \, \Ac^\lambda + A^d \, \Ac^d
  \quad\Rightarrow\quad
  \chi \Ac =
    \bigl(A^\lambda \chi_{\lambda\lambda} + A^d \chi_{d\lambda}\bigr)
    \, \Ac^\lambda
  + \bigl(A^\lambda \chi_{\lambda d} + A^d \chi_{dd}\bigr)
    \, \Ac^d
  \,.
\]

Der Koeffizient~$\chi^{(2)}$ wurde
aus dem masselosen
Zweischleifenergebnis~\cite{Anastasiou:2000ue,Glover:2004si} extrahiert.
Der vom Laufen der Kopplungskonstante unabh"angige Teil von $\chi^{(2)}$
ist proportional zum Einschleifenkoeffizienten~$\chi^{(1)}$:
\begin{align}
\label{eq:chi2erg1}
  \chi^{(2)}\Big|_{\beta_0=0} &=
    \frac{\gamma^{(2)}}{\gamma^{(1)}} \, \chi^{(1)}
    \,,
\end{align}
also
\begin{align*}
  \chi^{(2)}_{\lambda\lambda}\Big|_{\beta_0=0} &=
    \frac{\gamma^{(2)}}{\gamma^{(1)}} \chi^{(1)}_{\lambda\lambda}
    \,,&
  \chi^{(2)}_{\lambda d}\Big|_{\beta_0=0} &=
    \frac{\gamma^{(2)}}{\gamma^{(1)}} \chi^{(1)}_{\lambda d}
    \,,&
  \chi^{(2)}_{d\lambda}\Big|_{\beta_0=0} &=
    \frac{\gamma^{(2)}}{\gamma^{(1)}} \chi^{(1)}_{d\lambda}
    \,,&
  \chi^{(2)}_{dd}\Big|_{\beta_0=0} &= 0
    \,.
\end{align*}
Die Koeffizienten $\gamma^{(1)}$~(\ref{eq:g1z1}) und
$\gamma^{(2)}$~(\ref{eq:gamma2}) stammen aus der
Berechnung des Formfaktors und werden in
Abschnitt~\ref{sec:formfaktor} erl"autert.

Der andere Teil von $\chi^{(2)}$ kommt vom Laufen der Kopplungskonstante
in Gl.~(\ref{eq:A0exp}) und ist proportional zur $\beta$-Funktion.
"Uber~$\tilde\Ac_0^{(1)}$ liefert er einen isospin- und
chiralit"atsabh"angigen Beitrag:
\begin{align}
\label{eq:chi2ergbeta0}
  \chi^{(2)}\Big|_{\beta_0} \Ac^\lambda &\equiv
        \chi^{(2)}_{\lambda\lambda,LL}\Big|_{\beta_0} \Ac^\lambda_{LL}
      + \chi^{(2)}_{\lambda\lambda,LR}\Big|_{\beta_0} \Ac^\lambda_{LR}
      + \chi^{(2)}_{\lambda d,LL}\Big|_{\beta_0} \Ac^d_{LL}
      + \chi^{(2)}_{\lambda d,LR}\Big|_{\beta_0} \Ac^d_{LR}
\nonumber \\* &=
  -\beta_0 \left(
        \tilde A_0^{(1)}{}^\lambda_{LL} \, \Ac^\lambda_{LL}
      + \tilde A_0^{(1)}{}^\lambda_{LR} \, \Ac^\lambda_{LR}
      + \tilde A_0^{(1)}{}^d_{LL} \, \Ac^d_{LL}
      + \tilde A_0^{(1)}{}^d_{LR} \, \Ac^d_{LR} \right) .
\end{align}

Mit den angegebenen Ergebnissen f"ur $\tilde\Ac_0^{(1)}$~(\ref{eq:A01erg}),
$\chi^{(1)}$~(\ref{eq:chi1erg}) und
$\chi^{(2)}$~(\ref{eq:chi2erg1},\ref{eq:chi2ergbeta0})
ist die reduzierte Amplitude~$\tilde\Ac$ bis zur Ordnung~$\alpha^2$
in logarithmischer N"aherung bekannt.
In der Zweischleifenn"aherung fehlt lediglich die nichtlogarithmische
Konstante~$\tilde\Ac_0^{(2)}$, wie an Gl.~(\ref{eq:Atildechis})
zu sehen ist.

Der n"achste Schritt in der Auswertung der Vierfermionamplitude~$\Ac$ ist
also die Berechnung des Formfaktors~$F$, der entsprechend
Gl.~(\ref{eq:Afaktor}) quadratisch in~$\Ac$ eingeht.

\subsection{Der Formfaktor}
\label{sec:formfaktor}

Die Auswertung des im vorigen Abschnitt in Gl.~(\ref{eq:vertexformfaktor})
definierten Formfaktors~$F$ eines abelschen Vektorstroms
erfordert im Allgemeinen die Berechnung von
Vertexkorrekturen.
Das Interesse dieser Arbeit gilt jedoch dem Hochenergielimes
$Q^2 \gg M^2$, in dem die
Massen der Fer\-mionen vernachl"assigt werden.
Au"serdem verzichtet dieser sogenannte \emph{Sudakov-Limes} auf Terme,
die mit mindestens einem Faktor~$M^2/Q^2$ unterdr"uckt sind.
Das asymptotische Verhalten des Formfaktors ist dann durch die folgende
Evolutionsgleichung gegeben%
\cite{Mueller:1979ih,Collins:1980ih,Collins:1989bt,Sen:1981sd}:
\begin{equation}
\label{eq:formfaktorevol}
  \frac{\partial F(Q^2)}{\partial\ln Q^2} =
  \left[\,
    \int_{M^2}^{Q^2}\!\frac{\dd x}{x} \, \gamma\bigl(\alpha(x)\bigr)
    + \zeta\bigl(\alpha(Q^2)\bigr) + \xi\bigl(\alpha(M^2)\bigr)
  \right]
  F(Q^2)
  \,,
\end{equation}
mit den anomalen Dimension $\gamma$, $\zeta$ und $\xi$.
Die L"osung dieser Gleichung kann folgenderma"sen geschrieben werden:
\begin{equation}
\label{eq:formfaktorexpint}
  F(Q^2) = F_0\bigl(\alpha(M^2)\bigr) \, \exp\!\left\{\,
    \int_{M^2}^{Q^2}\!\frac{\dd x}{x} \left[\;
      \int_{M^2}^x\!\frac{\dd x'}{x'} \, \gamma\bigl(\alpha(x')\bigr)
      + \zeta\bigl(\alpha(x)\bigr) + \xi\bigl(\alpha(M^2)\bigr)
    \right] \right\}
  \,.
\end{equation}
Der Faktor~$F_0(\alpha(M^2)) = F(M^2)$ muss aus den Anfangsbedingungen
bestimmt werden.
Zur Auswertung der Integrale in~(\ref{eq:formfaktorexpint}) werden die
anomalen Dimensionen entsprechend Gl.~(\ref{eq:expchi}) in der
Kopplungskonstanten entwickelt, und f"ur die laufende Kopplungskonstante
selbst wird Gl.~(\ref{eq:alpharun}) eingesetzt.
Der Formfaktor l"asst sich dann mit den
Logarithmen~$\lqm(Q^2) = \ln(Q^2/M^2)$ und den Koeffizienten der anomalen
Dimensionen schreiben:
\begin{align}
\label{eq:formfaktorexp}
  F(Q^2) &=
    \left[ 1 + \frac{\alpha}{4\pi}\,F_0^{(1)}
      + \left(\frac{\alpha}{4\pi}\right)^2\,F_0^{(2)} \right]
    \, \exp\Bigg\{
    \frac{\alpha}{4\pi} \left[
      \frac12 \gamma^{(1)} \lqm^2
      + \Big(\zeta^{(1)}+\xi^{(1)}\Big) \lqm
      \right]
\nonumber \\* & \qquad\quad
    + \left(\frac{\alpha}{4\pi}\right)^2 \, \bigg[
      -\frac{\beta_0}{6} \gamma^{(1)} \lqm^3
      + \frac12 \Big( \gamma^{(2)} - \beta_0 \, \zeta^{(1)} \Big) \lqm^2
      + \Big(\zeta^{(2)}+\xi^{(2)}\Big) \lqm
      \bigg]
    \Bigg\}
\nonumber \\* & \quad
  + \Oc(\alpha^3)
  \,,
\end{align}
mit $\lqm = \ln(Q^2/M^2)$ und $\alpha \equiv \alpha(M^2)$.
Schreibt man die St"orungsreihe des Formfaktors entsprechend
\begin{equation} 
  F = 1 + \frac{\alpha}{4\pi} \, F^{(1)}
      + \left(\frac{\alpha}{4\pi}\right)^2 F^{(2)}
  \,,
\end{equation}
so erh"alt man:
\begin{align}
\label{eq:F1gzx}
  F^{(1)} &=
    \frac{1}{2} \gamma^{(1)} \lqm^2
    + \Bigl(\zeta^{(1)} + \xi^{(1)}\Bigr) \lqm
    + F_0^{(1)}
    \,,
\\
\label{eq:F2gzx}
  F^{(2)} &=
    \frac{1}{8} \Bigl(\gamma^{(1)}\Bigr)^2 \lqm^4
    + \frac{1}{2} \left(\zeta^{(1)} + \xi^{(1)} - \frac{\beta_0}{3}\right)
      \gamma^{(1)} \lqm^3
  \nonumber \\* & \qquad
    + \frac{1}{2} \left[ \gamma^{(2)} + \Bigl(\zeta^{(1)}+\xi^{(1)}\Bigr)^2
        - \beta_0 \zeta^{(1)} + F_0^{(1)} \gamma^{(1)} \right] \lqm^2
  \nonumber \\* & \qquad
    + \left[ \zeta^{(2)} + \xi^{(2)}
      + F_0^{(1)} \Bigl(\zeta^{(1)} + \xi^{(1)}\Bigr) \right] \lqm
    + F_0^{(2)}
  \,.
\end{align}

Im Gegensatz zur reduzierten Amplitude~$\tilde\Ac$ weist der Formfaktor
eine doppelt-logarithmische Struktur auf: Der
Einschleifenkoeffizient~$F^{(1)}$ besitzt Logarithmen bis zur quadratischen
Potenz, der Zweischleifenkoeffizient~$F^{(2)}$ sogar bis zur Potenz~4.
Diese Logarithmenstruktur kommt von den kollinearen und soften
Divergenzen im Fall verschwindender Eichbosonmasse.
Allgemein enth"alt jede Ordnung~$\alpha^n$ der St"orungstheorie im
Hochenergielimes Logarithmen $\ln^{2n-j}(Q^2/M^2)$ mit $j=0,\ldots,2n$.
Im Limes sehr gro"ser Impuls"ubertr"age~$Q^2$ wird der Formfaktor von den
f"uhrenden Logarithmen,
\[
  \alpha^n \, \ln^{2n}\left(\frac{Q^2}{M^2}\right) ,
\]
dominiert. Diese hei"sen auch \emph{Sudakov-Logarithmen}.
Man spricht von der f"uhrenden logarithmischen N"aherung (LL = leading
logarithmic approximation), wenn man sich in jeder Ordnung der
St"orungstheorie auf die Logarithmen der jeweils h"ochsten Potenz
beschr"ankt.
Bei mittleren Energien (also f"ur $M \approx 80\,\GeV$ im TeV-Bereich)
k"onnen jedoch auch die Logarithmen mit niedrigerer Potenz einen wichtigen
Beitrag zum Formfaktor liefern. Man spricht von der n"achstf"uhrenden
logarithmischen N"aherung (NLL = next-to-leading logarithmic), wenn die
Logarithmen $\alpha^n \, \ln^{2n-j}$ mit $j=0,1$ betrachtet werden.
Analog fortgesetzt bedeuteten NNLL=N$^2$LL die Logarithmen mit $j=0,1,2$,
NNNLL=N$^3$LL die Logarithmen mit $j=0,\ldots,3$ und so weiter.

Die anomalen Dimensionen $\gamma$ und $\zeta$ sind universal, "ahnlich wie
die Matrix~$\chi$ in der Evolutionsgleichung der Vierfermionamplitude.
Sie h"angen nicht davon ab, ob die Eichbosonen masselos oder massiv sind%
\cite{Mueller:1979ih,Collins:1980ih,Collins:1989bt,Sen:1981sd}.
Deshalb stimmen die aus $\gamma^{(1,2)}$ und
$\zeta^{(1,2)}$ gebildeten Koeffizienten vor den Logarithmen~$\ln(Q^2/M^2)$
in der Exponentialdarstellung~(\ref{eq:formfaktorexp}) des Formfaktors mit
den Koeffizienten vor den jeweiligen
Logarithmen~$\ln(Q^2/\mu^2)$ im masselosen Formfaktor "uberein.
Aufgrund des einfachen Zusammenhangs zwischen den Logarithmen und den
$\eps$-Polen im masselosen Fall k"onnen
$\gamma^{(1)}$ und $\gamma^{(2)}$ aus den Koeffizienten der
$1/\eps^2$-Pole der masselosen Ein- und Zweischleifenergebnisse extrahiert
werden, $\zeta^{(1)}$ und $\zeta^{(2)}$ aus den Koeffizienten der
$1/\eps$-Pole\cite{Sen:1981sd,Magnea:1990zb}.

Der masselose Einschleifenformfaktor f"ur eine $SU(N)$-Eichgruppe ist
bekannt:
\begin{equation}
\label{eq:F1herg}
  F^{(1)}_{M=0} = C_F \left(\frac{\mu^2}{Q^2}\right)^\eps S_\eps
  \left\{
    -\frac{2}{\eps^2} - \frac{3}{\eps} + \frac{\pi^2}{6} - 8
    + \eps \left( \frac{14}{3}\zeta_3 + \frac{\pi^2}{4} - 16 \right)
  \right\}
  + \Oc(\eps^2)
  \,,
\end{equation}
mit $S_\eps = (4\pi)^\eps \, e^{-\eps\gamma_E}$
und der Riemannschen $\zeta$-Funktion $\zeta_3 \approx 1{,}20206$
(\ref{eq:zetan}).
Daraus ergeben sich
\begin{equation}
\label{eq:g1z1}
  \gamma^{(1)} = -2 C_F \,,\quad
  \zeta^{(1)} = 3 C_F \,.
\end{equation}

Die Funktionen $\xi$ und $F_0$ sind nicht universal. Sie h"angen
insbesondere von den Anfangsbedingungen der Evolutionsgleichungen, von der
Regularisierung der IR-Divergenzen und von der Massengenerierung der
Eichbosonen durch den Higgs-Sektor des Modells ab.
Die Bestimmung von $\xi$ und $F_0$ erfordert also die Kenntnis des massiven
Formfaktors.
In Einschleifenn"aherung ist das Resultat bekannt:
\begin{equation}
\label{eq:F1SUN}
  F^{(1)} = -C_F \left(
    \lqm^2 - 3 \lqm + \frac{2}{3}\pi^2 + \frac{7}{2} \right) .
\end{equation}
Daraus lassen sich im Vergleich mit~(\ref{eq:F1gzx}) die
"ubrigen Einschleifenkoeffizienten bestimmen:
\begin{equation}
\label{eq:x1F01}
  \xi^{(1)} = 0 \,,\quad
  F_0^{(1)} = -C_F \left(\frac{2}{3}\pi^2 + \frac{7}{2} \right) .
\end{equation}

Die Kenntnis der Einschleifenkoeffizienten kombiniert mit~$\beta_0$ erlaubt
bereits die Vorhersage
der beiden f"uhrenden Zweischleifenlogarithmen $\alpha^2 (\lqm^4,\lqm^3)$
entsprechend Gl.~(\ref{eq:F2gzx}).
Auch f"ur die NNLL-N"aherung inklusive des quadratischen Logarithmus fehlt
nur ein Zweischleifenkoeffizient:~$\gamma^{(2)}$.
Dieser kann aus dem masselosen Zweischleifenergebnis%
\cite{Kramer:1986sg,Matsuura:1988sm}
bestimmt werden:
\begin{align}
\label{eq:F2hherg}
  F^{(2)}_{M=0} &= C_F \left(\frac{\mu^2}{Q^2}\right)^{2\eps} S_\eps^2 \,
  \biggl\{
  C_F \, \frac{1}{\eps} \left(-12\zeta_3 + \pi^2 - \frac{3}{4}\right)
\nonumber \\* & \qquad
  + C_A \left[ -\frac{11}{6\eps^3}
    + \frac{1}{\eps^2} \left(\frac{\pi^2}{6} - \frac{83}{9}\right)
    + \frac{1}{\eps} \left(13\zeta_3 - \frac{11}{36}\pi^2
    - \frac{4129}{108}\right)
    \right]
\nonumber \\* & \qquad
  + T_F n_f \left[ \frac{2}{3\eps^3} + \frac{28}{9\eps^2}
    + \frac{1}{\eps} \left(\frac{\pi^2}{9} + \frac{353}{27}\right)
    \right]
\nonumber \\* & \qquad
  + T_F n_s \left[ \frac{1}{6\eps^3} + \frac{17}{18\eps^2}
    + \frac{1}{\eps} \left(\frac{\pi^2}{36} + \frac{455}{108}\right)
    \right]
  \biggr\}
\nonumber \\* & \quad
  + \left(\frac{1}{2} F^{(1)}_{M=0} - \frac{\beta_0}{\eps}\right)
    F^{(1)}_{M=0}
  + \Oc(\eps^0)
  \,,
\end{align}
wobei das masselose Einschleifenergebnis $F^{(1)}_{M=0}$~(\ref{eq:F1herg})
eingesetzt werden muss.
Das Ergebnis~(\ref{eq:F2hherg}) erlaubt, wie oben beschrieben,
die Extraktion der
Zweischleifenkoeffizienten $\gamma^{(2)}$ und $\zeta^{(2)}$:
\begin{align}
\label{eq:gamma2}
  \gamma^{(2)} &= C_F \left[
    C_A \left(\frac{2}{3}\pi^2 - \frac{134}{9}\right)
    + \frac{40}{9} \, T_F n_f
    + \frac{16}{9} \, T_F n_s
  \right] ,
\\
\label{eq:zeta2}
  \zeta^{(2)} &= C_F \, \biggl[
    C_F \left(24\zeta_3 -2\pi^2 + \frac{3}{2}\right)
    + C_A \left(-26\zeta_3 + \frac{11}{9}\pi^2 + \frac{2545}{54}\right)
\nonumber \\* & \qquad\quad
    - T_F n_f \left(\frac{4}{9}\pi^2 + \frac{418}{27}\right)
    - T_F n_s \left(\frac{\pi^2}{9} + \frac{311}{54}\right)
  \biggr] \,.
\end{align}

Damit ist nun die NNLL-N"aherung von~$F^{(2)}$ komplett, ohne dass daf"ur
eine massive Zweischleifenrechnung n"otig gewesen w"are:
\begin{align}
\label{eq:F2NNLLSUN}
  F^{(2)} &= C_F \, \biggl\{
    C_F \left[ \frac12 \lqm^4 - 3 \lqm^3 +
      \left(\frac{2}{3}\pi^2 + 8\right) \lqm^2 \right]
    + C_A \left[ \frac{11}{9} \lqm^3
      + \left(\frac{\pi^2}{3} - \frac{233}{18}\right) \lqm^2 \right]
\nonumber \\* & \qquad\quad
    + T_F n_f \left[ -\frac{4}{9} \lqm^3 + \frac{38}{9} \lqm^2 \right]
    + T_F n_s \left[ -\frac{1}{9} \lqm^3 + \frac{25}{18} \lqm^2 \right]
  \biggr\}
  + \Oc(\lqm^1)
  \,.
\end{align}
Da $C_F$ weder in $\beta_0$ enthalten ist noch quadratisch in
$\gamma^{(2)}$ vorkommt, ist der $C_F^2$-Term in~(\ref{eq:F2NNLLSUN}) durch
$\frac12 (F^{(1)})^2$ gegeben.

Diese NNLL-N"aherung gilt mit $n_s=1$ auch f"ur ein $SU(2)$-Higgs-Modell.
Die Abh"angigkeit vom Mechanismus der spontanen Symmetriebrechung und von
der Higgs-Masse kommt erst im linearen Logarithmus "uber~$\xi^{(2)}$
ins Spiel.
Setzt man die $SU(2)$-Werte f"ur $C_F$, $C_A$ und
$T_F$~(\ref{eq:CasimirSU2}) sowie $n_s=1$ ein, so erh"alt man:
\begin{align}
  F^{(2)} = \frac{9}{32} \lqm^4
    + \left(\frac{5}{48} - \frac{n_f}{6}\right) \lqm^3
    + \left(\frac{7}{8}\pi^2 - \frac{691}{48} + \frac{19}{12} n_f\right)
      \lqm^2
    + \Oc(\lqm^1)
  \,.
\end{align}
Numerisch ergibt dies:
\begin{equation}
    F^{(2)} =
    \underbrace{0{,}28 \, \lqm^4}_{180}
    {}+ \underbrace{(0{,}10 - 0{,}17\,n_f) \, \lqm^3}_{13 - 21\,n_f}
    {}+ \underbrace{(-5{,}8 + 1{,}6\,n_f) \, \lqm^2}_{-150 + 40\,n_f}
    {}+ \Oc(\lqm^1)
  \,,
\end{equation}
wobei f"ur die Werte unterhalb der Klammern $M=80\,\GeV$ und $Q=1\,\TeV$
eingesetzt wurde.
Die Reihe der Logarithmenpotenzen scheint im TeV-Bereich nicht gut zu
"`konvergieren"'.
Die Gr"o"se und auch das Vorzeichen des zweiten und dritten Koeffizienten
h"angen von der Zahl der Fermionen ab.
F"ur $n_f=6$ (siehe Abschnitt~$\ref{eq:EWparam}$) tr"agt der kubische
Logarithmus den Wert~$-120$, der quadratische Logarithmus den Wert~$+95$
zu $F^{(2)}$ bei (alle Zahlen jeweils auf 2~g"ultige Stellen gerundet).
Alle drei logarithmischen Beitr"age sind von der gleichen Gr"o"senordnung,
und durch das wechselnde Vorzeichen treten gro"se gegenseitige
Kompensationen zwischen den logarithmischen Termen auf.

F"ur eine Pr"azisionsrechnung, die Vorhersagen f"ur den Formfaktor und die
Vier\-fermion\-amplitude mit einem Fehler relativ zum Born-Ergebnis von
weniger als 1\% treffen will,
reicht die NNLL-N"aherung des Zweischleifenformfaktors nicht aus.
Deshalb ist die Berechnung des Formfaktors mit Eichbosonmasse~$M$ und
Higgs-Mechanismus auf Zweischleifenniveau n"otig.

Die n"achsten Kapitel besch"aftigen sich mit verschiedenen Beitr"agen zum
massiven Zweischleifenformfaktor:
die fermionischen ($n_f$) und skalaren ($n_s$) Beitr"age in
Kapitel~\ref{chap:nfns},
die abelschen Beitr"age ($C_F^2$) in Kapitel~\ref{chap:abelsch} und
die nichtabelschen Beitr"age ($C_F C_A$) inklusive der Higgs-Beitr"age in
Kapitel~\ref{chap:nichtabelsch}.

Zun"achst aber werden im n"achsten Abschnitt die Einschleifenergebnisse,
insofern sie f"ur Beitr"age zum Zweischleifenformfaktor eine Rolle spielen,
aufgef"uhrt.

%
%
\section{Einschleifenergebnisse}
\label{sec:1loop}

F"ur den Formfaktor in Zweischleifenn"aherung werden auch Produkte von
Einschleifenergebnissen ben"otigt.
Nicht nur der gesamte Einschleifenformfaktor, sondern ebenso die einzelnen
Beitr"age dazu, die Vertexkorrektur und die Selbstenergiekorrektur,
werden verlangt.
Und weil die Einschleifenergebnisse mit Termen multipliziert werden, die
$1/\eps^2$-Pole enthalten, m"ussen sie bis einschlie"slich zur
Ordnung~$\eps^2$ bekannt sein.

\begin{figure}[ht]
  \centering
  \vcentergraphics{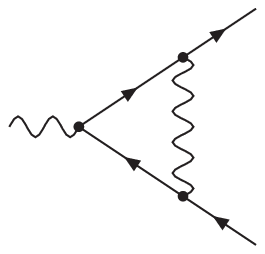}
  \hspace{2cm}
  \vcentergraphics{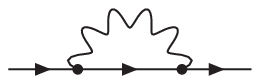}
  \caption{Einschleifen-Vertexdiagramm und -Selbstenergiediagramm}
  \label{fig:1loop}
\end{figure}

Die beiden Beitr"age zum Einschleifenergebnis sind in Abb.~\ref{fig:1loop}
dargestellt.

Der Beitrag der Vertexkorrektur zum Einschleifenformfaktor ist im
Hochenergielimes $Q^2 \gg M^2$ durch folgenden Ausdruck gegeben:
\begin{align}
\label{eq:Fv1erg}
  F_{v,1} &= C_F \, \frac{\alpha}{4\pi}
    \left(\frac{\mu^2}{M^2}\right)^\eps S_\eps \,
    \Biggl\{
    \frac{1}{\eps}
    - \lqm^2 + 3 \lqm - \frac{2}{3}\pi^2 - 4
\nonumber \\* & \qquad
    + \eps \left[
      \frac{1}{3} \lqm^3 - \frac{3}{2} \lqm^2
      + \left(-\frac{\pi^2}{3} + 8\right) \lqm
      + 2\zeta_3 + \frac{7}{12}\pi^2 - 12 \right]
\nonumber \\* & \qquad
    + \eps^2 \,\biggl[
      -\frac{1}{12} \lqm^4 + \frac{1}{2} \lqm^3
      + \left(\frac{\pi^2}{12} - 4\right) \lqm^2
      + \left(-4\zeta_3 - \frac{\pi^2}{4} + 16\right) \lqm
\nonumber \\* & \qquad\qquad
      - \frac{13}{180}\pi^4 + \frac{17}{3}\zeta_3 + \pi^2 - 28
      \biggr]
    \Biggr\}
    + \Oc(\eps^3) + \Oc\left(\frac{M^2}{Q^2}\right)
  ,
\end{align}
mit $\lqm = \ln(Q^2/M^2)$ und $S_\eps = (4\pi)^\eps \, e^{-\eps\gamma_E}$.
Zur Renormierung der Fermion-Feldst"arke wird zweimal der
Faktor~$\sqrt{Z_f}$ ben"otigt, also insgesamt gerade einmal der Faktor
$Z_f = 1 + \Sigma$ entsprechend~(\ref{eq:ZfSigma}).
Wenn $F_v$ der Beitrag der Vertexkorrekturen zum Formfaktor ist, gilt
in Einschleifenn"aherung:
\begin{align}
  F = F_v \cdot Z_f = 1 + \underbrace{F_{v,1} + \Sigma_1}_{\Oc(\alpha)}
  {}+ \Oc(\alpha^2)
  \,.
\end{align}
Die Einschleifen-Selbstenergiekorrektur betr"agt:
\begin{multline}
\label{eq:Sigma1erg}
  \Sigma_1 \equiv \Sigma_1(p^2=0) =
  C_F \, \frac{\alpha}{4\pi}
    \left(\frac{\mu^2}{M^2}\right)^\eps S_\eps \,
  \biggl\{ -\frac{1}{\eps} + \frac{1}{2}
    + \eps \left(-\frac{\pi^2}{12} + \frac{1}{4}\right)
  \\*
    + \eps^2 \left(\frac{1}{3}\zeta_3 + \frac{\pi^2}{24} + \frac{1}{8}\right)
  \biggr\}
  + \Oc(\eps^3)
  \,.
\end{multline}
Der gesamte Einschleifenformfaktor ist durch
\begin{align}
\label{eq:F1erg}
  F_1 &= F_{v,1} + \Sigma_1
\nonumber \\*
  &= C_F \, \frac{\alpha}{4\pi}
    \left(\frac{\mu^2}{M^2}\right)^\eps S_\eps \,
    \biggl\{
    - \lqm^2 + 3 \lqm - \frac{2}{3}\pi^2 - \frac{7}{2}
\nonumber \\* & \qquad
    + \eps \left[
      \frac{1}{3} \lqm^3 - \frac{3}{2} \lqm^2
      + \left(-\frac{\pi^2}{3} + 8\right) \lqm
      + 2\zeta_3 + \frac{\pi^2}{2} - \frac{47}{4} \right]
\nonumber \\* & \qquad
    + \eps^2 \,\biggl[
      -\frac{1}{12} \lqm^4 + \frac{1}{2} \lqm^3
      + \left(\frac{\pi^2}{12} - 4\right) \lqm^2
      + \left(-4\zeta_3 - \frac{\pi^2}{4} + 16\right) \lqm
\nonumber \\* & \qquad\qquad
      - \frac{13}{180}\pi^4 + 6\zeta_3 + \frac{25}{24}\pi^2 - \frac{223}{8}
      \biggr]
    \Biggr\}
    + \Oc(\eps^3) + \Oc\left(\frac{M^2}{Q^2}\right)
\end{align}
gegeben.
Dieser ist in $d=4$ Dimensionen endlich, die UV-Singularit"aten wurden
zwischen $F_{v,1}$ und $\Sigma_1$ aufgehoben.

Beim Renormieren der Kopplungskonstante gem"a"s Gl.~(\ref{eq:alpharen})
liefert $F_1$ einen Beitrag zum Zweischleifenformfaktor.
Der Parameter~$\alpha$ in~(\ref{eq:F1erg}) ist zun"achst eigentlich die
nackte Kopplung~$\alpha_0$.
Durch
\[
  \frac{\alpha_0}{4\pi} =
  \frac{\alpha}{4\pi}
  - \frac{\beta_0}{\eps} \, \left(\frac{\alpha}{4\pi}\right)^2
  + \Oc(\alpha^3)
\]
entsprechend (\ref{eq:alpharen}) und (\ref{eq:Zalpha})
ergibt sich der Zweischleifenbeitrag
\begin{align}
\label{eq:F2alpharenbeta0}
  \Delta F_2^\alpha &= -\beta_0 \, C_F \left(\frac{\alpha}{4\pi}\right)^2
    \left(\frac{\mu^2}{M^2}\right)^\eps S_\eps \,
    \biggl\{
    \frac{1}{\eps} \left[
      -\lqm^2 + 3 \lqm - \frac{2}{3}\pi^2 - \frac{7}{2} \right]
\nonumber \\* & \qquad
    + \frac{1}{3} \lqm^3 - \frac{3}{2} \lqm^2
      + \left(-\frac{\pi^2}{3} + 8\right) \lqm
      + 2\zeta_3 + \frac{\pi^2}{2} - \frac{47}{4}
    \biggr\}
    + \Oc(\eps) + \Oc\left(\frac{M^2}{Q^2}\right)
  .
\end{align}
Mit $\beta_0$ aus Gl.~(\ref{eq:beta0})
erh"alt man Terme, die, je nach Farbfaktor $C_A$, $T_F n_f$ oder $T_F n_s$,
zum nichtabelschen Formfaktor (Kapitel~\ref{chap:nichtabelsch}) oder zum
fermionischen Formfaktor (Kapitel~\ref{chap:nfns}) einen
Beitrag liefern.

Ein weiterer Renormierungsbeitrag ergibt sich aus~$F_1$ durch Einsetzen der
renormierten Masse entsprechend Gl.~(\ref{eq:Mren}).
Das f"uhrt zu den folgenden Transformationen:
\begin{align*} 
  \left(\frac{\mu^2}{M_0^2}\right)^\eps &=
    \left(\frac{\mu^2}{M^2}\right)^\eps \,
    \Bigl[ 1 + \eps \, \Rep\Pi_1(M^2) \Bigr]
    + \Oc(\alpha^2) \,,
\\
  \ln^n\!\left(\frac{Q^2}{M_0^2}\right) &=
    \ln^n\!\left(\frac{Q^2}{M^2}\right)
    + n \cdot \Rep\Pi_1(M^2) \, \ln^{n-1}\!\left(\frac{Q^2}{M^2}\right)
    + \Oc(\alpha^2) \,,
\end{align*}
mit der Eichboson-Selbstenergie~$\Pi_1$ entsprechend~(\ref{eq:Pimunu}).
Durch Einsetzen der Transformationen in $F_1$~(\ref{eq:F1erg}) ergibt sich
der folgende Renormierungsbeitrag in Zweischleifenordnung:
\begin{align}
\label{eq:F2MrenPi1}
  \Delta F_2^M &= \Rep\Pi_1(M^2) \,
    C_F \, \frac{\alpha}{4\pi}
    \left(\frac{\mu^2}{M^2}\right)^\eps S_\eps
    \left\{ - 2 \lqm + 3 + \eps \left(-\pi^2 + \frac{9}{2}\right)
    + \Oc(\eps^2) \right\}
\nonumber \\* & \quad
    + \Oc\left(\frac{M^2}{Q^2}\right)
  \,.
\end{align}
Da $\Pi_1$ in der Regel einen $1/\eps$-Pol enth"alt, werden die Terme
in~(\ref{eq:F2MrenPi1}) bis zur linearen Ordnung in~$\eps$
ben"otigt.
Die Eichboson-Selbstenergie~$\Pi_1$ erf"ahrt nichtabelsche ($C_A$) und
fermionische Beitr"age ($n_f$) sowie Higgs-Beitr"age.
Diese werden in den Kapiteln \ref{chap:nfns} und \ref{chap:nichtabelsch}
vorgestellt und in~(\ref{eq:F2MrenPi1}) eingesetzt.


\clearemptypage

\chapter{Fermionische und skalare Beitr"age zum Zweischleifenformfaktor}
\label{chap:nfns}

Dieses Kapitel referiert die Ergebnisse aus meiner
Diplomarbeit~\cite{Feucht:Diplom}. Dort ging es um die fermionischen
Beitr"age zum Zweischleifenformfaktor, also die Terme proportional zur
Anzahl~$n_f$ der Fermionen.
In Zusammenarbeit mit S.~Moch wurde zus"atzlich die G"ultigkeit des
Hochenergielimes getestet und das Ergebnis der Diplomarbeit um den Beitrag
masseloser skalarer Teilchen ($n_s$) erweitert.
Die Ergebnisse dieses Kapitels sind in \cite{Feucht:2003yx} ver"offentlicht.

%
%
\section{Beitr"age von Fermionschleifen}

\begin{figure}[ht]
  \centering
  \vcentergraphics{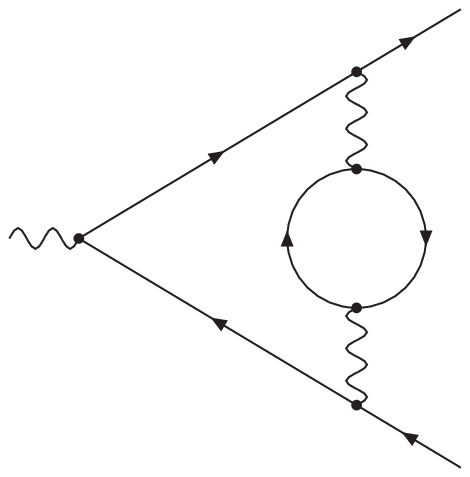}
  \hspace{1cm}
  \vcentergraphics{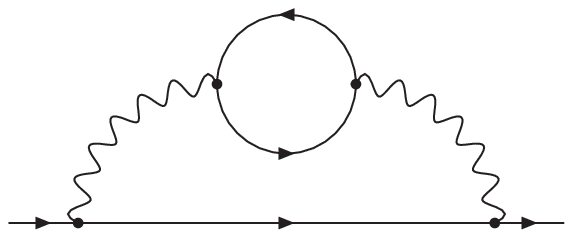}
  \caption{Vertex- und Selbstenergiediagramm der fermionischen
    Beitr"age}
  \label{fig:nf}
\end{figure}

Zu den fermionischen Beitr"agen des Formfaktors geh"oren das Vertexdiagramm
und das Selbstenergiediagramm in Abb.~\ref{fig:nf}.

Die Amplitude des Vertexdiagramms l"asst sich f"ur ein nichtabelsches
Modell folgenderma"sen schreiben (ohne die Spinoren der externen Fermionen):

\begin{multline}
  \Fc^\mu_{v,n_f} =
  \mu^{4\eps} \loopint dk \,
  ig \gamma^\nu t^a \, \frac{i}{\dslash k+\dslash p_1} \,
  \gamma^\mu
  \frac{i}{\dslash k+\dslash p_2} \, ig \gamma^\rho t^b \,
  \left(\frac{-i}{k^2-M^2}\right)^2
  \\* \times
  (-1) \, n_f \loopint d\ell \, \Tr\left(
    ig \gamma_\nu t^a \, \frac{i}{\dslash\ell} \,
    ig \gamma_\rho t^b \, \frac{i}{\dslash k+\dslash\ell}
  \right)
  \,,
\end{multline}
wobei die Spur sowohl "uber den Spinorraum der $\gamma$-Matrizen als
auch "uber den Isospinraum der $SU(2)$-Dubletts geht.
Gegen"uber der Diplomarbeit, wo die Rechnung f"ur ein abelsches
$U(1)$-Modell durchgef"uhrt wurde, erh"alt man hier
mittels Gl.~(\ref{eq:Casimir}) den $SU(2)$-Farbfaktor
\begin{equation}
\label{eq:CFTF}
  t^a t^b \, \Tr(t^a t^b) = C_F T_F \, \unity
  \,.
\end{equation}

Um diesen Farbfaktor wird das Ergebnis des Vertexdiagramms aus der
Diplomarbeit\cite{Feucht:Diplom} erweitert:
\begin{align}
\label{eq:Fvnferg}
  F_{v,n_f} &= C_F T_F n_f \left(\frac{\alpha}{4\pi}\right)^2
    \left(\frac{\mu^2}{M^2}\right)^{2\eps} S_\eps^2 \,
    \biggl\{
      \frac{1}{\eps} \left[ \frac{4}{3} \lqm^2 - \frac{20}{3} \lqm
        + \frac{8}{9}\pi^2 + \frac{29}{3} \right]
    - \frac{8}{9} \lqm^3
    + \frac{56}{9} \lqm^2
  \nonumber \\* & \qquad
    + \left(\frac{4}{9}\pi^2 - \frac{238}{9}\right) \lqm
    - \frac{8}{3}\zeta_3 - \frac{38}{27}\pi^2 + \frac{749}{18}
    \biggr\}
  + \Oc(\eps) + \Oc\!\left(\frac{M^2}{Q^2}\right) ,
\end{align}
mit $\lqm = \ln(Q^2/M^2)$ und $S_\eps = (4\pi)^\eps \, e^{-\eps\gamma_E}$.

Zur Renormierung der Fermion-Feldst"arke wird der Faktor~$Z_f$
entsprechend~(\ref{eq:ZfSigma}) ben"otigt.
Mit dem Formfaktorbeitrag der Vertexkorrekturen,
$F_v = 1 + F_{v,1} + F_{v,2} + \Oc(\alpha^3)$,
und der Fermion-Selbstenergie,
$\Sigma = \Sigma_1 + \Sigma_2 + \Oc(\alpha^3)$,
gilt in Zweischleifenn"aherung:
\begin{equation}
\label{eq:LSZ2loop}
  F = F_v \cdot Z_f =
  1 + \underbrace{F_{v,1} + \Sigma_1}_{\Oc(\alpha)}{}
  + \underbrace{F_{v,2} + \Sigma_2 + F_{v,1} \Sigma_1}_{\Oc(\alpha^2)}{}
  + \Oc(\alpha^3) \,.
\end{equation}
Der Zweischleifenformfaktor besteht aus den Beitr"agen
\begin{equation}
\label{eq:LSZF2}
  F_2 = F_{v,2} + \Sigma_2 + F_{v,1} \Sigma_1
  \,,
\end{equation}
also den Zweischleifen-Vertexkorrekturen, den
Zweischleifen"=Selbstenergiekorrekturen und dem Produkt aus
Einschleifen-Vertexkorrektur und Einschleifen-Selbst\-energie\-korrektur.

Da das Produkt $F_{v,1} \Sigma_1$ proportional zu $C_F^2$ ist (siehe
Abschnitt~\ref{sec:1loop}), spielt es nur f"ur die abelschen Beitr"age
(Kapitel~\ref{chap:abelsch}) eine Rolle.
F"ur den fermionischen Formfaktor fehlt neben der Vertexkorrektur nur die
Selbstenergiekorrektur in Abb.~\ref{fig:nf}.
Diese wurde ebenfalls in der Diplomarbeit berechnet:
\begin{align}
\label{eq:Sigmanferg}
  \Sigma_{n_f} &= C_F T_F n_f \left(\frac{\alpha}{4\pi}\right)^2
    \left(\frac{\mu^2}{M^2}\right)^{2\eps} S_\eps^2
    \left\{ -\frac{1}{\eps} - \frac{1}{2} \right\}
  + \Oc(\eps) \,.
\end{align}
Der Farbfaktor ist mit Gl.~(\ref{eq:CFTF}) der gleiche wie f"ur die
Vertexkorrektur.

\begin{figure}[ht]
  \centering
  \vcentergraphics{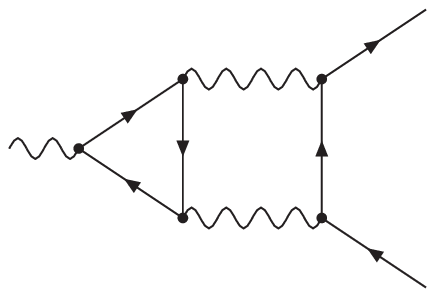}
  \hspace{1cm}
  \vcentergraphics{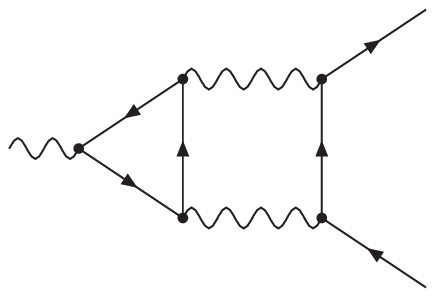}
  \caption{Vertexdiagramme mit Triangelschleife}
  \label{fig:triangel}
\end{figure}

Auch die beiden Vertexdiagramme in Abb.~\ref{fig:triangel} tragen
prinzipiell zum fermionischen Formfaktor bei.
Beide Diagramme enthalten eine Fermionschleife, an die drei Eichbosonen
koppeln. 
Die Diagramme unterscheiden sich in der Richtung des Fermionflusses durch
die Schleife (gekennzeichnet durch die Pfeilrichtung).
In einer abelschen Theorie 
verschwindet die Summe beider Beitr"age nach Furrys Theorem
(siehe z.B. \cite{Peskin:1995ev,Itzykson:1980rh}),
da eine ungerade Zahl von Eichbosonen an die Fermionschleife koppelt.
Aber auch im hier betrachteten Formfaktor des Vektorstroms, dessen
zentraler Vertex eine abelsche, parit"atserhaltende Kopplung darstellt,
verschwindet die Summe beider Diagramme, da nur zwei nichtabelsche
Eichbosonen an die Fermionschleife koppeln.

Neben den Zweischleifendiagrammen gibt es auch Renormierungsbeitr"age, die
proportional zu~$n_f$ sind.
Das Einsetzen des $n_f$-Terms von $\beta_0$~(\ref{eq:beta0}) in den
Renormierungsbeitrag der Kopplungskonstante~(\ref{eq:F2alpharenbeta0})
liefert:
\begin{align}
\label{eq:Fnfalpharenerg}
  \Delta F_{n_f}^\alpha &= C_F T_F n_f \left(\frac{\alpha}{4\pi}\right)^2
    \left(\frac{\mu^2}{M^2}\right)^\eps S_\eps \,
    \biggl\{
    \frac{1}{\eps} \left[
      -\frac{4}{3} \lqm^2 + 4 \lqm - \frac{8}{9}\pi^2 - \frac{14}{3} \right]
    + \frac{4}{9} \lqm^3 - 2 \lqm^2
\nonumber \\* & \qquad
      + \left(-\frac{4}{9}\pi^2 + \frac{32}{3}\right) \lqm
      + \frac{8}{3}\zeta_3 + \frac{2}{3}\pi^2 - \frac{47}{3}
    \biggr\}
    + \Oc(\eps) + \Oc\left(\frac{M^2}{Q^2}\right)
  .
\end{align}

F"ur die Renormierung der Eichbosonmasse~$M$ muss die
Einschleifenselbstenergie des Eichbosons berechnet werden.
Der fermionische ($n_f$) Beitrag dazu entspricht gerade der Fermionschleife im
Eichbosonpropagator der Diagramme in Abb.~\ref{fig:nf}.
Entsprechend der Notation~(\ref{eq:Pimunu}) ist diese Selbstenergie durch
\begin{equation}
  \tilde\Pi^{\mu\nu,ab}_{n_f}(k) =
  i \, \delta^{ab} \, (g^{\mu\nu} k^2 - k^\mu k^\nu) \, \Pi_{n_f}(k^2)
\end{equation}
mit
\begin{equation}
  \Pi_{n_f}(k^2) =
    T_F n_f \, \frac{\alpha}{4\pi}
    \left(\frac{\mu^2}{-k^2-i0}\right)^\eps S_\eps
    \left\{ -\frac{4}{3\eps} - \frac{20}{9} \right\}
    + \Oc(\eps)
\end{equation}
gegeben.
Ben"otigt wird der Realteil bei $k^2 = M^2$:
\begin{align}
  \Rep \Pi_{n_f}(M^2) =
    T_F n_f \, \frac{\alpha}{4\pi}
    \left(\frac{\mu^2}{M^2}\right)^\eps S_\eps
    \left\{ -\frac{4}{3\eps} - \frac{20}{9} \right\}
    + \Oc(\eps)
  \,.
\end{align}
Eingesetzt in den Beitrag der Massenrenormierung~(\ref{eq:F2MrenPi1})
erh"alt man:
\begin{align}
\label{eq:FnfMrenerg}
  \Delta F_{n_f}^M &= C_F T_F n_f \left(\frac{\alpha}{4\pi}\right)^2
    \left(\frac{\mu^2}{M^2}\right)^{2\eps} S_\eps^2
    \left\{ \frac{1}{\eps} \left[ \frac{8}{3} \lqm - 4 \right]
      + \frac{40}{9} \lqm + \frac{4}{3}\pi^2 - \frac{38}{3}
    \right\}
  \nonumber \\* & \quad
    + \Oc(\eps) + \Oc\left(\frac{M^2}{Q^2}\right)
  .
\end{align}

Der fermionische Formfaktor setzt sich folgenderma"sen zusammen:
\begin{equation} 
  F_{2,n_f} =
  F_{v,n_f} + \Sigma_{n_f} + \Delta F_{n_f}^\alpha + \Delta F_{n_f}^M
  \,.
\end{equation}
Mit den Ergebnissen aus (\ref{eq:Fvnferg}), (\ref{eq:Sigmanferg}),
(\ref{eq:Fnfalpharenerg}) und (\ref{eq:FnfMrenerg}) folgt,
zun"achst f"ur eine beliebige Renormierungsskala~$\mu$:
\begin{align}
  F_{2,n_f} &=
  C_F T_F n_f \left(\frac{\alpha}{4\pi}\right)^2 \, \biggl\{
    -\frac{4}{9} \lqm^3
    + \frac{38}{9} \lqm^2
    - \frac{34}{3} \lqm
    + \frac{16}{27}\pi^2 + \frac{115}{9}
  \nonumber \\* & \qquad
    + \ln\left(\frac{\mu^2}{M^2}\right) \left[
      \frac{4}{3} \lqm^2 - 4 \lqm + \frac{8}{9}\pi^2 + \frac{14}{3} \right]
    \biggr\}
    + \Oc(\eps) + \Oc\left(\frac{M^2}{Q^2}\right) .
\end{align}
Alle Pole in $\eps$ fallen heraus, das Ergebnis ist in $d=4$ Dimensionen
endlich.
Gem"a"s der \MSbar-Vorschrift~(\ref{eq:muMSbar}) wurde der
Faktor~$S_\eps$ in die Skala~$\mu$ absorbiert.
Der Logarithmus $\ln(\mu^2/M^2)$ kommt von der abweichenden
$(\mu^2/M^2)$-Abh"angigkeit im Beitrag der Kopplungskonstantenrenormierung
und entspricht dem Laufen der Kopplung im Einschleifenformfaktor.

Wenn die Kopplungskonstante~$\alpha$ bei $\mu=M$ renormiert wird, dann
lautet der fermionische Formfaktor in $d=4$ Dimensionen:
\begin{align}
\label{eq:F2nferg}
  F_{2,n_f} &=
  C_F T_F n_f \left(\frac{\alpha}{4\pi}\right)^2 \, \biggl\{
    -\frac{4}{9} \lqm^3
    + \frac{38}{9} \lqm^2
    - \frac{34}{3} \lqm
    + \frac{16}{27}\pi^2 + \frac{115}{9}
    \biggr\}
    + \Oc\left(\frac{M^2}{Q^2}\right) .
\end{align}

Dieses Ergebnis stimmt offensichtlich mit den $n_f$-Termen aus der
Evolutionsgleichung in~(\ref{eq:F2NNLLSUN}) "uberein.
Die fermionischen Beitr"age wurden also in NNLL-N"aherung richtig von der
Evolutionsgleichung vorhergesagt.
Der lineare Logarithmus und die Konstante in~(\ref{eq:F2nferg}) sind neu
dazugekommen.
Der Vergleich mit (\ref{eq:F2gzx}) liefert den $n_f$"~Beitrag zum 
Koeffizienten~$\xi^{(2)}$,
da alle anderen Terme im Koeffizienten des linearen Logarithmus bereits aus
(\ref{eq:g1z1}), (\ref{eq:x1F01}) und (\ref{eq:zeta2}) bekannt sind:
\begin{equation}
  \xi^{(2)}\Big|_{n_f} =
  C_F T_F n_f \left( \frac{4}{9}\pi^2 + \frac{112}{27} \right) .
\end{equation}

Die numerische Gr"o"se der Logarithmenkoeffizienten im
Ergebnis~(\ref{eq:F2nferg}) stellt sich wie folgt dar:
\begin{equation}
  F_{2,n_f} =
  C_F T_F n_f \left(\frac{\alpha}{4\pi}\right)^2
    \, \Bigl(
    - 0{,}\bar4 \,\lqm^3
    + 4{,}\bar2 \,\lqm^2
    - 11{,}\bar3 \,\lqm
    + 18{,}6264
    \Bigr)
    \,.
\end{equation}
Das Muster der alternierenden Koeffizienten zieht sich durch alle Terme:
Die Gr"o"se der Koeffizienten nimmt von $\lqm^3$ bis hin zur Konstanten zu
und wechselt von Term zu Term das Vorzeichen.

Um die Gr"o"se der Beitr"age einzelner Logarithmen abzusch"atzen, wird
$\alpha/(4\pi) = 0{,}003 \approx \alpha_\QED/(4\pi\sin^2\theta_W)$ gesetzt
sowie $C_F T_F = 3/8$ f"ur die schwache $SU(2)$-Eichgruppe
und $n_f = 6$ Fermion-Dubletts.
Bei $M = 80\,\GeV \approx M_W$ und dem Impuls"ubertrag $Q = 1\,\TeV$
lauten die Beitr"age der logarithmischen Terme zum Formfaktor in Promille
(d.h. mit $10^3$ multipliziert):
\begin{equation} 
  \lqm^3 \to -1{,}2 \,,\quad
  \lqm^2 \to +2{,}2 \,,\quad
  \lqm^1 \to -1{,}2 \,,\quad
  \lqm^0 \to +0{,}38
  \,.
\end{equation}
Der gr"o"ste Beitrag im TeV-Bereich stammt vom quadratischen Logarithmus.
Der lineare Logarithmus tr"agt bereits weniger bei, und die Konstante ist
relativ klein.
Es zeigt sich also, dass die logarithmische Reihe nach dem quadratischen
Term zu "`konvergieren"' beginnt.
Mindestens aber der lineare Logarithmus ist f"ur eine gute Genauigkeit des
fermionischen Beitrags n"otig.

Erstaunlich ist auch, dass die Summe aller Beitr"age bei $Q=1\,\TeV$ im
Vergleich zu den einzelnen Termen mit 0,24~Promille sehr klein ist.
Dieses Verhalten des fermionischen Formfaktors l"asst sich gut am Schaubild
in Abb.~\ref{fig:plotnf} ablesen,
wo die sukzessiven logarithmischen N"aherungen gezeichnet sind:
\begin{figure}[p]
  \centering
  \includegraphics{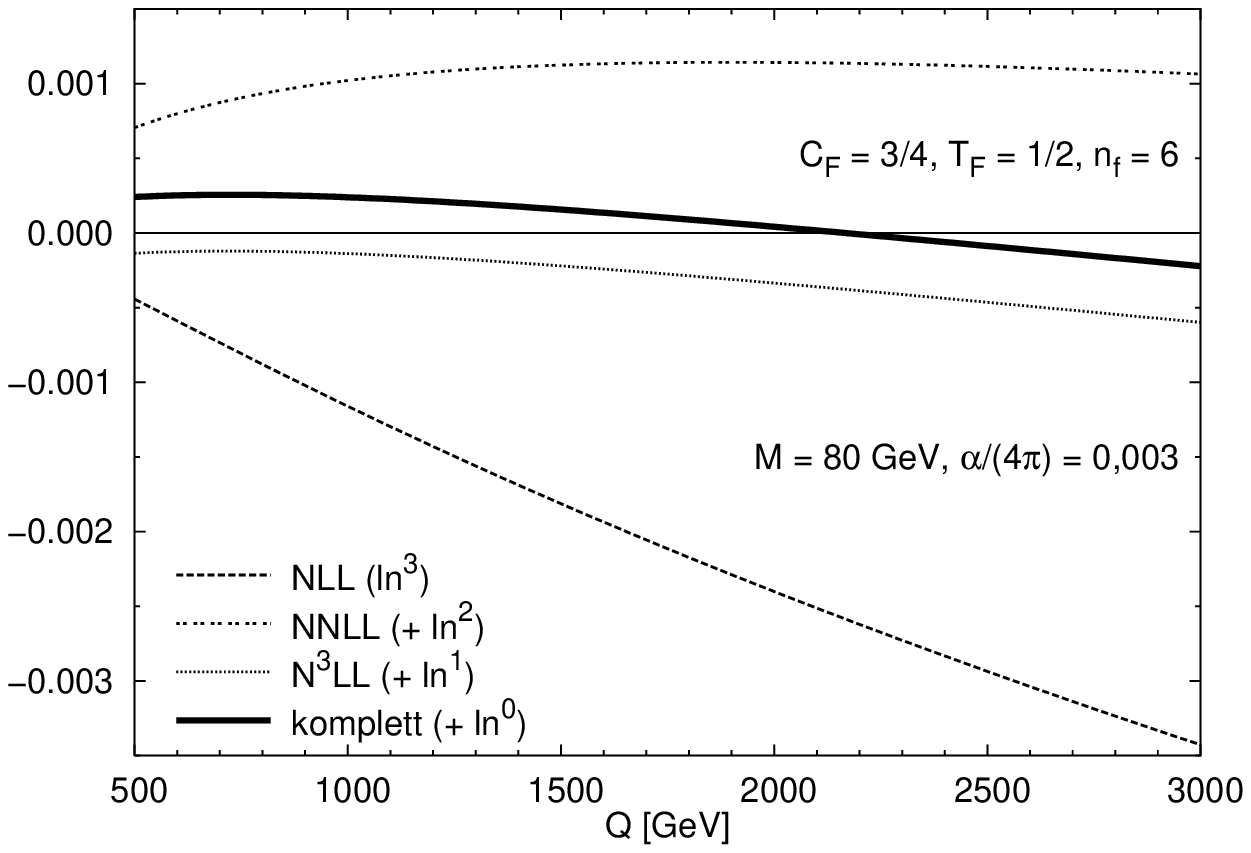}
  \caption{Fermionische Beitr"age zum Zweischleifenformfaktor in sukzessiven
    logarithmischen N"aherungen}
  \label{fig:plotnf}
\end{figure}
NLL mit $\lqm^3$,
NNLL mit $\lqm^3$ und $\lqm^2$,
$\NNNLL$ mit allen Logarithmen ohne die Konstante
und der komplette Formfaktor mit allen Beitr"agen im Hochenergielimes.

\begin{figure}[p]
  \centering
  \includegraphics{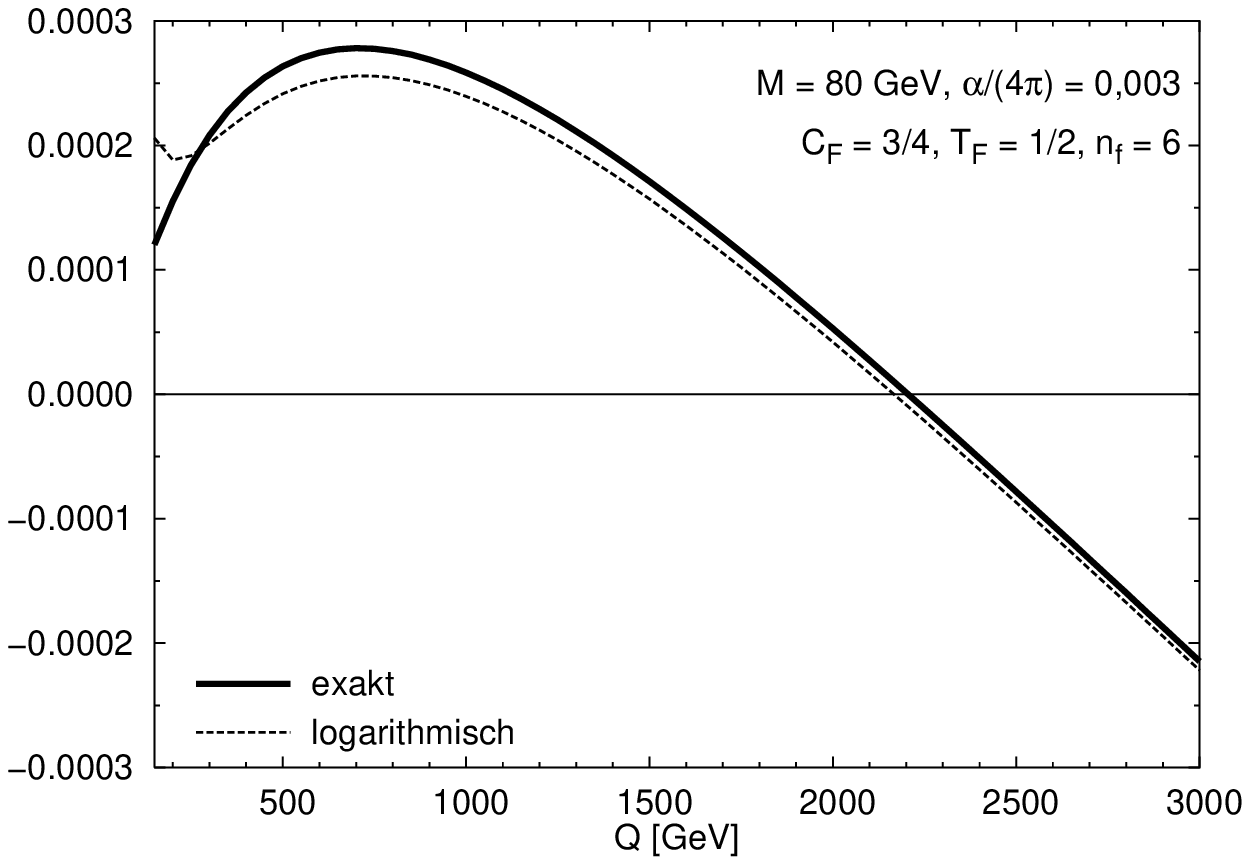}
  \caption{Fermionische Beitr"age zum Zweischleifenformfaktor, exakt und in der
    logarithmischen N"aherung des Hochenergielimes}
  \label{fig:plotnfexakt}
\end{figure}
Da die Diagramme mit Fermionschleife eine besonders einfache Topologie
besitzen und zudem in der inneren Schleife keine Massen auftreten,
konnte der fermionische Beitrag zum Zweischleifenformfaktor neben dem
Hochenergielimes auch exakt f"ur beliebige $z = M^2/Q^2$ berechnet werden:
\begin{align}
  F_{2,n_f} &= C_F T_F n_f \left(\frac{\alpha}{4\pi}\right)^2 \,
  \biggl\{
      (1-4z+3z^2) \biggl(
        \frac{8}{3}\,\Li3(z)
        + \frac{8}{3} \ln(z)\,\Li2(1-z)
        + \frac{4}{9} \ln^3(z)
\nonumber \\* & \qquad\qquad{}
        + \frac{4}{3} \ln^2(z) \ln(1-z)
        - \frac{4}{9}\pi^2 \ln(z)
        \biggr)
      + (1-z)^2 \biggl(
        \frac{16}{9}\,\Li2(1-z)
        + \frac{8}{27}\pi^2
        \biggr)
\nonumber \\* & \qquad
      + \biggl( \frac{38}{9} - \frac{52}{9}z + \frac{8}{9}z^2 \biggr) \ln^2(z)
      + \biggl( \frac{34}{3} - \frac{88}{9}z \biggr) \ln(z)
      + \frac{115}{9} - \frac{88}{9}z
      \biggr\}
  \,,
\end{align}
mit den Polylogarithmen $\Li2$ und $\Li3$~(\ref{eq:polylog}).

Der Vergleich des exakten Ergebnisses mit dem logarithmischen
Hochenergielimes (alle Terme inklusive der Konstanten) ist in
Abb.~\ref{fig:plotnfexakt} dargestellt.
Der Unterschied zwischen beiden Kurven liegt in den mit $M^2/Q^2$
unterdr"uckten Termen.
Das Schaubild zeigt die sehr gute "Ubereinstimmung der Hochenergien"aherung
mit dem exakten Ergebnis f"ur $Q > 300\,\GeV$. Unterhalb davon divergiert
die logarithmische N"aherung. Aber oberhalb dieser Schwelle betr"agt die
Abweichung maximal 0,02~Promille.
Der in den Evolutionsgleichungen und in den Schleifenrechnungen verwendete
Hochenergielimes reicht f"ur die gew"unschte Genauigkeit v"ollig aus.

%
%
\section{Beitr"age von masselosen skalaren Teilchen}

Wenn man im Hochenergielimes arbeitet, kann in erster N"aherung das
Higgs-Potential~$V(\Phi)$ und die damit verbundene spontane
Symmetriebrechung vernachl"assigt werden.
Das Higgs-Dublett verh"alt sich dann wie ein masseloses, geladenes Dublett
von Skalarfeldern.

In der Evolutionsgleichung~(\ref{eq:formfaktorevol}) f"ur den Formfaktor
sind die Funktionen $\gamma$ und $\zeta$ universal, sie h"angen nicht vom
Higgs-Potential ab. Effekte der spontanen Symmetriebrechung tauchen
nur in $\xi^{(2)}$ und $F_0^{(2)}$ und damit beim
Zweischleifenformfaktor erst im linearen Logarithmus auf.
Die NNLL-N"aherung kann also mit einem masselosen skalaren Dublett
anstelle des tats"achlichen Higgs-Teilchens berechnet werden.

Weil das Ergebnis f"ur den fermionischen Formfaktor aus dem vorigen
Abschnitt leicht auf masselose skalare Teilchen "ubertragen werden kann,
soll diese Rechnung hier kurz skizziert werden.

Wir betrachten $n_s$~masselose, geladene Dubletts von Skalarfeldern, die
sich in der fundamentalen Darstellung der $SU(2)$ transformieren.
Ihre Lagrange-Dichte ist durch
\begin{equation}
  \Lc_s = \sum_{i=1}^{n_s} \, (D_\mu \Phi_i)^\dag (D^\mu \Phi_i)
\end{equation}
mit der kovarianten Ableitung
\begin{equation}
  D_\mu = \partial_\mu - ig \, t^a \, W^a_\mu
\end{equation}
gegeben.
Die Feynman-Regeln lauten:
\begin{equation}
  \vcentergraphics{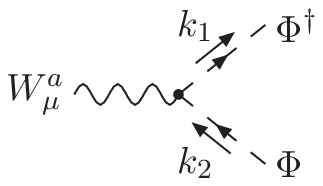} =
    ig\,t^a (k_1+k_2)_\mu \,,
  \qquad
  \vcentergraphics{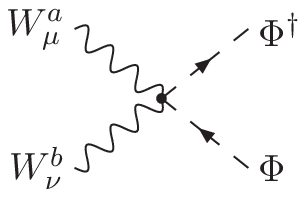} =
    \frac{i}{2} g^2 g_{\mu\nu} \delta^{ab} \,.
\end{equation}

\begin{figure}[ht]
  \centering
  \vcentergraphics{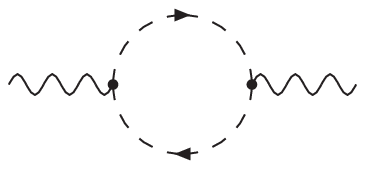}
  \hspace{1cm}
  \vcentergraphics{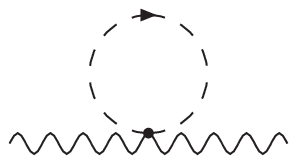}
  \caption{Eichboson-Selbstenergiediagramme mit skalaren Teilchen}
  \label{fig:bcorrns}
\end{figure}

Die beiden Vertizes steuern die Feynman-Diagramme in Abb.~\ref{fig:bcorrns}
zur Eichboson-Selbstenergie bei.
Das rechte Diagramm enth"alt jedoch eine sogenannte Tadpole-Schleife;
diese Selbstenergie ist unabh"angig vom Impuls des Eichbosons. Deshalb wird
ihr Beitrag exakt von der Renormierung der Eichbosonmasse kompensiert und
kann weggelassen werden.

Die Amplitude des linken Diagramms in Abb.~\ref{fig:bcorrns} lautet:
\begin{align}
  \tilde\Pi_{n_s}^{\mu\nu,ab}(k) &=
    n_s \, \mu^{2\eps} \loopint d\ell \,
    \Tr\left( ig\,t^a (k+2\ell)^\mu \, \frac{i}{\ell^2} \,
    ig\,t^b (k+2\ell)^\nu \, \frac{i}{(k+\ell)^2} \right)
  \\* &=
    i \, \delta^{ab} \, (g^{\mu\nu} k^2 - k^\mu k^\nu) \, \Pi_{n_s}(k^2)
    \,. \nonumber
\end{align}
Es stellt sich heraus, dass
\begin{equation}
\label{eq:PinsPinf}
  \Pi_{n_s}(k^2) =
  \frac{n_s}{n_f} \, \frac{1}{4(1-\eps)} \, \Pi_{n_f}(k^2)
\end{equation}
gilt.
Die Relation~(\ref{eq:PinsPinf}) zwischen der skalaren und der fermionischen
Eichboson-Selbstenergie ist eine exakte Beziehung, die nicht nur im Limes
$\eps\to0$ g"ultig ist.
Der Vorfaktor in~(\ref{eq:PinsPinf}) ist eine $\eps$-abh"angige Konstante,
mit der die Beitr"age des fermionischen Formfaktors aus dem
Vertexdiagramm, der Fermion-Selbstenergie und der Massenrenormierung
multipliziert werden k"onnen, um die entsprechenden Beitr"age des skalaren
Formfaktors zu erhalten:
\begin{align}
\label{eq:Fvnserg}
  F_{v,n_s} &= \frac{n_s}{n_f} \, \frac{1}{4(1-\eps)} \, F_{v,n_f}
  \nonumber \\*
  &= C_F T_F n_s \left(\frac{\alpha}{4\pi}\right)^2
    \left(\frac{\mu^2}{M^2}\right)^{2\eps} S_\eps^2 \,
    \biggl\{
      \frac{1}{\eps} \left[ \frac{1}{3} \lqm^2 - \frac{5}{3} \lqm
        + \frac{2}{9}\pi^2 + \frac{29}{12} \right]
    - \frac{2}{9} \lqm^3
    + \frac{17}{9} \lqm^2
  \nonumber \\* & \qquad
    + \left(\frac{\pi^2}{9} - \frac{149}{18}\right) \lqm
    - \frac{2}{3}\zeta_3 - \frac{7}{54}\pi^2 + \frac{923}{72}
    \biggr\}
  + \Oc(\eps) + \Oc\!\left(\frac{M^2}{Q^2}\right) ,
\\
\label{eq:Sigmanserg}
  \Sigma_{n_s} &= \frac{n_s}{n_f} \, \frac{\Sigma_{n_f}}{4(1-\eps)}
  = C_F T_F n_s \left(\frac{\alpha}{4\pi}\right)^2
    \left(\frac{\mu^2}{M^2}\right)^{2\eps} S_\eps^2
    \left\{ -\frac{1}{4\eps} - \frac{3}{8} \right\}
  + \Oc(\eps) \,,
\\
\label{eq:FnsMrenerg}
  \Delta F_{n_s}^M &= \frac{n_s}{n_f} \, \frac{1}{4(1-\eps)} \,
    \Delta F_{n_f}^M
  \nonumber \\
  &= C_F T_F n_s \left(\frac{\alpha}{4\pi}\right)^2
    \left(\frac{\mu^2}{M^2}\right)^{2\eps} S_\eps^2
    \left\{ \frac{1}{\eps} \left[ \frac{2}{3} \lqm - 1 \right]
      + \frac{16}{9} \lqm + \frac{\pi^2}{3} - \frac{25}{6}
    \right\}
  \nonumber \\* & \quad
    + \Oc(\eps) + \Oc\left(\frac{M^2}{Q^2}\right) .
\end{align}

Da die \MSbar-Renormierung nur den $\eps$-Pol ohne endliche Konstanten (mit
Ausnahme von $\ln(4\pi)$ und $\gamma_E$) entfernt, besteht zwischen den
Beitr"agen der Fermionschleife und der skalaren Schleife in der
Kopplungskonstantenrenormierung ein Faktor von $4\,n_f/n_s$, wie auch an
$\beta_0$ in Gl.~(\ref{eq:beta0}) zu sehen ist:
\begin{align}
\label{eq:Fnsalpharenerg}
  \Delta F_{n_s}^\alpha &= \frac{n_s}{n_f} \, \frac{1}{4} \,
    \Delta F_{n_f}^\alpha
  \nonumber \\*
  &= C_F T_F n_s \left(\frac{\alpha}{4\pi}\right)^2
    \left(\frac{\mu^2}{M^2}\right)^\eps S_\eps \,
    \biggl\{
    \frac{1}{\eps} \left[
      -\frac{1}{3} \lqm^2 + \lqm - \frac{2}{9}\pi^2 - \frac{7}{6} \right]
    + \frac{1}{9} \lqm^3 - \frac{1}{2} \lqm^2
\nonumber \\* & \qquad
      + \left(-\frac{\pi^2}{9} + \frac{8}{3}\right) \lqm
      + \frac{2}{3}\zeta_3 + \frac{\pi^2}{6} - \frac{47}{12}
    \biggr\}
    + \Oc(\eps) + \Oc\left(\frac{M^2}{Q^2}\right) .
\end{align}

In der Summe dieser vier Beitr"age ergibt sich der Formfaktor mit der
skalaren Schleife in $d=4$ Dimensionen:
\begin{multline}
\label{F2nserg}
  F_{2,n_s} =
  C_F T_F n_s \left(\frac{\alpha}{4\pi}\right)^2 \, \biggl\{
    -\frac{1}{9} \lqm^3
    + \frac{25}{18} \lqm^2
    - \frac{23}{6} \lqm
    + \frac{10}{27}\pi^2 + \frac{157}{36}
  \\*
    + \ln\left(\frac{\mu^2}{M^2}\right) \left[
      \frac{1}{3} \lqm^2 - \lqm + \frac{2}{9}\pi^2 + \frac{7}{6} \right]
    \biggr\}
    + \Oc\left(\frac{M^2}{Q^2}\right) .
\end{multline}
Im speziellen Fall, wenn f"ur $\alpha$ die Renormierungsskala $\mu=M$
gew"ahlt wird, fallen die Beitr"age der zweiten Zeile weg.
Dieses Ergebnis best"atigt die Vorhersage der Evolutionsgleichung
in~(\ref{eq:F2NNLLSUN}).
Der $n_s$-Beitrag zu $\xi^{(2)}$,
\begin{equation}
  \xi^{(2)}\Big|_{n_s} =
  C_F T_F n_s \left( \frac{\pi^2}{9} + \frac{52}{27} \right) ,
\end{equation}
gilt jedoch -- genau wie der Koeffizient des linearen Logarithmus
und die Konstante in~(\ref{F2nserg}) -- nur f"ur masselose skalare
Teilchen, nicht f"ur eine Theorie mit spontaner Symmetriebrechung und einem
massiven Higgs-Teilchen.


\clearemptypage

\chapter{Abelsche Beitr"age zum Zweischleifenformfaktor}
\label{chap:abelsch}

Der $SU(2)$-Formfaktor des abelschen Vektorstroms
wurde in Abschnitt~\ref{sec:Sudakov} eingef"uhrt.
In diesem Kapitel werden die abelschen Korrekturen zu diesem
Formfaktor betrachtet.
Das sind Beitr"age, die auch in einer abelschen $U(1)$-Eichtheorie
vorkommen. Deshalb k"onnen sie zun"achst separat von den "ubrigen,
nichtabelschen Beitr"agen betrachtet werden.
Im allgemeineren Fall eines $SU(N)$-Modells besitzen die entsprechenden
Feynman-Diagramme den
Farbfaktor $C_F^2$ oder $C_F^2 - \frac{1}{2} C_F C_A$,
der sich aus den $SU(N)$-Generatoren~$t^a$
entsprechend Gl.~(\ref{eq:Casimir}) ergibt.
F"ur dieses Kapitel sind nur die Anteile proportional zu $C_F^2$ relevant,
die Terme mit $C_F C_A$ werden zu den nichtabelschen Beitr"agen in
Kapitel~\ref{chap:nichtabelsch} gerechnet.

Die Resultate dieses Kapitels wurden in \cite{Feucht:2004rp} ver"offentlicht.
Eine separate Ver"offentlichung von Details aus den Rechnungen in
Zusammenarbeit mit V.A.~Smirnov ist geplant.

Da der Formfaktor die Streuung eines Fermions im externen abelschen
Feld beschreibt, kann die Kopplungskonstante des zentralen Vertex als
externe Gr"o"se betrachtet werden, die nicht zum Formfaktor geh"ort und
nicht mit diesem renormiert wird.

\begin{figure}[pht]
  \centering
  \valignbox[b]{\includegraphics{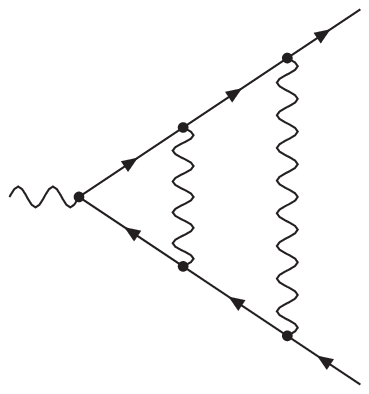}\\planar (LA)}
  \hspace{1.5cm}
  \valignbox[b]{\includegraphics{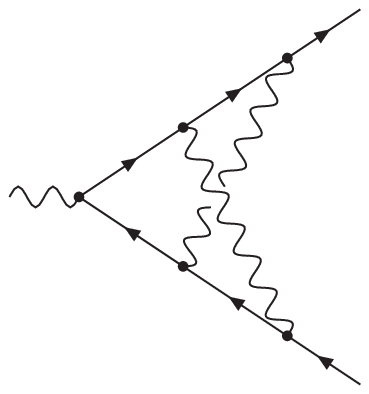}\\nichtplanar (NP)}
  \\[2ex]
  \valignbox[b]{\includegraphics{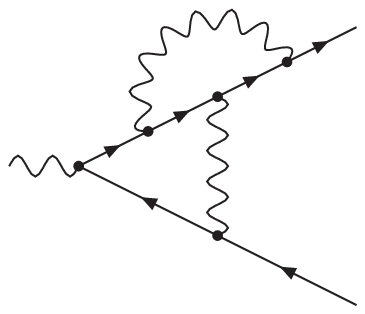}\\Benz-Topologie (BE, $2{\times}$)}
  \hspace{1.5cm}
  \valignbox[b]{\includegraphics{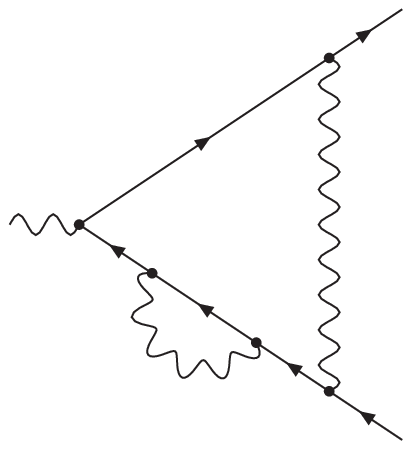}\\mit Fermion-Selbstenergie
    (fc, $2{\times}$)}
  \\[2ex]
  \caption{Abelsche Zweischleifen-Vertexdiagramme}
  \label{fig:abelschvertex}
\end{figure}
\begin{figure}[pht]
  \centering
  \valignbox[b]{\includegraphics{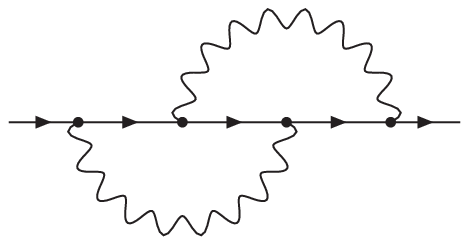}\\T1-Topologie}
  \hspace{1.5cm}
  \valignbox[b]{\includegraphics{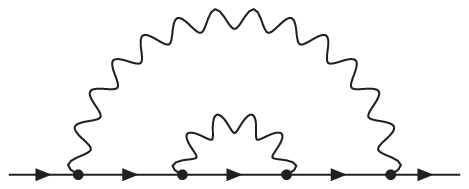}\\T2-Topologie}
  \\[2ex]
  \valignbox[b]{\includegraphics{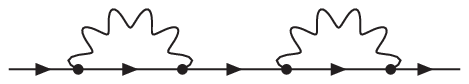}\\
    (Einschleifen-Selbstenergie)$^2$}
  \\[2ex]
  \caption{Abelsche Zweischleifen-Selbstenergiediagramme}
  \label{fig:abelschfcorr}
\end{figure}
F"ur den abelschen Zweischleifenformfaktor m"ussen
die Vertexdiagramme in Abb.~\ref{fig:abelschvertex} betrachtet werden.
Deren Berechnung wird in Abschnitt~\ref{sec:abelschvertex} vorgestellt.
Zwei Diagramme (BE und fc) kommen jeweils zweifach vor: so wie abgebildet
und horizontal gespiegelt.

F"ur die Renormierung der Fermion-Feldst"arke sind au"serdem die
Zwei"-schleifen-Selbstenergiediagramme in Abb.~\ref{fig:abelschfcorr}
erforderlich, die somit ebenfalls
zum abelschen Formfaktor beitragen.
Sie werden in Abschnitt~\ref{sec:abelschfcorr} vorgestellt,
zusammen mit weiteren Beitr"agen zur Feldst"arkerenormierung aus
Einschleifenkorrekturen.

Die abelschen Diagramme enthalten alle nur zwei massive Propagatoren.
Deshalb war ihre Berechnung im Vergleich zu den
nichtabelschen Beitr"agen mit drei massiven Propagatoren (siehe
Kapitel~\ref{chap:nichtabelsch}) noch relativ einfach und konnte im
Hochenergielimes vollst"andig durchgef"uhrt werden, d.h. alle Terme der
logarithmischen Reihe inklusive der Konstanten wurden ermittelt.


%
%
\section{Vertexdiagramme}
\label{sec:abelschvertex}

Die Amplitude der Vertexkorrekturen hat die Form
\begin{equation}
\label{eq:formfaktorv}
  \bar u(p_1) \, \Fc_v^\mu \, v(-p_2) =
  F_v \cdot \bar u(p_1) \, \gamma^\mu \, v(-p_2)
  \,.
\end{equation}
Die Vertexamplitude~$\Fc_v^\mu$ wurde ohne die Spinoren der beiden
Fermionen definiert.
Es gilt $p_1^2 = p_2^2 = 0$, $q = p_1 - p_2$
und $Q^2 = -q^2 = 2 p_1 \cdot p_2$.
Eigentlich ist $\Fc_v^\mu$ eine \mbox{Matrix} nicht nur im Spinorraum, sondern
auch im Isospinraum der
Fermion-Dubletts. Aber da der zentrale Vertex keinen $SU(2)$-Generator
enth"alt (vgl. die Definition des Formfaktors in
Abschnitt~\ref{sec:Sudakov}), sind auch alle Korrekturen proportional zur
Einheitsmatrix im Isospinraum.

Unter Ausnutzung der Antivertauschungsrelationen f"ur Dirac-Matrizen,
\[
  \{ \gamma^\mu, \gamma^\nu \} = 2 g^{\mu\nu}
  \,,
\]
und der kinematischen Eigenschaften der Spinoren von masselosen Fermionen,
\[
  \bar u(p_1) \, \dslash p_1 = 0
  \,,\quad
  \dslash p_2 \, v(-p_2) = 0
  \,,
\]
kann die Amplitude eines Vertexdiagramms auf die Form~(\ref{eq:formfaktorv})
mit dem Beitrag $F_v$ zum Formfaktor gebracht werden.
Dazu ist jedoch immer auch eine Tensorreduktion der Integrale mit offenen
Lorentz-Indizes auf skalare Integrale notwendig.
Eleganter ist die Extraktion des Formfaktors aus der Vertexamplitude mit einer
Projektionstechnik (siehe z.B. \cite{Bernreuther:2004ih}):
\begin{equation}
\label{eq:projformfaktor}
  F_v =
    \frac{\Tr(\gamma_\mu \dslash p_1 \Fc_v^\mu \dslash p_2)}
      {2 (d-2) \, q^2}
  \,.
\end{equation}
Dabei bezeichnet $d = 4-2\eps$ die Zahl der Raum-Zeit-Dimensionen.
Die Einheitsmatrix im $SU(2)$-Isospinraum wurde hier nicht
ber"ucksichtigt. Wenn die Spur auch dar"uber l"auft, muss zus"atzlich durch
$\Tr\unity = N = 2$ geteilt werden.

Die Projektion~(\ref{eq:projformfaktor}) kann auf einzelne Vertexdiagramme
angewendet werden, um deren Beitrag zum Formfaktor zu bestimmen.
Das Ergebnis der Projektion sind skalare Schleifen\-integrale, die neben
Propagatoren im Nenner lediglich Skalarprodukte im Z"ahler enthalten.

%
%
\subsection{Planares Vertexdiagramm}
\label{sec:LA}

\begin{figure}[ht]
  \center
  \includegraphics{vertex-LA}
  \hspace{1cm}
  \includegraphics{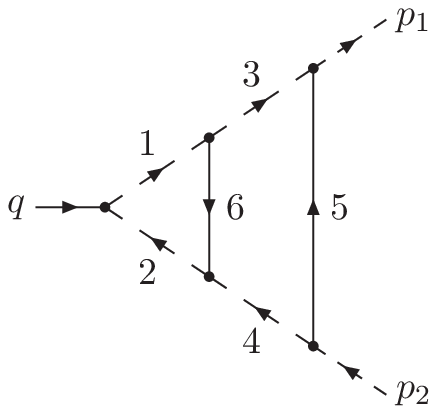}
  \caption{Planares Vertexdiagramm (mit skalarem Diagramm)}
  \label{fig:LA}
\end{figure}

Das planare Vertexdiagramm (Bezeichnung LA = ladder)
ist in Abb.~\ref{fig:LA} dargestellt,
links das Feynman-Diagramm mit Pfeilen f"ur Fermionen und Wellenlinien
f"ur Eichbosonen,
rechts das zugeh"orige skalare Diagramm mit durchgezogenen Linien f"ur
massive Propagatoren und gestrichelten Linien f"ur masselose
Propagatoren.

Im skalaren Diagramm sind die inneren Impulse $k_i$, $i=1,\ldots,6$,
nur durch die Angabe des Index~$i$ gekennzeichnet,
die Pfeilrichtung gibt hier die Richtung des Impulsflusses an.

In der Feynman-'t~Hooft-Eichung ($\xi=1$) lautet die Amplitude des planaren
Vertexdiagramms (ohne die Spinoren der externen Fermionen):
\begin{multline}
\label{eq:LAamplitude}
  \Fc_{v,\LA}^\mu =
  \mu^{4\eps} \loopint dk \loopint d\ell \,
  ig \gamma^\rho t^b \, \frac{i\dslash k_3}{k_3^2} \,
  ig \gamma^\nu t^a \, \frac{i\dslash k_1}{k_1^2} \,
  \gamma^\mu
  \frac{i\dslash k_2}{k_2^2} \, ig \gamma_\nu t^a \,
  \frac{i\dslash k_4}{k_4^2} \, ig \gamma_\rho t^b
  \\* \times
  \frac{-i}{k_5^2-M^2} \, \frac{-i}{k_6^2-M^2}
  \,,
\end{multline}
mit den Impulsen
\begin{equation}
\label{eq:LAmom}
  k_1 = p_1 - \ell ,\quad k_2 = p_2 - \ell ,\quad
  k_3 = p_1 - k ,\quad k_4 = p_2 - k ,\quad
  k_5 = k ,\quad k_6 = k - \ell .
\end{equation}

Der Farbfaktor des planaren Vertexdiagramms lautet
\[
  t^b t^a t^a t^b = C_F^2 \, \unity
  \,.
\]
Aus der Amplitude~$\Fc_{v,\LA}^\mu$ kann der Beitrag~$F_{v,\LA}$ zum
Formfaktor durch Projektion entsprechend~(\ref{eq:projformfaktor})
extrahiert werden.
Anschlie"send werden die Skalarprodukte im Z"ahler des
Schleifenintegrals mit Propagatornennern gek"urzt.
Daf"ur sind folgende Umformungen hilfreich:
\begin{equation}
\label{eq:LAredskalar}
  \begin{aligned}
    p_1 \cdot k_5 &= -\tfrac12 (k_3^2 - k_5^2) \,,\quad &
    k_3 \cdot k_6 &= +\tfrac12 (k_1^2 - k_3^2 - k_6^2) \,, \\
    p_2 \cdot k_5 &= -\tfrac12 (k_4^2 - k_5^2) \,,\quad &
    k_4 \cdot k_6 &= +\tfrac12 (k_2^2 - k_4^2 - k_6^2) \,.
  \end{aligned}
\end{equation}
Mit Hilfe von (\ref{eq:LAmom}) und (\ref{eq:LAredskalar}) k"onnen alle
Skalarprodukte gek"urzt werden mit Ausnahme eines sogennanten
\emph{irreduziblen} Skalarprodukts, das als
$2k \cdot \ell$ gew"ahlt wurde.
Die skalaren Zweischleifenintegrale der planaren Topologie wurden
folgenderma"sen definiert:
\begin{multline}
\label{eq:LAskalar}
  F_\LA(n_1,n_2,n_3,n_4,n_5,n_6,n_7) =
  e^{2\eps\gamma_E} \, (M^2)^{2\eps} \,
  (Q^2)^{n-n_7-4}
  \\*
  \times \loopintf dk \loopintf d\ell \,
  \frac{(2k \cdot \ell)^{n_7}}
    {(\ell^2 - 2p_1\cdot\ell)^{n_1} \, (\ell^2 - 2p_2\cdot\ell)^{n_2} \,
    (k^2 - 2p_1\cdot k)^{n_3} \, (k^2 - 2p_2\cdot k)^{n_4}}
  \\*
  \times \frac{1}{(k^2 - M^2)^{n_5} \, ((k-\ell)^2 - M^2)^{n_6}}
  \,,
\end{multline}
mit der Notation $n_{ij\ldots} = n_i + n_j + \ldots$ und
$n = n_{123456}$ (ohne $n_7$).
Die skalaren Integrale wurden so definiert, dass sie in $d=4-2\eps$
Raum-Zeit-Dimensionen keine Massendimension aufweisen.
Damit kann $F_{v,\LA}$ als Linearkombination von skalaren
Integralen~(\ref{eq:LAskalar}) geschrieben werden.
Diese Zerlegung findet sich im Anhang~\ref{sec:SkalarLA}
in Gl.~(\ref{eq:LAzerlegung}).
Da das Feynman-Diagramm jeden Propagator nur einmal enth"alt und bei der
Reduktion auf skalare Integrale Propagatoren gek"urzt werden, aber keine
hinzukommen, sind die Parameter $n_i$ ($i=1,\ldots,6$) nach oben
durch $n_i\le1$ beschr"ankt.
In der Zerlegung des planaren Diagramms kommen die Werte 1, 0 und -1 f"ur
die $n_i$ vor.
Durch Ausnutzung der Symmetrie
\begin{equation} 
  F_\LA(n_1,n_2,n_3,n_4,n_5,n_6,n_7) = F_\LA(n_2,n_1,n_4,n_3,n_5,n_6,n_7)
\end{equation}
wurden skalare Integrale mit $n_1<n_2$ oder $n_1=n_2$ und $n_3<n_4$
durch ihre symmetrischen Partner ersetzt.
Die Auswertung der Spur "uber die Amplitude zur Projektion auf den
Formfaktor, das K"urzen der Skalarprodukte und die Zerlegung in skalare
Integrale wurden mit dem Computeralgebraprogramm
FORM\cite{Vermaseren:2000nd} durchgef"uhrt.

Die Auswertung der skalaren Integrale erfolgte mit der Methode der
\emph{Expansion by Regions} (siehe Anhang~\ref{sec:ExpReg}).
Nichtverschwindende Beitr"age zur skalaren Topologie liefern die
folgenden Regionen~\cite{Smirnov:2002pj}:
\[
  \renewcommand{\arraystretch}{1.2}
  \left.
  \begin{array}{rll}
    \text{(h-h):} & k \sim Q, & \ell \sim Q \\
    \text{(1c-h):} & k \;\|\; p_1, & \ell \sim Q \\
    \text{(2c-h):} & k \;\|\; p_2, & \ell \sim Q
  \end{array}
  \qquad
  \right|
  \qquad
  \begin{array}{rll}
    \text{(1c-1c):} & k \;\|\; p_1, & \ell \;\|\; p_1 \\
    \text{(2c-2c):} & k \;\|\; p_2, & \ell \;\|\; p_2 \\
    \text{(h-s'):} & k \sim Q, & k_6 = k-\ell \sim M
  \end{array}
\]
Dabei bedeutet $k \sim Q$, dass jede Komponente des Vektors~$k$ von der
Gr"o"senordnung~$Q$ ist.
Und $k \;\|\; p_i$ bezeichnet eine Region, in welcher der Impuls $k$
kollinear zum externen Impuls~$p_i$ steht (analog auch f"ur $\ell$):
\begin{align*}
  k \;\|\; p_1 &\iff k_+ \sim \frac{M^2}{Q} \,,\quad
    k_- \sim Q \,,\quad k_\bot \sim M
    \,, \\
  k \;\|\; p_2 &\iff k_+ \sim Q \,,\quad
    k_- \sim \frac{M^2}{Q} \,,\quad k_\bot \sim M
  \,,
\end{align*}
wobei $k_\pm \equiv (2p_{1,2}\cdot k)/Q$ die Komponenten von $k$ in
Richtung von $p_2$ bzw. $p_1$ sind
und $k_\bot \equiv k - (k_-/Q)p_1 - (k_+/Q)p_2$ den Vektor mit den Komponenten
von $k$ bezeichnet, die senkrecht auf $p_{1,2}$ stehen.
Damit lassen sich die Skalarprodukte der Schleifenimpulse als
$k^2 = k_+ k_- + k_\bot^2$ und
$k\cdot\ell = \frac12(k_+ \ell_- + k_- \ell_+) + k_\bot\cdot\ell_\bot$
schreiben.

Der f"uhrende Beitrag jeder Region hat eine charakteristische Abh"angigkeit
vom Parameter~$M^2/Q^2$.
Im Folgenden wird f"ur eine Region~(x-y) mit der Abh"angigkeit
$F_\LA^{\text{(x-y)}} \propto (M^2/Q^2)^p$
die Potenz~$p$ angegeben:
\[
  \left.
  \begin{array}{rl}
    \text{(h-h):} & 2\eps \\
    \text{(1c-h):} & 2-n_{35}+\eps \\
    \text{(2c-h):} & 2-n_{45}+\eps
  \end{array}
  \qquad
  \right|
  \qquad
  \begin{array}{rl}
    \text{(1c-1c):} & 4-n_{1356}+n_7 \\
    \text{(2c-2c):} & 4-n_{2456}+n_7 \\
    \text{(h-s'):} & 2-n_6+\eps
  \end{array}
\]
Da Beitr"age, die gegen"uber anderen mit dem kleinen Faktor
$M^2/Q^2$ unterdr"uckt sind, vernachl"assigt werden sollen,
wird die (h-s')-Region nicht ben"otigt.
Alle Propagatoren kommen in den skalaren Integralen h"ochstens einmal
vor, so dass $n_i \le 1$ f"ur $i=1,\ldots,6$ ist.
Im physikalischen Limes $d \to 4$ ($\eps \to 0$)
ist $F_\LA^{\text{(h-s')}}$ daher
mit mindestens einem Faktor~$M^2/Q^2$ gegen"uber der (h-h)-Region
unterdr"uckt.
Ebenso gilt, dass die rein kollinearen (c-c)-Regionen, also (1c-1c) und
(2c-2c), nur f"ur $n_7 = 0$ ber"ucksichtigt werden m"ussen. Bei Integralen mit
dem irreduziblen Skalarprodukt im Z"ahler ($n_7 > 0$) sind beide (c-c)-Regionen
mit mindestens einem Faktor~$M^2/Q^2$ unterdr"uckt.
Im Allgemeinen pr"uft man f"ur jedes ben"otigte skalare Integral zuerst,
welche Regionen den f"uhrenden Beitrag liefern, und kann sich dann auf die
Auswertung dieser Regionen beschr"anken.

Die harte Region (h-h), die dem masselosen Diagramm entspricht,
l"asst sich am einfachsten mit der Methode der partiellen Integration
(Integration by Parts \cite{Tkachov:1981wb,Chetyrkin:1981qh})
auf einfachere Topologien reduzieren, die alle
mit Feynman-Parametern ausgewertet werden k"onnen
und lediglich $\Gamma$-Funktionen liefern.
Mit den Beitr"agen der harten Region l"asst sich das Ergebnis des
masselosen planaren Feynman-Diagramms angeben:
\begin{multline}
  F_{v,\LA}^{\text{(h-h)}} =
    C_F^2
    \left(\frac{\alpha}{4\pi}\right)^2
    \left(\frac{\mu^2}{Q^2}\right)^{2\eps} S_\eps^2
    \, \Biggl\{
  \frac{1}{\eps^4}
  + \frac{2}{\eps^3}
  + \frac{1}{\eps^2} \left( \frac{\pi^2}{6} + \frac{17}{2} \right)
  + \frac{1}{\eps} \left( \frac{46}{3}\zeta_3 - \frac{\pi^2}{3}
    + \frac{101}{4} \right)
  \\*
  + \frac{103}{360}\pi^4 + \frac{152}{3}\zeta_3 - \frac{35}{12}\pi^2
    + \frac{631}{8}
  \Biggr\}
  + \Oc(\eps)
  \,,
\end{multline}
mit $S_\eps = (4\pi)^\eps \, e^{-\eps\gamma_E}$.
Dieses Ergebnis stimmt mit \cite{Kramer:1986sg,Matsuura:1988sm} "uberein.

Die (c-h)- und (c-c)-Regionen wurden dagegen mit Hilfe von
Schwinger-Parametern (siehe Anhang~\ref{sec:loopparam})
und Mellin-Barnes-Darstellungen (siehe Anhang~\ref{sec:MB})
berechnet.
Ihre Beitr"age lassen sich durch ein- bzw. zweifache
Mellin-Barnes-Integrale ausdr"ucken:
\begin{align}
\label{eq:FLA1chgen}
  \lefteqn{F_\LA^{\text{(1c-h)}}(n_1,n_2,n_3,n_4,n_5,n_6,n_7) =} 
  \nonumber \\* &
  \left(\frac{M^2}{Q^2}\right)^{2-n_{35}+\eps}
  e^{-n i\pi} \, e^{2\eps\gamma_E} \,
  \frac{\Gamma(\frac d2-n_3) \Gamma(\frac d2-n_{16}+n_7)
    \Gamma(n_{35}-\frac d2)}{
    \Gamma(n_1) \Gamma(n_2) \Gamma(n_3) \Gamma(n_5) \Gamma(n_6)
    \Gamma(d-n_{126}+n_7)}
  \nonumber \\* & \times
  \MBint z \,
    \frac{\Gamma(-z) \Gamma(\frac d2-n_{26}-z)
      \Gamma(n_6+z) \Gamma(n_{37}-n_4+z) \Gamma(n_{126}-\frac d2+z)}{
      \Gamma(\frac d2-n_4+n_7+z)}
  \,,
\\
\label{eq:FLA1c1cgen}
  \lefteqn{F_\LA^{\text{(1c-1c)}}(n_1,n_2,n_3,n_4,n_5,n_6,0) =
    \left(\frac{M^2}{Q^2}\right)^{4-n_{1356}}
    \frac{e^{-n i\pi} \, e^{2\eps\gamma_E}}{
    \Gamma(n_1) \Gamma(n_3) \Gamma(n_5) \Gamma(n_6) \Gamma(\frac d2-n_{24})}
  } 
  \nonumber \\ & \times
  \MBint{z_1} \MBint{z_2} \,
  \frac{\Gamma(-z_1) \Gamma(n_{13}-\frac d2-z_1)
    \Gamma(\frac d2-n_1+z_1) \Gamma(\frac d2-n_{24}+z_1)}{
    \Gamma(\frac d2-n_4+z_1)}
  \qquad\qquad
  \nonumber \\* & \times
  \frac{\Gamma(-z_2) \Gamma(\frac d2-n_{35}-z_2) \Gamma(\frac d2-n_{45}-z_2)
    \Gamma(n_{1356}-d+z_2)}{
    \Gamma(\frac d2-n_5-z_2)} \,
  \Gamma(n_5+z_1+z_2)
  \,.
\end{align}

Der Integrationsweg der Mellin-Barnes-Integrale verl"auft so von $-i\infty$
nach $+i\infty$, dass Pole aus $\Gamma$-Funktionen der Form
$\Gamma(\ldots+z)$ links des Integrationswegs liegen ("`IR-Pole"')
und Pole der Form $\Gamma(\ldots-z)$ rechts des Integrationswegs liegen
("`UV-Pole"').
Aus Symmetriegr"unden gilt:
\begin{align}
  F_\LA^{\text{(2c-h)}}(n_1,n_2,n_3,n_4,n_5,n_6,n_7) &=
    F_\LA^{\text{(1c-h)}}(n_2,n_1,n_4,n_3,n_5,n_6,n_7) \,, \\
  F_\LA^{\text{(2c-2c)}}(n_1,n_2,n_3,n_4,n_5,n_6,0) &=
    F_\LA^{\text{(1c-1c)}}(n_2,n_1,n_4,n_3,n_5,n_6,0) \,.
\end{align}

Bei der Auswertung der Mellin-Barnes-Integrale ist zu beachten, dass sowohl
der Z"ahler als auch der Nenner an manchen Stellen des Parameterraums
Singularit"aten aufweisen, die sich teilweise gegenseitig aufheben.

In Gl.~(\ref{eq:FLA1chgen}) f"ur $F_\LA^{\text{(1c-h)}}$
ist das Integral singul"ar bei $n_6=0$, weil dann der IR-Pol bei $z=-n_6$
und der UV-Pol bei $z=0$, zwischen denen der Integrationsweg verl"auft,
auf einen Punkt zusammenr"ucken.
Diese Singularit"at wird durch die entsprechende
Singularit"at von $\Gamma(n_6)$ im Nenner des Vorfaktors gek"urzt.
Das Ergebnis ist hier durch den Limes $n_6 \to 0$ gegeben, zu dem nur das
Residuum des Integranden bei $z=-n_6$ bzw. $z=0$ beitr"agt.
Entsprechendes gilt f"ur ganzzahlige Werte $n_6 \le 0$ und $n_1 \le 0$.

Etwas anders ist die Situation bei $n_3+n_7 \le n_4$.  Hier wird die
Singularit"at im Integral nur f"ur $n_3 \le 0$ durch den Nenner
aufgehoben (wenn dieser Beitrag sowieso durch $M^2/Q^2$ unterdr"uckt ist).
Im Fall $n_3=n_4=1$, $n_7=0$ bleibt die Singularit"at als Pol
$1/(n_3-n_4)$ bestehen. Sie wird durch die entsprechende Singularit"at mit
umgekehrtem Vorzeichen in $F_\LA^{\text{(2c-h)}}$ aufgeboben.
Kollineare Regionen weisen h"aufig solche zus"atzlichen
Singularit"aten auf, die
analytisch regularisiert werden k"onnen, d.h. die Parameter~$n_i$ werden
als komplexe Parameter nahe dem ben"otigten reellen Wert verstanden.
Die Summe mehrerer oder aller kollinearen Regionen ist endlich im Rahmen
der dimensionalen Regularisierung.

Der einzige singul"are Beitrag von Gl.~(\ref{eq:FLA1c1cgen}) f"ur
$F_\LA^{\text{1c-1c}}$, der nicht durch $M^2/Q^2$ unterdr"uckt ist,
tritt bei $n_1=n_2=n_3=n_4=n_5=n_6=1$ auf.
Er "au"sert sich in einem Pol $1/(n_{13}-n_{24})$, der durch einen
entsprechenden Pol mit umgekehrtem Vorzeichen in $F_\LA^{\text{2c-2c}}$
kompensiert wird.

Die Auswertung von denjenigen Mellin-Barnes-Integralen,
bei denen nicht nur einzelne
Residuen ben"otigt werden, erfolgt durch Aufsummation aller Residuen
auf einer Seite des Integrationswegs.
In einfachen F"allen von Gl.~(\ref{eq:FLA1c1cgen}),
bei denen sich $\Gamma$"~Funktionen k"urzen,
kann f"ur die erste Integration das erste Barnsche
Lemma~(\ref{eq:BarnesLemma1}) angewandt werden (siehe Anhang~\ref{sec:MB}).
In den anderen F"allen werden die Residuen, die Singularit"aten im
Parameterraum der $n_i$ oder in $\eps$ produzieren, separat betrachtet.
F"ur die restlichen Residuen k"onnen alle Limites vollzogen werden
(inklusive $\eps\to0$).
Dadurch vereinfachen sich die Summen "uber die Residuen,
so dass diese von \textsc{Mathematica}\cite{Wolfram:Mathematica4.2}
gel"ost oder in einer Summationstabelle (z.B.~\cite{Smirnov:2004ym})
nachgeschlagen werden k"onnen.

Die Ergebnisse der einzelnen skalaren Integrale~$F_\LA$ finden sich im
Anhang~\ref{sec:SkalarLA} in Gl.~(\ref{eq:LAergskalar}).
Der gesamte Beitrag des planaren Vertexdiagramms zum Formfaktor lautet
in f"uhrender Ordnung in $M^2/Q^2$:
\begin{align}
\label{eq:LAerg}
  F_{v,\LA} &=
    C_F^2
    \left(\frac{\alpha}{4\pi}\right)^2
    \left(\frac{\mu^2}{M^2}\right)^{2\eps} S_\eps^2
    \, \Biggl\{
  \frac{1}{2\eps^2}
  + \frac{1}{\eps} \left[
    - \lqm^2 + 3 \lqm - \frac{2}{3}\pi^2 - \frac{11}{4} \right]
  \nonumber \\* & \qquad
  + \frac{1}{6} \lqm^4
  + \left(\frac{2}{3}\pi^2 - 1\right) \lqm^2
  + \left(-32\zeta_3 - \pi^2 + \frac{11}{2}\right) \lqm
  \nonumber \\* & \qquad
  + \frac{8}{15}\pi^4 + 62\zeta_3 + \frac{13}{12}\pi^2 - \frac{41}{8}
  \Biggr\}
  + \Oc(\eps) + \Oc\!\left(\frac{M^2}{Q^2}\right)
  ,
\end{align}
mit $\lqm = \ln(Q^2/M^2)$.
Das Vertexdiagramm mit massiven Eichbosonen enth"alt keine infraroten
Singularit"aten, deshalb ist der f"uhrende Pol in~$\eps$
lediglich eine zweifache ultraviolette Singularit"at der
Ordnung~$\eps^{-2}$.

%
%
\subsection{Nichtplanares Vertexdiagramm}
\label{sec:NP}

\begin{figure}[ht]
  \center
  \includegraphics{vertex-NP}
  \hspace{1cm}
  \includegraphics{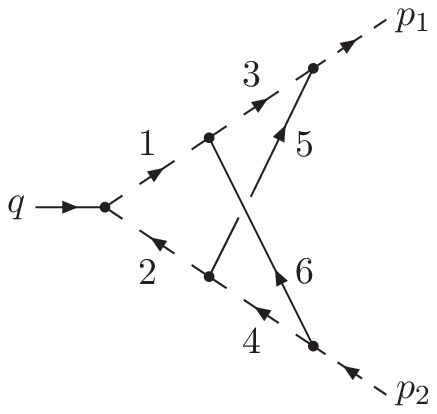}
  \caption{Nichtplanares Vertexdiagramm (mit skalarem Diagramm)}
  \label{fig:NP}
\end{figure}

Das nichtplanare Vertexdiagramm (Bezeichnung NP)
ist in Abb.~\ref{fig:NP} dargestellt, links das Feynman-Diagramm, rechts
das skalare Diagramm mit Impulsbezeichnungen.
Die Notationen entsprechen denen aus Abschnitt~\ref{sec:LA}.

Die Amplitude des nichtplanaren Vertexdiagramms lautet:
\begin{multline}
\label{eq:NPamplitude}
  \Fc_{v,\NP}^\mu =
  \mu^{4\eps} \loopint dk \loopint d\ell \,
  ig \gamma^\rho t^b \, \frac{i\dslash k_3}{k_3^2} \,
  ig \gamma^\nu t^a \, \frac{i\dslash k_1}{k_1^2} \,
  \gamma^\mu
  \frac{i\dslash k_2}{k_2^2} \, ig \gamma_\rho t^b \,
  \frac{i\dslash k_4}{k_4^2} \, ig \gamma_\nu t^a
  \\* \times
  \frac{-i}{k_5^2-M^2} \, \frac{-i}{k_6^2-M^2}
  \,,
\end{multline}
mit den Impulsen
\begin{equation}
\label{eq:NPmom}
  k_1 = p_1 - k - \ell ,\quad k_2 = p_2 - k - \ell ,\quad
  k_3 = p_1 - k ,\quad k_4 = p_2 - \ell ,\quad
  k_5 = k ,\quad k_6 = \ell .
\end{equation}
Der Farbfaktor des nichtplanaren Vertexdiagramms ergibt sich zu
\[
  t^b t^a t^b t^a = \left( C_F^2 - \frac{1}{2} C_F C_A \right) \unity
  \,.
\]

Der Beitrag der Amplitude~$\Fc_{v,\NP}^\mu$ zum Formfaktor ist
$F_{v,\NP}$.
Zum K"urzen von Skalarprodukten im Z"ahler von~$F_{v,\NP}$ mit
Propagatornennern dienen die Umformungen
\begin{equation}
\label{eq:NPredskalar}
  \begin{aligned}
    p_1 \cdot k_5 &= -\tfrac12 (k_3^2 - k_5^2) \,,\quad &
    k_3 \cdot k_6 &= -\tfrac12 (k_1^2 - k_3^2 - k_6^2) \,, \\
    p_2 \cdot k_6 &= -\tfrac12 (k_4^2 - k_6^2) \,,\quad &
    k_4 \cdot k_5 &= -\tfrac12 (k_2^2 - k_4^2 - k_5^2) \,.
  \end{aligned}
\end{equation}

Als irreduzibles Skalarprodukt bleibt $2k_5\cdot k_6 = 2k\cdot\ell$
"ubrig.
Damit lautet die Definition der skalaren Zweischleifenintegrale der
nichtplanaren Topologie:
\begin{multline}
\label{eq:NPskalar}
  F_\NP(n_1,n_2,n_3,n_4,n_5,n_6,n_7) =
  e^{2\eps\gamma_E} \, (M^2)^{2\eps} \,
  (Q^2)^{n-n_7-4}
  \\*
  \times \loopintf dk \loopintf d\ell \,
  \frac{(2k \cdot \ell)^{n_7}}
    {((k+\ell)^2-2p_1\cdot(k+\ell))^{n_1} \,
      ((k+\ell)^2-2p_2\cdot(k+\ell))^{n_2}}
  \\*
  \times \frac{1}{(k^2 - 2p_1\cdot k)^{n_3} \, (\ell^2 - 2p_2\cdot \ell)^{n_4}
    \, (k^2 - M^2)^{n_5} \, (\ell^2 - M^2)^{n_6}}
  \,,
\end{multline}
mit der Notation aus Gl.~(\ref{eq:LAskalar})
und $n = n_{123456}$ (ohne $n_7$).
Die Zerlegung von~$F_{v,\NP}$ in skalare Integrale ist im
Anhang~\ref{sec:SkalarNP} in Gl.~(\ref{eq:NPzerlegung}) aufgef"uhrt.
Durch Ausnutzung der Symmetrie
\begin{equation}
\label{eq:NPsym}
  F_\NP(n_1,n_2,n_3,n_4,n_5,n_6,n_7) = F_\NP(n_2,n_1,n_4,n_3,n_6,n_5,n_7)
\end{equation}
wurden skalare Integrale mit $n_1<n_2$ oder $n_1=n_2$ und $n_3<n_4$
oder $n_1=n_2$ und $n_3=n_4$ und $n_5<n_6$
durch ihre symmetrischen Partner ersetzt.

Zur Auswertung der skalaren Integrale tragen im Rahmen der \emph{Expansion by
Regions} folgende Regionen bei\cite{Smirnov:2002pj}:
\[
  \renewcommand{\arraystretch}{1.2}
  \left.
  \begin{array}{rll}
    \text{(h-h):} & k \sim Q, & \ell \sim Q \\
    \text{(1c-h):} & k \;\|\; p_1, & \ell \sim Q \\
    \text{(h-2c):} & k \sim Q, & \ell \;\|\; p_2 \\
    \text{(1c-1c):} & k \;\|\; p_1, & \ell \;\|\; p_1 \\
    \text{(2c-2c):} & k \;\|\; p_2, & \ell \;\|\; p_2 \\
    \text{(1c-2c):} & k \;\|\; p_1, & \ell \;\|\; p_2
  \end{array}
  \qquad
  \right|
  \qquad
  \begin{array}{rll}
    \text{(1c-1c'):} & k \;\|\; p_1, & k_4 \;\|\; p_1 \\
    \text{(2c'-2c):} & k_3 \;\|\; p_2, & \ell \;\|\; p_2 \\
    \text{(us'-us'):} & k_3 \sim M^2/Q, & k_4 \sim M^2/Q \\
    \text{(1c-us'):} & k \;\|\; p_1, & k_4 \sim M^2/Q \\
    \text{(us'-2c):} & k_3 \sim M^2/Q, & \ell \;\|\; p_2
  \end{array}
\]
In der Regel bezeichnet der erste Eintrag im Namen der jeweiligen Region
den Schleifenimpuls~$k$, der zweite Eintrag den Schleifenimpuls~$\ell$.
Abweichend davon steht ein gestrichener Eintrag (z.B. 1c') f"ur den
Impuls $k_3$ bzw. $k_4$.
Es ist unerheblich, ob $k_5=k$ oder $k_3$ kollinear zu $p_1$ ist,
da mit dem einen Impuls auch der andere kollinear zu $p_1$ wird.
Es macht aber einen Unterschied, ob $k_5$ oder $k_3$ kollinear zu $p_2$
wird, wie in den Regionen (2c-2c) bzw. (2c'-2c).
Gleiches gilt umgekehrt f"ur die Kollinearit"at von $k_6=\ell$ oder $k_4$.

Der f"uhrende Beitrag der jeweiligen Region h"angt "uber die folgende
Potenz von $M^2/Q^2$ ab (Notation wie in Abschnitt~\ref{sec:LA}):
\[
  \left.
  \begin{array}{rl}
    \text{(h-h):} & 2\eps \\
    \text{(1c-h):} & 2-n_{35}+\eps \\
    \text{(h-2c):} & 2-n_{46}+\eps \\
    \text{(1c-1c):} & 4-n_{1356}+n_7 \\
    \text{(2c-2c):} & 4-n_{2456}+n_7 \\
    \text{(1c-2c):} & 4-n_{3456}
  \end{array}
  \qquad
  \right|
  \qquad
  \begin{array}{rl}
    \text{(1c-1c'):} & 4-n_{2345} \\
    \text{(2c'-2c):} & 4-n_{1346} \\
    \text{(us'-us'):} & 8-n_{1256}-2n_{34}-2\eps \\
    \text{(1c-us'):} & 6-n_{2356}-2n_4-\eps \\
    \text{(us'-2c):} & 6-n_{1456}-2n_3-\eps
  \end{array}
\]
Der Beitrag jeder Region au"ser (h-h) kann durch den Faktor $M^2/Q^2$
unterdr"uckt sein, wenn gewisse Parameter~$n_i$ nicht ihren Maximalwert
$n_i=1$ haben.
Au"serdem tragen die Regionen (1c-1c) und (2c-2c) nur f"ur $n_7=0$ bei\
und wurden nur f"ur diesen Fall berechnet.

Die Beitr"age der einzelnen Regionen wurden mit Hilfe von
Schwinger-Parametern und Mellin-Barnes-Transformationen berechnet.
F"ur die harte Region~(h-h) ergab sich im Spezialfall $n_7=0$ das folgende
zweifache Mellin-Barnes-Integral:
\begin{align}
  \lefteqn{F_\NP^{\text{(h-h)}}(n_1,n_2,n_3,n_4,n_5,n_6,0) = } \quad
  \nonumber \\* &
  \left(\frac{M^2}{Q^2}\right)^{2\eps}
  \frac{e^{-n i \pi} \, e^{2\eps\gamma_E} \,
    \Gamma(\frac d2-n_{35}) \Gamma(\frac d2-n_{46})}
    {\Gamma(n_1) \Gamma(n_2) \Gamma(n_3) \Gamma(n_4) \Gamma(n_5) \Gamma(n_6)
    \Gamma(d-n_{3456}) \Gamma(\frac32d-n_{123456})}
  \nonumber \\ & \times
  \MBint{z_1} \,
  \Gamma(d-n_{12345}-z_1) \Gamma(d-n_{12456}-z_1)
  \Gamma(n_1+z_1) \Gamma(n_{123456}-d+z_1)
  \nonumber \\* & \times
  \MBint{z_2} \,
  \Gamma(-z_2) \Gamma(d-n_{13456}-z_2)
  \Gamma(n_4+z_2) \Gamma(n_5+z_2)
  \nonumber \\* & \times
  \frac{\Gamma(-z_1+z_2) \Gamma(\frac d2-n_{12}-z_1+z_2)}
    {\Gamma(d-n_{1235}-z_1+z_2) \Gamma(d-n_{1246}-z_1+z_2)}
  \,.
\end{align}

Die Auswertung dieses zweifachen Mellin-Barnes-Integrals, das zudem nur die
skalaren Integrale mit $n_7=0$ abdeckt, w"are vergleichsweise aufw"andig
gewesen.
Deshalb wurden die masselosen Diagramme, die der harten Region entsprechen,
mit der Methode der partiellen Integration berechnet.
Das Ergebnis f"ur das Integral mit der vollen Topologie ohne Z"ahler,
$F_\NP^{\text{(h-h)}}(1,1,1,1,1,1,0)$,
wurde aus \cite{Gonsalves:1983nq} "ubernommen.
Alle anderen masselosen skalaren Integrale konnten mittels partieller
Integration entweder auf dieses Integral oder auf einfachere Topologien
zur"uckgef"uhrt werden.
Der gesamte Beitrag der harten Region liefert das Ergebnis des
masselosen nichtplanaren Feynman-Diagramms:
\begin{multline}
  F_{v,\NP}^{\text{(h-h)}} =
    \left( C_F^2 - \frac{1}{2} C_F C_A \right)
    \left(\frac{\alpha}{4\pi}\right)^2
    \left(\frac{\mu^2}{Q^2}\right)^{2\eps} S_\eps^2
    \, \Biggl\{
  \frac{1}{\eps^4}
  + \frac{4}{\eps^3}
  + \frac{1}{\eps^2} \left( -\frac{7}{6}\pi^2 + 16 \right)
  \\*
  + \frac{1}{\eps} \left( -\frac{122}{3}\zeta_3 - \frac{8}{3}\pi^2
    + 58 \right)
  - \frac{53}{72}\pi^4 - \frac{380}{3}\zeta_3 - \frac{29}{3}\pi^2
    + 204
  \Biggr\}
  + \Oc(\eps)
  \,.
\end{multline}
Dieses Ergebnis stimmt mit \cite{Kramer:1986sg,Matsuura:1988sm} "uberein.

Die Beitr"age der anderen Regionen wurden mit Hilfe von
Schwinger-Parametern und Mellin-Barnes-Darstellungen berechnet.
Ihre f"uhrenden Beitr"age lassen sich durch einfache
Mellin-Barnes-Integrale oder durch schlichte Br"uche mit
$\Gamma$"~Funk"-tionen ausdr"ucken:
\begin{align}
\label{eq:FNP1chgen}
  \lefteqn{F_\NP^{\text{(1c-h)}} (n_1,n_2,n_3,n_4,n_5,n_6,n_7) =} \quad
  \nonumber \\* &
  \left(\frac{M^2}{Q^2}\right)^{2-n_{35}+\eps}
  e^{-n i\pi} \, e^{2\eps\gamma_E} \,
  \frac{\Gamma(\frac d2-n_{24}) \Gamma(\frac d2-n_{16}+n_7)
    \Gamma(n_{35}-\frac d2)}
    {\Gamma(n_1) \Gamma(n_2) \Gamma(n_3) \Gamma(n_5)
    \Gamma^2(d-n_{1246}+n_7)}
  \nonumber \\* & \times
  \MBint z \,
  \frac{\Gamma(-z) \Gamma(\frac d2-n_{146}-z)
    \Gamma(\frac d2-n_{1246}+n_{37}-z)}
    {\Gamma(\frac d2-n_{16}-z)}
  \nonumber \\* & \times
  \frac{\Gamma(n_1+z) \Gamma(\frac d2-n_3+z) \Gamma(n_{1246}-\frac d2+z)}
    {\Gamma(n_{16}+z)}
  \,,
\\
\label{eq:FNP1c1cgen}
  \lefteqn{F_\NP^{\text{(1c-1c)}} (n_1,n_2,n_3,n_4,n_5,n_6,0) =} \quad
  \nonumber \\* &
  \left(\frac{M^2}{Q^2}\right)^{4-n_{1356}} \,
  \frac{e^{-n i\pi} \, e^{2\eps\gamma_E} \,
    \Gamma(n_{16}-\frac d2) \Gamma(n_{1356}-d)}
    {\Gamma(n_1) \Gamma(n_3) \Gamma(n_5) \Gamma(n_6) \Gamma(\frac d2-n_{24})}
  \MBint z \,
  \Gamma(-z) \Gamma(n_{13}-\tfrac d2-z)
  \nonumber \\* & \times
  \frac{\Gamma(n_5-n_4+z) \Gamma(\frac d2-n_1+z)
    \Gamma(\frac d2-n_{24}+z) \Gamma(\frac d2-n_{34}+z)}
    {\Gamma(\frac d2-n_4+z) \Gamma(n_{156}-n_4-\frac d2+z)}
  \,,
\\
\label{eq:FNP1c2cgen}
  \lefteqn{F_\NP^{\text{(1c-2c)}} (n_1,n_2,n_3,n_4,n_5,n_6,n_7) =
  \left(\frac{M^2}{Q^2}\right)^{4-n_{3456}}
  e^{-n i\pi} \, e^{2\eps\gamma_E}
  } \quad
  \nonumber \\* & \times
  \frac{\Gamma(n_{37}-n_2) \Gamma(n_{47}-n_1)
    \Gamma(\frac d2-n_{13}) \Gamma(\frac d2-n_{24})
    \Gamma(n_{35}-\frac d2) \Gamma(n_{46}-\frac d2)}
    {\Gamma(n_3) \Gamma(n_4) \Gamma(n_5) \Gamma(n_6)
    \Gamma^2(\frac d2-n_{12}+n_7)}
  \,,
\\
\label{eq:FNP1c1cPgen}
  \lefteqn{F_\NP^{\text{(1c-1c')}} (n_1,n_2,n_3,n_4,n_5,n_6,n_7) =} \quad
  \nonumber \\* &
  \left(\frac{M^2}{Q^2}\right)^{4-n_{2345}} \,
  \frac{e^{-n i\pi} \, e^{2\eps\gamma_E} \,
    \Gamma(d-n_{234}) \Gamma(n_{24}-\frac d2) \Gamma(n_{2345}-d)}
    {\Gamma(n_2) \Gamma(n_3) \Gamma(n_4) \Gamma(n_5)
    \Gamma(\frac d2-n_{16}+n_7) \Gamma(d-n_{2347})}
  \MBint z
  \nonumber \\* & \times
  \frac{\Gamma(-z) \Gamma(\frac d2-n_{347}-z)
    \Gamma(n_{37}-n_6+z) \Gamma(\frac d2-n_2+z) \Gamma(\frac d2-n_{16}+n_7+z)}
    {\Gamma(\frac d2-n_6+n_7+z)}
  \,,
\\
\label{eq:FNPusPusPgen}
  \lefteqn{F_\NP^{\text{(us'-us')}} (n_1,n_2,n_3,n_4,n_5,n_6,n_7) =
  \left(\frac{M^2}{Q^2}\right)^{8-n_{1256}-2n_{34}-2\eps}
  e^{-n i\pi} \, e^{2\eps\gamma_E}
  } \quad
  \nonumber \\* & \times
  \frac{\Gamma(d-n_{134}) \Gamma(d-n_{234})
    \Gamma(n_{13}-\frac d2) \Gamma(n_{24}-\frac d2)
    \Gamma(n_{2345}-d) \Gamma(n_{1346}-d)}
    {\Gamma(n_1) \Gamma(n_2) \Gamma(n_3) \Gamma(n_4) \Gamma(n_5) \Gamma(n_6)}
  \,,
\\
\label{eq:FNP1cusPgen}
  \lefteqn{F_\NP^{\text{(1c-us')}} (n_1,n_2,n_3,n_4,n_5,n_6,n_7) =
  \left(\frac{M^2}{Q^2}\right)^{6-n_{2356}-2n_4-\eps}
  e^{-n i\pi} \, e^{2\eps\gamma_E} \,
  \Gamma(\tfrac d2-n_4)
  } \quad
  \nonumber \\* & \times
  \frac{\Gamma(\frac d2-n_{13}) \Gamma(d-n_{234})
    \Gamma(n_{24}-\frac d2) \Gamma(n_{46}-\frac d2) \Gamma(n_{347}-\frac d2)
    \Gamma(n_{2345}-d)}
    {\Gamma(n_2) \Gamma(n_3) \Gamma(n_4) \Gamma(n_5) \Gamma(n_6)
    \Gamma(n_{47}-n_1) \Gamma(\frac d2-n_3)}
  \,.
\end{align}

Der Ausdruck f"ur die (us'-us')-Region gilt tats"achlich f"ur
beliebige~$n_7$, auch wenn $n_7$ darin nicht mehr vorkommt.
Der f"uhrende Beitr"ag des Z"ahlers im Schleifenintegral von
$F_\NP^{\text{(us'-us')}}$ enth"alt lediglich die Konstante $(Q^2)^{n_7}$,
die durch den Vorfaktor $(Q^2)^{-n_7}$ in~(\ref{eq:NPskalar}) gek"urzt wird.

Wegen der Symmetrie des Diagramms gilt:
\begin{align}
  F_\NP^{\text{(h-2c)}}(n_1,n_2,n_3,n_4,n_5,n_6,n_7) &=
    F_\NP^{\text{(1c-h)}}(n_2,n_1,n_4,n_3,n_6,n_5,n_7) \,, \\
  F_\NP^{\text{(2c-2c)}}(n_1,n_2,n_3,n_4,n_5,n_6,0) &=
    F_\NP^{\text{(1c-1c)}}(n_2,n_1,n_4,n_3,n_6,n_5,0) \,, \\
  F_\NP^{\text{(2c'-2c)}}(n_1,n_2,n_3,n_4,n_5,n_6,n_7) &=
    F_\NP^{\text{(1c-1c')}}(n_2,n_1,n_4,n_3,n_6,n_5,n_7) \,, \\
  F_\NP^{\text{(us'-2c)}}(n_1,n_2,n_3,n_4,n_5,n_6,n_7) &=
    F_\NP^{\text{(1c-us')}}(n_2,n_1,n_4,n_3,n_6,n_5,n_7) \,.
\end{align}

Bei der gro"sen Zahl von Regionen, die zum nichtplanaren Diagramm
beitragen, stellt sich die berechtigte Frage nach einer "Uberpr"ufung,
ob alle relevanten Regionen ber"ucksichtigt worden sind.
F"ur das nichtplanare Diagramm, wie auch f"ur viele andere
Diagramme, konnte die Vollst"andigkeit der Regionen folgenderma"sen
gepr"uft werden.
Das gesamte skalare Diagramm mit allgemeinen Parametern ohne Z"ahler
wurde mit Hilfe von Schwinger-Parametern in die folgende Form
mit einem vier\-fachen Mellin-Barnes-Integral gebracht:
\begin{align}
  & \! F_\NP(n_1,n_2,n_3,n_4,n_5,n_6,0) =
  \frac{e^{-n i\pi} \, e^{2\eps\gamma_E}}
    {\Gamma(n_1) \Gamma(n_2) \Gamma(n_3) \Gamma(n_4) \Gamma(n_5) \Gamma(n_6)}
  \MBint{z_1} \left(\frac{M^2}{Q^2}\right)^{2\eps+z_1}
  \nonumber \\*[-1ex] & \times
  \MBint{z_2} \MBint{z_3} \MBint{z_4} \,
  \frac{\Gamma(n_1+z_2) \Gamma(-z_3) \Gamma(n_4+z_3)
    \Gamma(-z_4) \Gamma(n_{46}-\frac d2-z_4)}
    {\Gamma(d-n_{3456}-z_1) \Gamma(\frac32d-n_{123456}-z_1)}
  \nonumber \\* & \times
  \Gamma(d-n_{12456}-z_1-z_2) \Gamma(d-n_{13456}-z_1-z_3)
    \Gamma(\tfrac d2-n_{46}-z_1+z_4)
  \nonumber \\* & \times
  \Gamma(d-n_{3456}-z_1+z_4)
    \Gamma(n_{123456}-d+z_1+z_2)
    \Gamma(z_3-z_2) \Gamma(\tfrac d2-n_{12}-z_2+z_3)
  \nonumber \\* & \times
  \frac{\Gamma(\frac32d-n_{12356}-2n_4-z_1-z_2+z_4)
    \Gamma(n_{456}-\frac d2+z_1+z_3-z_4)}
    {\Gamma(\frac d2-n_{12}-z_2+z_3-z_4)
    \Gamma(\frac32d-n_{123456}-z_1-z_2+z_3+z_4)}
  \,.
\end{align}
Aus dieser Darstellung konnten die f"uhrenden Residuen, d.h. diejenigen
Beitr"age, die nicht mit $M^2/Q^2$ gegen"uber anderen unterdr"uckt sind,
extrahiert werden.
Dies ergab 11~Terme, deren Abh"angigkeit von $M^2/Q^2$ exakt derjenigen
der 11~Regionen entspricht. Manche der Terme sind identisch mit den
Ausdr"ucken f"ur die jeweiligen Regionen, andere sind durch zus"atzliche
Mellin-Barnes-Integrale wesentlich komplizierter.
Da jedoch keine zus"atzlichen Terme mit anderen als den aus den
ber"ucksichtigen Regionen bekannten Abh"angigkeiten von $M^2/Q^2$
auftraten, ist davon auszugehen, dass keine weiteren Regionen zum
nichtplanaren Diagramm beitragen.

Die Struktur der Singularit"aten in den Mellin-Barnes-Integralen der
einzelnen Regionen ist komplizierter als im Fall des planaren Diagramms.
Neben Singularit"aten, die dimensional regularisiert sind (durch
$\eps\ne0$) und Singularit"aten im Z"ahler der Integrale, die durch
entsprechende Singularit"aten im Nenner der Vorfaktoren kompensiert werden,
treten Singularit"aten im Parameterraum der $n_i$ auf, die nur zwischen den
kollinearen Regionen (1c-1c), (2c-2c), (1c-2c), (1c-1c') und (2c'-2c)
kompensiert werden.
So wird der Pol $1/(n_4-n_5)$, der f"ur $n_1=n_3=n_4=n_5=n_6=1$, $n_7=0$
auftritt, zwischen den Regionen (1c-1c) und (2c'-2c) kompensiert.
Entsprechendes gilt f"ur den Pol $1/(n_3-n_6)$, der f"ur
$n_2=n_3=n_4=n_5=n_6=1$, $n_7=0$ in den Regionen (2c-2c) und (1c-1c')
auftritt.
Auch von den Polen $1/(n_1-n_4)$, $1/(n_2-n_3)$ und
$1/(n_{13}-n_{24})$, die in unterschiedlicher Kombination in den Regionen
(1c-1c), (2c-2c) und (1c-2c) vorkommen, bleibt nach Addition dieser drei
Regionen nichts mehr "ubrig.

Die Auswertung der Mellin-Barnes-Integrale erfolgt wie f"ur das planare
Diagramm beschrieben.
Die Ergebnisse der einzelnen skalaren Integrale $F_\NP$ finden sich im
Anhang~\ref{sec:SkalarNP} in Gl.~(\ref{eq:NPergskalar}).
Der gesamte Beitrag des nichtplanaren Vertexdiagramms zum Formfaktor lautet
in f"uhrender Ordnung in $M^2/Q^2$:
\begin{align}
\label{eq:NPerg}
  F_{v,\NP} &=
    \left( C_F^2 - \frac{1}{2} C_F C_A \right)
    \left(\frac{\alpha}{4\pi}\right)^2
    \left(\frac{\mu^2}{M^2}\right)^{2\eps} S_\eps^2
    \, \Biggl\{
  - \frac{2}{\eps}
  \nonumber \\* & \qquad
  + \frac{1}{3} \lqm^4
  - \frac{8}{3} \lqm^3
  + \left(-\frac{2}{3}\pi^2 + 12\right) \lqm^2
  + \left(40\zeta_3 + \frac{2}{3}\pi^2 - 28\right) \lqm
  \nonumber \\* & \qquad
  - \frac{4}{15}\pi^4 - 72\zeta_3 - \pi^2 + 28
  \Biggr\}
  + \Oc(\eps) + \Oc\!\left(\frac{M^2}{Q^2}\right)
  \,.
\end{align}
Das Ergebnis des Vertexdiagramms enth"alt nur einen einfachen Pol in
$\eps$. Dies ist darauf zur"uckzuf"uhren, dass jede Schleife des
Zweischleifendiagramms vier Propagatoren enth"alt und f"ur sich gesehen
ultraviolett-konvergent ist. Erst die zweite Schleifenintegration
produziert eine ultraviolette Singularit"at.

%
%
\subsection{Vertexdiagramm mit Benz-Topologie}
\label{sec:BE}

\begin{figure}[ht]
  \center
  \includegraphics{vertex-BE}
  \hspace{1cm}
  \includegraphics{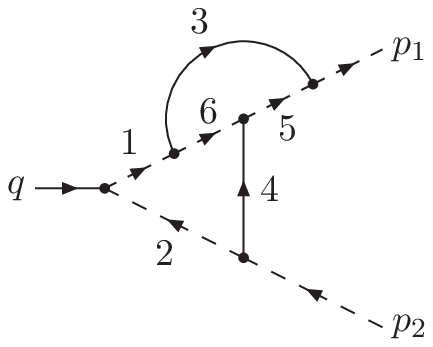}
  \caption{Vertexdiagramm mit Benz-Topologie (mit skalarem Diagramm)}
  \label{fig:BE}
\end{figure}

Das Vertexdiagramm mit Benz-Topologie (Bezeichnung~BE)
ist in Abb.~\ref{fig:BE} als Feynman-Diagramm und als skalares Diagramm mit
Impulsbezeichnungen dargestellt.
Der Name \emph{Benz-Topologie} leitet sich von der bekannten Form ab, die
man erh"alt, wenn die Linien 2, 1, 3, $p_1$, $p_2$ einen Kreis bilden,
von dessen Rand aus "aquidistant die Linien 4, 6, 5 weggehen, um sich in
der Kreismitte zu treffen.
Das horizontal gespiegelte Feynman-Diagramm liefert aus Symmetriegr"unden
denselben Beitrag zum Formfaktor, so dass das hier dargestellte
Benz-Diagramm zweifach gez"ahlt werden muss.

Die Amplitude des Benz-Vertexdiagramms lautet:
\begin{multline}
\label{eq:BEamplitude}
  \Fc_{v,\BE}^\mu =
  \mu^{4\eps} \loopint dk \loopint d\ell \,
  ig \gamma^\rho t^b \, \frac{i\dslash k_5}{k_5^2} \,
  ig \gamma^\nu t^a \, \frac{i\dslash k_6}{k_6^2} \,
  ig \gamma_\rho t^b \, \frac{i\dslash k_1}{k_1^2} \,
  \gamma^\mu
  \frac{i\dslash k_2}{k_2^2} \, ig \gamma_\nu t^a
  \\* \times
  \frac{-i}{k_3^2-M^2} \, \frac{-i}{k_4^2-M^2}
  \,,
\end{multline}
mit den Impulsen
\begin{equation}
\label{eq:BEmom}
  k_1 = p_1 - \ell ,\quad k_2 = p_2 - \ell ,\quad
  k_3 = p_1 - k ,\quad k_4 = \ell ,\quad
  k_5 = k ,\quad k_6 = k - \ell .
\end{equation}
Der Farbfaktor des Benz-Vertexdiagramms ist der gleiche wie der des
nichtplanaren Vertexdiagramms:
\[
  t^b t^a t^b t^a = \left( C_F^2 - \frac{1}{2} C_F C_A \right) \unity
  \,.
\]

Der Beitrag der Amplitude~$\Fc_{v,\BE}^\mu$ zum Formfaktor ist
$F_{v,\BE}$.
Das K"urzen der Skalarprodukte im Z"ahler von~$F_{v,\BE}$ mit
Propagatornennern erledigen die folgenden Umformungen:
\begin{equation}
\label{eq:BEredskalar}
  \begin{aligned}
    p_1 \cdot k_5 &= -\tfrac12 (k_3^2 - k_5^2) \,,\quad &
    p_1 \cdot k_4 &= -\tfrac12 (k_1^2 - k_4^2) \,, \\
    p_2 \cdot k_4 &= -\tfrac12 (k_2^2 - k_4^2) \,,\quad &
    k_4 \cdot k_5 &= -\tfrac12 (k_6^2 - k_4^2 - k_5^2) \,.
  \end{aligned}
\end{equation}

Als irreduzibles Skalarprodukt verbleibt $2p_2\cdot k_5 = 2p_2\cdot k$.
Die skalaren Zweischleifenintegrale der Benz-Topologie
sind folgenderma"sen definiert:
\begin{multline}
\label{eq:BEskalar}
  F_\BE(n_1,n_2,n_3,n_4,n_5,n_6,n_7) =
  e^{2\eps\gamma_E} \, (M^2)^{2\eps} \,
  (Q^2)^{n-n_7-4}
  \\* \times
  \loopintf dk \loopintf d\ell \,
  \frac{(2p_2 \cdot k)^{n_7}}
    {(\ell^2-2p_1\cdot\ell)^{n_1} \, (\ell^2-2p_2\cdot\ell)^{n_2} \,
      (k^2 - 2p_1\cdot k - M^2)^{n_3}}
  \\* \times
  \frac{1}{(\ell^2 - M^2)^{n_4} \, (k^2)^{n_5} \, ((k-\ell)^2)^{n_6}}
  \,,
\end{multline}
mit der Notation aus Gl.~(\ref{eq:LAskalar})
und $n = n_{123456}$ (ohne $n_7$).
Die Zerlegung von~$F_{v,\BE}$ in skalare Integrale findet sich im
Anhang~\ref{sec:SkalarBE} in Gl.~(\ref{eq:BEzerlegung}).
Das Benz-Diagramm weist keine Symmetrie auf, mit der analog zum planaren
und nichtplanaren Diagramm verschiedene skalare Integrale zueinander
in Beziehung gesetzt werden k"onnten.

Zur \emph{Expansion by Regions} f"ur die Auswertung der skalaren Integrale
tragen folgende Regionen bei:
\[
  \renewcommand{\arraystretch}{1.2}
  \left.
  \begin{array}{rll}
    \text{(h-h):} & k \sim Q, & \ell \sim Q \\
    \text{(1c-h):} & k \;\|\; p_1, & \ell \sim Q \\
    \text{(h-2c):} & k \sim Q, & \ell \;\|\; p_2 \\
    \text{(us-2c):} & k \sim M^2/Q, & \ell \;\|\; p_2
  \end{array}
  \qquad
  \right|
  \qquad
  \begin{array}{rll}
    \text{(1c-1c):} & k \;\|\; p_1, & \ell \;\|\; p_1 \\
    \text{(2c-2c):} & k \;\|\; p_2, & \ell \;\|\; p_2 \\
    \text{(1c-2c):} & k \;\|\; p_1, & \ell \;\|\; p_2
  \end{array}
\]

Der f"uhrende Beitrag der jeweiligen Region h"angt "uber die folgende
Potenz von $M^2/Q^2$ ab (Notation wie in Abschnitt~\ref{sec:LA}):
\[
  \left.
  \begin{array}{rl}
    \text{(h-h):} & 2\eps \\
    \text{(1c-h):} & 2-n_{35}+\eps \\
    \text{(h-2c):} & 2-n_{24}+\eps \\
    \text{(us-2c):} & 6-n_{2346}-2n_5+n_7-\eps
  \end{array}
  \qquad
  \right|
  \qquad
  \begin{array}{rl}
    \text{(1c-1c):} & 4-n_{13456} \\
    \text{(2c-2c):} & 4-n_{2456}+n_7 \\
    \text{(1c-2c):} & 4-n_{2345}
  \end{array}
\]
Die Regionen (us-2c) und (2c-2c) sind gegen"uber (h-h) unterdr"uckt,
wenn $n_7 > 0$ ist.
Au"serdem tritt f"ur $n_1=n_3=n_4=n_5=n_6=1$ der Fall ein,
dass der Beitrag der Region (1c-1c) von der Ordnung~$(M^2/Q^2)^{-1}$ ist,
also einen Pol in $M^2$ aufweist. Dieser Pol wird jedoch vom Vorfaktor des
skalaren Diagramms in der Zerlegung~(\ref{eq:BEzerlegung}) gek"urzt, der
in all diesen F"allen proportional zu $M^2/Q^2$ ist.

Die Beitr"age aller Regionen wurden mit Hilfe von Schwinger-Parametern und
Mellin-Barnes-Darstellungen f"ur allgemeine Parameter~$n_i$ als
maximal zweifache Mellin-Barnes-Integrale dargestellt.
Zur Vereinfachung von Summen aus mehreren Termen mit ganzzahligen Potenzen
im Z"ahler der Parameterintegrale wurde zus"atzlich die binomische Formel
verwendet:
\begin{equation}
\label{eq:binom}
  (a+b)^n = \sum_{i=0}^n \frac{n!}{i! \, (n-i)!} \, a^i \, b^{n-i}
\end{equation}
Wiederholte Anwendung von (\ref{eq:binom}) f"uhrt auf Summen der Form
\[
  \sum_{i_1,i_2,i_3\ge0}^{i_{123}\le n} \equiv
  \sum_{i_1=0}^n \, \sum_{i_2=0}^{n-i_1} \,\sum_{i_3=0}^{n-i_1-i_2}
  \,,
\]
wobei wie f"ur die Parameter~$n_i$ die Notation
$i_{12\cdots} = i_1 + i_2 + \ldots$ eingef"uhrt wird.
Die Beitr"age der Regionen lauten:
\begin{align}
\label{eq:FBEhhgen}
  \lefteqn{F_\BE^{\text{(h-h)}}(n_1,n_2,n_3,n_4,n_5,n_6,n_7) =
  \left(\frac{M^2}{Q^2}\right)^{2\eps}
  e^{-n i\pi} \, e^{2\eps\gamma_E} \,
  \sum_{i_1,i_2,i_3\ge0}^{i_{123}\le n_7} \,
    \frac{n_7!}{i_1!\,i_2!\, i_3!\,(n_7-i_{123})!}
  } \quad
  \nonumber \\* & \times
  \frac{\Gamma(n_1+i_3) \Gamma(n_4+i_2)
    \Gamma(\frac d2-n_{35}) \Gamma(d-n_{13456}+i_1) \Gamma(n_{123456}-d)}
    {\Gamma(n_1) \Gamma(n_2) \Gamma(n_3) \Gamma(n_4) \Gamma(n_5) \Gamma(n_6)
    \Gamma(d-n_{356}+i_{123}) \Gamma(\frac32d-n_{123456}+n_7)}
  \nonumber \\* & \times
  \MBint z \,
  \frac{\Gamma(-z) \Gamma(\frac d2-n_{56}+i_{12}-z)
    \Gamma(d-n_{36}-2n_5+i_{123}-z)}
    {\Gamma(d-n_{36}-2n_5+i_{12}-z)}
  \nonumber \\* & \times
  \frac{\Gamma(n_5+z) \Gamma(\frac d2-n_{24}+n_7-i_{12}+z)
    \Gamma(n_{3567}-i_{123}-\frac d2+z)}
    {\Gamma(n_{13567}-i_{12}-\frac d2+z)}
  \,, \qquad\qquad\quad
\\
\label{eq:FBE1chgen}
  \lefteqn{F_\BE^{\text{(1c-h)}}(n_1,n_2,n_3,n_4,n_5,n_6,n_7) =} \quad
  \nonumber \\* &
  \left(\frac{M^2}{Q^2}\right)^{2-n_{35}+\eps}
  e^{-n i\pi} \, e^{2\eps\gamma_E} \,
  \frac{\Gamma(n_{35}-\frac d2) \Gamma(\frac d2-n_{146})}
    {\Gamma(n_1) \Gamma(n_2) \Gamma(n_3) \Gamma(n_6) \Gamma(d-n_{1246})}
  \nonumber \\ & \times
  \MBint z \,
  \frac{\Gamma(-z) \Gamma(\frac d2-n_{246}-z)
    \Gamma(n_6+z) \Gamma(\frac d2-n_5+n_7+z) \Gamma(n_{1246}-\frac d2+z)}
    {\Gamma(\frac d2+n_7+z)}
  \,,
\\
\label{eq:FBEh2cgen}
  \lefteqn{F_\BE^{\text{(h-2c)}}(n_1,n_2,n_3,n_4,n_5,n_6,n_7) =
  \left(\frac{M^2}{Q^2}\right)^{2-n_{24}+\eps}
  e^{-n i\pi} \, e^{2\eps\gamma_E}
  } \quad
  \nonumber \\* & \times
  \frac{\Gamma(\frac d2-n_2) \Gamma(\frac d2-n_{35})
    \Gamma(\frac d2-n_{1356}+n_2) \Gamma(\frac d2-n_{56}+n_7)
    \Gamma(n_{24}-\frac d2) \Gamma(n_{356}-\frac d2)}
    {\Gamma(n_2) \Gamma(n_3) \Gamma(n_4) \Gamma(n_6)
    \Gamma(d-n_{1356}) \Gamma(d-n_{356}+n_7)}
  \,,
\\
\label{eq:FBEus2cgen}
  \lefteqn{F_\BE^{\text{(us-2c)}}(n_1,n_2,n_3,n_4,n_5,n_6,n_7) =} \quad
  \nonumber \\* &
  \left(\frac{M^2}{Q^2}\right)^{6-n_{2346}-2n_5+n_7-\eps}
  e^{-n i\pi} \, e^{2\eps\gamma_E} \,
  \frac{\Gamma(\frac d2-n_5+n_7) \Gamma(d-n_{256}+n_7)}
    {\Gamma(n_2) \Gamma(n_3) \Gamma(n_4) \Gamma(n_5) \Gamma(n_6)
    \Gamma(n_5-n_{17})}
  \nonumber \\* & \times
  \Gamma(n_{35}-\tfrac d2) \Gamma(n_{25}-n_{17}-\tfrac d2)
    \Gamma(n_{56}-n_7-\tfrac d2) \Gamma(n_{2456}-n_7-d)
  \,,
\\
\label{eq:FBE1c1cgen}
  \lefteqn{F_\BE^{\text{(1c-1c)}}(n_1,n_2,n_3,n_4,n_5,n_6,n_7) =
  \left(\frac{M^2}{Q^2}\right)^{4-n_{13456}}
  e^{-n i\pi} \, e^{2\eps\gamma_E}
  } \quad
  \nonumber \\ & \times
  \sum_{i_1,i_2\ge0}^{i_{12}\le n_7} \,
    \frac{n_7!}{i_1!\,i_2!\,(n_7-i_{12})!} \,
  \frac{\Gamma(n_1+i_2)}
    {\Gamma(n_1) \Gamma(n_3) \Gamma(n_4) \Gamma(n_5) \Gamma(n_6)
    \Gamma(\frac d2-n_2+n_7)}
  \nonumber \\ & \times
  \MBint{z_1} \,
  \frac{\Gamma(-z_1) \Gamma(n_{16}-i_1-\frac d2-z_1)
    \Gamma(n_{1356}-i_1-d-z_1)
    \Gamma(n_4+i_1+z_1)}
    {\Gamma(n_1+i_2-z_1)}
  \nonumber \\ & \times
  \MBint{z_2} \,
  \Gamma(-z_2) \Gamma(\tfrac d2-n_{56}+i_1-z_2) \Gamma(n_5+z_2) \,
  \frac{\Gamma(n_1-n_5+i_2-z_1-z_2)}
    {\Gamma(n_1-n_5-z_1-z_2)}
  \nonumber \\ & \times
  \frac{\Gamma(\frac d2-n_1+n_7-i_2+z_1+z_2) \Gamma(\frac d2-n_2+n_7+z_1+z_2)}
    {\Gamma(\frac d2+n_7+z_1+z_2)}
  \,,
\\
\label{eq:FBE2c2cgen}
  \lefteqn{F_\BE^{\text{(2c-2c)}}(n_1,n_2,n_3,n_4,n_5,n_6,n_7) =
  \left(\frac{M^2}{Q^2}\right)^{4-n_{2456}+n_7}
  e^{-n i\pi} \, e^{2\eps\gamma_E} \,
  \sum_{i_1,i_2,i_3\ge0}^{i_{123}\le n_7}
  } \quad
  \nonumber \\* & \times
  \frac{n_7!}{i_1!\,i_2!\, i_3!\,(n_7-i_{123})!} \,
  \frac{\Gamma(n_1+i_1) \Gamma(n_{37}-i_1) \Gamma(n_2-n_{137}+i_2)}
    {\Gamma(n_1) \Gamma(n_2) \Gamma(n_3) \Gamma(n_4) \Gamma(n_5) \Gamma(n_6)}
  \, \Gamma(\tfrac d2-n_{35}+i_1)
  \nonumber \\* & \times
  \frac{\Gamma(\frac d2-n_6+i_{23})
    \Gamma(d-n_{256}+n_7+i_3)
    \Gamma(n_{56}-i_{23}-\frac d2) \Gamma(n_{2456}-n_7-d)}
    {\Gamma(\frac d2-n_{13}) \Gamma(d-n_{356}+i_{123})}
  \,,\!\!
\\
\label{eq:FBE1c2cgen}
  \lefteqn{F_\BE^{\text{(1c-2c)}}(n_1,n_2,n_3,n_4,n_5,n_6,n_7) =
  \left(\frac{M^2}{Q^2}\right)^{4-n_{2345}}
  e^{-n i\pi} \, e^{2\eps\gamma_E}
  } \quad
  \nonumber \\* & \times
  \frac{\Gamma(n_2-n_{16})
    \Gamma(\frac d2-n_2) \Gamma(\frac d2-n_{56}+n_7)
    \Gamma(n_{24}-\frac d2) \Gamma(n_{35}-\frac d2)}
    {\Gamma(n_2) \Gamma(n_3) \Gamma(n_4)
    \Gamma(\frac d2-n_{16}) \Gamma(\frac d2-n_6+n_7)}
  \,.
\end{align}

Diese Ausdr"ucke f"ur die 7~Regionen konnten, "ahnlich wie in
Abschnitt~\ref{sec:NP} f"ur das nichtplanare Diagramm,
auch durch eine Berechnung des gesamten skalaren Diagramms ohne Entwicklung
in den Regionen erzielt werden.
Das skalare Benz-Diagramm mit allgemeinen Parametern (inklusive Z"ahler)
wurde mit Hilfe von Schwinger-Parametern in die folgende Form mit einem
dreifachen Mellin-Barnes-Integral und einer dreifachen Summe gebracht:
\begin{align}
  \lefteqn{F_\BE(n_1,n_2,n_3,n_4,n_5,n_6,n_7) =
  \frac{e^{-n i\pi} \, e^{2\eps\gamma_E}}
    {\Gamma(n_2) \Gamma(n_3) \Gamma(n_4) \Gamma(n_5) \Gamma(n_6)}
  } \quad
  \nonumber \\* & \times
  \sum_{i_1,i_2,i_3\ge0}^{i_{123}\le n_7} \,
    \frac{n_7!}{i_1!\,i_2!\, i_3!\,(n_7-i_{123})!} \,
  \frac{\Gamma(n_1+i_3)}{\Gamma(n_1)}
  \MBint{z_1} \MBint{z_2} \MBint{z_3} \,
  \left(\frac{M^2}{Q^2}\right)^{2\eps+z1}
  \nonumber \\* & \times
  \frac{\Gamma(d-n_{13456}+i_1-z_1) \Gamma(n_{123456}-d+z_1)}
    {\Gamma(\frac32d-n_{123456}+n_7-z_1)} \,
  \Gamma(-z_2) \Gamma(n_4+i_2+z_2)
  \nonumber \\* & \times
  \Gamma(-z_3) \Gamma(\tfrac d2-n_{56}+i_{12}-z_3) \Gamma(n_5+z_3)
  \Gamma(\tfrac d2-n_2+n_7-i_1+z_2+z_3)
  \nonumber \\* & \times
  \frac{\Gamma(-n_4-i_2-z_1-z_2) \Gamma(\frac d2-n_{345}-i_2-z_1-z_2)}
    {\Gamma(d-n_{3456}+i_{13}-z_1-z_2)}
  \nonumber \\* & \times
  \frac{\Gamma(d-n_{346}-2n_5+i_{13}-z_1-z_2-z_3)
    \Gamma(n_{34567}-i_{13}-\frac d2+z_1+z_2+z_3)}
    {\Gamma(d-n_{346}-2n_5+i_1-z_1-z_2-z_3)
    \Gamma(n_{134567}-i_1-\frac d2+z_1+z_2+z_3)}
  \,.
\end{align}
Die Extraktion der f"uhrenden Beitr"age in $M^2/Q^2$ ergab 7~Terme.
Diese waren zwar teilweise komplizierter als die obigen Ausdr"ucke f"ur die
7~Regionen, aber nach Umformungen und Vereinfachungen konnten alle Terme
mit dem Ausdruck je einer Region identifiziert werden.

Die harte (h-h)-Region wurde au"ser durch Berechnung der
Mellin-Barnes-Integrale auch durch partielle Integration der masselosen
Integrale ausgewertet. Die Ergebnisse der beiden Methoden stimmen "uberein.
Das Ergebnis des masselosen Feynman-Diagramms mit Benz-Topologie,
oder anders ausgedr"uckt, der gesamte Beitrag der harten Region, lautet:
\begin{multline}
  F_{v,\BE}^{\text{(h-h)}} =
    \left( C_F^2 - \frac{1}{2} C_F C_A \right)
    \left(\frac{\alpha}{4\pi}\right)^2
    \left(\frac{\mu^2}{Q^2}\right)^{2\eps} S_\eps^2
    \, \Biggl\{
  -\frac{1}{\eps^3}
  + \frac{1}{\eps^2} \left( \frac{\pi^2}{3} - \frac{11}{2} \right)
  \\*
  + \frac{1}{\eps} \left( 2\zeta_3 + \frac{5}{3}\pi^2
    - \frac{109}{4} \right)
  + \frac{2}{45}\pi^4 + \frac{59}{3}\zeta_3 + \frac{91}{12}\pi^2
    - \frac{911}{8}
  \Biggr\}
  + \Oc(\eps)
  \,.
\end{multline}
Dieses Ergebnis stimmt mit \cite{Kramer:1986sg,Matsuura:1988sm} "uberein.

Die Auswertung der Mellin-Barnes-Integrale erfolgte im wesentlichen wie
f"ur das planare Diagramm beschrieben.
Singularit"aten, die nicht dimensional regularisiert sind,
treten ausschlie"slich zwischen den (c-c)-Regionen (1c-1c), (2c-2c) und
(1c"~2c) auf.
In den F"allen $F_\BE(n_1,1,n_3,1,1,1,0)$ mit $n_1+n_3=1$
wird der Pol \mbox{$1/(n_{13}-n_2)$} zwischen den Regionen (1c-1c) und (2c-2c)
kompensiert.
Und in den F"allen $F_\BE(n_1,1,1,1,1,n_6,n_7)$ mit beliebigem $n_7$ und
$n_1+n_6=1$ wird der Pol $1/(n_{16}-n_2)$ zwischen den Regionen (1c-1c)
und (1c-2c) kompensiert.

Am kompliziertesten ist die Auswertung der (1c-1c)-Region mit dem zweifachen
Mellin-Barnes-Integral.
Wenn einer der Parameter $n_i$, $i = 3, 4, 5$ oder $6$, kleiner
als~1 ist, geht der Nenner des Vorfaktors mit $1/\Gamma(n_i)$ gegen Null.
Dann liefern nur diejenigen Residuen des Integrals einen Beitrag,
die mit einer entsprechenden Singularit"at $\Gamma(n_i)$ die Nullstelle
des Vorfaktors k"urzen.
Im Fall $n_1=0$ verschwinden Summanden mit $i_2>0$. F"ur $n_1=i_2=0$ aber
k"urzen sich je zwei $\Gamma$-Funktionen $\Gamma(-z_1)$ und
$\Gamma(\frac d2+n_7+z_1+z_2)$, so dass der Integrand einfacher wird.
Auch wenn $n_2\le0$ ist, kann $\Gamma(\frac d2+n_7+z_1+z_2)$ gek"urzt
werden.
Der Nenner $\Gamma(n_1-n_5-z_1-z_2)$ kann immer gek"urzt werden,
falls n"otig mit der Umformung $\Gamma(x+1) = x \Gamma(x)$ im Z"ahler.
Deswegen kann das $z_2$-Integral f"ur $n_1=0$ oder $n_2\le0$ mit dem ersten
Barnschen Lemma~(\ref{eq:BarnesLemma1}) gel"ost werden.

Die Schwierigkeit der (1c-1c)-Region zeigt sich bei der vollen
Topologie, also f"ur $F_\BE^{\text{(1c-1c)}}(1,1,1,1,1,1,n_7)$.
Dann ist der f"uhrende Beitrag, wie bereits vorher analysiert,
von der Ordnung~$(M^2/Q^2)^{-1}$, alle
anderen Regionen sind gegen"uber (1c"~1c) mit $M^2/Q^2$ unterdr"uckt.
(Letzteres gilt auch f"ur den einfacheren Fall $n_2\le0$.)
Die analytische Auswertung des doppelten Mellin-Barnes-Integrals f"ur die
volle Topologie zeigte sich als zu schwierig.
Allerdings sind diese Integrale in $d=4$ Dimensionen endlich, so dass sie
bis auf den Faktor $Q^2/M^2$ lediglich Konstanten liefern.
Diese wurden durch einen Ansatz in Kombination mit
einer numerischen Auswertung der Integrale bestimmt.
Neben den bisherigen analytischen Konstanten $\zeta_2 = \pi^2/6$,
$\zeta_3 \approx 1{,}202057$ und $\zeta_4 = \pi^4/90$ tauchen in diesen
Ergebnissen neue analytische Konstanten auf:
$\ln^4{2}$, $\pi^2\ln^2{2}$ sowie der Polygarithmus
$\Li4(\tfrac12) \approx 0{,}517479$ (\ref{eq:polylog}).

Die skalaren Integrale $F_\BE(1,1,1,1,1,1,n_7)$ mit der vollen Topologie
und Z"ahler, $n_7>0$, wurden nicht direkt durch Auswertung von
Gl.~(\ref{eq:FBE1c1cgen}) gel"ost, sondern auf Integrale ohne Z"ahler
zur"uckgef"uhrt.
Dazu wurde mit Hilfe des in \cite{Anastasiou:2000bn} beschriebenen
Algorithmus, der teilweise auf \cite{Tarasov:1996br,Tarasov:1997kx}
zur"uckgeht, das Zweischleifenintegral mit Schwinger-Parametern
geschrieben und die Tensorreduktion in dieser Form durchgef"uhrt.
Als Ergebnis erh"alt man Integrale ohne Z"ahler mit h"oheren
Propagatorpotenzen und einer erh"ohten Zahl von Raum-Zeit-Dimensionen.
Das Ergebnis wurde anhand einer direkten numerischen Auswertung
von~(\ref{eq:FBE1c1cgen}) "uberpr"uft.

Die Ergebnisse der einzelnen skalaren Integrale~$F_\BE$ finden sich im
Anhang~\ref{sec:SkalarBE} in Gl.~(\ref{eq:BEergskalar}).
Der gesamte Beitrag des Benz-Diagramms zum Formfaktor in f"uhrender Ordnung
in~$M^2/Q^2$ lautet wie folgt:
\begin{align}
\label{eq:BEerg}
  F_{v,\BE} &=
    \left( C_F^2 - \frac{1}{2} C_F C_A \right)
    \left(\frac{\alpha}{4\pi}\right)^2
    \left(\frac{\mu^2}{M^2}\right)^{2\eps} S_\eps^2
    \, \Biggl\{
  \frac{1}{2\eps^2}
  + \frac{1}{\eps} \left[
    - \lqm^2 + 3 \lqm - \frac{2}{3}\pi^2 - \frac{13}{4} \right]
  \nonumber \\* & \quad
  + \lqm^3
  + \left(\frac{\pi^2}{3} - 7\right) \lqm^2
  + \left(8\zeta_3 - 2\pi^2 + \frac{53}{2}\right) \lqm
  + 128\,\Li4\!\left(\frac12\right)
  + \frac{16}{3}\ln^4{2}
  \nonumber \\* & \quad
  - \frac{16}{3}\pi^2\ln^2{2}
  - \frac{28}{15}\pi^4 + 54\zeta_3 + \frac{115}{12}\pi^2 - \frac{263}{8}
  \Biggr\}
  + \Oc(\eps) + \Oc\!\left(\frac{M^2}{Q^2}\right)
  .
\end{align}

%
%
\subsection{Vertexdiagramm mit Fermion-Selbstenergie}
\label{sec:fc}

\begin{figure}[ht]
  \center
  \includegraphics{vertex-fcorr}
  \hspace{1cm}
  \includegraphics{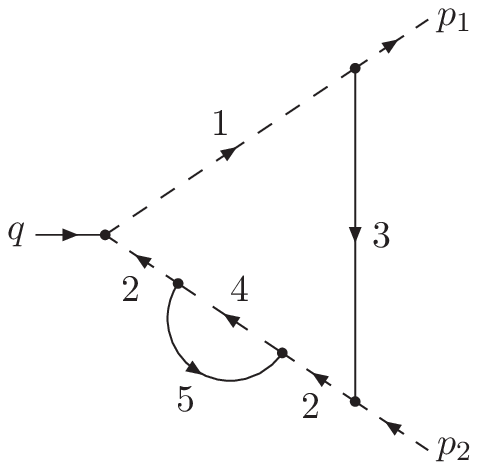}
  \caption{Vertexdiagramm mit Fermion-Selbstenergie (mit skalarem Diagramm)}
  \label{fig:fc}
\end{figure}

Das Vertexdiagramm mit Selbstenergie-Einsetzung in einen der
Fermionpropagatoren (Bezeichnung~fc)
ist in Abb.~\ref{fig:fc} mit dem skalaren Diagramm dargestellt.
Das horizontal gespiegelte Feynman-Diagramm mit der Selbstenergie"=Einsetzung
im anderen Fermionpropagator liefert aus Symmetriegr"unden
denselben Beitrag zum Formfaktor. Das hier berechnete
Diagramm muss also zweifach gez"ahlt werden.

Die Amplitude des Vertexdiagramms mit Fermion-Selbstenergie lautet:
\begin{multline}
\label{eq:fcamplitude}
  \Fc_{v,\fc}^\mu =
  \mu^{4\eps} \loopint dk \loopint d\ell \,
  ig \gamma^\nu t^a \, \frac{i\dslash k_1}{k_1^2} \,
  \gamma^\mu
  \frac{i\dslash k_2}{k_2^2} \, ig \gamma^\rho t^b \,
  \frac{i\dslash k_4}{k_4^2} \, ig \gamma_\rho t^b \,
  \frac{i\dslash k_2}{k_2^2} \, ig \gamma_\nu t^a
  \\* \times
  \frac{-i}{k_3^2-M^2} \, \frac{-i}{k_5^2-M^2}
  \,,
\end{multline}
mit den Impulsen
\begin{equation}
\label{eq:fcmom}
  k_1 = p_1 - p_2 + k ,\quad k_2 = k ,\quad
  k_3 = k - p_2 ,\quad k_4 = k + \ell ,\quad
  k_5 = \ell .
\end{equation}
Der Farbfaktor dieses Vertexdiagramms ist der gleiche wie der des
planaren Vertexdiagramms:
\[
  t^a t^b t^b t^a = C_F^2 \, \unity
  \,.
\]

Der Beitrag der Amplitude~$\Fc_{v,\fc}^\mu$ zum Formfaktor ist
$F_{v,\fc}$.
Zum K"urzen der meisten Skalarprodukte im Z"ahler von~$F_{v,\fc}$ mit
Propagatornennern dienen die folgenden Umformungen:
\begin{equation}
\label{eq:fcredskalar}
  \begin{aligned}
    p_2 \cdot k_2 &= -\tfrac12 (k_3^2 - k_2^2) \,,\quad &
    k_1 \cdot k_2 &= -\tfrac12 (q^2 - k_1^2 - k_2^2) \,, \\
    k_3 \cdot k_2 &= -\tfrac12 (k_2^2 - k_4^2) \,,\quad &
    k_2 \cdot k_5 &= +\tfrac12 (k_4^2 - k_2^2 - k_5^2) \,.
  \end{aligned}
\end{equation}

Die vorliegende Topologie hat nur 5~Propagatoren (anstelle von
6~Propagatoren bei den vorhergehenden Topologien). Von den 7~unabh"angigen
Skalarprodukten, die aus den externen Impulsen und den Schleifenimpulsen
gebildet werden k"onnen, bleiben also formal zwei irreduzible
Skalarprodukte $p_{1,2}\cdot k_5$ "ubrig, die nicht gek"urzt werden k"onnen.
Jedoch erlaubt die einfache Struktur der Topologie eine Tensorreduktion
(siehe Anhang~\ref{sec:tensorred})
f"ur die innere Schleifenintegration, indem man
$p_{1,2}\cdot k_5 = p_{1,2}^\nu \ell_\nu$ schreibt.
Die Tensorintegrale werden mit Hilfe von (\ref{eq:2punkttensorred1}) und
(\ref{eq:2punkttensorred2}) auf skalare Integrale zur"uckgef"uhrt.
Die bei der Tensorreduktion entstehenden Skalarprodukte
$p_{1,2}\cdot k = p_{1,2}\cdot k_2$, $\ell^2 = k_5^2$
und $\ell\cdot k = k_5\cdot k_2$ k"onnen mittels (\ref{eq:fcmom}) und
(\ref{eq:fcredskalar}) gek"urzt werden,
so dass f"ur diese Topologie keine irreduziblen Skalarprodukte verbleiben.
Die skalaren Zweischleifenintegrale
sind folgenderma"sen definiert:
\begin{multline}
\label{eq:fcskalar}
  F_\fc(n_1,n_2,n_3,n_4,n_5) =
  e^{2\eps\gamma_E} \, (M^2)^{2\eps} \,
  (Q^2)^{n-4}
  \loopintf dk \loopintf d\ell
  \\* \times
  \frac{1}
    {((p_1-p_2+k)^2)^{n_1} \, (k^2)^{n_2} \,
    (k^2 - 2p_2\cdot k - M^2)^{n_3} \,
    ((k+\ell)^2)^{n_4} \, (\ell^2 - M^2)^{n_5}}
  \,,
\end{multline}
mit der Notation aus Gl.~(\ref{eq:LAskalar})
und $n = n_{12345}$.
Der zweite Propagator kommt im urspr"unglichen
Integral~(\ref{eq:fcamplitude}) bereits zweifach vor.
Seine Potenz~$n_2$ wird durch Faktoren $1/k^2$ aus der Tensorreduktion
weiter erh"oht.
Durch das K"urzen von Skalarprodukten tritt jedoch nur $n_2\le2$ auf.
Die Zerlegung von $F_{v,\fc}$ in skalare Integrale ist im
Anhang~\ref{sec:Skalarfc} in Gl.~(\ref{eq:fczerlegung}) aufgef"uhrt.

Mit Hilfe von Schwinger-Parametern und Mellin-Barnes-Darstellungen kann das
skalare Integral~(\ref{eq:fcskalar}) in eine Form mit einem zweifachem
Mellin-Barnes-Integral gebracht werden:
\begin{align}
  \lefteqn{F_\fc(n_1,n_2,n_3,n_4,n_5) =
  \frac{e^{-n i\pi} \, e^{2\eps\gamma_E} \,
    \Gamma(\frac d2-n_4)}
    {\Gamma(n_1) \Gamma(n_3) \Gamma(n_4) \Gamma(n_5)}
  \MBint{z_1} \MBint{z_2} \,
  \left(\frac{M^2}{Q^2}\right)^{2\eps+z1}
  } \quad
  \nonumber \\* & \times
  \frac{\Gamma(d-n_{2345}-z_1) \Gamma(n_{12345}-d+z_1)}
    {\Gamma(\frac32d-n_{12345}-z_1)} \,
  \Gamma(-z_2) \Gamma(n_3+z_2) \Gamma(\tfrac d2-n_1+z_2) \,
  \nonumber \\* & \times
  \frac{\Gamma(-n_3-z_1-z_2) \Gamma(\frac d2-n_{35}-z_1-z_2)
    \Gamma(n_{345}-\frac d2+z_1+z_2)}
    {\Gamma(d-n_{345}-z_1-z_2) \Gamma(n_{2345}-\frac d2+z_1+z_2)}
  \,.
\end{align}

Die Extraktion der f"uhrenden Terme in $M^2/Q^2$ liefert folgende
Beitr"age:
\begin{align}
\label{eq:Ffchhgen}
  \lefteqn{F_\fc^{\text{(h-h)}}(n_1,n_2,n_3,n_4,n_5) =
  \left(\frac{M^2}{Q^2}\right)^{2\eps}
  e^{-n i\pi} \, e^{2\eps\gamma_E}
  } \quad
  \nonumber \\* & \times
  \frac{\Gamma(\frac d2-n_{13}) \Gamma(\frac d2-n_4) \Gamma(\frac d2-n_5)
    \Gamma(d-n_{2345})
    \Gamma(n_{45}-\frac d2) \Gamma(n_{12345}-d)}
    {\Gamma(n_1) \Gamma(n_4) \Gamma(n_5)
    \Gamma(d-n_{45}) \Gamma(\frac32d-n_{12345})
    \Gamma(n_{245}-\frac d2)}
  \,,
\\
\label{eq:Ffc1chgen}
  \lefteqn{F_\fc^{\text{(1c'-h)}}(n_1,n_2,n_3,n_4,n_5) =
  \left(\frac{M^2}{Q^2}\right)^{2-n_{13}+\eps}
  e^{-n i\pi} \, e^{2\eps\gamma_E}
  } \quad
  \nonumber \\* & \times
  \frac{\Gamma(\frac d2-n_1) \Gamma(\frac d2-n_4) \Gamma(\frac d2-n_5)
    \Gamma(\frac d2+n_1-n_{245})
    \Gamma(n_{13}-\frac d2) \Gamma(n_{45}-\frac d2)}
    {\Gamma(n_1) \Gamma(n_3) \Gamma(n_4) \Gamma(n_5)
    \Gamma(d-n_{45}) \Gamma(d-n_{245})}
  \,,
\\
\label{eq:Ffc2c2cgen}
  \lefteqn{F_\fc^{\text{(2c-2c)}}(n_1,n_2,n_3,n_4,n_5) =
  \left(\frac{M^2}{Q^2}\right)^{4-n_{2345}}
  \frac{e^{-n i\pi} \, e^{2\eps\gamma_E} \,
    \Gamma(\frac d2-n_4)}
    {\Gamma(n_3) \Gamma(n_4) \Gamma(n_5) \Gamma(\frac d2-n_1)}
  \MBint z
  } \quad
  \nonumber \\* & \times
  \frac{\Gamma(-z) \Gamma(n_{24}-\frac d2-z) \Gamma(n_{245}-d-z)
    \Gamma(n_3+z) \Gamma(\frac d2-n_1+z) \Gamma(\frac d2-n_2+z)}
    {\Gamma(n_2-z) \Gamma(\frac d2+z)}
  \,,
\\
\label{eq:Ffchsgen}
  \lefteqn{F_\fc^{\text{(h-s)}}(n_1,n_2,n_3,n_4,n_5) =
  } \quad
  \nonumber \\* &
  \left(\frac{M^2}{Q^2}\right)^{2-n_5+\eps}
  e^{-n i\pi} \, e^{2\eps\gamma_E} \,
  \frac{\Gamma(\frac d2-n_{13}) \Gamma(\frac d2-n_{234})
    \Gamma(n_{1234}-\frac d2) \Gamma(n_5-\frac d2)}
    {\Gamma(n_1) \Gamma(n_5) \Gamma(n_{24})
    \Gamma(d-n_{1234})}
  \,,
\\
\label{eq:Ffc1csgen}
  \lefteqn{F_\fc^{\text{(1c'-s)}}(n_1,n_2,n_3,n_4,n_5) =
  } \quad
  \nonumber \\* &
  \left(\frac{M^2}{Q^2}\right)^{4-n_{135}}
  e^{-n i\pi} \, e^{2\eps\gamma_E} \,
  \frac{\Gamma(n_1-n_{24})
    \Gamma(\frac d2-n_1)
    \Gamma(n_{13}-\frac d2) \Gamma(n_5-\frac d2)}
    {\Gamma(n_1) \Gamma(n_3) \Gamma(n_5)
    \Gamma(\frac d2-n_{24})}
  \,.
\end{align}

Diese 5~Beitr"age entsprechen exakt denen, die man im Rahmen der
\emph{Expansion by Regions} bekommt:
\[
  \renewcommand{\arraystretch}{1.2}
  \left.
  \begin{array}{rll}
    \text{(h-h):} & k \sim Q, & \ell \sim Q \\
    \text{(1c'-h):} & k_3 \;\|\; p_1, & \ell \sim Q \\
    \text{(2c-2c):} & k \;\|\; p_2, & \ell \;\|\; p_2
  \end{array}
  \qquad
  \right|
  \qquad
  \begin{array}{rll}
    \text{(h-s):} & k \sim Q, & \ell \sim M \\
    \text{(1c'-s):} & k_3 \;\|\; p_1, & \ell \sim M
  \end{array}
\]
Wie bereits aus den Gleichungen (\ref{eq:Ffchhgen})--(\ref{eq:Ffc1csgen})
ersichtlich, h"angt der f"uhrende Beitrag der jeweiligen Region "uber die
folgende Potenz von $M^2/Q^2$ ab:
\[
  \left.
  \begin{array}{rl}
    \text{(h-h):} & 2\eps \\
    \text{(1c'-h):} & 2-n_{13}+\eps \\
    \text{(2c-2c):} & 4-n_{2345}
  \end{array}
  \qquad
  \right|
  \qquad
  \begin{array}{rl}
    \text{(h-s):} & 2-n_5+\eps \\
    \text{(1c'-s):} & 4-n_{135}
  \end{array}
\]
Nur $n_2$ kann den Wert~2 annehmen, die anderen Parameter $n_1$, $n_3$,
$n_4$ und $n_5$ sind nach oben durch den Wert~1 beschr"ankt.
Deshalb sind die Regionen (h-s) und (1c'"~s) gegen"uber der (h-h)-Region
unterdr"uckt.
F"ur $n_2=2$ und $n_3=n_4=n_5=1$ ist der Beitrag der
(2c-2c)-Region von der Ordnung $(M^2/Q^2)^{-1}$.
Dieser Pol in $M^2$ wird jedoch vom jeweiligen Vorfaktor in der
Zerlegung~(\ref{eq:fczerlegung}) gek"urzt.

Die Beitr"age der harten (h-h)-Region k"onnen zum masselosen
Feynman-Diagramm zusammengesetzt werden:
\begin{multline}
  F_{v,\fc}^{\text{(h-h)}} =
    C_F^2 \left(\frac{\alpha}{4\pi}\right)^2
    \left(\frac{\mu^2}{Q^2}\right)^{2\eps} S_\eps^2
    \, \Biggl\{
  \frac{1}{\eps^3}
  + \frac{7}{2\eps^2}
  + \frac{1}{\eps} \left( -\frac{\pi^2}{6} + \frac{53}{4} \right)
  \\*
  - \frac{32}{3}\zeta_3 - \frac{7}{12}\pi^2 + \frac{355}{8}
  \Biggr\}
  + \Oc(\eps)
  \,.
\end{multline}
Dieses Ergebnis stimmt mit
\cite{Gonsalves:1983nq,Kramer:1986sg,Matsuura:1988sm} "uberein.

Von allen Beitr"agen zum massiven Diagramm weist nur die (2c-2c)-Region ein
Mellin-Barnes-Integral auf, das jedoch auf die "ubliche Weise berechnet
werden kann.
Singularit"aten im Parameterraum zwischen den Regionen treten in der
f"uhrenden Ordnung in $M^2/Q^2$ nicht auf.
Die (2c-2c)-Region besitzt einen Beitrag, der zu $\Gamma(n_{24}-n_1)$
proportional ist. Dieser kann nur f"ur $n_{24}\le1$ singul"ar sein.
Dann aber ist die (2c-2c)-Region durch $M^2/Q^2$ unterdr"uckt, ebenso wie
die (1c'-s)-Region, die diese Singularit"at kompensiert.

Neben der Auswertung der obigen Regionen wurden sowohl das massive als auch
das masselose Diagramm mit Fermion-Selbstenergie zus"atzlich durch
Reduktion der Integrale auf einfachere Topologien mittels partieller
Integration berechnet.
Die Resultate stimmen jeweils "uberein.

Die Ergebnisse der einzelnen skalaren Integrale~$F_\fc$ sind im
Anhang~\ref{sec:Skalarfc} in Gl.~(\ref{eq:fcergskalar}) aufgelistet.
Der gesamte Beitrag des Diagramms mit Fermion-Selbstenergie
ergibt sich in f"uhrender Ordnung in $M^2/Q^2$ wie folgt:
\begin{align}
\label{eq:fcerg}
  F_{v,\fc} &=
    C_F^2
    \left(\frac{\alpha}{4\pi}\right)^2
    \left(\frac{\mu^2}{M^2}\right)^{2\eps} S_\eps^2
    \, \Biggl\{
  -\frac{1}{2\eps^2}
  + \frac{1}{\eps} \left[
    \lqm^2 - 3 \lqm + \frac{2}{3}\pi^2 + \frac{13}{4} \right]
  \nonumber \\* & \qquad
  - \lqm^3
  + 5 \lqm^2
  - \frac{33}{2} \lqm
  - 8\zeta_3 - \frac{\pi^2}{4} + \frac{171}{8}
  \Biggr\}
  + \Oc(\eps) + \Oc\!\left(\frac{M^2}{Q^2}\right)
  .
\end{align}

%
%
\section{Selbstenergiekorrekturen}
\label{sec:abelschfcorr}

F"ur die Renormierung der Fermion-Feldst"arke gem"a"s dem LSZ-Theorem
werden die Selbstenergiekorrekturen an den "au"seren Fermionlinien
ben"otigt.
Diese Korrekturen haben die Form
\begin{equation}
  \tilde\Sigma = -i \dslash p \, \Sigma(p^2)
  \,,
\end{equation}
wobei $p = p_{1,2}$ mit $p^2=0$ der externe Fermionimpuls ist.
Aus der Amplitude~$\tilde\Sigma$ der Selbstenergiekorrektur l"asst sich die
skalare Funktion~$\Sigma$ durch Projektion extrahieren
(ohne Ber"ucksichtigung der Einheitsmatrix im $SU(2)$-Isospinraum):
\begin{equation}
\label{eq:fcorrproj}
  \Sigma = \frac{i}{4p^2} \, \Tr(\dslash p \, \tilde\Sigma)
  \,.
\end{equation}
Die Selbstenergiekorrektur wird f"ur $p^2=0$ ben"otigt. Allerdings tritt in
der Projektion~(\ref{eq:fcorrproj}) der Faktor $1/p^2$ auf.
Deshalb muss die Auswertung zun"achst bei einem infinitesimalen, aber
endlichen Wert von $p^2$ geschehen. Die Pole in $p^2$ fallen im Gesamtergebnis
heraus.
Der Limes $p^2 \to 0$ muss vor dem Limes $\eps \to 0$ der dimensionalen
Regularisierung vollzogen werden, um die Selbstenergie beim Wert $p^2=0$ zu
erhalten.
Die Limites $p^2 \to 0$ und $\eps \to 0$ vertauschen bei Feynman-Diagrammen,
die f"ur $p^2=0$ infrarot-konvergent sind, was f"ur die
Selbstenergiediagramme der Fall ist. Allerdings k"onnen einzelne skalare
Diagramme infrarote Singularit"aten aufweisen, so dass die beiden Limites
f"ur sie nicht vertauschen und die Unterschiede zwischen den Reihenfolgen
der Limites nur in der Summe aller Beitr"age herausfallen.

In der vorliegenden Arbeit wurde konsequent die Entwicklung um $p^2=0$
vor der Entwicklung um $\eps=0$ durchgef"uhrt.
Dann entspricht die Entwicklung in $p^2$ einer naiven Taylor-Entwicklung
mit lediglich ganzzahligen Potenzen von $p^2$ und ohne Logarithmen
$\ln(p^2)$.

%
%
\subsection{Selbstenergiekorrektur mit T1-Topologie}
\label{sec:T1}

\begin{figure}[ht]
  \center
  \vcentergraphics{fcorr-T1}
  \hspace{1cm}
  \vcentergraphics{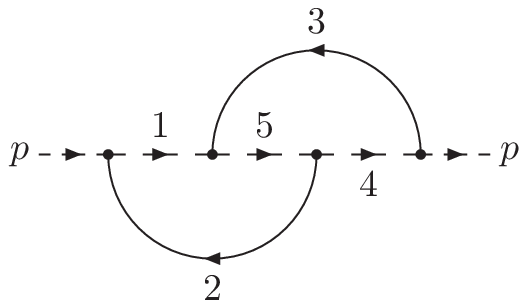}
  \caption{Selbstenergiekorrektur mit T1-Topologie (mit skalarem Diagramm)}
  \label{fig:T1}
\end{figure}
Die Selbstenergiekorrektur mit T1-Topologie ist in Abb.~\ref{fig:T1}
zusammen mit dem skalaren Diagramm dargestellt.
Die Amplitude dieses Diagramms lautet wie folgt:
\begin{multline}
\label{eq:T1amplitude}
  \tilde\Sigma_\Tone =
  \mu^{4\eps} \loopint dk \loopint d\ell \,
  ig \gamma^\rho t^b \, \frac{i\dslash k_4}{k_4^2} \,
  ig \gamma^\nu t^a \, \frac{i\dslash k_5}{k_5^2} \,
  ig \gamma_\rho t^b \, \frac{i\dslash k_1}{k_1^2} \,
  ig \gamma_\nu t^a
  \\* \times
  \frac{-i}{k_2^2-M^2} \, \frac{-i}{k_3^2-M^2}
  \,,
\end{multline}
mit den Impulsen
\begin{equation}
\label{eq:T1mom}
  k_1 = k + p ,\quad k_2 = k ,\quad
  k_3 = \ell ,\quad k_4 = \ell + p ,\quad
  k_5 = k + \ell + p .
\end{equation}
Der Farbfaktor dieses Diagramms ist der gleiche wie der des
nichtplanaren Vertexdiagramms:
\[
  t^b t^a t^b t^a = \left( C_F^2 - \frac{1}{2} C_F C_A \right) \unity
  \,.
\]

Die skalare Funktion~$\Sigma_\Tone$ ergibt sich aus der
Amplitude~$\tilde\Sigma_\Tone$ entsprechend Gl.~(\ref{eq:fcorrproj}).
Zum K"urzen von Skalarprodukten im Z"ahler von $\Sigma_\Tone$ mit
Propagatornennern dienen die Umformungen
\begin{equation}
\label{eq:T1redskalar}
  \begin{aligned}
    k_2 \cdot p &= +\tfrac12 (k_1^2 - k_2^2 - p^2) \,,\quad &
    k_2 \cdot k_4 &= +\tfrac12 (k_5^2 - k_2^2 - k_4^2) \,, \\
    k_3 \cdot p &= +\tfrac12 (k_4^2 - k_3^2 - p^2) \,,\quad &
    k_3 \cdot k_1 &= +\tfrac12 (k_5^2 - k_3^2 - k_1^2) \,.
  \end{aligned}
\end{equation}
Mit (\ref{eq:T1mom}) und (\ref{eq:T1redskalar}) k"onnen alle Skalarprodukte
gek"urzt werden.
Die skalaren Zweischleifenintegrale wurden folgenderma"sen definiert:
\begin{multline}
\label{eq:T1skalar}
  B_\Tone(p^2;n_1,n_2,n_3,n_4,n_5) =
  e^{2\eps\gamma_E} \, (M^2)^{n-4+2\eps}
  \loopintf dk \loopintf d\ell
  \\* \times
  \frac{1}
    {((k+p)^2)^{n_1} \, (k^2 - M^2)^{n_2} \,
    (\ell^2 - M^2)^{n_3} \,
    ((\ell+p)^2)^{n_4} \, ((k+\ell+p)^2)^{n_5}}
  \,,
\end{multline}
mit der Notation aus Gl.~(\ref{eq:LAskalar})
und $n = n_{12345}$.
Die Integrale wurden so definiert, dass sie in $d=4-2\eps$
Raum-Zeit-Dimensionen keine Massendimension aufweisen.
F"ur die Auswertung des Selbstenergiediagramms bei infinitesimalem $p^2$
wurde
\[
  B_\Tone(p^2;n_1,n_2,n_3,n_4,n_5) =
  \Bm_\Tone(n_1,n_2,n_3,n_4,n_5)
  + \frac{p^2}{M^2} \, \DBm_\Tone(n_1,n_2,n_3,n_4,n_5)
  + \ldots
\]
mit
\begin{align}
\label{eq:T1skalar0}
  \Bm_\Tone(n_1,n_2,n_3,n_4,n_5) &=
    B_\Tone(p^2;n_1,n_2,n_3,n_4,n_5) \Big|_{p^2=0}
  \,,\\
\label{eq:T1skalarD0}
  \DBm_\Tone(n_1,n_2,n_3,n_4,n_5) &=
    M^2 \, \frac{\partial}{\partial p^2} B_\Tone(p^2;n_1,n_2,n_3,n_4,n_5)
    \Big|_{p^2=0}
\end{align}
definiert.
Die Zerlegung von $\Sigma_\Tone$ in skalare Integrale (\ref{eq:T1skalar0})
und (\ref{eq:T1skalarD0}) ist im Anhang~\ref{sec:SkalarT1} in
Gl.~(\ref{eq:T1zerlegung}) dargestellt.
Durch Ausnutzung der Symmetrie
\begin{equation}
\label{eq:T1sym}
  B_\Tone(p^2;n_1,n_2,n_3,n_4,n_5) = B_\Tone(p^2;n_4,n_3,n_2,n_1,n_5)
\end{equation}
wurden skalare Integrale mit $n_1<n_4$ oder $n_1=n_4$ und $n_2<n_3$
durch ihre symmetrischen Partner ersetzt.

Zur Auswertung der skalaren Integrale wurde (\ref{eq:T1skalar}) zun"achst
f"ur allgemeines~$p^2$ mit Schwinger-Parametern~$\alpha_i$ und
Mellin-Barnes-Darstellungen geschrieben:
\begin{align}
  \lefteqn{B_\Tone(p^2;n_1,n_2,n_3,n_4,n_5) =
  \frac{e^{-n i\pi} \, e^{2\eps\gamma_E}}
    {\Gamma(n_1) \Gamma(n_2) \Gamma(n_3) \Gamma(n_4) \Gamma(n_5)}
  } \quad
  \nonumber \\* & \times
  \MBint{z_1} \MBint{z_2}
  \left(\frac{-p^2}{M^2}\right)^{z_1} \,
  \Gamma(-z_1) \Gamma(-z_2) \Gamma(n_{12345}-d+z_1+z_2)
  \nonumber \\* & \times
  \left(\prod_{i=1}^5\!\int_0^\infty\!\dd\alpha_i\right)
    \delta\!\left(\sum_{j\in S} \alpha_j - 1\right)
    \alpha_1^{n_1-1} \alpha_2^{n_2+z_2-1}
    \alpha_3^{d-n_{1245}-z_1-z_2-1}
    \alpha_4^{n_4-1} \alpha_5^{n_5-1}
  \nonumber \\* & \times
  \Bigl( (\alpha_1+\alpha_2)(\alpha_3+\alpha_4+\alpha_5)
    + (\alpha_3+\alpha_4)\alpha_5 \Bigr)^{-\frac d2-z_1}
  \,
  \Bigl( \alpha_1 \alpha_2 \alpha_3 + \alpha_1 \alpha_2 \alpha_4
  \nonumber \\* & \qquad\quad
    + \alpha_1 \alpha_3 \alpha_4 + \alpha_2 \alpha_3 \alpha_4
    + \alpha_1 \alpha_2 \alpha_5 + \alpha_2 \alpha_3 \alpha_5
    + \alpha_1 \alpha_4 \alpha_5 + \alpha_3 \alpha_4 \alpha_5 \Bigr)^{z_1}
  \,,
\end{align}
wobei die Menge~$S$ der Indizes, "uber die in der
$\delta$-Funktion summiert wird, eine beliebige, nicht leere Teilmenge der
5~Indizes sein darf: $\emptyset \ne S \subseteq \{1,2,3,4,5\}$.
Die Entwicklung um $p^2=0$ erh"alt man aus den Residuen des
$z_1$-Integrals:
\[
  \Res\big|_{z_1=0} \longrightarrow (-2\pi i) \, \Bm_\Tone
  \,,\quad
  \Res\big|_{z_1=1} \longrightarrow (-2\pi i) \, \frac{p^2}{M^2} \,
    \DBm_\Tone \,.
\]
Die Integrale "uber die Schwinger-Parameter k"onnen anschlie"send gel"ost
werden.
F"ur $\Bm_\Tone$ ergibt sich
\begin{align}
\label{eq:B0T1gen}
  \lefteqn{\Bm_\Tone(n_1,n_2,n_3,n_4,n_5) =
  \frac{e^{-n i\pi} \, e^{2\eps\gamma_E} \,
    \Gamma(\frac d2-n_5)}
    {\Gamma(n_2) \Gamma(n_3) \Gamma(n_5) \Gamma(\frac d2)}
  \, \MBint z \,
  \Gamma(-z) \Gamma(\tfrac d2-n_{12}-z) 
  } \quad
  \nonumber \\* & \times
  \frac{\Gamma(d-n_{1245}-z)}
    {\Gamma(d-n_{125}-z)}
  \,
  \frac{\Gamma(n_2+z) \Gamma(n_{125}-\frac d2+z) \Gamma(n_{12345}-d+z)}
    {\Gamma(n_{12}+z)}
  \,.
\end{align}

Das Ergebnis f"ur $\DBm_\Tone$ kann durch die Funktion $\Bm_\Tone$ mit
erh"ohter Raum-Zeit-Dimension und erh"ohten Propagator-Potenzen
ausgedr"uckt werden:
\begin{multline}
\label{eq:DBT1gen}
  \DBm_\Tone(n_1,n_2,n_3,n_4,n_5) =
  - \mathbf{d^{++}} \,
  \Bigl( \mathbf{1^+} \mathbf{2^+} \mathbf{3^+}
    + \mathbf{1^+} \mathbf{2^+} \mathbf{4^+}
    + \mathbf{1^+} \mathbf{3^+} \mathbf{4^+}
    + \mathbf{2^+} \mathbf{3^+} \mathbf{4^+}
  \\*
    + \mathbf{1^+} \mathbf{2^+} \mathbf{5^+}
    + \mathbf{2^+} \mathbf{3^+} \mathbf{5^+}
    + \mathbf{1^+} \mathbf{4^+} \mathbf{5^+}
    + \mathbf{3^+} \mathbf{4^+} \mathbf{5^+} \Bigr)
  \, \Bm_\Tone(n_1,n_2,n_3,n_4,n_5)
  \,,
\end{multline}
wobei die Operatoren $\mathbf{d^{++}}$ und $\mathbf{i^+}$ ($i = 1,\ldots,5$)
entsprechend folgender Definition wirken:
\begin{align}
  \mathbf{d^{++}} \, \Bm_\Tone(n_1,n_2,n_3,n_4,n_5) &=
    \Bm_\Tone(n_1,n_2,n_3,n_4,n_5) \big|_{d \to d+2}
  \,, \\
  \mathbf{i^+} \, \Bm_\Tone(\ldots,n_i,\ldots) &=
    n_i \, \Bm_\Tone(\ldots,n_i+1,\ldots)
  \,.
\end{align}

Die Auswertung des Mellin-Barnes-Integrals in $\Bm_\Tone$ und $\DBm_\Tone$
erfolgte mit den bereits beschriebenen Methoden.
Zus"atzlich wurde die Selbstenergiekorrektur auch durch Reduktion auf
einfachere Topologien mittels partieller Integration berechnet.
Beide Methoden liefern identische Resultate.
Die Ergebnisse der einzelnen skalaren Integrale $\Bm_\Tone$ und
$\DBm_\Tone$ finden sich im Anhang~\ref{sec:SkalarT1} in
Gl.~(\ref{eq:T1ergskalar}).
Der gesamte Beitrag des Selbstenergiediagramms mit T1-Topologie lautet:
\begin{align}
\label{eq:T1erg}
  \Sigma_\Tone &=
    \left( C_F^2 - \frac{1}{2} C_F C_A \right)
    \left(\frac{\alpha}{4\pi}\right)^2
    \left(\frac{\mu^2}{M^2}\right)^{2\eps} S_\eps^2
    \left(
  -\frac{1}{\eps^2}
  + \frac{3}{2\eps}
  - \frac{\pi^2}{2} + \frac{7}{4}
  \right)
  + \Oc(\eps)
  \,.
\end{align}

%
%
\subsection{Selbstenergiekorrektur mit T2-Topologie}
\label{sec:T2}

\begin{figure}[ht]
  \center
  \includegraphics{fcorr-T2}
  \hspace{1cm}
  \includegraphics{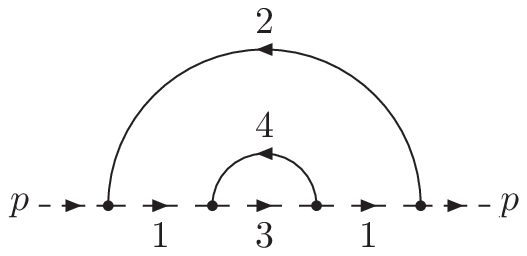}
  \caption{Selbstenergiekorrektur mit T2-Topologie (mit skalarem Diagramm)}
  \label{fig:T2}
\end{figure}
Die Selbstenergiekorrektur mit T2-Topologie ist in Abb.~\ref{fig:T2}
zusammen mit dem skalaren Diagramm dargestellt.
Die Amplitude des Diagramms lautet wie folgt:
\begin{multline}
\label{eq:T2amplitude}
  \tilde\Sigma_\Ttwo =
  \mu^{4\eps} \loopint dk \loopint d\ell \,
  ig \gamma^\nu t^a \, \frac{i\dslash k_1}{k_1^2} \,
  ig \gamma^\rho t^b \, \frac{i\dslash k_3}{k_3^2} \,
  ig \gamma_\rho t^b \, \frac{i\dslash k_1}{k_1^2} \,
  ig \gamma_\nu t^a
  \\* \times
  \frac{-i}{k_2^2-M^2} \, \frac{-i}{k_4^2-M^2}
  \,,
\end{multline}
mit den Impulsen
\begin{equation}
\label{eq:T2mom}
  k_1 = k + p ,\quad k_2 = k ,\quad
  k_3 = k + \ell + p ,\quad k_4 = \ell .
\end{equation}
Der Farbfaktor dieses Diagramms ist der gleiche wie der des
planaren Vertexdiagramms:
\[
  t^a t^b t^b t^a = C_F^2 \, \unity
  \,.
\]

Die skalare Funktion~$\Sigma_\Ttwo$ ergibt sich aus der
Amplitude~$\tilde\Sigma_\Ttwo$ entsprechend Gl.~(\ref{eq:fcorrproj}).
Zum K"urzen von Skalarprodukten im Z"ahler von $\Sigma_\Ttwo$ mit
Propagatornennern k"onnen die Relationen
\begin{equation}
\label{eq:T2redskalar}
  \begin{aligned}
    k_1 \cdot k_3 &= -\tfrac12 (k_4^2 - k_1^2 - k_3^2) \,,\quad &
    k_1 \cdot p &= -\tfrac12 (k_2^2 - k_1^2 - p^2)
  \end{aligned}
\end{equation}
verwendet werden.
Das Skalarprodukt $k_3 \cdot p$ bleibt formal als irreduzibel zur"uck.
Es kann jedoch mittels Tensorreduktion entsprechend
(\ref{eq:2punkttensorred1}) und (\ref{eq:2punkttensorred2}) eliminiert
werden.
Die Definition der skalaren Zweischleifenintegrale lautet folgenderma"sen:
\begin{multline}
\label{eq:T2skalar}
  B_\Ttwo(p^2;n_1,n_2,n_3,n_4) =
  e^{2\eps\gamma_E} \, (M^2)^{n-4+2\eps}
  \\* \times
  \loopintf dk \loopintf d\ell \,
  \frac{1}
    {((k+p)^2)^{n_1} \, (k^2 - M^2)^{n_2} \,
    ((k+\ell+p)^2)^{n_3} \, (\ell^2 - M^2)^{n_4}} \,
  \,,
\end{multline}
mit der Notation aus Gl.~(\ref{eq:LAskalar})
und $n = n_{1234}$.
F"ur die Entwicklung um $p^2=0$ wurden $\Bm_\Ttwo(n_1,n_2,n_3,n_4)$
und $\DBm_\Ttwo(n_1,n_2,n_3,n_4)$ in Analogie zu (\ref{eq:T1skalar0}) und
(\ref{eq:T1skalarD0}) definiert.
Die Zerlegung von $\Sigma_\Ttwo$ in skalare Integrale $\Bm_\Ttwo$ und
$\DBm_\Ttwo$ findet sich im Anhang~\ref{sec:SkalarT2} in
Gl.~(\ref{eq:T2zerlegung}).

F"ur allgemeines $p^2$ l"asst sich (\ref{eq:T2skalar}) als zweifaches
Mellin-Barnes-Integral schreiben:
\begin{align}
  \lefteqn{B_\Ttwo(p^2;n_1,n_2,n_3,n_4) =
  e^{-n i\pi} \, e^{2\eps\gamma_E} \,
  \frac{\Gamma(\frac d2-n_3)}
    {\Gamma(n_2) \Gamma(n_3) \Gamma(n_4)}
  \MBint{z_1} \MBint{z_2}
  \left(\frac{-p^2}{M^2}\right)^{z_1}
  } \quad
  \nonumber \\* & \times
  \frac{\Gamma(-z_1)}{\Gamma(\frac d2+z_1)} \,
  \Gamma(-z_2) \Gamma(\tfrac d2-n_2-z_2) \,
  \frac{\Gamma(\frac d2-n_{12}-z_1-z_2)}
    {\Gamma(\frac d2-n_2-z_1-z_2)}
  \nonumber \\* & \times
  \frac{\Gamma(n_2+z_1+z_2) \Gamma(n_{123}-\frac d2+z_1+z_2)
    \Gamma(n_{1234}-d+z_1+z_2)}
    {\Gamma(n_{12}+z_1+z_2)}
  \,.\qquad\quad
\end{align}
Wie in Abschnitt~\ref{sec:T1} erh"alt man die Entwicklung um $p^2=0$
aus den Residuen des $z_1$-Integrals bei $z_1=0$ und $z_1=1$:
\begin{align}
\label{eq:B0T2gen}
  \lefteqn{\Bm_\Ttwo(n_1,n_2,n_3,n_4) =
  \frac{e^{-n i\pi} \, e^{2\eps\gamma_E} \,
    \Gamma(\frac d2-n_3)}
    {\Gamma(n_2) \Gamma(n_3) \Gamma(n_4) \Gamma(\frac d2)}
  \, \MBint z \,
  \Gamma(-z) \Gamma(\tfrac d2-n_{12}-z)
  } \quad
  \nonumber \\* & \times
  \frac{\Gamma(n_2+z) \Gamma(n_{123}-\frac d2+z) \Gamma(n_{1234}-d+z)}
    {\Gamma(n_{12}+z)}
  \,,
\\
\label{eq:DBT2gen}
  \lefteqn{\DBm_\Ttwo(n_1,n_2,n_3,n_4) =
  \frac{e^{-n i\pi} \, e^{2\eps\gamma_E} \,
    \Gamma(\frac d2-n_3)}
    {\Gamma(n_2) \Gamma(n_3) \Gamma(n_4) \Gamma(1+\frac d2)}
  \, \MBint z \,
  \Gamma(-z) \Gamma(\tfrac d2-n_{12}-1-z)
  } \quad
  \nonumber \\* & \times
  \frac{\Gamma(\frac d2-n_2-z)}{\Gamma(\frac d2-n_2-1-z)} \,
  \frac{\Gamma(n_2+1+z) \Gamma(n_{123}+1-\frac d2+z) \Gamma(n_{1234}+1-d+z)}
    {\Gamma(n_{12}+1+z)}
  \,.
\end{align}

Neben der Auswertung dieser Mellin-Barnes-Integrale wurde die
Selbstenergiekorrektur mit T2-Topologie auch durch Reduktion auf einfachere
Topologien mittels partieller Integration berechnet.
Die Resultate der beiden Methoden stimmen "uber\-ein, und
die Ergebnisse der einzelnen skalaren Integrale sind im
Anhang~\ref{sec:SkalarT2} in Gl.~(\ref{eq:T2ergskalar}) aufgelistet.
Die gesamte Selbstenergiekorrektur des Diagramms mit T2"~Topologie lautet:
\begin{align}
\label{eq:T2erg}
  \Sigma_\Ttwo &=
    C_F^2 \left(\frac{\alpha}{4\pi}\right)^2
    \left(\frac{\mu^2}{M^2}\right)^{2\eps} S_\eps^2
    \left(
  \frac{1}{2\eps^2}
  - \frac{1}{4\eps}
  - \frac{\pi^2}{12} + \frac{7}{8}
  \right)
  + \Oc(\eps)
  \,.
\end{align}

%
%
\subsection{Sonstige Beitr"age mit Selbstenergiekorrekturen}
\label{sec:ecorrsonst}

\begin{figure}[ht]
  \center
  \includegraphics{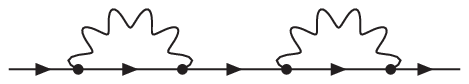}
  \caption{Selbstenergiekorrektur mit Einschleifentopologie}
  \label{fig:fcorr1x1}
\end{figure}

Ein weiteres Selbstenergiediagramm tr"agt zum Zweischleifenformfaktor bei.
Es ist in Abb.~\ref{fig:fcorr1x1} dargestellt.
Dieses Diagramm ist jedoch lediglich ein Produkt aus
Einschleifendiagrammen.
Seine Amplitude ist
\begin{equation}
  \tilde\Sigma_\oxo
  = \tilde\Sigma_1 \, \frac{i \dslash p}{p^2} \, \tilde\Sigma_1
  = -i \dslash p \, (\Sigma_1)^2
  \,,
\end{equation}
mit der Einschleifen-Selbstenergiekorrektur
$\tilde\Sigma_1 = -i \dslash p \, \Sigma_1$ (\ref{eq:Sigma1erg}).
Der Beitrag zum Formfaktor ist also
\begin{equation}
  \Sigma_\oxo = (\Sigma_1)^2
  \,.
\end{equation}

Aus der Feldst"arkerenormierung ergibt sich gem"a"s Gl.~(\ref{eq:LSZF2})
ein weiterer Beitrag:
das Produkt aus der Einschleifen-Vertexkorrektur $F_{1,v}$~(\ref{eq:Fv1erg})
mit der Einschleifen"=Selbstenergiekorrektur $\Sigma_1$.
Insgesamt erh"alt man folgende weitere Beitr"age:
\begin{equation}
  (\Sigma_1)^2 + F_{1,v} \, \Sigma_1
  = \bigl( F_{1,v} + \Sigma_1 \bigr) \, \Sigma_1
  = F_1 \, \Sigma_1
  \,,
\end{equation}
da sich die beiden Einschleifenkorrekturen gerade zum
Einschleifenformfaktor $F_1$ addieren.
Weil $F_1$ in $d=4$ Raum-Zeit-Dimensionen endlich ist, gen"ugt f"ur
$\Sigma_1$ im zweiten Faktor des Produkts die Ordnung~$\eps^0$.
Aber da $\Sigma_1$ einen $1/\eps$-Pol enth"alt, wird von $F_1$ auch die
lineare Ordnung in $\eps$ ben"otigt.
Mit den Ergebnissen aus (\ref{eq:F1erg}) und (\ref{eq:Sigma1erg})
erh"alt man:
\begin{multline}
\label{eq:F1Sigma1erg}
  F_1 \, \Sigma_1 =
    C_F^2
    \left(\frac{\alpha}{4\pi}\right)^2
    \left(\frac{\mu^2}{M^2}\right)^{2\eps} S_\eps^2
    \, \Biggl\{
  \frac{1}{\eps} \left[
    \lqm^2 - 3 \lqm + \frac{2}{3}\pi^2 + \frac{7}{2} \right]
  \\*
  - \frac{1}{3} \lqm^3
  + \lqm^2
  + \left(\frac{\pi^2}{3} - \frac{13}{2}\right) \lqm
  - 2\zeta_3 - \frac{5}{6}\pi^2 + 10
  \Biggr\}
  + \Oc(\eps) + \Oc\!\left(\frac{M^2}{Q^2}\right)
  .
\end{multline}

%
%
\section{Zusammenfassung der abelschen Beitr"age}

Die abelschen Zweischleifenbeitr"age sind nun komplett.
Dies sind die Beitr"age zum Formfaktor mit dem Farbfaktor $C_F^2$,
der in einem abelschen $U(1)$-Modell mit $C_F=1$ gerade eins ist.
Drei der vorgestellten Diagramme, das nichtplanare Vertexdiagramm
(Abschnitt~\ref{sec:NP}), das Vertexdiagramm mit Benz-Topologie
(Abschnitt~\ref{sec:BE}) und die Selbstenergiekorrektur mit T1-Topologie
(Abschnitt~\ref{sec:T1}) haben einen Farbfaktor, in dem auch $C_A$
vorkommt.
Wie in einer abelschen Theorie wird zun"achst $C_A=0$ gesetzt, um die
abelschen Beitr"age separat zu betrachten.
Die nichtabelschen Terme mit dem Faktor $C_F C_A$ werden in
Kapitel~\ref{chap:nichtabelsch} durch weitere Beitr"age erg"anzt.

Die abelschen Beitr"age zum Zweischleifenformfaktor setzen sich
folgenderma"sen zusammen:
\begin{equation}
  F_{2,C_F^2} =
  F_{v,\LA} + F_{v,\NP}|_{C_F^2} + 2\,F_{v,\BE}|_{C_F^2} + 2\,F_{v,\fc}
  + \Sigma_\Tone|_{C_F^2} + \Sigma_\Ttwo
  + F_1 \, \Sigma_1
  \,,
\end{equation}
wobei ber"ucksichtigt wurde, dass die Vertexdiagramme mit Benz-Topologie
und mit Fermion-Selbstenergie zweifach vorkommen.
Nach Einsetzen der Ergebnisse aus den Gleichungen (\ref{eq:LAerg}),
(\ref{eq:NPerg}), (\ref{eq:BEerg}), (\ref{eq:fcerg}), (\ref{eq:T1erg}),
(\ref{eq:T2erg}) und (\ref{eq:F1Sigma1erg})
heben sich die Pole in $\eps$ gegenseitig auf,
und der Limes $\eps \to 0$ kann vollzogen werden.
Man erh"alt den folgenden abelschen Beitrag zum Zweischleifenformfaktor im
Limes $M^2 \ll Q^2$:
\begin{multline}
\label{eq:F2CFerg}
  F_{2,C_F^2} =
  C_F^2
  \left(\frac{\alpha}{4\pi}\right)^2
  \, \Biggl\{
  \frac{1}{2} \lqm^4
  - 3 \lqm^3
  + \left(\frac{2}{3}\pi^2 + 8\right) \lqm^2
  - \Bigl(-24\zeta_3 + 4\pi^2 + 9\Bigr) \, \lqm
  \\*
  + 256\,\Li4\!\left(\frac{1}{2}\right)
  + \frac{32}{3}\ln^4{2} - \frac{32}{3}\pi^2\ln^2{2}
  - \frac{52}{15}\pi^4 + 80\zeta_3 + \frac{52}{3}\pi^2
  + \frac{25}{2}
  \Biggr\}
  \,,
\end{multline}
mit $\lqm = \ln(Q^2/M^2)$.

Die "Ubereinstimmung von (\ref{eq:F2CFerg}) mit den $C_F^2$-Termen aus der
Evolutionsgleichung in~(\ref{eq:F2NNLLSUN}) ist offensichtlich.
Die Vorhersagen der Evolutionsgleichung wurden also f"ur den abelschen Fall
best"atigt.
Der Koeffizient des linearen Logarithmus und die Konstante
in~(\ref{eq:F2CFerg}) sind ein neues Ergebnis.
Aus dem Vergleich mit der Darstellung~(\ref{eq:F2gzx}) l"asst sich zeigen,
dass der abelsche Beitrag zum Koeffizienten~$\xi^{(2)}$ verschwindet:
\begin{equation} 
  \xi^{(2)}\Big|_{C_F^2} = 0 \,.
\end{equation}
Alle anderen Terme im Koeffizienten des linearen Logarithmus
in~(\ref{eq:F2gzx}) sind bereits aus
(\ref{eq:g1z1}), (\ref{eq:x1F01}) und (\ref{eq:zeta2}) bekannt.

Die numerische Gr"o"se der Logarithmenkoeffizienten im
Ergebnis~(\ref{eq:F2CFerg}) ist wie folgt:
\begin{equation}
  F_{2,C_F^2} =
  C_F^2 \left(\frac{\alpha}{4\pi}\right)^2
    \, \Bigl(
    + 0{,}5\,\lqm^4
    - 3\,\lqm^3
    + 14{,}5797\,\lqm^2
    - 19{,}6291\,\lqm
    + 26{,}4097
    \Bigr)
    \,.
\end{equation}
Das Schema der alternierenden Koeffizienten setzt sich fort:
Vom f"uhrenden $\lqm^4$ "uber die n"achstf"uhrenden Logarithmen bis hin zur
Konstanten nimmt die Gr"o"se der Koeffizienten -- bei wechselnden
Vorzeichen -- zu.

Zur Gr"o"senabsch"atzung der Einzelterme wird wieder
$\alpha/(4\pi) = 0{,}003$, $C_F = 3/4$ f"ur die $SU(2)$-Gruppe,
$M = 80\,\GeV$ und $Q = 1000\,\GeV$ gesetzt.
Die Beitr"age der logarithmischen Terme zum Formfaktor lauten in Promille:
\begin{equation} 
  \lqm^4 \to +1{,}6 \,,\quad
  \lqm^3 \to -2{,}0 \,,\quad
  \lqm^2 \to +1{,}9 \,,\quad
  \lqm^1 \to -0{,}5 \,,\quad
  \lqm^0 \to +0{,}1
  \,.
\end{equation}
W"ahrend die Beitr"age der drei aus der Evolutionsgleichung bekannten
Logarithmen $\lqm^4$, $\lqm^3$ und $\lqm^2$ von der gleichen Gr"o"senordnung
sind, beginnt die logarithmische Reihe danach zu "`konvergieren"':
Der lineare $\lqm^1$ liefert einen wesentlich kleineren Beitrag als die
drei h"oheren Logarithmenpotenzen.
Und die Konstante $\lqm^0$, die nur noch 0,1~Promille zum Formfaktor
beitr"agt, ist vernachl"assigbar.

\begin{figure}[ht]
  \centering
  \includegraphics{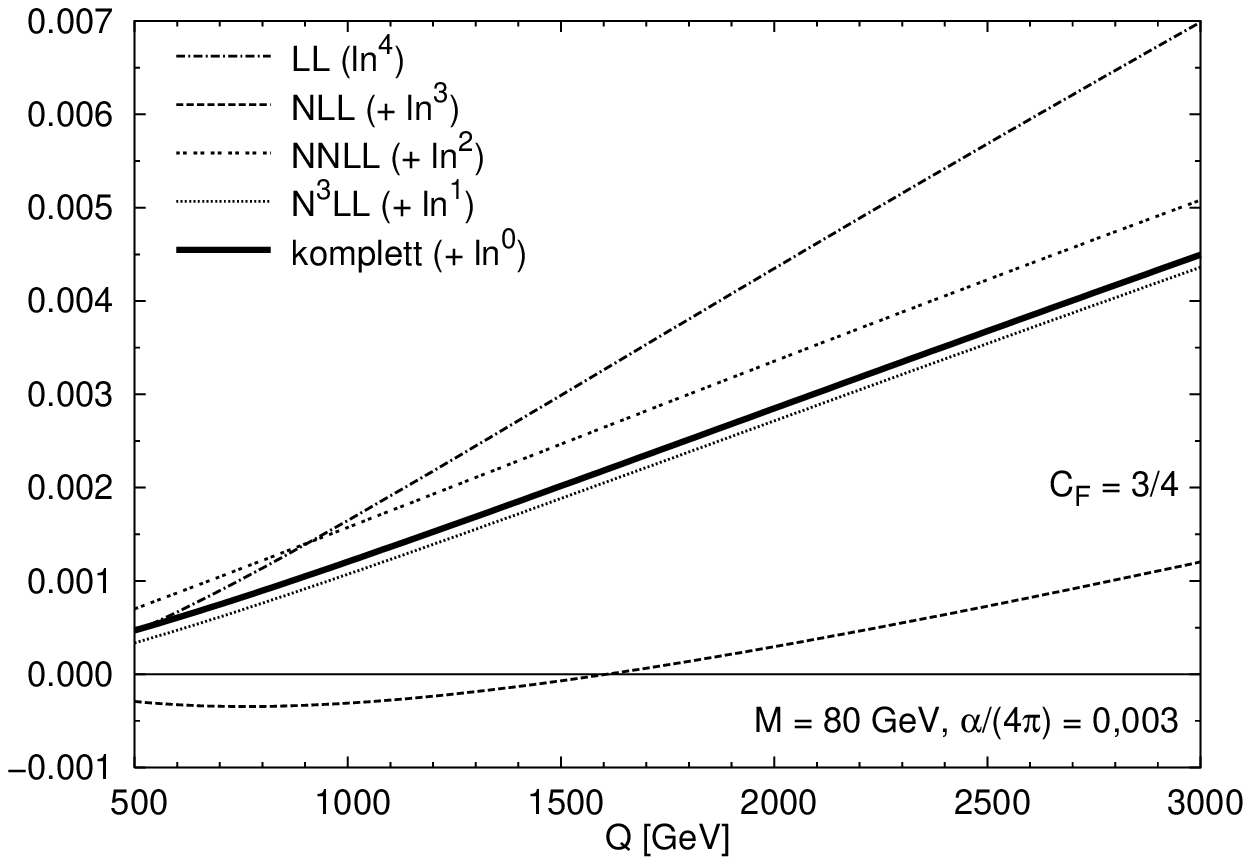}
  \caption{Abelscher Zweischleifenformfaktor in sukzessiven
    logarithmischen N"aherungen}
  \label{fig:plotabelsch}
\end{figure}

Der abelsche Zweischleifenformfaktor ist in Abb.~\ref{fig:plotabelsch}
graphisch als Funktion des Impuls"ubertrags~$Q$ dargestellt.
Dabei sind die sukzessiven logarithmischen N"aherungen gezeichnet:
die f"uhrende logarithmische N"aherung (LL) mit $\lqm^4$,
die n"achstf"uhrende (NLL) mit $\lqm^4$ und $\lqm^3$,
NNLL mit $\lqm^4$ bis $\lqm^2$,
$\NNNLL$ mit allen Logarithmen ohne die Konstante
und der kompletten abelsche Zweischleifenformfaktor mit allen Beitr"agen.

Aufgrund der gro"sen Koeffizienten vor den n"achstf"uhrenden Logarithmen
mit alternierenden Vorzeichen setzt sich das asymptotische
$\lqm^4$-Verhalten des Zweischleifenformfaktors im TeV-Energiebereich nicht
durch.
Auch der lineare Logarithmus liefert noch einen signifikanten Beitrag,
der f"ur ein Ergebnis mit Promille-Genauigkeit n"otig ist.
Allerdings kann die Konstante vernachl"assigt werden, so dass bereits die
$\NNNLL$-N"aherung, die alle logarithmischen Beitr"age umfasst, ein
ausreichend pr"azises Resultat liefert.


\clearemptypage

\chapter{Nichtabelsche Beitr"age zum Zweischleifenformfaktor}
\label{chap:nichtabelsch}

In diesem Kapitel geht es um die nichtabelschen Beitr"age zum
Zweischleifenformfaktor, der in Abschnitt~\ref{sec:Sudakov} eingef"uhrt
wurde.
Dies sind Beitr"age, die in einer nicht\-abel\-schen spontan gebrochenen
$SU(2)$-Eichtheorie
zus"atzlich zu denen eines abelschen $U(1)$-Modells (siehe
Kapitel~\ref{chap:abelsch}) auftreten.
Eine Ver"offentlichung der Ergebnisse dieses Kapitels
wurde eingereicht\cite{Jantzen:2005xi}.

Zum einen handelt es sich dabei um Diagramme, die in einer $SU(N)$-Theorie
den Farbfaktor $C_F C_A$ besitzen.
Diese sind jedoch f"ur sich gesehen nicht eichinvariant.
Die Eichbosonmasse wird "uber den Higgs-Mechanismus der spontanen
Symmetrie"-brechung eingef"uhrt. Nur wenn Diagramme mit dem Higgs-Boson und
den Goldstone-Bosonen hinzugenommen werden, ist das Gesamtergebnis
eichinvariant, d.h. es h"angt nicht mehr von der "uber den Parameter
$\xi$~(\ref{eq:Lfix}) gew"ahlten Eichung ab.
Die folgenden Rechnungen werden in der Feynman-'t~Hooft-Eichung, also mit
$\xi=1$ durchgef"uhrt.

Das nichtplanare Vertexdiagramm (Abschnitt~\ref{sec:NP}), das abelsche
Diagramm mit Benz-Topologie (Abschnitt~\ref{sec:BE}) und die abelsche
Selbstenergie mit T1-Topologie (Abschnitt~\ref{sec:T1}) besitzen den
Farbfaktor $C_F^2-\frac12 C_F C_A$, der neben dem abelschen Anteil~$C_F^2$
auch den nichtabelschen Anteil~$C_F C_A$ aufweist. Die Beitr"age von diesen
Diagrammen m"ussen hier mit ber"ucksichtigt werden (siehe
Abschnitt~\ref{sec:sumCA}).

\begin{figure}[pht]
  \centering
  \valignbox[b]{\includegraphics[scale=0.9]{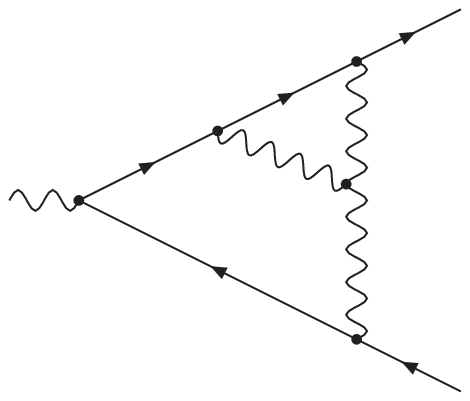}\\
    Benz-Topologie (\BECA, $2{\times}$)}
  \hfill
  \valignbox[b]{\includegraphics[scale=0.9]{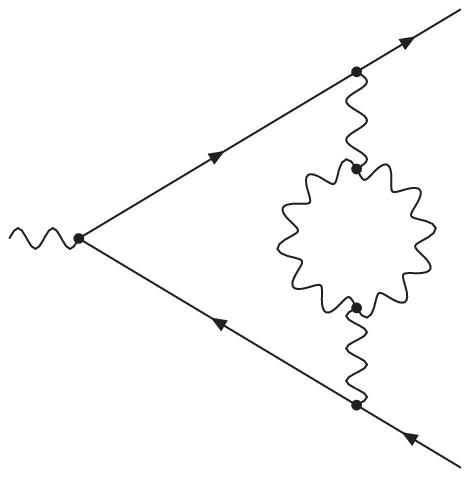}\\
    $W$-Schleife (\WW)}
  \hfill
  \valignbox[b]{\includegraphics[scale=0.9]{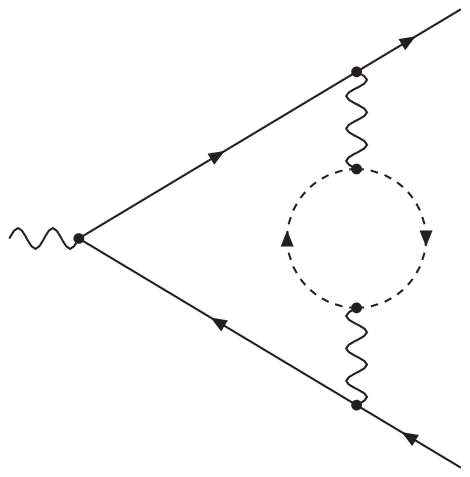}\\
    Geist-Schleife (\cc)}
  \\[2ex]
  \valignbox[b]{\includegraphics[scale=0.9]{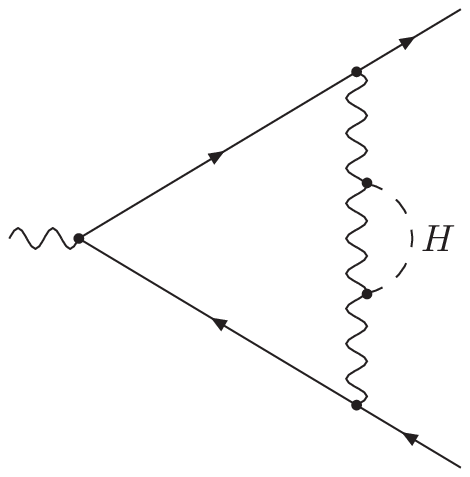}\\
    $W$-Higgs-Schleife (\WH)}
  \hfill
  \valignbox[b]{\includegraphics[scale=0.9]{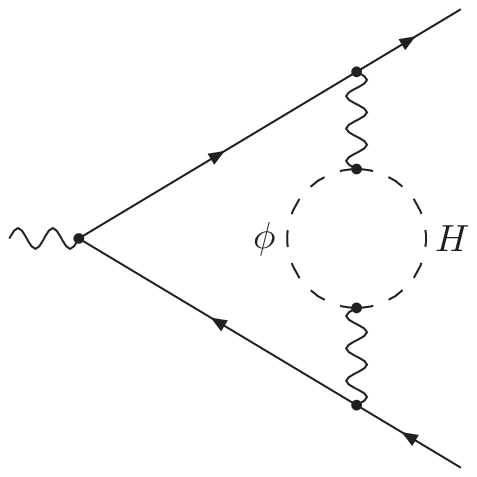}\\
    Higgs-Goldstone-Schleife (\Hphi)}
  \hfill
  \valignbox[b]{\includegraphics[scale=0.9]{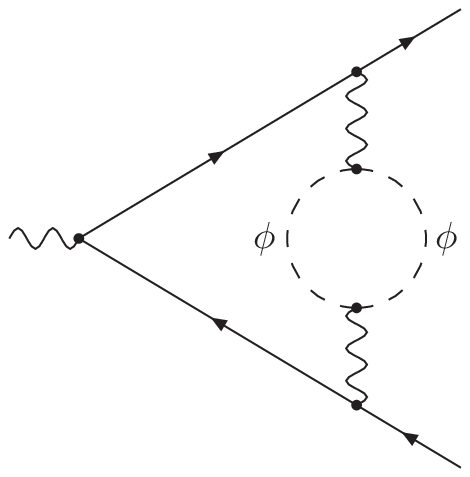}\\
    Goldstone-Schleife (\phiphi)}
  \\[2ex]
  \caption{Nichtabelsche Zweischleifen-Vertexdiagramme}
  \label{fig:nichtabelschvertex}
\end{figure}
\begin{figure}[pht]
  \centering
  \valignbox[b]{\includegraphics[scale=0.85]{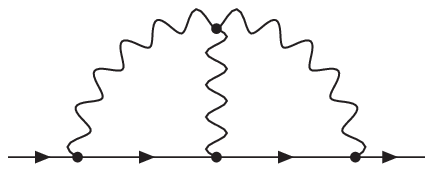}\\[-1ex]
    T1-Topologie (\ToneCA)}
  \hfill
  \valignbox[b]{\includegraphics[scale=0.85]{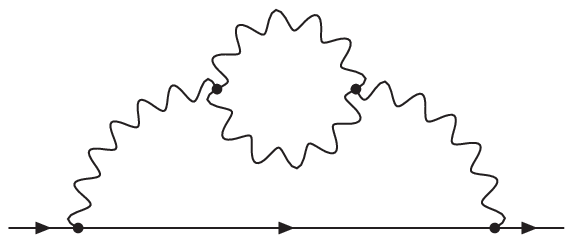}\\[-1ex]
    $W$-Schleife (\WW)}
  \hfill
  \valignbox[b]{\includegraphics[scale=0.85]{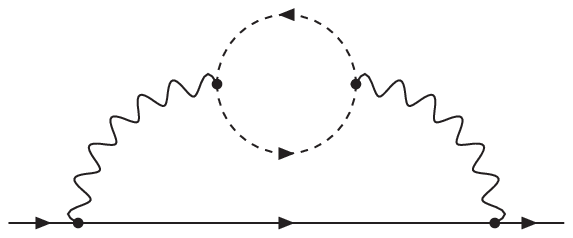}\\[-1ex]
    Geist-Schleife (\cc)}
  \\[2ex]
  \valignbox[b]{\includegraphics[scale=0.85]{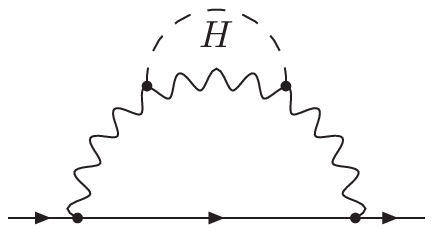}\\[-1ex]
    $W$-Higgs-Schleife (\WH)}
  \hfill
  \valignbox[b]{\includegraphics[scale=0.85]{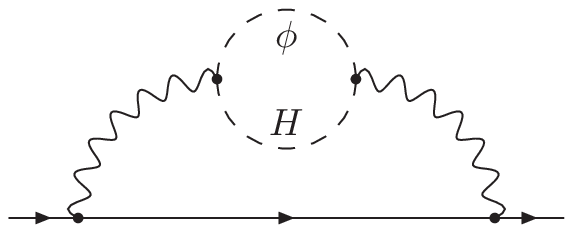}
    \\[-1ex]
    Higgs-Goldst.-Schleife (\Hphi)}
  \hfill
  \valignbox[b]{\includegraphics[scale=0.85]{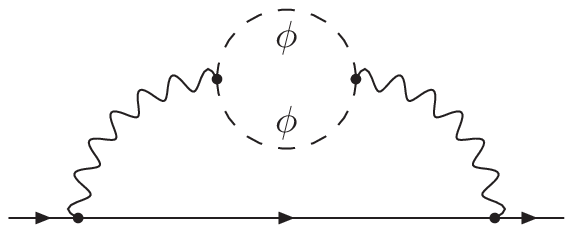}\\[-1ex]
    Goldstone-Schleife (\phiphi)}
  \\[2ex]
  \caption{Nichtabelsche Zweischleifen-Selbstenergiediagramme}
  \label{fig:nichtabelschfcorr}
\end{figure}
Dar"uber hinaus tragen die Vertexdiagramme in
Abb.~\ref{fig:nichtabelschvertex} zum nichtabelschen Formfaktor bei.
Die drei Diagramme in der ersten Reihe sind proportional zu $C_F C_A$,
w"ahrend die drei Diagramme in der zweiten Reihe Higgs-Beitr"age sind.
Die f"ur den nichtabelschen Formfaktor relevanten Selbstenergiediagramme
sind in Abb.~\ref{fig:nichtabelschfcorr} dargestellt.

Jedem Vertexdiagramm in Abb.~\ref{fig:nichtabelschvertex} entspricht genau
ein Selbstenergiediagramm in Abb.~\ref{fig:nichtabelschfcorr}. Den
Zusammenhang sieht man durch Herausnehmen des zentralen Vertex im
Vertexdiagramm.
Deshalb ist dieses Kapitel so gegliedert, dass jedes Vertexdiagramm
zusammen mit dem zugeh"origen Selbstenergiediagramm behandelt wird.
Die beiden Diagramme mit jeweils einer Drei-Eichboson-Kopplung
(\BECA{} und \ToneCA) werden in Abschnitt~\ref{sec:CA3W} vorgestellt.
Die Berechnung der Diagramme mit Eichboson- (\WW) oder Geistschleife (\cc),
die ebenfalls proportional zu $C_F C_A$ sind, wird in
Abschnitt~\ref{sec:CAWcorr} beschrieben.
In Abschnitt~\ref{sec:sumCA} werden die $C_F C_A$-Beitr"age
zusammengefasst.
Anschlie"send folgen in Abschnitt~\ref{sec:Higgs} die Beitr"age von Higgs-
und Goldstone-Bosonen (\WH, \Hphi{} und \phiphi).
Die Summe aller nichtabelschen Beitr"age wird in
Abschnitt~\ref{sec:sumnichtabelsch} diskutiert.

Zus"atzlich gibt es Vertex- und Selbstenergiediagramme, die eine
Tadpole-Schleife im Eichbosonpropagator haben, "ahnlich wie das rechte
Diagramm in Abb.~\ref{fig:bcorrns}.
Der Beitrag dieser Diagramme wird jedoch vollst"andig von der Renormierung
der Eichbosonmasse kompensiert, so dass er hier nicht betrachtet werden
muss.

Die Berechnung der Feynman-Diagramme wurde weitgehend wie im abelschen Fall
(Kapitel~\ref{chap:abelsch}) durchgef"uhrt.
Jedoch ist die Komplexit"at der nichtabelschen Diagramme dadurch erh"oht,
dass sie drei massive Propagatoren haben (Eichbosonen, Higgs- oder
Goldstone-Bosonen) und teilweise auch einer dieser drei Propagatoren
doppelt vorkommt.
Dadurch konnte die Rechnung nicht im vollst"andigen Hochenergielimes
durchgef"uhrt werden, auf die nichtlogarithmische Konstante musste
verzichtet werden.
Es hat sich jedoch bei den fermionischen und den abelschen Beitr"agen
(Kapitel \ref{chap:nfns} und \ref{chap:abelsch}) gezeigt, dass die
\NNNLL-N"aherung mit allen Logarithmen ausreicht.
Der Anteil der Konstante ist dort vernachl"assigbar.

%
%
\section[$C_F C_A$-Beitr"age mit einer Drei-Eichboson-Kopplung]
  {\boldmath $C_F C_A$-Beitr"age mit einer\\ Drei-Eichboson-Kopplung}
\label{sec:CA3W}

\subsection{Nichtabelsches Vertexdiagramm mit Benz-Topologie}

\begin{figure}[ht]
  \center
  \includegraphics{vertex-BE-CA}
  \hspace{1cm}
  \includegraphics{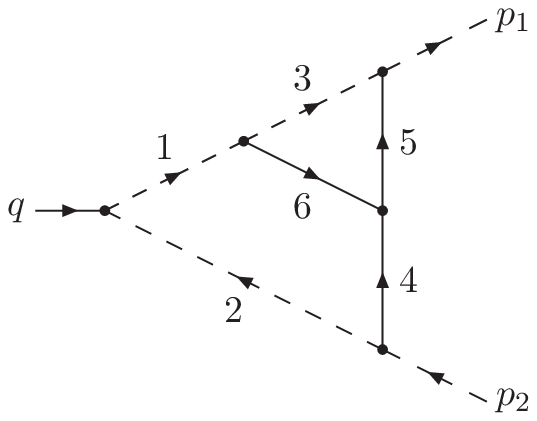}
  \caption{Nichtabelsches Vertexdiagramm mit Benz-Topologie
    (mit skalarem Diagramm)}
  \label{fig:BECA}
\end{figure}

Das nichtabelsche Vertexdiagramm mit Benz-Topologie (Bezeichnung \BECA) ist
in Abb.~\ref{fig:BECA} dargestellt.
Es kommt auch horizontal gespiegelt vor, so dass es zweifach gez"ahlt
werden muss.
Die Amplitude des Diagramms lautet:
\begin{align}
\label{eq:BECAamplitude}
  \lefteqn{\Fc_{v,\BECA}^\mu =
  \mu^{4\eps} \loopint dk \loopint d\ell \,
  ig \gamma^\nu t^a \, \frac{i\dslash k_3}{k_3^2} \,
  ig \gamma^\rho t^b \, \frac{i\dslash k_1}{k_1^2} \,
  \gamma^\mu
  \frac{i\dslash k_2}{k_2^2} \, ig \gamma^\sigma t^c \,
  \frac{-i}{k_4^2-M^2}
  } \quad
  \nonumber \\* & \times
  \frac{-i}{k_5^2-M^2} \, \frac{-i}{k_6^2-M^2} \,
  g \, f^{abc} \Bigl[
    g_{\nu\rho}(-k_5-k_6)_\sigma + g_{\rho\sigma}(k_6-k_4)_\nu
    + g_{\sigma\nu}(k_4+k_5)_\rho \Bigr]
  \,,
\end{align}
mit den Impulsen
\begin{equation}
\label{eq:BECAmom}
  k_1 = p_1 - \ell ,\quad k_2 = p_2 - \ell ,\quad
  k_3 = p_1 - k ,\quad k_4 = \ell ,\quad
  k_5 = k ,\quad k_6 = k - \ell
\end{equation}
wie in der abelschen Benz-Topologie.
Der Farbfaktor dieses Diagramms ist
\[
  f^{abc} \, t^a t^b t^c = \frac{i}{2} \, C_F C_A \, \unity
  \,.
\]

Der Beitrag der Amplitude $\Fc_{v,\BECA}^\mu$ zum Formfaktor ist
$F_{v,\BECA}$. Er wird
mit den gleichen Umformungen wie f"ur die abelsche Benz-Topologie in
Abschnitt~\ref{sec:BE} als Linearkombination von skalaren Integralen der
folgenden Definition geschrieben:
\begin{multline}
\label{eq:BECAskalar}
  F_\BECA(n_1,n_2,n_3,n_4,n_5,n_6,n_7) =
  e^{2\eps\gamma_E} \, (M^2)^{2\eps} \,
  (Q^2)^{n-n_7-4}
  \\* \times
  \loopintf dk \loopintf d\ell \,
  \frac{(2p_2 \cdot k)^{n_7}}
    {(\ell^2-2p_1\cdot\ell)^{n_1} \, (\ell^2-2p_2\cdot\ell)^{n_2} \,
      (k^2 - 2p_1\cdot k)^{n_3}}
  \\* \times
  \frac{1}{(\ell^2 - M^2)^{n_4} \, (k^2-M^2)^{n_5} \, ((k-\ell)^2-M^2)^{n_6}}
  \,,
\end{multline}
mit der Notation aus Gl.~(\ref{eq:LAskalar})
und $n = n_{123456}$ (ohne $n_7$).
Die Zerlegung von $F_{v,\BECA}$ in skalare Integrale ist im
Anhang~\ref{sec:SkalarBECA} in Gl.~(\ref{eq:BECAzerlegung}) aufgelistet.

Folgende Regionen treten im Rahmen der \emph{Expansion by Regions} auf:
\[
  \renewcommand{\arraystretch}{1.2}
  \left.
  \begin{array}{rll}
    \text{(h-h):} & k \sim Q, & \ell \sim Q \\
    \text{(1c-h):} & k \;\|\; p_1, & \ell \sim Q \\
    \text{(h-2c):} & k \sim Q, & \ell \;\|\; p_2 \\
    \text{(s'-h):} & k_6 \sim M, & \ell \sim Q
  \end{array}
  \qquad
  \right|
  \qquad
  \begin{array}{rll}
    \text{(1c-1c):} & k \;\|\; p_1, & \ell \;\|\; p_1 \\
    \text{(2c-2c):} & k \;\|\; p_2, & \ell \;\|\; p_2 \\
    \text{(1c-2c):} & k \;\|\; p_1, & \ell \;\|\; p_2
  \end{array}
\]

Die Potenz in $(M^2/Q^2)$ des f"uhrenden Beitrags jeder Region lautet
(Notation wie in Abschnitt~\ref{sec:LA}):
\[
  \left.
  \begin{array}{rl}
    \text{(h-h):} & 2\eps \\
    \text{(1c-h):} & 2-n_{35}+\eps \\
    \text{(h-2c):} & 2-n_{24}+\eps \\
    \text{(s'-h):} & 2-n_6+\eps
  \end{array}
  \qquad
  \right|
  \qquad
  \begin{array}{rl}
    \text{(1c-1c):} & 4-n_{13456} \\
    \text{(2c-2c):} & 4-n_{2456}+n_7 \\
    \text{(1c-2c):} & 4-n_{2345}
  \end{array}
\]

Die Region (s'-h) ist immer mit $M^2/Q^2$ unterdr"uckt, und die Region
(2c-2c) muss nur f"ur $n_7=0$ betrachtet werden.
Wie in der abelschen Benz-Topologie tritt f"ur $n_1=n_3=n_4=n_5=n_6=1$ der
Fall ein, dass der (1c-1c)-Beitrag von der Ordnung $(M^2/Q^2)^{-1}$ ist.
Dieser Pol in $M^2$ wird aber jeweils vom Vorfaktor der Zerlegung in skalare
Integrale gek"urzt.

Die Beitr"age der Regionen (h-h) und (h-2c) entsprechen denjenigen
des abelschen Benz-Diagramms in den Gleichungen
(\ref{eq:FBEhhgen}) und (\ref{eq:FBEh2cgen}):
\begin{align}
  F_\BECA^{\text{(h-h)}}(n_1,n_2,n_3,n_4,n_5,n_6,n_7) &=
    F_\BE^{\text{(h-h)}}(n_1,n_2,n_3,n_4,n_5,n_6,n_7) \,,
\\
  F_\BECA^{\text{(h-2c)}}(n_1,n_2,n_3,n_4,n_5,n_6,n_7) &=
    F_\BE^{\text{(h-2c)}}(n_1,n_2,n_3,n_4,n_5,n_6,n_7) \,.
\end{align}
Die f"uhrenden Beitr"age der anderen Regionen lauten:
\begin{align}
\label{eq:FBECA1chgen}
  \lefteqn{F_\BECA^{\text{(1c-h)}}(n_1,n_2,n_3,n_4,n_5,n_6,n_7) =} \quad
  \nonumber \\* &
  \left(\frac{M^2}{Q^2}\right)^{2-n_{35}+\eps}
  e^{-n i\pi} \, e^{2\eps\gamma_E} \,
  \frac{\Gamma(\frac d2-n_3) \Gamma(\frac d2-n_{146}) \Gamma(n_{35}-\frac d2)}
    {\Gamma(n_1) \Gamma(n_2) \Gamma(n_3) \Gamma(n_5) \Gamma(n_6)
    \Gamma(d-n_{1246})}
  \nonumber \\* & \times
  \MBint z \,
  \frac{\Gamma(-z) \Gamma(\frac d2-n_{246}-z)
    \Gamma(n_6+z) \Gamma(n_{37}+z) \Gamma(n_{1246}-\frac d2+z)}
    {\Gamma(\frac d2+n_7+z)}
  \,,
\\
\label{eq:FBECA1c1cgen}
  \lefteqn{F_\BECA^{\text{(1c-1c)}}(n_1,n_2,n_3,n_4,n_5,n_6,n_7) =
  \left(\frac{M^2}{Q^2}\right)^{4-n_{13456}}
  e^{-n i\pi} \, e^{2\eps\gamma_E} \,
  \sum_{i_1,i_2,i_3\ge0}^{i_{123}\le n_7}
  } \quad
  \nonumber \\* & \times
    \frac{n_7!}{i_1!\,i_2!\,i_3!\,(n_7-i_{123})!} \,
  \frac{\Gamma(n_{17}-i_{12})}
    {\Gamma(n_1) \Gamma(n_3) \Gamma(n_4) \Gamma(n_5) \Gamma(n_6)
    \Gamma(\frac d2-n_2+n_7)}
  \MBint{z_1} \MBint{z_2}
  \nonumber \\* & \times
    \Gamma(-z_1)
    \Gamma(n_5+z_1) \Gamma(\tfrac d2-n_3+n_7-i_{123}+z_1)
    \cdot
    \Gamma(n_{134}+i_{13}-\tfrac d2-z_1+z_2)
  \nonumber \\* & \times
  \frac{\Gamma(-z_2)
    \Gamma(n_4+i_1+z_2) \Gamma(n_{1467}-i_{123}-\tfrac d2+z_2)
    \Gamma(n_{13456}-d+z_2)}
    {\Gamma(n_{147}-i_2+z_2)}
  \nonumber \\* & \times
  \frac{\Gamma(\frac d2-n_{14}+i_2+z_1-z_2)
    \Gamma(\frac d2-n_{24}+n_7-i_1+z_1-z_2)}
    {\Gamma(\frac d2-n_4+n_7-i_1+z_1-z_2)
    \Gamma(n_{14567}-i_{123}-\frac d2+z_1+z_2)}
  \,,
\\
\label{eq:FBECA2c2cgen}
  \lefteqn{F_\BECA^{\text{(2c-2c)}}(n_1,n_2,n_3,n_4,n_5,n_6,0) =
  } \quad
  \nonumber \\* & \times
  \left(\frac{M^2}{Q^2}\right)^{4-n_{2456}} \,
  \frac{e^{-n i\pi} \, e^{2\eps\gamma_E} \,
    \Gamma(n_2-n_{13})}
    {\Gamma(n_2) \Gamma(n_4) \Gamma(n_5) \Gamma(n_6) \Gamma(\frac d2-n_{13})}
  \MBint z
  \nonumber \\* & \times
  \frac{\Gamma(-z) \Gamma(n_{24}-\frac d2-z)
    \Gamma(n_5+z) \Gamma(n_6-n_3+z)
    \Gamma(\frac d2-n_2+z) \Gamma(n_{56}-\frac d2+z)}
    {\Gamma(n_{56}-n_3+2z)}
  \,,
\\
\label{eq:FBECA1c2cgen}
  \lefteqn{F_\BECA^{\text{(1c-2c)}}(n_1,n_2,n_3,n_4,n_5,n_6,n_7) =
  \left(\frac{M^2}{Q^2}\right)^{4-n_{2345}}
  e^{-n i\pi} \, e^{2\eps\gamma_E}
  } \quad
  \nonumber \\* & \times
  \frac{\Gamma(n_2-n_{16}) \Gamma(n_{37}-n_6)
    \Gamma(\frac d2-n_2) \Gamma(\frac d2-n_3)
    \Gamma(n_{24}-\frac d2) \Gamma(n_{35}-\frac d2)}
    {\Gamma(n_2) \Gamma(n_3) \Gamma(n_4) \Gamma(n_5)
    \Gamma(\frac d2-n_{16}) \Gamma(\frac d2-n_6+n_7)}
  \,.
\end{align}

Die Auswertung der (1c-h)-Region erfolgt wie f"ur die abelschen Diagramme
beschrieben.
Die drei (c-c)-Regionen (1c-1c), (2c-2c) und (1c-2c) m"ussen gemeinsam
berechnet werden, da sie Singularit"aten im Parameterraum der $n_i$
aufweisen, die sich nur zwischen diesen drei Regionen aufheben.

An der Region (2c-2c) sieht man die neue Schwierigkeit des nichtabelschen
Diagramms mit drei massiven Propagatoren: Die Funktion
$\Gamma(n_{56}-n_3+2z)$ im Nenner besitzt die Abh"angigkeit $2z$ vom
Integrationsparameter. Dadurch lassen sich die Residuen, die zum
Mellin-Barnes-Integral beitragen, nicht als Br"uche mit gleichartigen
$\Gamma$"~Funktionen im Z"ahler und Nenner schreiben,
und die Aufsummierung unendlich vieler solcher Residuen gestaltet sich
schwierig.
In der (1c-1c)-Region verh"alt sich nach dem Aufl"osen einer der beiden
Mellin-Barnes-Integrationen das zweite Integral "ahnlich.

Allerdings ist es m"oglich, die in den (c-c)-Regionen auftretenden
Logarithmen $\ln(Q^2/M^2)$ zu isolieren. Da die Potenz des
$(M^2/Q^2$)-Vorfaktors in diesen drei Regionen nicht von $\eps$ abh"angt,
f"uhren nur Pole in den Parametern~$n_i$ zu Logarithmen $\ln(Q^2/M^2)$.
Eine genaue Untersuchung der Mellin"=Barnes"=Integrale zeigt, dass f"ur
$\eps\ne0$ singul"are Beitr"age nur in den folgenden 7~Integralen
auftreten:
$F_\BECA^{\text{(c-c)}}(-1,1,1,1,1,1,0)$,
$F_\BECA^{\text{(c-c)}}(0,1,1,1,1,1,n_7)$ mit $n_7=0,1,2$,
$F_\BECA^{\text{(c-c)}}(1,1,0,1,1,1,0)$
und $F_\BECA^{\text{(c-c)}}(1,1,1,1,1,0,n_7)$ mit $n_7=0,1$.
Es gen"ugt dann jeweils, diejenigen der Residuen in den
Mellin-Barnes-Integralen zu betrachten, die f"ur die Singularit"aten
verantwortlich sind.
In f"unf der sieben Integrale waren dies jeweils nur eine endliche Zahl von
Termen.

Lediglich f"ur die Integrale $F_\BECA^{\text{(c-c)}}(0,1,1,1,1,1,0)$ und
$F_\BECA^{\text{(c-c)}}(1,1,0,1,1,1,0)$, bei denen auch der Vorfaktor in
der (2c-2c)-Region singul"ar wird, musste jeweils eine unendliche Reihe von
Residuen aufsummiert werden.
Beispielsweise war ein Beitrag zu $F_\BECA^{\text{(c-c)}}(1,1,0,1,1,1,0)$
die folgende Summe:
\begin{equation}
\label{eq:suminvbinomS1}
  \sum_{j=0}^\infty \frac{1}{\binom{2j}{j}}
    \left(\frac{1}{3+2j} - \frac{1}{1+2j}\right)
    \left(\frac{1}{3+2j} + \frac{1}{1+2j} + S_1(2j) - S_1(j)\right) ,
\end{equation}
wobei
$S_m(j) = \sum_{i=1}^j \, i^{-m}$
die harmonische Summe~(\ref{eq:HarmonicSum}) ist.
Zwar wurden in j"ungster Zeit einige Fortschritte im Aufsummieren solcher
Reihen gemacht (siehe z.B. \cite{Moch:2001zr,Weinzierl:2004bn}), auch
Reihen mit inversen Binomialkoeffizienten wie im obigen Fall k"onnen
teilweise gel"ost werden. Jedoch ist ein allgemeiner Algorithmus, der alle
vorkommenden F"alle abdeckt, noch nicht bekannt. Und es ist nicht immer
klar, ob eine gegebene Reihe auf einen der l"osbaren F"alle
transformiert werden kann.

Eine alternative Methode (siehe z.B.\cite{Kalmykov:2000qe})
besteht darin, die unendlichen Reihen numerisch mit
einer Genauigkeit von vielen Dezimalstellen zu berechnen.
Anschlie"send macht man einen Ansatz mit einer Basis von analytischen
Konstanten wie $\pi^2$, $\zeta_3$, $\ln^4{2}$ usw. und versucht, rationale
Koeffizienten zu finden, die den gesuchten Wert als Linearkombination dieser
analytischen Konstanten darstellen.
Zum Aufsp"uren dieser rationalen Koeffizienten ist 
der \emph{PSLQ-Algorithmus}\cite{Ferguson:1991,Bailey:1993,Ferguson:1996}
geeignet.
F"ur die hier beschriebenen Rechnungen wurde eine PSLQ-Implementierung
in Fortran von O.~Veretin mit
Multipr"azisionsarithmetik\cite{Bailey:1990,Bailey:1991} verwendet.
Die unendlichen Reihen wurden mit einer Genauigkeit von 100~Stellen
berechnet, wozu das einfache Aufsummieren der ersten 300~Reihenglieder mit
\textsc{Mathematica} gen"ugte.
Die obige Summe in Gl.~(\ref{eq:suminvbinomS1}) f"uhrte dabei zum
analytischen Ausdruck $4\sqrt{3}\,\Cl2\!\left(\frac{\pi}{3}\right) - 8$
mit der Clausen-Funktion $\Cl2(\frac{\pi}{3}) \approx 1{,}014942$
(\ref{eq:Clausen}).

Zus"atzlich zu den Logarithmen in den Ergebnissen der skalaren Integrale
sind auch diejenigen Pole in~$\eps$ wichtig, die nicht in Verbindung mit
Logarithmen auftreten, damit am Ende
"uberpr"uft werden kann, ob das Gesamtergebnis in $d=4$ Dimensionen endlich
ist.
Da in den Mellin-Barnes-Integralen der Regionen (1c-1c) und (2c-2c) Pole
in~$\eps$ nur als Singularit"aten des Integrals auftreten, gen"ugt auch
hier die Betrachtung von denjenigen Residuen, die f"ur solche
Singularit"aten verantwortlich sind.
Alle diese Beitr"age beschr"anken sich entweder auf einzelne Residuen, oder
auf Mellin-Barnes-Integrale, die mit dem ersten Barnschen
Lemma~(\ref{eq:BarnesLemma1}) oder Ableitungen davon gel"ost werden
k"onnen.

Die Ergebnisse der einzelnen skalaren Integrale sind im
Anhang~\ref{sec:SkalarBECA} in Gl.~(\ref{eq:BECAergskalar}) aufgelistet.
Die oben angesprochenen Integrale der Form
$F_\BECA(1,n_2,1,1,1,1,n_7)$, bei denen der f"uhrende Beitrag der Ordnung
$(M^2/Q^2)^{-1}$ von der (1c-1c)-Region herr"uhrt, weisen weder
Logarithmen~$\ln(Q^2/M^2)$ noch Pole in~$\eps$ auf. Sie kommen deshalb in
der Ergebnisliste~(\ref{eq:BECAergskalar}) nicht vor.

Der Beitrag des gesamten nichtabelschen Feynman-Diagramms mit
Benz-Topologie lautet:
\begin{align}
\label{eq:BECAerg}
  F_{v,\BECA} &=
    C_F C_A
    \left(\frac{\alpha}{4\pi}\right)^2
    \left(\frac{\mu^2}{M^2}\right)^{2\eps} S_\eps^2
    \, \Biggl\{
  \frac{3}{4\eps^2}
  + \frac{1}{\eps} \left[
    - \frac{3}{2} \lqm^2 + \frac{9}{2} \lqm - \pi^2 - \frac{37}{8} \right]
  + \frac{1}{12} \lqm^4
  \nonumber \\* & \qquad
  + \frac{1}{2} \lqm^3
  + \left(\frac{\pi^2}{6} - \frac{11}{2}\right) \lqm^2
  + \left(4\sqrt{3}\,\Cl2\!\left(\frac{\pi}{3}\right)
    - \frac{2}{3}\zeta_3 - \frac{5}{6}\pi^2 + \frac{89}{4}\right) \lqm
  \Biggr\}
  \nonumber \\* & \quad
  + \Oc(\eps^0 \lqm^0) + \Oc(\eps) + \Oc\!\left(\frac{M^2}{Q^2}\right)
  ,
\end{align}
mit $\lqm = \ln(Q^2/M^2)$.

\subsection[Nichtabelsche Selbstenergiekorrektur mit T1-Topologie]
  {Nichtabelsche Selbstenergiekorrektur mit\\ T1-Topologie}

\begin{figure}[ht]
  \center
  \vcentergraphics{fcorr-T1-CA}
  \hspace{1cm}
  \vcentergraphics{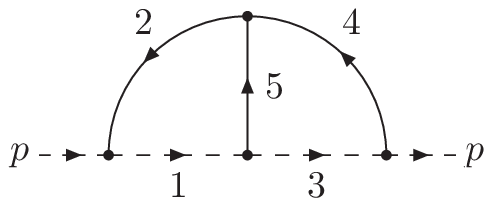}
  \caption{Nichtabelsche Selbstenergiekorrektur mit T1-Topologie
    (mit skalarem Diagramm)}
  \label{fig:T1CA}
\end{figure}
Das nichtabelsche Selbstenergiediagramm mit T1-Topologie ist in
Abb.~\ref{fig:T1CA} dargestellt.
Die Amplitude dieses Diagramms lautet wie folgt:
\begin{multline}
\label{eq:T1CAamplitude}
  \tilde\Sigma_\ToneCA =
  \mu^{4\eps} \loopint dk \loopint d\ell \,
  ig \gamma^\mu t^a \, \frac{i\dslash k_3}{k_3^2} \,
  ig \gamma^\nu t^b \, \frac{i\dslash k_1}{k_1^2} \,
  ig \gamma^\rho t^c \,
  \frac{-i}{k_2^2-M^2} \, \frac{-i}{k_4^2-M^2}
  \\* \times
  \frac{-i}{k_5^2-M^2} \,
  g\,f^{abc} \Bigl[ g_{\mu\nu}(k_4-k_5)_\rho + g_{\nu\rho}(k_5+k_2)_\mu
    + g_{\rho\mu}(-k_2-k_4)_\nu \Bigr]
  \,,
\end{multline}
mit den Impulsen
\begin{equation}
\label{eq:T1CAmom}
  k_1 = k + p ,\quad k_2 = k ,\quad
  k_3 = \ell + p ,\quad k_4 = \ell ,\quad
  k_5 = k - \ell .
\end{equation}
Der Farbfaktor ist der gleiche wie der des nichtabelschen Benz-Diagramms:
\[
  f^{abc} \, t^a t^b t^c = \frac{i}{2} \, C_F C_A \, \unity
  \,.
\]

Mit~(\ref{eq:fcorrproj}) wird aus der Amplitude~$\tilde\Sigma_\ToneCA$
der skalare Beitrag~$\Sigma_\ToneCA$ zum Formfaktor extrahiert.
Dieser wird wie beim abelschen Selbstenergiediagramm mit T1-Topologie
in skalare Integrale zerlegt, die folgenderma"sen definiert sind:
\begin{multline}
\label{eq:T1CAskalar}
  B_\ToneCA(p^2;n_1,n_2,n_3,n_4,n_5) =
  e^{2\eps\gamma_E} \, (M^2)^{n-4+2\eps}
  \loopintf dk \loopintf d\ell
  \\* \times
  \frac{1}
    {((k+p)^2)^{n_1} \, (k^2 - M^2)^{n_2} \,
    ((\ell+p)^2)^{n_3} \,
    (\ell^2 - M^2)^{n_4} \, ((k-\ell)^2 - M^2)^{n_5}}
  \,,
\end{multline}
mit $n = n_{12345}$.
Analog zu den abelschen
Selbstenergiekorrekturen~(\ref{eq:T1skalar0},\ref{eq:T1skalarD0}) ist dann
$B_\ToneCA^0$ der Wert des skalaren Integrals bei $p^2=0$ und
$B_\ToneCA'$ die mit $M^2$ multiplizierte Ableitung nach $p^2$ bei $p^2=0$.
Durch Ausnutzung der Symmetrie
\begin{equation} 
  B_\ToneCA(p^2;n_1,n_2,n_3,n_4,n_5) = B_\ToneCA(p^2;n_3,n_4,n_1,n_2,n_5)
\end{equation}
wurden skalare Integrale mit $n_1<n_3$ oder $n_1=n_3$ und $n_2<n_4$
durch ihre symmetrischen Partner ersetzt.
Die Zerlegung von $\Sigma_\ToneCA$ in skalare Integrale $B_\ToneCA^0$
und $B_\ToneCA'$ ist im Anhang~\ref{sec:SkalarT1CA} in
Gl.~(\ref{eq:T1CAzerlegung}) dargestellt.

F"ur allgemeines~$p^2$ l"asst sich (\ref{eq:T1CAskalar}) mit
Schwinger-Parametern~$\alpha_i$ und einer Mellin-Barnes-Darstellung
folgenderma"sen schreiben:
\begin{align}
\label{eq:BT1CAgen}
  \lefteqn{B_\ToneCA(p^2;n_1,n_2,n_3,n_4,n_5) =
  \frac{e^{-n i\pi} \, e^{2\eps\gamma_E}}
    {\Gamma(n_1) \Gamma(n_2) \Gamma(n_3) \Gamma(n_4) \Gamma(n_5)}
  \MBint z
  \left(\frac{-p^2}{M^2}\right)^z \, \Gamma(-z)
  } \quad
  \nonumber \\* & \times
  \Gamma(n_{12345}-d+z)
  \left(\prod_{i=1}^5\!\int_0^\infty\!\dd\alpha_i \, \alpha_i^{n_i-1}\right)
    \delta\!\left(\sum_{j\in S} \alpha_j - 1\right)
  (\alpha_2+\alpha_4+\alpha_5)^{d-n_{12345}-z} \quad
  \nonumber \\* & \times
  \Bigl( (\alpha_1+\alpha_2)(\alpha_3+\alpha_4+\alpha_5)
    + (\alpha_3+\alpha_4)\alpha_5 \Bigr)^{-\frac d2-z}
  \,
  \Bigl( \alpha_1 \alpha_2 \alpha_3 + \alpha_1 \alpha_2 \alpha_4
  \nonumber \\* & \qquad\quad
    + \alpha_1 \alpha_3 \alpha_4 + \alpha_2 \alpha_3 \alpha_4
    + \alpha_1 \alpha_2 \alpha_5 + \alpha_2 \alpha_3 \alpha_5
    + \alpha_1 \alpha_4 \alpha_5 + \alpha_3 \alpha_4 \alpha_5 \Bigr)^z
  \,,
\end{align}
mit $\emptyset \ne S \subseteq \{1,2,3,4,5\}$.
Der Wert dieses Ausdrucks f"ur $p^2=0$ ergibt sich aus dem Residuum des
Mellin-Barnes-Integrals bei $z=0$.
Nach der Einf"uhrung von zwei weiteren Mellin-Barnes-Integralen erh"alt
man:

\begin{align}
\label{eq:B0T1CAgen}
  \lefteqn{\Bm_\ToneCA(n_1,n_2,n_3,n_4,n_5) =
  } \quad
  \nonumber \\* &
  \frac{e^{-n i\pi} \, e^{2\eps\gamma_E}}
    {\Gamma(n_2) \Gamma(n_4) \Gamma(n_5) \Gamma(\frac d2)}
  \, \MBint{z_1} \MBint{z_2} \,
  \Gamma(n_5+z_1+z_2) \Gamma(\tfrac d2+z_1+z_2)
  \nonumber \\* & \times
  \frac{\Gamma(-z_1) \Gamma(n_{12}-\frac d2-z_1)
    \Gamma(\frac d2-n_1+z_1)}
    {\Gamma(\frac d2+z_1)} \,
  \frac{\Gamma(-z_2) \Gamma(n_{34}-\frac d2-z_2)
    \Gamma(\frac d2-n_3+z_2)}
    {\Gamma(\frac d2+z_2)}
  \,.
\end{align}

Analog ergibt sich $\DBm_\ToneCA$ aus dem Residuum des $z$-Integrals
in~(\ref{eq:BT1CAgen}) bei $z=1$. Der Ausdruck f"ur $\DBm_\ToneCA$ besteht
aus 8~Termen von der Form~(\ref{eq:B0T1CAgen}).

Das zweifache Mellin-Barnes-Integral in~(\ref{eq:B0T1CAgen}) muss nicht
allgemein gel"ost werden.
Bei den nichtabelschen Vertexdiagrammen wird auf die Berechnung der
nichtlogarithmischen, in $d=4$ Dimensionen endlichen Konstante verzichtet.
Die Selbstenergiekorrekturen enthalten keine Logarithmen $\ln(Q^2/M^2)$,
deshalb gen"ugt hier die Berechnung der Pole in~$\eps$.
Diese entstehen nur als Singularit"aten der Mellin-Barnes-Integrale.
Eine endliche Zahl von Residuen liefert singul"are Beitr"age,
diese k"onnen leicht isoliert und addiert werden.

Die Ergebnisse der einzelnen skalaren Integrale sind im
Anhang~\ref{sec:SkalarT1CA} in Gl.~(\ref{eq:T1CAergskalar}) aufgelistet.
Das nichtabelsche Selbstenergiediagramm mit T1-Topologie liefert insgesamt
den folgenden Beitrag:
\begin{align}
\label{eq:T1CAerg}
  \Sigma_\ToneCA &=
    C_F C_A
    \left(\frac{\alpha}{4\pi}\right)^2
    \left(\frac{\mu^2}{M^2}\right)^{2\eps} S_\eps^2
    \left(
  -\frac{3}{2\eps^2}
  - \frac{5}{4\eps}
  \right)
  + \Oc(\eps^0)
  \,.
\end{align}

Nun k"onnen die beiden Beitr"age mit einer Drei-Eichboson-Kopplung
aus den Gleichungen (\ref{eq:BECAerg}) und (\ref{eq:T1CAerg})
addiert werden:
\begin{align}
\label{eq:3Werg}
  F_\tW &= 2\,F_{v,\BECA} + \Sigma_\ToneCA
  \nonumber \\* &=
    C_F C_A
    \left(\frac{\alpha}{4\pi}\right)^2
    \left(\frac{\mu^2}{M^2}\right)^{2\eps} S_\eps^2
    \, \Biggl\{
  \frac{1}{\eps} \left[
    - 3 \lqm^2 + 9 \lqm - 2\pi^2 - \frac{21}{2} \right]
  + \frac{1}{6} \lqm^4
  + \lqm^3
  \nonumber \\* & \qquad
  + \left(\frac{\pi^2}{3} - 11\right) \lqm^2
  + \left(8\sqrt{3}\,\Cl2\!\left(\frac{\pi}{3}\right)
    - \frac{4}{3}\zeta_3 - \frac{5}{3}\pi^2 + \frac{89}{2}\right) \lqm
  \Biggr\}
  \nonumber \\* & \quad
  + \Oc(\eps^0 \lqm^0) + \Oc(\eps) + \Oc\!\left(\frac{M^2}{Q^2}\right)
  ,
\end{align}
wobei das Vertexdiagramm doppelt ber"ucksichtigt wurde.
Das Ergebnis enth"alt nur noch einen einfachen $1/\eps$-Pol, der zudem
bis auf den Faktor $3 C_A \frac{\alpha}{4\pi}$ dem
Einschleifenformfaktor in Gl.~(\ref{eq:F1erg}) entspricht. Der $1/\eps$-Pol
kann somit durch die Kopplungskonstantenrenormierung aufgehoben werden.

%
%
\section[$C_F C_A$-Beitr"age mit Eichboson- oder Geistschleife]
  {\boldmath $C_F C_A$-Beitr"age mit Eichboson- oder\\ Geistschleife}
\label{sec:CAWcorr}

\subsection{Vertexdiagramme mit Eichboson- oder Geistschleife}
\label{sec:WWccvertex}

\begin{figure}[ht]
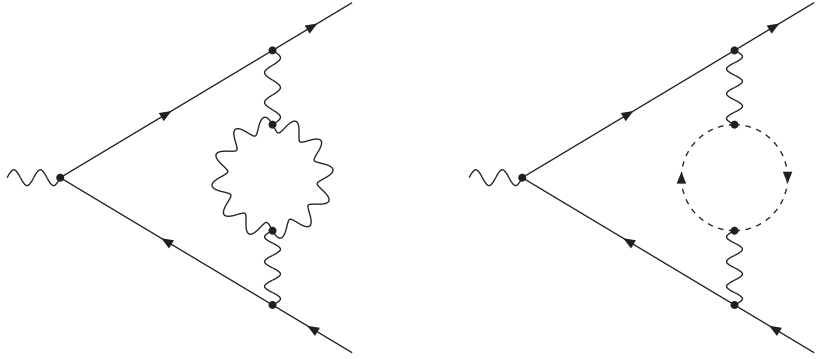

  \center
  \vcentergraphics{vertex-Wcorr-CA}
  \hspace{1cm}
  \vcentergraphics{vertex-Wcorr-cc}
  \caption{Vertexdiagramme mit Eichboson- bzw. Geistschleife}
  \label{fig:WWcc}
\end{figure}

In diesem Abschnitt wird die Berechnung der beiden Vertexdiagramme in
Abb.~\ref{fig:WWcc} mit einer Eichboson- bzw. Geistschleife im
Eichbosonpropagator vorgestellt.
\begin{figure}[ht]
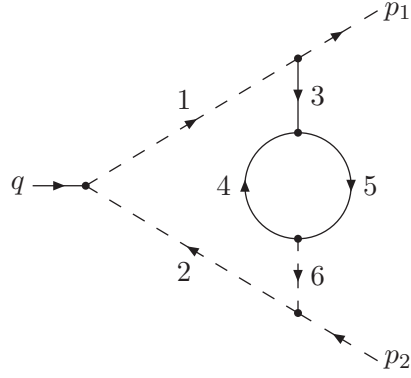

  \center
  \vcentergraphics{vertex-Wcorr-skalar}
  \caption{Skalares Vertexdiagramm mit Schleife im Eichbosonpropagator}
  \label{fig:Wcorrskalar}
\end{figure}
Beiden Feynman-Diagrammen entspricht das gemeinsame skalare Diagramm in
Abb.~\ref{fig:Wcorrskalar}, da in der Feynman-'t~Hooft-Eichung sowohl das
Eichboson als auch das Geistfeld die gleiche Masse~$M$ besitzen.
Der massive Propagator~3 kommt in den Feynman-Diagrammen jeweils zweifach
vor. Dagegen fehlt der masselose Propagator~6 (mit $k_6 = k_3$) in den
Feynman-Diagrammen; er kommt durch die Tensorreduktion (siehe unten) ins
Spiel.

Die Summe der Amplituden beider Feynman-Diagramme in Abb.~\ref{fig:WWcc}
ergibt:
\begin{align}
\label{eq:WWccamplitude}
  \lefteqn{\Fc_{v,\WWcc}^\mu =
  \mu^{4\eps} \loopint dk \loopint d\ell \,
  ig \gamma^\nu t^a \, \frac{i\dslash k_1}{k_1^2} \,
  \gamma^\mu
  \frac{i\dslash k_2}{k_2^2} \, ig \gamma_\rho t^b
  \left(\frac{-i}{k_3^2-M^2}\right)^2
  } \;
\nonumber \\* & \times
  \biggl\{
  \frac{1}{2} \,
    g \, f^{acd} \Bigl[
      g_{\nu\sigma}(k_3+k_5)_\tau + g_{\sigma\tau}(-k_5-k_4)_\nu
      + g_{\tau\nu}(k_4-k_3)_\sigma \Bigr] \,
    \frac{-i}{k_4^2-M^2}
\nonumber \\* & \qquad\quad \times
    g \, f^{bcd} \Bigl[
      g^{\rho\sigma}(-k_3-k_5)^\tau + g^{\sigma\tau}(k_5+k_4)^\rho
      + g^{\tau\rho}(-k_4+k_3)^\sigma \Bigr] \,
    \frac{-i}{k_5^2-M^2}
\nonumber \\* & \qquad
  + (-1) \,
    (-g) f^{acd} \, {k_5}_\nu \, \frac{i}{k_4^2-M^2} \,
    (-g) f^{bdc} \, k_4^\rho \, \frac{i}{k_5^2-M^2}
  \biggr\}
  \,,
\end{align}
mit den Impulsen
\begin{equation}
\label{eq:WWccmom}
  k_1 = p_1 + k ,\quad k_2 = p_2 + k ,\quad
  k_3 = k ,\quad k_4 = \ell ,\quad
  k_5 = k + \ell .
\end{equation}
Bei der Eichbosonschleife steht der Symmetriefaktor~$1/2$, da die
Impulsintegration "uber~$\ell$ virtuelle Korrekturen doppelt z"ahlt, die
physikalisch identisch sind.
Die Geistschleife erh"alt ein Minuszeichen, wie in den Feynman-Regeln
(Anhang~\ref{chap:feynman}) beschrieben.
Die Farbfaktoren der Diagramme sind
\[
  f^{acd} f^{bcd} t^a t^b = C_F C_A \, \unity
  \quad \text{und} \quad
  f^{acd} f^{bdc} t^a t^b = -C_F C_A \, \unity
  \,.
\]

Die Indizes $\nu$, $\rho$, $\sigma$ und $\tau$ in der
Amplitude~$\Fc^\mu_{v,\WWcc}$ wurden mit FORM kontrahiert. Man erh"alt
den Beitrag~$F_{v,\WWcc}$ beider Diagramme zum Formfaktor.
Dessen Z"ahler enth"alt Skalarprodukte, die mit den folgenden
Beziehungen umgeformt werden:
\begin{equation}
\label{eq:Wcorrredskalar}
  \begin{aligned}
    p_1 \cdot k &= \tfrac12 (k_1^2 - k_3^2) \,,\quad &
    p_2 \cdot k &= \tfrac12 (k_2^2 - k_3^2) \,,\quad &
    k \cdot \ell &= \tfrac12 (k_5^2 - k_3^2 - k_4^2) \,.
  \end{aligned}
\end{equation}
Formal bleiben zwei Skalarprodukte als irreduzibel zur"uck und k"onnen
nicht mit Propagatornennern gek"urzt werden.
Wie beim Vertexdiagramm mit Fermion-Selbstenergie k"onnen jedoch
Skalarprodukte zwischen $\ell$ und $p_{1,2}$ durch Tensorreduktion
entsprechend (\ref{eq:2punkttensorred1}) und (\ref{eq:2punkttensorred2})
in k"urzbare Skalarprodukte aus (\ref{eq:Wcorrredskalar}) umgewandelt werden.

Allerdings entstehen bei dieser Tensorreduktion neue Faktoren
$1/k^2=1/k_3^2$, also Propagatornenner mit Impuls~$k_3$, aber ohne die
Masse des Propagators~3. Man muss somit einen sechsten, masselosen
Propagator mit Impuls $k_6=k_3$ einf"uhren.
Die skalaren Integrale sind dann folgenderma"sen definiert:
\begin{multline}
\label{eq:Wcorrskalar}
  F_\Wc(n_1,n_2,n_3,n_4,n_5,n_6) =
  e^{2\eps\gamma_E} \, (M^2)^{2\eps} \,
  (Q^2)^{n-4}
  \\* \times
  \loopintf dk \loopintf d\ell \,
  \frac{1}
    {(k^2+2p_1\cdot k)^{n_1} \, (k^2+2p_2\cdot k)^{n_2} \,
    (k^2 - M^2)^{n_3} \, (\ell^2 - M^2)^{n_4}}
  \\* \times
  \frac{1}{((k+\ell)^2 - M^2)^{n_5} \, (k^2)^{n_6}}
  \,,
\end{multline}
mit $n = n_{123456}$.
Zwar w"are es m"oglich, durch Partialbruchzerlegung die Propagatoren 3 und
6 zu jeweils einem Propagator zu kombinieren, der dann entweder massiv
oder masselos ist. Dabei entstehen jedoch Faktoren $1/M^2$, so dass die
Berechnung im Limes $M^2\ll Q^2$ eine h"ohere Entwicklung der Integrale in
$M^2/Q^2$ n"otig machen w"urde.
Bel"asst man beide Propagatoren, den massiven dritten und den masselosen
sechsten, in der Definition des skalaren Integrals~(\ref{eq:Wcorrskalar}),
so ist die Zerlegung der Amplitude in skalare Integrale nicht eindeutig.
F"ur die folgende Rechnung wurde die Wahl getroffen, den dritten Propagator
nicht zu k"urzen, sondern alle auftretenden Faktoren~$k^2$ mit dem sechsten
Propagator zu k"urzen.
Der Parameter~$n_3$ hat also f"ur alle auftretenden Integrale den Wert
$n_3=2$ entsprechend dem Feynman-Diagramm.

Die Zerlegung von $F_{v,\WWcc}$ in skalare Integrale $F_\Wc$ findet sich im
Anhang~\ref{sec:SkalarWWcc} in Gl.~(\ref{eq:WWcczerlegung}).
Durch Ausnutzung der beiden Symmetrien
\begin{align}
  F_\Wc(n_1,n_2,n_3,n_4,n_5,n_6) &= F_\Wc(n_2,n_1,n_3,n_4,n_5,n_6) \,,
  \\
  F_\Wc(n_1,n_2,n_3,n_4,n_5,n_6) &= F_\Wc(n_1,n_2,n_3,n_5,n_4,n_6)
\end{align}
wurden skalare Integrale mit $n_1<n_2$ oder mit $n_4<n_5$ durch ihre
symmetrischen Partner ersetzt.

Folgende Regionen liefern im Rahmen der \emph{Expansion by Regions}
nichtverschwindende Beitr"age:
\[
  \renewcommand{\arraystretch}{1.2}
  \left.
  \begin{array}{rll}
    \text{(h-h):} & k \sim Q, & \ell \sim Q \\
    \text{(h-s):} & k \sim Q, & \ell=k_4 \sim M \\
    \text{(h-s'):} & k \sim Q, & k_5 \sim M
  \end{array}
  \qquad
  \right|
  \qquad
  \begin{array}{rll}
    \text{(1c-1c):} & k \;\|\; p_1, & \ell \;\|\; p_1 \\
    \text{(2c-2c):} & k \;\|\; p_2, & \ell \;\|\; p_2
  \end{array}
\]

Die Potenz in $(M^2/Q^2)$ des f"uhrenden Beitrags jeder Region lautet:
\[
  \left.
  \begin{array}{rl}
    \text{(h-h):} & 2\eps \\
    \text{(h-s):} & 2-n_4+\eps \\
    \text{(h-s'):} & 2-n_5+\eps
  \end{array}
  \qquad
  \right|
  \qquad
  \begin{array}{rl}
    \text{(1c-1c):} & 4-n_{13456} \\
    \text{(2c-2c):} & 4-n_{23456}
  \end{array}
\]
Wegen $n_4\le1$ und $n_5\le1$ sind die Regionen (h-s) und (h-s') immer mit
mindestens einem Faktor $M^2/Q^2$ unterdr"uckt.
Auf der anderen Seite ist es m"oglich, dass der f"uhrende Beitrag der
Regionen (1c-1c) und (2c-2c) eine negative Potenz von $(M^2/Q^2)$ hat.
Dies tritt nur in den drei F"allen
$F_\Wc(0,0,2,1,1,1)$, $F_\Wc(1,0,2,1,1,0)$ und $F_\Wc(1,1,2,1,1,0)$
auf. Hier ist der f"uhrende Beitrag von der Ordnung $(M^2/Q^2)^{-1}$, und
der Pol in~$M^2$ wird vom jeweiligen Vorfaktor aus der Zerlegung in skalare
Integrale gek"urzt.

Die f"uhrenden Beitr"age der nichtunterdr"uckten Regionen lauten:
\begin{align}
\label{eq:FWchhgen}
  \lefteqn{F_\Wc^{\text{(h-h)}}(n_1,n_2,n_3,n_4,n_5,n_6) =
  \left(\frac{M^2}{Q^2}\right)^{2\eps}
  e^{-n i\pi} \, e^{2\eps\gamma_E}
  } \quad
  \nonumber \\* & \times
  \frac{\Gamma(\frac d2-n_4) \Gamma(\frac d2-n_5)
    \Gamma(d-n_{13456}) \Gamma(d-n_{23456})
    \Gamma(n_{45}-\frac d2)
    \Gamma(n_{123456}-d)}
    {\Gamma(n_1) \Gamma(n_2) \Gamma(n_4) \Gamma(n_5)
    \Gamma(d-n_{45})
    \Gamma(\frac32d-n_{123456})}
  \,,
\\
\label{eq:FWc1c1cgen}
  \lefteqn{F_\Wc^{\text{(1c-1c)}}(n_1,n_2,n_3,n_4,n_5,n_6) =
  } \quad
  \nonumber \\* &
  \left(\frac{M^2}{Q^2}\right)^{4-n_{13456}}
  e^{-n i\pi} \, e^{2\eps\gamma_E} \,
  \frac{\Gamma(n_1-n_2)}
    {\Gamma(n_1) \Gamma(n_3) \Gamma(n_4) \Gamma(n_5)
    \Gamma(\frac d2-n_2)} \,
  \MBint z
  \nonumber \\* & \times
  \frac{\Gamma(-z) \Gamma(n_{136}-\frac d2-z)
    \Gamma(n_4+z) \Gamma(n_5+z)
    \Gamma(\frac d2-n_{16}+z) \Gamma(n_{45}-\frac d2+z)}
    {\Gamma(n_{45}+2z)}
  \,,
\end{align}
und aus Gr"unden der Symmetrie:
\begin{equation} 
  F_\Wc^{\text{(2c-2c)}}(n_1,n_2,n_3,n_4,n_5,n_6) =
  F_\Wc^{\text{(1c-1c)}}(n_2,n_1,n_3,n_4,n_5,n_6) \,.
\end{equation}

Logarithmen $\ln(Q^2/M^2)$ werden von den (c-c)-Regionen (1c-1c) und
(2c-2c) nur f"ur $n_1=n_2=1$ produziert.
Wenn dabei $n_4\le0$ oder $n_5\le0$ ist, dann tragen nur diejenigen
Residuen des Mellin-Barnes-Integrals zum Ergebnis bei, die durch eine
Singularit"at im Integral den Vorfaktor $1/\Gamma(n_4)/\Gamma(n_5)$
kompensieren, also nur endlich viele.
Lediglich f"ur die beiden Integrale
$F_\Wc^{\text{(c-c)}}(1,1,2,1,1,-1)$ und
$F_\Wc^{\text{(c-c)}}(1,1,2,1,1,0)$
muss die ganze unendliche Reihe der Residuen f"ur die logarithmischen
Beitr"age aufsummiert werden.
In diesen beiden F"allen wurde die Summe numerisch mit jeweils
100~Dezimalstellen
Genauigkeit berechnet und das zugeh"orige analytische Ergebnis mit dem
PSLQ-Algorithmus gefunden.

F"ur die nichtlogarithmischen Pole in~$\eps$ war in jedem Fall nur eine
endliche Zahl von Residuen n"otig.

Die Ergebnisse der einzelnen skalaren Integrale sind im
Anhang~\ref{sec:SkalarWWcc} in Gl.~(\ref{eq:WWccergskalar}) aufgelistet.
Die beiden Feynman-Diagramme mit Eichboson- und Geistschleife liefern
insgesamt den folgenden Beitrag zum Formfaktor:
\begin{align}
\label{eq:WWccerg}
  F_{v,\WWcc} &=
    C_F C_A
    \left(\frac{\alpha}{4\pi}\right)^2
    \left(\frac{\mu^2}{M^2}\right)^{2\eps} S_\eps^2
    \, \Biggl\{
  \frac{1}{\eps} \left[
    - \frac{5}{3} \lqm^2 + \frac{49}{3} \lqm - \frac{10}{9}\pi^2
    - \frac{337}{12} \right]
  \nonumber \\* & \qquad
  + \frac{10}{9} \lqm^3
  - \frac{76}{9} \lqm^2
  + \left(-4\sqrt{3}\,\Cl2\!\left(\frac{\pi}{3}\right)
    + \frac{859}{18}\right) \lqm
  \Biggr\}
  \nonumber \\* & \quad
  + \Oc(\eps^0 \lqm^0) + \Oc(\eps) + \Oc\!\left(\frac{M^2}{Q^2}\right)
  .
\end{align}

\subsection[Selbstenergiekorrekturen mit Eichboson- oder Geistschleife]
  {Selbstenergiekorrekturen mit Eichboson- oder\\ Geistschleife}
\label{sec:WWccfcorr}

Die beiden Selbstenergiediagramme mit einer Eichboson- bzw. Geistschleife
im Eichbosonpropagator in Abb.~\ref{fig:fcorrWWcc} sind dem gleichen
skalaren Diagramm in Abb.~\ref{fig:fcorrWWccskalar} zugeordnet.
\begin{figure}[ht]
  \center
  \vcentergraphics{fcorr-Wcorr-CA}
  \hspace{1cm}
  \vcentergraphics{fcorr-Wcorr-cc}
  \caption{Selbstenergiediagramme mit Eichboson- bzw. Geistschleife}
  \label{fig:fcorrWWcc}
\end{figure}%
\begin{figure}[ht]
  \center
  \vcentergraphics{fcorr-Wcorr-skalar}
  \caption{Skalares Selbstenergiediagramm mit Schleife im Eichbosonpropagator}
  \label{fig:fcorrWWccskalar}
\end{figure}%
Analog zu den Vertexdiagrammen aus dem vorigen Abschnitt kommt der
Propagator~2 in den Selbstenergiediagrammen doppelt vor,
und der masselose Propagator~5 (mit $k_5 = k_2$) entsteht durch die
Tensorreduktion.

Die Summe beider Feynman-Diagramme in Abb.~\ref{fig:fcorrWWcc} ergibt
die folgende Amplitude:
\begin{align}
\label{eq:fcorrWWccamplitude}
  \lefteqn{\tilde\Sigma_\WWcc =
  \mu^{4\eps} \loopint dk \loopint d\ell \,
  ig \gamma^\mu t^a \, \frac{i\dslash k_1}{k_1^2} \,
  ig \gamma_\nu t^b
  \left(\frac{-i}{k_2^2-M^2}\right)^2
  } \;
\nonumber \\* & \times
  \biggl\{
  \frac{1}{2} \,
    g \, f^{acd} \Bigl[
      g_{\mu\rho}(k_2+k_4)_\sigma + g_{\rho\sigma}(-k_4-k_3)_\mu
      + g_{\sigma\mu}(k_3-k_2)_\rho \Bigr] \,
    \frac{-i}{k_3^2-M^2}
\nonumber \\* & \qquad\quad \times
    g \, f^{bcd} \Bigl[
      g^{\nu\rho}(-k_2-k_4)^\sigma + g^{\rho\sigma}(k_4+k_3)^\nu
      + g^{\sigma\nu}(-k_3+k_2)^\rho \Bigr] \,
    \frac{-i}{k_4^2-M^2}
\nonumber \\* & \qquad
  + (-1) \,
    (-g) f^{acd} \, {k_4}_\mu \, \frac{i}{k_3^2-M^2} \,
    (-g) f^{bdc} \, k_3^\nu \, \frac{i}{k_4^2-M^2}
  \biggr\}
  \,,
\end{align}
mit den Impulsen
\begin{equation}
\label{eq:fcorrWWccmom}
  k_1 = k + p ,\quad
  k_2 = k ,\quad k_3 = \ell ,\quad
  k_4 = k + \ell .
\end{equation}
Die Farbfaktoren der Diagramme sind die gleichen wie f"ur die
Vertexdiagramme im vorigen Abschnitt:
\[
  f^{acd} f^{bcd} t^a t^b = C_F C_A \, \unity
  \quad \text{und} \quad
  f^{acd} f^{bdc} t^a t^b = -C_F C_A \, \unity
  \,.
\]

Entsprechend Gl.~(\ref{eq:fcorrproj}) erh"alt man aus der
Amplitude~$\tilde\Sigma_\WWcc$ den skalaren Beitrag~$\Sigma_\WWcc$ zum
Formfaktor.
Zum K"urzen von Skalarprodukten dienen die folgenden Umformungen:
\begin{equation}
\label{eq:fcorrWcorrredskalar}
  \begin{aligned}
    p \cdot k &= \tfrac12 (k_1^2 - k_2^2 - p^2) \,,\quad &
    k \cdot \ell &= \tfrac12 (k_4^2 - k_2^2 - k_3^2) \,.
  \end{aligned}
\end{equation}
Als formal irreduzibles Skalarprodukt verbleibt $p\cdot\ell$.
Es kann aber durch Tensorreduktion entsprechend
(\ref{eq:2punkttensorred1}) und (\ref{eq:2punkttensorred2}) eliminiert
werden. Stattdessen kommt durch die Tensorreduktion der
masselose Propagator mit $k_5=k_2$ hinzu.
Die Definition der skalaren Integrale lautet:
\begin{multline}
\label{eq:fcorrWcorrskalar}
  B_\Wc(p^2;n_1,n_2,n_3,n_4,n_5) =
  e^{2\eps\gamma_E} \, (M^2)^{n-4+2\eps}
  \loopintf dk \loopintf d\ell
  \\* \times
  \frac{1}
    {((k+p)^2)^{n_1} \, (k^2 - M^2)^{n_2} \,
    (\ell^2 - M^2)^{n_3} \, ((k+\ell)^2 - M^2)^{n_4} \,
    (k^2)^{n_5}}
  \,,
\end{multline}
mit $n = n_{12345}$.
Wie bei den vorherigen Selbstenergiekorrekturen werden $\Bm_\Wc$ und
$\DBm_\Wc$ f"ur das skalare Integral bei $p^2=0$ respektive seine (mit
$M^2$ multiplizierte) Ableitung nach~$p^2$ bei $p^2=0$ eingef"uhrt.
Die Symmetrie
\begin{equation} 
  B_\Wc(p^2;n_1,n_2,n_3,n_4,n_5) = B_\Wc(p^2;n_1,n_2,n_4,n_3,n_5)
\end{equation}
wird dazu benutzt, um Integrale mit $n_3<n_4$ durch ihre symmetrischen
Partner zu ersetzen.
Die Zerlegung von $\Sigma_\WWcc$ in skalare Integrale $\Bm_\Wc$
und $\DBm_\Wc$ ist im Anhang~\ref{sec:SkalarfcorrWWcc} in
Gl.~(\ref{eq:fcorrWWcczerlegung}) aufgelistet.

F"ur allgemeines $p^2$ l"asst sich (\ref{eq:fcorrWcorrskalar}) als
zweifaches Mellin-Barnes-Integral darstellen:
\begin{align}
  \lefteqn{B_\Wc(p^2;n_1,n_2,n_3,n_4.n_5) =
  \frac{e^{-n i\pi} \, e^{2\eps\gamma_E} \,
    \Gamma(\frac d2-n_1)}
    {\Gamma(n_1) \Gamma(n_2) \Gamma(n_3) \Gamma(n_4)}
  \MBint{z_1} \MBint{z_2}
  \left(\frac{-p^2}{M^2}\right)^{z_1}
  } \quad
  \nonumber \\* & \times
  \frac{\Gamma(-z_1) \Gamma(n_1+z_1)}
    {\Gamma(\frac d2-n_1-z_1) \Gamma(\frac d2+z_1)} \,
  \frac{\Gamma(-z_2) \Gamma(n_3+z_2) \Gamma(n_4+z_2)
    \Gamma(n_{34}-\tfrac d2+z_2)}
    {\Gamma(n_{34}+2z_2)}
  \quad \nonumber \\* & \times
  \Gamma(\tfrac d2-n_{15}-z_1+z_2) \Gamma(n_{125}-\tfrac d2+z_1-z_2)
  \,.
\end{align}
Die Entwicklung um $p^2=0$ erh"alt man aus den Residuen des $z_1$-Integrals
bei $z_1=0$ und $z_1=1$:
\begin{align}
\label{eq:B0Wcgen}
  \lefteqn{\Bm_\Wc(n_1,n_2,n_3,n_4,n_5) =
  \frac{e^{-n i\pi} \, e^{2\eps\gamma_E}}
    {\Gamma(n_2) \Gamma(n_3) \Gamma(n_4) \Gamma(\frac d2)}
  \, \MBint z \,
  \Gamma(-z)
  } \quad
  \nonumber \\* & \times
  \Gamma(n_{125}-\tfrac d2-z) \,
  \frac{\Gamma(n_3+z) \Gamma(n_4+z)
    \Gamma(\frac d2-n_{15}+z) \Gamma(n_{34}-\frac d2+z)}
    {\Gamma(n_{34}+2z)}
  \,,
\\
\label{eq:DBWcgen}
  \lefteqn{\DBm_\Wc(n_1,n_2,n_3,n_4,n_5) =
  \frac{e^{-n i\pi} \, e^{2\eps\gamma_E} \,
    n_1 \, (\frac d2-n_1-1)}
    {\Gamma(n_2) \Gamma(n_3) \Gamma(n_4) \Gamma(1+\frac d2)}
  \, \MBint z \,
  \Gamma(-z)
  } \quad
  \nonumber \\* & \times
  \Gamma(n_{125}+1-\tfrac d2-z) \,
  \frac{\Gamma(n_3+z) \Gamma(n_4+z)
    \Gamma(\frac d2-n_{15}-1+z) \Gamma(n_{34}-\frac d2+z)}
    {\Gamma(n_{34}+2z)}
  \,.
\end{align}

Weil von der Selbstenergiekorrektur nur die Pole in~$\eps$ ben"otigt werden,
gen"ugt die Auswertung endlich vieler Residuen der Mellin-Barnes-Integrale
in (\ref{eq:B0Wcgen}) und (\ref{eq:DBWcgen}).
Die Ergebnisse der einzelnen skalaren Integrale sind im
Anhang~\ref{sec:SkalarfcorrWWcc} in Gl.~(\ref{eq:fcorrWWccergskalar})
aufgef"uhrt.
Die beiden Selbstenergiediagramme mit Eichboson- und Geistschleife liefern
insgesamt den folgenden Beitrag:
\begin{align}
\label{eq:fcorrWWccerg}
  \Sigma_\WWcc &=
    C_F C_A
    \left(\frac{\alpha}{4\pi}\right)^2
    \left(\frac{\mu^2}{M^2}\right)^{2\eps} S_\eps^2
  \cdot \frac{21}{4\eps}
  + \Oc(\eps^0)
  \,.
\end{align}

\subsection[Massenrenormierung mit Eichboson- oder Geistschleife]
  {Massenrenormierung mit Eichboson- oder\\ Geistschleife}
\label{sec:WWccMren}

\begin{figure}[ht]
  \center
  \vcentergraphics{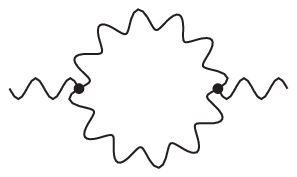}
  \hspace{1cm}
  \vcentergraphics{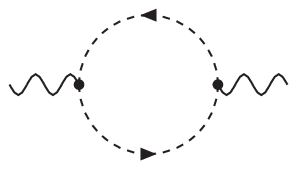}
  \caption{Beitr"age zur Massenrenormierung mit Eichboson- bzw. Geistschleife}
  \label{fig:MrenWWcc}
\end{figure}

Zur Renormierung der Eichbosonmasse~$M$ muss die Einschleifenselbstenergie
des Eichbosons berechnet werden (siehe Abschnitt~\ref{sec:Renormierung}).
Die zwei Diagramme mit Eichboson- oder Geistschleife sind in
Abb.~\ref{fig:MrenWWcc} dargestellt.
Diagramme mit Tadpole-Schleifen werden hier genauso wenig betrachtet wie bei
den Vertex- oder Selbstenergiekorrekturen, da deren
Schleifenkorrekturen durch die Massenrenormierung kompensiert werden.
Die Summe der Amplituden beider Feynman-Diagramme in
Abb.~\ref{fig:MrenWWcc} lautet:
\begin{align}
\label{eq:MrenWWccamplitude}
  \lefteqn{\tilde\Pi_{\WWcc}^{\mu\nu,ab}(k) =
  \mu^{2\eps} \loopint d\ell \,
  \biggl\{
  } \quad
\nonumber \\* &
  \frac{1}{2} \,
    g \, f^{acd} \Bigl[
      g^\mu_\rho(2k+\ell)_\sigma + g_{\rho\sigma}(-k-2\ell)^\mu
      + g_\sigma^\mu(\ell-k)_\rho \Bigr] \,
    \frac{-i}{\ell^2-M^2}
\nonumber \\* & \quad \times
    g \, f^{bcd} \Bigl[
      g^{\nu\rho}(-2k-\ell)^\sigma + g^{\rho\sigma}(k+2\ell)^\nu
      + g^{\sigma\nu}(-\ell+k)^\rho \Bigr] \,
    \frac{-i}{(k+\ell)^2-M^2}
\nonumber \\* &
  + (-1) \,
    (-g) f^{acd} \, (k+\ell)^\mu \, \frac{i}{\ell^2-M^2} \,
    (-g) f^{bdc} \, \ell^\nu \, \frac{i}{(k+\ell)^2-M^2}
  \biggr\}
  \,.
\end{align}
Die Farbfaktoren der Diagramme sind
\[
  f^{acd} f^{bcd} = C_A \, \delta^{ab}
  \quad \text{und} \quad
  f^{acd} f^{bdc} = -C_A \, \delta^{ab}
  \,.
\]

Die Amplitude~$\tilde\Pi_{\WWcc}^{\mu\nu,ab}$ ist von der Form
\begin{equation}
  \tilde\Pi_\WWcc^{\mu\nu,ab}(k) =
  i \, \delta^{ab} \Bigl[
    g^{\mu\nu} k^2 \, \Pi_\WWcc(k^2) - k^\mu k^\nu \, \Pi'_\WWcc(k^2)
    \Bigr]
  \,,
\end{equation}
mit den skalaren Funktionen $\Pi_\WWcc$ und $\Pi'_\WWcc$.
Entsprechend den Gleichungen~(\ref{eq:Mren}) und (\ref{eq:F2MrenPi1})
wird zur Massenrenormierung nur die Funktion~$\Pi_\WWcc$ ben"otigt.
Diese kann aus der Amplitude durch Projektion gewonnen werden, analog zur
Tensorreduktion:
\begin{equation}
\label{eq:bcorrproj}
  \Pi_\WWcc(k^2) =
    \frac{-i}{d-1} \, \frac{1}{k^2} \, \left(
      g_{\mu\nu} - \frac{k_\mu k_\nu}{k^2}\right)
  \tilde\Pi_\WWcc^{\mu\nu}(k)
  \,,
\end{equation}
wobei $\tilde\Pi_\WWcc^{\mu\nu}(k)$ die Amplitude ohne den
Faktor~$\delta^{ab}$ (die Identit"at in der adjungierten
$SU(2)$-Darstellung) bezeichnet.

Von der mit Gl.~(\ref{eq:bcorrproj}) projizierten Funktion~$\Pi_\WWcc(k^2)$
wird entsprechend~(\ref{eq:Mren}) und (\ref{eq:F2MrenPi1}) der Wert
(bzw. der Realteil) bei $k^2=M^2$ ben"otigt.
Er wird als Linearkombination von Integralen mit der Definition
\begin{equation}
\label{eq:MrenWcorrskalar}
  B_1^M(n_1,n_2) =
  e^{\eps\gamma_E} \, (M^2)^{n_{12}-2+\eps}
  \left. \loopintf d\ell \,
  \frac{1}{(\ell^2 - M^2)^{n_1} \, ((k+\ell)^2 - M^2)^{n_2}}
  \, \right|_{k^2=M^2}
\end{equation}
geschrieben.
Diese Zerlegung lautet:
\begin{multline}
  \Pi_\WWcc(M^2) =
  C_A \, \frac{\alpha}{4\pi} \left(\frac{\mu^2}{M^2}\right)^\eps S_\eps \,
  \biggl\{
    \left(-1 + \frac{1}{d-1}\right) B_1^M(1,-1)
  \\*
    + \left(3 - \frac{2}{d-1}\right) B_1^M(1,0)
    + \left(\frac{9}{2} - \frac{3}{2(d-1)}\right) B_1^M(1,1)
  \biggr\}
  \,,
\end{multline}
wobei $B_1^M(0,0) = 0$ und die Symmetrie
$B_1^M(n_1,n_2) = B_1^M(n_2,n_1)$ benutzt wurden.
Die skalaren Zweipunktintegrale~$B_1^M$ sind in der Literatur bekannt
(siehe z.B.\cite{Passarino:1979jh}):
\begin{align}
\label{eq:B1Mred}
  B_1^M(1,-1) &= B_1^M(1,0) = \frac{1}{\eps} + 1 + \Oc(\eps) \,, \\
\label{eq:B1Mfull}
  B_1^M(1,1) &= \frac{1}{\eps} - \frac{\pi}{\sqrt3} + 2 + \Oc(\eps) \,.
\end{align}
Daraus ergibt sich:
\begin{equation}
  \Pi_\WWcc(M^2) =
  C_A \, \frac{\alpha}{4\pi} \left(\frac{\mu^2}{M^2}\right)^\eps S_\eps
  \left( \frac{17}{3\eps} - 4\frac{\pi}{\sqrt3} + \frac{82}{9} \right)
  + \Oc(\eps)
  \,.
\end{equation}
Nach Einsetzen von $\Pi_\WWcc(M^2)$ in den allgemeinen
Ausdruck~(\ref{eq:F2MrenPi1}) f"ur die Renormierung der Masse im
Einschleifenergebnis erh"alt man den Beitrag der Eichboson- und
Geistschleife dazu:
\begin{align}
\label{eq:MrenWWccerg}
  \Delta F_\WWcc^M &=
    C_F C_A
    \left(\frac{\alpha}{4\pi}\right)^2
    \left(\frac{\mu^2}{M^2}\right)^{2\eps} S_\eps^2
    \, \Biggl\{
  \frac{1}{\eps} \left[
    - \frac{34}{3} \lqm + 17 \right]
  + \left(8\frac{\pi}{\sqrt3} - \frac{164}{9}\right) \lqm
  \nonumber \\* & \qquad
  - 4\sqrt{3}\pi - \frac{17}{3}\pi^2 + \frac{317}{6}
  \Biggr\}
  + \Oc(\eps) + \Oc\!\left(\frac{M^2}{Q^2}\right)
  .
\end{align}
Die nichtlogarithmische Konstante in der Ordnung~$\eps^0$ wird hier
allerdings nicht ben"otigt.

Alle Beitr"age mit Eichboson- und Geistschleife aus den Gleichungen
(\ref{eq:WWccerg}), (\ref{eq:fcorrWWccerg}) und (\ref{eq:MrenWWccerg})
k"onnen nun addiert werden:
\begin{align}
\label{eq:WWcctoterg}
  F_\WWcc &= F_{v,\WWcc} + \Sigma_\WWcc + \Delta F_\WWcc^M
  \nonumber \\* &=
    C_F C_A
    \left(\frac{\alpha}{4\pi}\right)^2
    \left(\frac{\mu^2}{M^2}\right)^{2\eps} S_\eps^2
    \, \Biggl\{
  \frac{1}{\eps} \left[
    - \frac{5}{3} \lqm^2 + 5 \lqm - \frac{10}{9}\pi^2 - \frac{35}{6} \right]
  + \frac{10}{9} \lqm^3
  \nonumber \\* & \qquad
  - \frac{76}{9} \lqm^2
  + \left(-4\sqrt{3}\,\Cl2\!\left(\frac{\pi}{3}\right)
    + 8\frac{\pi}{\sqrt3} + \frac{59}{2}\right) \lqm
  \Biggr\}
  \nonumber \\* & \quad
  + \Oc(\eps^0 \lqm^0) + \Oc(\eps) + \Oc\!\left(\frac{M^2}{Q^2}\right)
  .
\end{align}
Auch dieses Ergebnis enth"alt nur einen einfachen $1/\eps$-Pol, der
bis auf den Faktor $\frac{5}{3} C_A \frac{\alpha}{4\pi}$ dem
Einschleifenformfaktor in Gl.~(\ref{eq:F1erg}) entspricht.
Die $1/\eps$-Pole aus den Gleichungen (\ref{eq:3Werg}) und
(\ref{eq:WWcctoterg}) k"onnen also durch die
Kopplungskonstantenrenormierung (siehe n"achster Abschnitt) aufgehoben
werden.

%
%
\section[Zusammenfassung der $C_F C_A$-Beitr"age]
  {\boldmath Zusammenfassung der $C_F C_A$-Beitr"age}
\label{sec:sumCA}

Bevor alle Beitr"age zum Zweischleifenformfaktor, die proportional zu
$C_F C_A$ sind, zusammengefasst werden k"onnen, muss noch die Renormierung
der Kopplungskonstante ber"ucksichtigt werden.
Wenn man in die allgemeine Form~(\ref{eq:F2alpharenbeta0})
den $C_A$"~Anteil von $\beta_0$~(\ref{eq:beta0}) einsetzt, erh"alt man:
\begin{align}
\label{eq:alpharenCAerg}
  \Delta F_{C_F C_A}^\alpha &=
    C_F C_A
    \left(\frac{\alpha}{4\pi}\right)^2
    \left(\frac{\mu^2}{M^2}\right)^\eps S_\eps
    \, \Biggl\{
  \frac{1}{\eps} \left[
    \frac{11}{3} \lqm^2 - 11 \lqm + \frac{22}{9}\pi^2 + \frac{77}{6} \right]
  - \frac{11}{9} \lqm^3
  \nonumber \\* & \qquad
  + \frac{11}{2} \lqm^2
  + \left(\frac{11}{9}\pi^2 - \frac{88}{3}\right) \lqm
  - \frac{22}{3}\zeta_3 - \frac{11}{6}\pi^2 + \frac{517}{12}
  \Biggr\}
  \nonumber \\* & \quad
  + \Oc(\eps) + \Oc\!\left(\frac{M^2}{Q^2}\right)
  ,
\end{align}
wobei die nichtlogarithmische Konstante in $\Oc(\eps^0)$ hier nicht
ben"otigt wird.

Zu den $C_F C_A$-Beitr"agen im Zweischleifenformfaktor geh"oren auch die
$C_F C_A$-Anteile aus den Ergebnissen
des nichtplanaren Vertexdiagramms~(\ref{eq:NPerg}),
des abelschen Vertexdiagramms mit Benz-Topologie~(\ref{eq:BEerg})
und der abelschen Selbstenergiekorrektur mit T1-Topologie~(\ref{eq:T1erg}).
Die anderen Beitr"age finden sich in den Gleichungen (\ref{eq:3Werg}),
(\ref{eq:WWcctoterg}) und (\ref{eq:alpharenCAerg}) zusammengefasst:
\begin{equation}
  F_{2,C_F C_A} =
  \Bigl[ F_{v,\NP} + 2\,F_{v,\BE} + \Sigma_\Tone
    \Bigr]_{C_F C_A}
  + F_\tW + F_\WWcc + \Delta F_{C_F C_A}^\alpha
  \,.
\end{equation}
Die Summe lautet f"ur eine allgemeine Renormierungsskala~$\mu$ (in die der
Faktor $S_\eps$ absorbiert wird) in $d=4$ Dimensionen im Hochenergielimes:
\begin{multline} 
\label{eq:F2CFCAerg}
  F_{2,C_F C_A} =
    C_F C_A
    \left(\frac{\alpha}{4\pi}\right)^2
    \, \Biggl\{
  \\*
  \frac{11}{9} \lqm^3
  - \left(-\frac{\pi^2}{3} + \frac{233}{18}\right) \lqm^2
  + \left( 4\sqrt{3}\,\Cl2\!\left(\frac{\pi}{3}\right)
    + 8\frac{\pi}{\sqrt3} - \frac{88}{3}\zeta_3 + \frac{11}{9}\pi^2
    + \frac{193}{6} \right) \lqm
  \\*
  + \ln\left(\frac{\mu^2}{M^2}\right) \left[
    -\frac{11}{3} \lqm^2 + 11 \lqm - \frac{22}{9}\pi^2 - \frac{77}{6}
    \right]
  \Biggr\}
  + \Oc(\lqm^0)
  \,,
\end{multline}
mit $\lqm = \ln(Q^2/M^2)$.
Dieses Ergebnis ist bereits in $d=4$ Dimensionen endlich, aber es ist nicht
eichinvariant und darum nur in der Feynman-'t~Hooft-Eichung g"ultig.
Im n"achsten Abschnitt werden die Higgs-Beitr"age vorgestellt, mit denen
zusammen der nichtabelsche Beitrag dann eichunabh"angig wird.

Die Koeffizienten der Logarithmen $\lqm^3$ und $\lqm^2$ im
Ergebnis~(\ref{eq:F2CFCAerg}) sind jedoch getrennt von den Higgs-Beitr"agen
eichinvariant und stimmen mit der Vorhersage der Evolutionsgleichung
in~(\ref{eq:F2NNLLSUN}) "uberein.

%
%
\section{Higgs-Beitr"age}
\label{sec:Higgs}

Die Masse des Higgs-Bosons ist ein freier Parameter des Standardmodells.
Durch die direkte experimentelle Suche ist die Higgs-Masse nach unten
beschr"ankt: $M_H > 114{,}4\,\GeV$ (mit 95\% statistischer Sicherheit).
Fits der experimentellen Daten an das Standardmodell sind damit konsistent:
$M_H = 113^{+56}_{-40}\,\GeV$
(\emph{Review of Particle Physics}\cite{Eidelman:2004wy}).

Grunds"atzlich sind die Strahlungskorrekturen, die ein Higgs-Teilchen
beinhalten, neben der Eichbosonmasse~$M$ auch von der Higgs-Masse~$M_H$
abh"angig.
Es bedeutet jedoch eine erhebliche Komplikation der Schleifenrechnungen,
wenn in den massiven Propagatoren zwei verschiedene Massen auftauchen.
Deshalb wurde in dieser Arbeit die Higgs-Masse gleich der Eichbosonmasse
gesetzt:
\[
  M_H = M \,.
\]
Dies stellt eine nicht unwesentliche "Anderung der physikalischen
Parameter dar, da die tats"achliche Higgs-Masse deutlich oberhalb der
Eichbosonmassen von $M_Z \approx 91{,}2\,\GeV$ und
$M_W \approx 80{,}4\,\GeV$ liegt.
Jedoch beeinflusst die "Anderung der Higgs-Masse das Ergebnis f"ur den
Formfaktor oder die Vierfermionamplitude im Hochenergielimes nur wenig.
Die f"uhrenden drei Logarithmen in Zweischleifenordnung, $\lqm^4$, $\lqm^3$
und $\lqm^2$, h"angen nicht von der Higgs-Masse ab.
Erst in den Koeffizienten des linearen Logarithmus geht der Wert der
Higgs-Masse ein.
Am Ergebnis wird zu sehen sein, dass die Higgs-Beitr"age insgesamt nur eine
untergeordnete Rolle spielen, so dass die Variation mit~$M_H$ im gesamten
Formfaktor vernachl"assigbar ist.

Die Goldstone-Bosonen haben in der Feynman-'t~Hooft-Eichung mit $\xi=1$ die
gleiche Masse wie die Eichbosonen, so dass in den Schleifendiagrammen der
Higgs-Beitr"age mit $M_H=M$ nur noch eine von Null verschiedene Masse
auftaucht.

Die bisherigen Rechnungen f"ur den fermionischen, abelschen und
nichtabelschen Beitrag zum Formfaktor sind allgemein f"ur $SU(N)$- oder
$U(1)$-Eichgruppen g"ultig, wenn man f"ur $C_F$, $C_A$ und $T_F$ die
entsprechenden Werte einsetzt.
Der Higgs-Formalismus mit der spontanen Symmetriebrechung funktioniert in
der Form des Standardmodells jedoch nur f"ur Eichgruppen,
bei denen die Zahl der Goldstone-Bosonen mit der Zahl der Eichbosonen
"ubereinstimmt.
Hier wird, wie in Abschnitt~\ref{sec:SU2U1} beschrieben, ein
$SU(2)$-Higgs-Modell verwendet.

\subsection{Vertexdiagramme mit Higgs- und Goldstone-Bosonen}

\begin{figure}[ht]
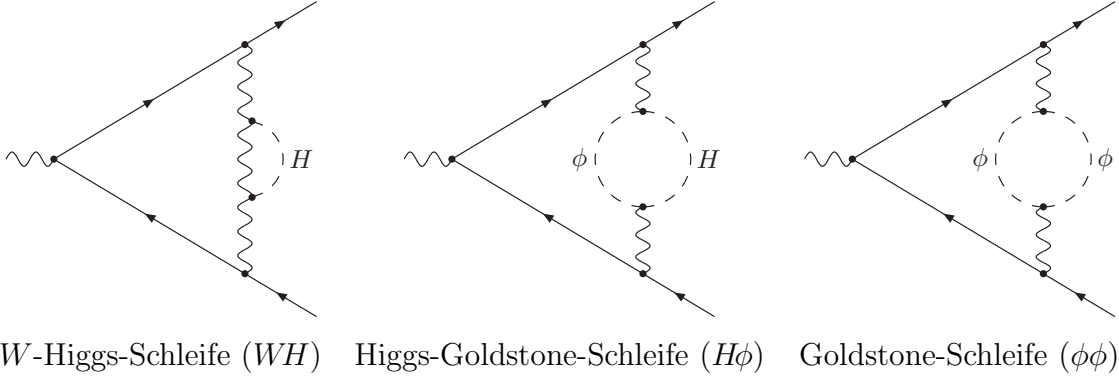

  \centering
  \valignbox[b]{\includegraphics[scale=0.9]{vertex-Wcorr-Higgs}\\
    $W$-Higgs-Schleife (\WH)}
  \hfill
  \valignbox[b]{\includegraphics[scale=0.9]{vertex-Wcorr-HiggsGoldstone}\\
    Higgs-Goldstone-Schleife (\Hphi)}
  \hfill
  \valignbox[b]{\includegraphics[scale=0.9]{vertex-Wcorr-Goldstones}\\
    Goldstone-Schleife (\phiphi)}
  \caption{Vertexdiagramme mit Higgs- und Goldstone-Bosonen}
  \label{fig:Higgsvertex}
\end{figure}

Zu den Beitr"agen mit Higgs- und Goldstone-Bosonen geh"oren die
Vertexdiagramme in Abb.~\ref{fig:Higgsvertex}.
Diagramme mit Tadpole-Schleifen im Eichbosonpropagator werden durch die
Massenrenormierung kompensiert und daher nicht betrachtet.
\begin{figure}[ht]
  \centering
  \includegraphics{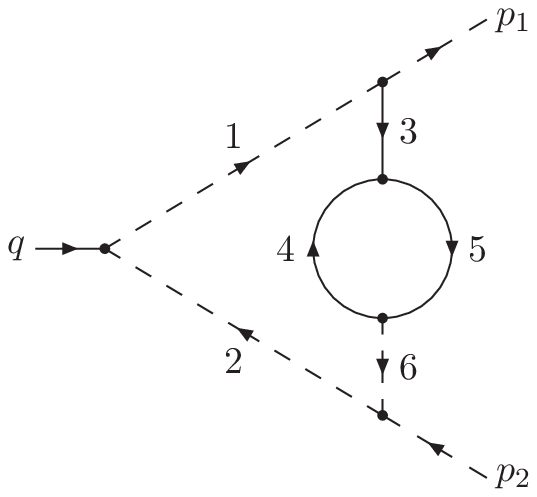}
  \caption{Skalares Vertexdiagramm mit Schleife im Eichbosonpropagator}
  \label{fig:Higgsvertexskalar}
\end{figure}

Die Struktur der skalaren Integrale ist bei allen drei Feynman-Diagrammen
die gleiche und entspricht derjenigen der Diagramme mit Eichboson- und
Geistschleife in Abb.~\ref{fig:Wcorrskalar}, hier nochmals dargestellt in
Abb.~\ref{fig:Higgsvertexskalar}.

Die Amplituden dieser drei Diagramme haben die folgende Form:
\begin{align}
\label{eq:Higgsvertexamplitude}
  \Fc_{v,\Higgs(i)}^\mu =
  \mu^{2\eps} \loopint dk \,
  ig \gamma_\nu t^a \, \frac{i\dslash k_1}{k_1^2} \,
  \gamma^\mu
  \frac{i\dslash k_2}{k_2^2} \, ig \gamma_\rho t^b
  \left(\frac{-i}{k_3^2-M^2}\right)^2 \,
  \tilde\Pi^{\nu\rho,ab}_{\Higgs(i)}(k_3)
  \,,
\end{align}
wobei $\Higgs(i)$ f"ur einen der drei Beitr"age \WH, \Hphi{} und \phiphi{}
steht,
mit den Impulsen
\begin{equation}
\label{eq:Higgsvertexmom}
  k_1 = p_1 + k ,\quad k_2 = p_2 + k ,\quad
  k_3 = k ,\quad k_4 = \ell ,\quad
  k_5 = k + \ell .
\end{equation}
Die Selbstenergie-Einsetzungen im Eichbosonpropagator lauten:
\begin{align}
\label{eq:WcorrWHamplitude}
  \tilde\Pi^{\nu\rho,ab}_\WH(k_3) &=
    \delta^{ab} \, g^{\nu\rho} \,
    \mu^{2\eps} \loopint d\ell \,
    igM \, \frac{-i}{k_4^2 - M^2} \, igM \, \frac{i}{k_5^2 - M^2}
  \,,
  \\
\label{eq:WcorrHphiamplitude}
  \tilde\Pi^{\nu\rho,ab}_\Hphi(k_3) &=
    \delta^{ab} \,
    \mu^{2\eps} \loopint d\ell \,
    \frac{g}{2} \, (-k_5-k_4)^\nu \, \frac{i}{k_4^2 - M^2} \,
    \frac{g}{2} \, (k_5+k_4)^\rho \, \frac{i}{k_5^2 - M^2}
  \,,
  \\
\label{eq:Wcorrphiphiamplitude}
  \tilde\Pi^{\nu\rho,ab}_\phiphi(k_3) &=
    \frac{1}{2} \, \mu^{2\eps} \loopint d\ell
    \left(-\frac{g}{2}\right) f^{acd} \, (-k_5-k_4)^\nu \,
      \frac{i}{k_4^2 - M^2}
    \nonumber \\* & \qquad\qquad\qquad\qquad \times
    \left(-\frac{g}{2}\right) f^{bcd} \, (k_5+k_4)^\rho \,
      \frac{i}{k_5^2 - M^2}
  \,.
\end{align}
Der Faktor~$1/2$ vor der $\phiphi$-Schleife ist ein Symmetriefaktor, genau
wie f"ur die Eichbosonschleife in Gl.~(\ref{eq:WWccamplitude}).

Die Farbfaktoren der Vertexdiagramme $\Fc_{v,\WH}^\mu$ und
$\Fc_{v,\Hphi}^\mu$ sind jeweils
\[
  \delta^{ab} \, t^a t^b = C_F \, \unity = \frac{3}{4} \, \unity \,,
\]
der Farbfaktor des Diagramms $\Fc_{v,\phiphi}$ ist
\[
  f^{acd} f^{bcd} \, t^a t^b = C_A C_F \, \unity = \frac{3}{2} \, \unity \,,
\]
f"ur ein $SU(2)$-Higgs-Modell, das hier zugrunde liegt.
Die Kombination des Farbfaktors und des Symmetriefaktors~$1/2$ zeigt, dass
die Amplituden der Diagramme mit Goldstone-Schleife und mit
Higgs-Goldstone-Schleife gleich sind:
\[
  \Fc_{v,\Hphi}^\mu = \Fc_{v,\phiphi} \,.
\]

Die Beitr"age der drei Vertexdiagramme zum Formfaktor sind jeweils
$F_{v,\WH}$ und $F_{v,\Hphi} = F_{v,\phiphi}$.
Sie werden in die gleichen skalaren Integrale~(\ref{eq:Wcorrskalar})
aufgeteilt wie die Diagramme mit Eichboson- oder Geistschleife in
Abschnitt~\ref{sec:WWccvertex}.
Diese Zerlegung ist im Anhang~\ref{sec:SkalarHiggsvertex} in den
Gleichungen (\ref{eq:WHvertexzerlegung}) und
(\ref{eq:Hphiphivertexzerlegung}) angegeben.

Die Koeffizienten des Diagramms mit Eichboson-Higgs-Schleife~(\WH) sind
aufgrund der zweifachen Kopplung~$igM$ alle proportional zu $M^2/Q^2$.  Man
k"onnte naiv vermuten, dass deshalb der Beitrag dieses Diagramms im
Hochenergielimes komplett mit $M^2/Q^2$ unterdr"uckt w"are.  Dies ist
jedoch nicht der Fall, da mehrere skalare Integrale von der
Ordnung~$(M^2/Q^2)^{-1}$ sind und den Faktor $M^2/Q^2$ in den Koeffizienten
k"urzen.

Die Ergebnisse der skalaren Integrale sind aus
Abschnitt~\ref{sec:WWccvertex} bzw. aus Gleichung~(\ref{eq:WWccergskalar})
im Anhang bekannt.
Die Beitr"age der Feynman-Diagramme zum Formfaktor sind:
\begin{align}
\label{eq:WHvertexerg}
  F_{v,\WH} &=
    \left(\frac{\alpha}{4\pi}\right)^2
    \left(\frac{\mu^2}{M^2}\right)^{2\eps} S_\eps^2
    \, \Biggl\{
  \frac{1}{\eps} \left[ -\frac{3}{2} \lqm + 3 \right]
  + \biggl( 2\sqrt{3}\,\Cl2\!\left(\frac{\pi}{3}\right)
    - 3 \biggr) \, \lqm
  \Biggr\}
  \nonumber \\* & \quad
  + \Oc(\eps^0 \lqm^0) + \Oc(\eps) + \Oc\!\left(\frac{M^2}{Q^2}\right)
  ,
\\
\label{eq:Hphiphivertexerg}
  F_{v,\Hphi} &= F_{v,\phiphi} =
    \left(\frac{\alpha}{4\pi}\right)^2
    \left(\frac{\mu^2}{M^2}\right)^{2\eps} S_\eps^2
    \, \Biggl\{
  \frac{1}{\eps} \left[
    \frac{1}{16} \lqm^2 + \frac{7}{16} \lqm
     + \frac{\pi^2}{24} - \frac{67}{64} \right]
  - \frac{1}{24} \lqm^3 + \frac{17}{48} \lqm^2
  \nonumber \\* & \qquad
  + \biggl( -\frac{3}{4}\sqrt{3}\,\Cl2\!\left(\frac{\pi}{3}\right)
    + \frac{19}{96} \biggr) \, \lqm
  \Biggr\}
  + \Oc(\eps^0 \lqm^0) + \Oc(\eps) + \Oc\!\left(\frac{M^2}{Q^2}\right)
  .
\end{align}

\subsection[Selbstenergiekorrekturen mit Higgs- und Goldstone-Bosonen]
  {Selbstenergiekorrekturen mit Higgs- und\\ Goldstone-Bosonen}

\begin{figure}[ht]
  \centering
  \valignbox[b]{\includegraphics[scale=0.85]{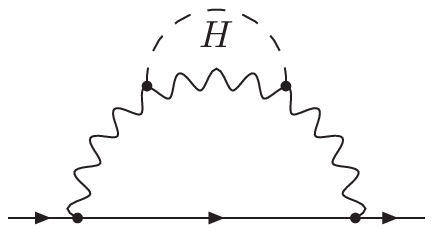}\\
    $W$-Higgs-Schleife (\WH)}
  \hfill
  \valignbox[b]{\includegraphics[scale=0.85]{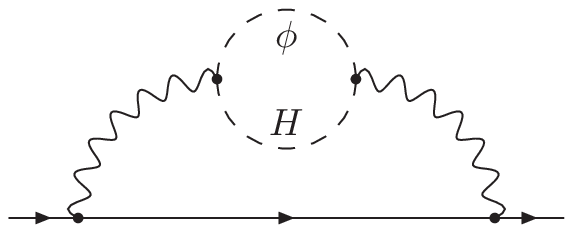}\\
    Higgs-Goldst.-Schleife (\Hphi)}
  \hfill
  \valignbox[b]{\includegraphics[scale=0.85]{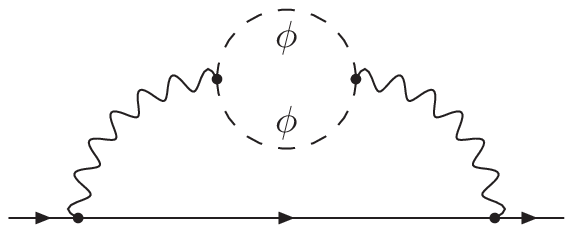}\\
    Goldstone-Schleife (\phiphi)}
  \caption{Selbstenergiediagramme mit Higgs- und Goldstone-Bosonen}
  \label{fig:Higgsfcorr}
\end{figure}
\begin{figure}[ht]
  \centering
  \includegraphics{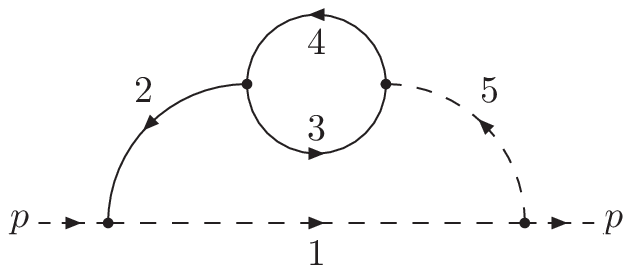}
  \caption{Skalares Selbstenergiediagramm mit Schleife im Eichbosonpropagator}
  \label{fig:Higgsfcorrskalar}
\end{figure}
Die Selbstenergiediagramme mit Higgs- und Goldstone-Bosonen sind in
Abb.~\ref{fig:Higgsfcorr} dargestellt.
Das zugeh"orige skalare Diagramm ist bei allen drei Feynman"=Diagrammen
durch dasjenige mit Eichboson- und Geistschleife in
Abb.~\ref{fig:fcorrWWccskalar} gegeben, das hier nochmals in
Abb.~\ref{fig:Higgsfcorrskalar} zu sehen ist.

Die Amplituden der drei Selbstenergiediagramme haben die folgende Form:
\begin{align}
\label{eq:Higgsfcorramplitude}
  \tilde\Sigma_{\Higgs(i)} =
  \mu^{2\eps} \loopint dk \,
  ig \gamma_\mu t^a \, \frac{i\dslash k_1}{k_1^2} \,
  ig \gamma_\nu t^b
  \left(\frac{-i}{k_2^2-M^2}\right)^2 \,
  \tilde\Pi^{\mu\nu,ab}_{\Higgs(i)}(k_2)
  \,,
\end{align}
wobei $\Higgs(i)$ f"ur einen der drei Beitr"age \WH, \Hphi{} und \phiphi{}
steht,
mit den Impulsen
\begin{equation}
\label{eq:Higgsfcorrmom}
  k_1 = p + k ,\quad
  k_2 = k ,\quad k_3 = \ell ,\quad
  k_4 = k + \ell .
\end{equation}
Die Selbstenergie-Einsetzungen
$\tilde\Pi^{\mu\nu,ab}_\WH(k_2)$, $\tilde\Pi^{\mu\nu,ab}_\Hphi(k_2)$
und $\tilde\Pi^{\mu\nu,ab}_\phiphi(k_2)$
im Eichbosonpropagator sind aus den Gleichungen
(\ref{eq:WcorrWHamplitude}), (\ref{eq:WcorrHphiamplitude}) und
(\ref{eq:Wcorrphiphiamplitude}) bekannt.
Die Farbfaktoren entsprechen denen der Vertexdiagramme im vorigen Abschnitt,
und auch hier gilt die Identit"at
\begin{equation}
  \tilde\Sigma_\Hphi = \tilde\Sigma_\phiphi \,.
\end{equation}

Die Beitr"age der drei Feynman-Diagramme zum Formfaktor,
$\Sigma_\WH$ und $\Sigma_\Hphi = \Sigma_\phiphi$,
werden als Linearkombinationen der skalaren
Integrale $\Bm_\Wc$ und $\DBm_\Wc$ geschrieben,
die bereits von den Selbstenergiediagrammen mit Eichboson- oder
Geistschleife in Abschnitt~\ref{sec:WWccfcorr} bekannt sind.
Diese Zerlegung ist im Anhang~\ref{sec:SkalarHiggsfcorr} in den
Gleichungen (\ref{eq:WHfcorrzerlegung}) und
(\ref{eq:Hphiphifcorrzerlegung}) aufgelistet.

Die Ergebnisse der einzelnen skalaren Integrale wurden in
Abschnitt~\ref{sec:WWccfcorr} berechnet und sind im
Anhang~\ref{sec:SkalarfcorrWWcc} in Gl.~(\ref{eq:fcorrWWccergskalar})
angegeben.
Die Selbstenergiekorrekturen mit Higgs- und Goldstone-Bosonen lauten:
\begin{align}
\label{eq:WHfcorrerg}
  \Sigma_\WH &=
    \left(\frac{\alpha}{4\pi}\right)^2
    \left(\frac{\mu^2}{M^2}\right)^{2\eps} S_\eps^2
  \left( -\frac{3}{4\eps} \right)
  + \Oc(\eps^0)
  \,,
\\
\label{eq:Hphiphifcorrerg}
  \Sigma_\Hphi &= \Sigma_\phiphi =
    \left(\frac{\alpha}{4\pi}\right)^2
    \left(\frac{\mu^2}{M^2}\right)^{2\eps} S_\eps^2
  \cdot \frac{21}{64\eps}
  + \Oc(\eps^0)
  \,.
\end{align}

\subsection[Massenrenormierung mit Higgs- und Goldstone-Bosonen]
  {Massenrenormierung mit Higgs- und\\ Goldstone-Bosonen}

\begin{figure}[ht]
  \centering
  \valignbox[b]{\includegraphics{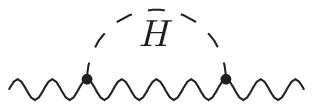}\\
    $W$-Higgs-Schleife (\WH)}
  \hfill
  \valignbox[b]{\includegraphics{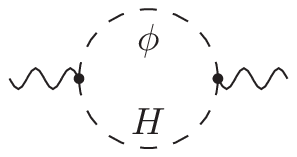}\\
    Higgs-Goldstone-Schleife (\Hphi)}
  \hfill
  \valignbox[b]{\includegraphics{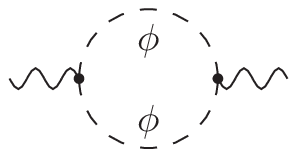}\\
    Goldstone-Schleife (\phiphi)}
  \caption{Higgs-Beitr"age zur Massenrenormierung}
  \label{fig:HiggsMren}
\end{figure}

Zur Renormierung der Eichbosonmasse tragen die drei Selbstenergiediagramme
in Abb.~\ref{fig:HiggsMren} bei.

Die Amplituden dieser Feynman-Diagramme,
$\tilde\Pi^{\mu\nu,ab}_\WH$, $\tilde\Pi^{\mu\nu,ab}_\Hphi$
und $\tilde\Pi^{\mu\nu,ab}_\phiphi$,
sind bereits als Selbstenergie-Einsetzungen aus den vorigen Abschnitten
bekannt, sie sind in den Gleichungen
(\ref{eq:WcorrWHamplitude}), (\ref{eq:WcorrHphiamplitude}) und
(\ref{eq:Wcorrphiphiamplitude}) definiert.
Die Farbfaktoren der Diagramme sind
\[
  \delta^{ab} \;(\text{f"ur }\WH,\Hphi) \quad \text{und} \quad
  f^{acd} f^{bcd} = C_A \, \delta^{ab} = 2\,\delta^{ab}
    \;(\text{f"ur }\phiphi) \,.
\]

Analog zum Abschnitt~\ref{sec:WWccMren} werden die skalaren Funktionen
$\Pi_\WH(k^2)$, $\Pi_\Hphi(k^2)$ und $\Pi_\phiphi(k^2)$
aus den Amplituden extrahiert.
F"ur $k^2=M^2$ lassen sie sich als Linearkombinationen der skalaren
Integrale $B_1^M(n_1,n_2)$ aus Gl.~(\ref{eq:MrenWcorrskalar}) schreiben:
\begin{align}
  \Pi_\WH(M^2) &=
    \frac{\alpha}{4\pi} \left(\frac{\mu^2}{M^2}\right)^\eps S_\eps \,
    \biggl\{ - B_1^M(1,1) \biggr\}
  \,,
\\
  \Pi_\Hphi(M^2) &= \Pi_\phiphi(M^2) =
    \frac{\alpha}{4\pi} \left(\frac{\mu^2}{M^2}\right)^\eps S_\eps \,
    \biggl\{
      - \frac{1}{2(d-1)} \, B_1^M(1,-1)
      + \frac{1}{d-1} \, B_1^M(1,0)
  \nonumber \\* & \qquad\qquad\qquad\qquad\qquad\qquad\qquad
      + \frac{3}{4(d-1)} \, B_1^M(1,1)
    \biggr\}
  \,.  
\end{align}
Mit den Ergebnissen f"ur die skalaren Integrale~$B_1^M(n_1,n_2)$ aus
den Gleichungen (\ref{eq:B1Mred}) und (\ref{eq:B1Mfull}) erh"alt man:
\begin{align}
  \Pi_\WH(M^2) &=
    \frac{\alpha}{4\pi} \left(\frac{\mu^2}{M^2}\right)^\eps S_\eps
    \left( -\frac{1}{\eps} + \frac{\pi}{\sqrt3} - 2 \right)
    + \Oc(\eps)
  \,,
\\
  \Pi_\Hphi(M^2) &= \Pi_\phiphi(M^2) =
    \frac{\alpha}{4\pi} \left(\frac{\mu^2}{M^2}\right)^\eps S_\eps
    \left( \frac{5}{12\eps} - \frac{\pi}{4\sqrt3} + \frac{17}{18} \right)
    + \Oc(\eps)
  \,.
\end{align}

Durch Einsetzen dieser Selbstenergiebeitr"age in die
Gleichung~(\ref{eq:F2MrenPi1}) ergeben sich die Beitr"age zum Formfaktor:
\begin{align}
\label{eq:MrenWHerg}
  \Delta F_\WH^M &=
    \left(\frac{\alpha}{4\pi}\right)^2
    \left(\frac{\mu^2}{M^2}\right)^{2\eps} S_\eps^2
    \, \Biggl\{
  \frac{1}{\eps} \left[
    \frac{3}{2} \lqm - \frac{9}{4} \right]
  + \left( -\frac{1}{2}\sqrt{3}\pi + 3 \right) \lqm
  \nonumber \\* & \qquad
  + \frac{3}{4}\sqrt{3}\pi + \frac{3}{4}\pi^2 - \frac{63}{8}
  \Biggr\}
  + \Oc(\eps) + \Oc\!\left(\frac{M^2}{Q^2}\right)
  ,
\\
\label{eq:MrenHphiphierg}
  \Delta F_\Hphi^M &= \Delta F_\phiphi^M =
    \left(\frac{\alpha}{4\pi}\right)^2
    \left(\frac{\mu^2}{M^2}\right)^{2\eps} S_\eps^2
    \, \Biggl\{
  \frac{1}{\eps} \left[
    -\frac{5}{8} \lqm + \frac{15}{16} \right]
  + \left( \frac{1}{8}\sqrt{3}\pi - \frac{17}{12} \right) \lqm
  \nonumber \\* & \qquad
  - \frac{3}{16}\sqrt{3}\pi - \frac{5}{16}\pi^2 + \frac{113}{32}
  \Biggr\}
  + \Oc(\eps) + \Oc\!\left(\frac{M^2}{Q^2}\right)
  ,
\end{align}
wobei $C_F = 3/4$ gesetzt wurde.
Die nichtlogarithmische Konstante in der Ordnung~$\eps^0$ wird f"ur die
Higgs-Beitr"age nicht ben"otigt.

\subsection{Zusammenfassung der Higgs-Beitr"age}

Alle Higgs-Beitr"age, die jeweils von derselben
Selbstenergie-Einsetzung \WH, \Hphi{}
oder \phiphi{} herr"uhren, werden nun zusammengefasst:
\begin{equation}
  F_{\Higgs(i)} =
    F_{v,\Higgs(i)} + \Sigma_{\Higgs(i)} + \Delta F^M_{\Higgs(i)}
  \,.
\end{equation}
Mit den Ergebnissen aus
(\ref{eq:WHvertexerg}), (\ref{eq:Hphiphivertexerg}),
(\ref{eq:WHfcorrerg}), (\ref{eq:Hphiphifcorrerg}),
(\ref{eq:MrenWHerg}) und (\ref{eq:MrenHphiphierg})
erh"alt man:
\begin{align}
\label{eq:WHerg}
  F_\WH  &=
    \left(\frac{\alpha}{4\pi}\right)^2
    \left(\frac{\mu^2}{M^2}\right)^{2\eps} S_\eps^2
  \left( 2\sqrt{3}\,\Cl2\!\left(\frac{\pi}{3}\right)
    - \frac{1}{2}\sqrt{3}\pi \right) \lqm
  \nonumber \\* & \quad
  + \Oc(\eps^0 \lqm^0) + \Oc(\eps) + \Oc\!\left(\frac{M^2}{Q^2}\right)
  ,
\\
\label{eq:Hphiphierg}
  F_\Hphi  &= F_\phiphi =
    \left(\frac{\alpha}{4\pi}\right)^2
    \left(\frac{\mu^2}{M^2}\right)^{2\eps} S_\eps^2 \,
  \Biggl\{
    \frac{1}{\eps} \left[
      \frac{1}{16} \lqm^2 - \frac{3}{16} \lqm + \frac{\pi^2}{24}
      + \frac{7}{32} \right]
  - \frac{1}{24} \lqm^3 + \frac{17}{48} \lqm^2
  \nonumber \\* & \qquad
  + \left( -\frac{3}{4}\sqrt{3}\,\Cl2\!\left(\frac{\pi}{3}\right)
    + \frac{1}{8}\sqrt{3}\pi - \frac{39}{32} \right) \lqm
  \Biggr\}
  + \Oc(\eps^0 \lqm^0) + \Oc(\eps) + \Oc\!\left(\frac{M^2}{Q^2}\right)
  .
\end{align}
Der Beitrag $F_\WH$ ist bereits in $d=4$ Dimensionen endlich.
Der $\eps$-Pol der Beitr"age $F_\Hphi = F_\phiphi$ ist proportional
zum Einschleifenformfaktor und wird durch die Renormierung der
Kopplungskonstante eliminiert.
Mit dem $n_s$-Anteil von $\beta_0$ aus~(\ref{eq:beta0}),
eingesetzt in die allgemeine Form~(\ref{eq:F2alpharenbeta0}), ergibt sich:
\begin{align}
\label{eq:Higgsalpharenerg}
  \Delta F_\Higgs^\alpha &=
    \left(\frac{\alpha}{4\pi}\right)^2
    \left(\frac{\mu^2}{M^2}\right)^\eps S_\eps \,
    \Biggl\{
    \frac{1}{\eps} \left[
      -\frac{1}{8} \lqm^2 + \frac{3}{8} \lqm
      - \frac{\pi^2}{12} - \frac{7}{16} \right]
    + \frac{1}{24} \lqm^3 - \frac{3}{16} \lqm^2
\nonumber \\* & \qquad
      + \left(-\frac{\pi^2}{24} + 1\right) \lqm
      + \frac{1}{4}\zeta_3 + \frac{\pi^2}{16} - \frac{47}{32}
    \Biggr\}
    + \Oc(\eps) + \Oc\left(\frac{M^2}{Q^2}\right)
  ,
\end{align}
mit $C_F = 3/4$, $T_F = 1/2$ und $n_s=1$,
wobei die nichtlogarithmische Konstante in
der Ordnung~$\eps^0$ hier nicht ben"otigt wird.

Jetzt k"onnen alle Higgs-Beitr"age aus
(\ref{eq:WHerg}), (\ref{eq:Hphiphierg}) und (\ref{eq:Higgsalpharenerg})
zusammengesetzt werden.
In $d=4$ Dimensionen erh"alt man das folgende endliche Ergebnis im
Hochenergielimes:
\begin{align}
\label{eq:F2Higgserg}
  F_{2,\Higgs} &=
    F_\WH + F_\Hphi + F_\phiphi + \Delta F_\Higgs^\alpha
  \nonumber \\* &=
    \left(\frac{\alpha}{4\pi}\right)^2 \,
    \Biggl\{
    -\frac{1}{24} \lqm^3 + \frac{25}{48} \lqm^2
    - \left( -\frac{1}{2}\sqrt{3}\,\Cl2\!\left(\frac{\pi}{3}\right)
    + \frac{1}{4}\sqrt{3}\pi + \frac{\pi^2}{24}
    + \frac{23}{16} \right) \lqm
\nonumber \\* & \qquad\qquad\qquad
    + \ln\left(\frac{\mu^2}{M^2}\right) \left[
      \frac{1}{8} \lqm^2 - \frac{3}{8} \lqm + \frac{\pi^2}{12} + \frac{7}{16}
      \right]
    \Biggr\}
    + \Oc(\lqm^0)
  \,.      
\end{align}
Dieses Ergebnis ist genauso wie die $C_F C_A$-Beitr"age nicht separat
eichinvariant und darum nur in der Feynman-'t~Hooft-Eichung g"ultig.
Es ist zudem auf ein $SU(2)$-Higgs-Modell beschr"ankt.
F"ur ein $U(1)$-Higgs-Modell (in dem das Higgs die Hyperladung~1 hat)
"andert sich folgendes:
Alle Beitr"age werden durch die Anpassung von~$C_F$ mit $4/3$
multipliziert. 
Au"serdem erhalten die Vertizes zwischen zwei Eichbosonen und dem
Higgs-Boson sowie zwischen Eichboson, Higgs- und Goldstone-Boson den
Faktor~2. Der Vertex zwischen dem Eichboson und zwei Goldstone-Bosonen
verschwindet in einer $U(1)$-Theorie.
Bei der Kopplungskonstantenrenormierung "andert sich neben $C_F$ auch
$T_F$, was zu einem weiteren Faktor~2 f"uhrt.
Insgesamt m"ussen die Beitr"age wie folgt transformiert werden:
\begin{equation}
  F_\WH^{U(1)} = \frac{16}{3} \, F_\WH \,,\quad
  F_\Hphi^{U(1)} = \frac{16}{3} \, F_\Hphi \,,\quad
  F_\phiphi^{U(1)} = 0 \,,\quad
  \Delta F_\Higgs^{\alpha \, U(1)} = \frac{8}{3} \, \Delta F_\Higgs^\alpha \,.
\end{equation}
Der Beitrag zum $U(1)$-Formfaktor lautet:
\begin{align}
\label{eq:F2HiggsU1erg}
  F_{2,\Higgs}^{U(1)} &=
    \left(\frac{\alpha}{4\pi}\right)^2 \,
    \Biggl\{
    -\frac{1}{9} \lqm^3 + \frac{25}{18} \lqm^2
    - \left( -\frac{20}{\sqrt{3}}\,\Cl2\!\left(\frac{\pi}{3}\right)
    + 2\sqrt{3}\pi + \frac{\pi^2}{9}
    + \frac{23}{6} \right) \lqm
\nonumber \\* & \qquad\qquad\qquad
    + \ln\left(\frac{\mu^2}{M^2}\right) \left[
      \frac{1}{3} \lqm^2 - \lqm + \frac{2}{9}\pi^2 + \frac{7}{6}
      \right]
    \Biggr\}
    + \Oc(\lqm^0)
  \,.      
\end{align}
Sowohl im $SU(2)$-Ergebnis~(\ref{eq:F2Higgserg}) als auch im
$U(1)$-Ergebnis~(\ref{eq:F2HiggsU1erg}) stimmen die Koeffizienten der
Logarithmen $\lqm^3$ und $\lqm^2$ mit den Vorhersagen der
Evolutionsgleichung in~(\ref{eq:F2NNLLSUN}) "uberein, wenn man f"ur
$C_F$ und $T_F$ jeweils die entsprechenden Werte einsetzt.

Um beurteilen zu k"onnen, wie sich eine "Anderung im Parameter der
Higgs-Masse auf das Ergebnis auswirkt, wurde die Berechnung der
Higgs-Beitr"age neben der bisherigen Annahme $M_H = M$ auch f"ur den
hypothetischen Fall eines
masselosen Higgs-Bosons, $M_H = 0$, durchgef"uhrt.
Dann w"are der Beitrag zum Formfaktor in einer $SU(2)$-Theorie der folgende:
\begin{align}
\label{eq:F2HiggsMH0erg}
  F_{2,\Higgs}^{M_H=0} &=
    \left(\frac{\alpha}{4\pi}\right)^2 \,
    \Biggl\{
    -\frac{1}{24} \lqm^3 + \frac{25}{48} \lqm^2
    - \left( \frac{3}{4}\sqrt{3}\,\Cl2\!\left(\frac{\pi}{3}\right)
      - \frac{1}{8}\sqrt{3}\pi -  \frac{3}{16}\pi^2
      + \frac{25}{16} \right) \lqm
\nonumber \\* & \qquad\qquad\qquad
    + \ln\left(\frac{\mu^2}{M^2}\right) \left[
      \frac{1}{8} \lqm^2 - \frac{3}{8} \lqm + \frac{\pi^2}{12} + \frac{7}{16}
      \right]
    \Biggr\}
    + \Oc(\lqm^0)
  \,.      
\end{align}
Nur der Koeffizient des linearen Logarithmus hat sich mit der Higgs-Masse
ge"andert, die f"uhrenden Logarithmen $\lqm^3$ und $\lqm^2$ sind gleich
geblieben.
Dies erkl"art sich dadurch, dass im Fall $M_H=M$ die Logarithmen zwar von
dieser gemeinsamen Masse abh"angen. Da das Higgs-Boson und die
Goldstone-Bosonen jedoch nur an das Eichboson und nicht an die Fermionen
koppeln, entstehen im Limes $M_H\to0$ keine kollinearen oder weichen
Divergenzen, so dass auch die Masse~$M$ in den Logarithmen~$\ln(Q^2/M^2)$
nicht mit der Higgs-Masse in Verbindung steht.
Der Limes $M_H\to0$ ist also ein stetiger Limes, der nur die Koeffizienten
der Logarithmen beeinflusst. Und weil die Koeffizienten der h"oheren
Logarithmenpotenzen nur massenunabh"angige Terme aufweisen, "andert sich
f"ur $M_H=0$ lediglich der Beitrag des linearen Logarithmus.

Die numerische Gr"o"se dieser "Anderung wird im n"achsten Abschnitt
diskutiert, wenn die Higgs-Beitr"age mit den zu $C_F C_A$ proportionalen
Beitr"agen zum nichtabelschen Anteil des Formfaktors zusammengesetzt
werden.

%
%
\section[Zusammenfassung der nichtabelschen Beitr"age]
  {Zusammenfassung der nichtabelschen\\ Beitr"age}
\label{sec:sumnichtabelsch}

Im Folgenden wird der Formfaktor nur noch f"ur eine an der
Skala $\mu=M$ renormierte Kopplung~$\alpha$ betrachtet, wie es auch f"ur
die Analyse der Evolutionsgleichung in Abschnitt~\ref{sec:Sudakov}
geschehen ist.
Die Renormierungsgruppenabh"angigkeit des Formfaktors kann leicht mittels
des Einschleifenergebnisses~(\ref{eq:F1erg}) und $\beta_0$~(\ref{eq:beta0})
wieder eingef"uhrt werden.

Die Beitr"age proportional zu $C_F C_A$~(\ref{eq:F2CFCAerg})
und die Higgs-Beitr"age~(\ref{eq:F2Higgserg}) addieren
sich zum eichinvarianten nichtabelschen Anteil des Formfaktors:
\begin{align}
\label{eq:F2NAerg}
  F_{2,\NA} &= F_{2,C_F C_A} + F_{2,\Higgs}
  \nonumber \\* &=
    \left(\frac{\alpha}{4\pi}\right)^2 \,
    \Biggl\{
    \frac{43}{24} \lqm^3
    - \left( -\frac{\pi^2}{2} + \frac{907}{48} \right) \lqm^2
  \nonumber \\* & \quad
    + \left( \frac{13}{2}\sqrt{3}\,\Cl2\!\left(\frac{\pi}{3}\right)
      + \frac{15}{4}\sqrt{3}\pi
      - 44\zeta_3
      + \frac{43}{24}\pi^2
      + \frac{749}{16} \right) \lqm
    \Biggr\}
    + \Oc(\lqm^0)
  \,,
\end{align}
mit $\lqm = \ln(Q^2/M^2)$ sowie
$C_F = 3/4$ und $C_A = 2$ f"ur das $SU(2)$-Higgs-Modell.

Die "Ubereinstimmung der f"uhrenden Logarithmen $\lqm^3$ und $\lqm^2$ mit
der Vorhersage der Evolutionsgleichung~(\ref{eq:F2NNLLSUN}) wurde bereits
festgestellt.
Der Koeffizient des linearen Logarithmus ist ein neues Ergebnis.
Aus ihm kann im Vergleich mit Gl.~(\ref{eq:F2gzx}) der nichtabelsche
Anteil des Koeffizienten~$\xi^{(2)}$ bestimmt werden,
da alle anderen Terme im Koeffizienten des linearen Logarithmus bereits aus
(\ref{eq:g1z1}), (\ref{eq:x1F01}) und (\ref{eq:zeta2}) bekannt sind:
\begin{equation}
  \xi^{(2)}\Big|_\NA =
    \frac{13}{2}\sqrt{3}\,\Cl2\!\left(\frac{\pi}{3}\right)
    + \frac{15}{4}\sqrt{3}\pi
    - 5\zeta_3
    - \frac{391}{18}
  \,.
\end{equation}
Damit sind die anomalen Dimensionen $\gamma$, $\zeta$ und $\xi$ aus der
Evolutionsgleichung~(\ref{eq:formfaktorevol}) f"ur den Formfaktor bis
in Zweischleifenn"aherung vollst"andig bekannt.

Die Koeffizienten der Logarithmen im Ergebnis~(\ref{eq:F2NAerg}) haben die
folgenden numerischen Werte:
\begin{equation}
  F_{2,\NA} = \left(\frac{\alpha}{4\pi}\right)^2 \Bigl(
    +1{,}7917\,\lqm^3
    - 13{,}9610\,\lqm^2
    + 43{,}4368\,\lqm
  \Bigr)
  + \Oc(\lqm^0)
  \,.
\end{equation}
Auch im nichtabelschen Anteil des Formfaktors gilt, dass die Koeffizienten
mit alternierendem Vorzeichen vom f"uhrenden~$\lqm^3$ bis hin zum
linearen~$\lqm$ ansteigen.

Wie bisher werden zur Gr"o"senabsch"atzung
$\alpha/(4\pi) = 0{,}003$ sowie $M=80\,\GeV$ und $Q=1000\,\GeV$ eingesetzt.
Die logarithmischen Beitr"age werden in Promille angegeben:
\begin{equation}
  \lqm^3 \to +2{,}0 \,,\quad
  \lqm^2 \to -3{,}2 \,,\quad
  \lqm^1 \to +1{,}97 \,.
\end{equation}
Wie man sieht, gibt es durch das wechselnde Vorzeichen auch im
nichtabelschen Anteil gro"se Kompensationen zwischen den logarithmischen
Termen. Aber ebenso gilt hier, dass der Beitrag des linearen Logarithmus
schon wesentlich kleiner als der Beitrag des quadratischen Logarithmus ist,
so dass sich die beginnende "`Konvergenz"' der logarithmischen Reihe
abzeichnet.

Zum Vergleich ist im Folgenden das nichtabelsche Ergebnis f"ur verschwindende
Higgs-Masse, $M_H=0$, aus (\ref{eq:F2CFCAerg}) und (\ref{eq:F2HiggsMH0erg})
angegeben:
\begin{align}
\label{eq:F2NAMH0erg}
  F_{2,\NA}^{M_H=0} &= F_{2,C_F C_A} + F_{2,\Higgs}^{M_H=0}
  \nonumber \\* &=
    \left(\frac{\alpha}{4\pi}\right)^2 \,
    \Biggl\{
    \frac{43}{24} \lqm^3
    - \left( -\frac{\pi^2}{2} + \frac{907}{48} \right) \lqm^2
  \nonumber \\* & \quad
    + \left( \frac{21}{4}\sqrt{3}\,\Cl2\!\left(\frac{\pi}{3}\right)
      + \frac{33}{8}\sqrt{3}\pi
      - 44\zeta_3
      + \frac{97}{48}\pi^2
      + \frac{747}{16} \right) \lqm
    \Biggr\}
    + \Oc(\lqm^0)
  \,.
\end{align}
Nur der Koeffizient des linearen Logarithmus hat jetzt einen anderen
numerischen Wert:
\begin{equation}
  F_{2,\NA}^{M_H=0} = \left(\frac{\alpha}{4\pi}\right)^2 \Bigl(
    +1{,}7917\,\lqm^3
    - 13{,}9610\,\lqm^2
    + 45{,}4167\,\lqm
  \Bigr)
  + \Oc(\lqm^0)
  \,.
\end{equation}
Der Beitrag des linearen Logarithmus "andert sich durch die verschwindende
Higgs-Masse von $1{,}97$ auf $2{,}06$~Promille. Der Unterschied betr"agt also
nur vernachl"assigbare $0{,}1$~Promille.
Dies ist eine "Anderung in der Gr"o"senordnung der nichtlogarithmischen
Konstante, auf deren Berechnung wir verzichtet haben.
Durch die Annahme, dass die Higgs-Masse der Eichbosonmasse entspricht,
wurde also ein Fehler von vernachl"assigbarer Gr"o"senordnung gemacht.

Zum Abschluss soll nun der gesamte Zweischleifenformfaktor betrachtet
werden, d.h. die Summe aus den fermionischen~(\ref{eq:F2nferg}),
abelschen~(\ref{eq:F2CFerg}) und nichtabelschen Beitr"agen~(\ref{eq:F2NAerg}).
Es gilt wieder $M_H=M$.
Der $SU(2)$"=Zweischleifenformfaktor lautet:
\begin{align}
\label{eq:F2erg}
  F_2 &= F_{2,n_f} + F_{2,C_F^2} + F_{2,\NA}
  \nonumber \\* &=
  \left(\frac{\alpha}{4\pi}\right)^2 \Biggl\{
    \frac{9}{32} \lqm^4
    + \left(\frac{5}{48} - \frac{n_f}{6}\right) \lqm^3
    + \left(\frac{7}{8}\pi^2 - \frac{691}{48} + \frac{19}{12} n_f\right)
      \lqm^2
  \nonumber \\* & \quad
    + \left( \frac{13}{2}\sqrt{3}\,\Cl2\!\left(\frac{\pi}{3}\right)
      + \frac{15}{4}\sqrt{3}\pi
      - \frac{61}{2}\zeta_3
      - \frac{11}{24}\pi^2
      + \frac{167}{4}
      - \frac{17}{4} n_f \right) \lqm
    \Biggr\}
    + \Oc(\lqm^0)
  \,.
\end{align}

\begin{figure}[pt]
  \centering
  \includegraphics{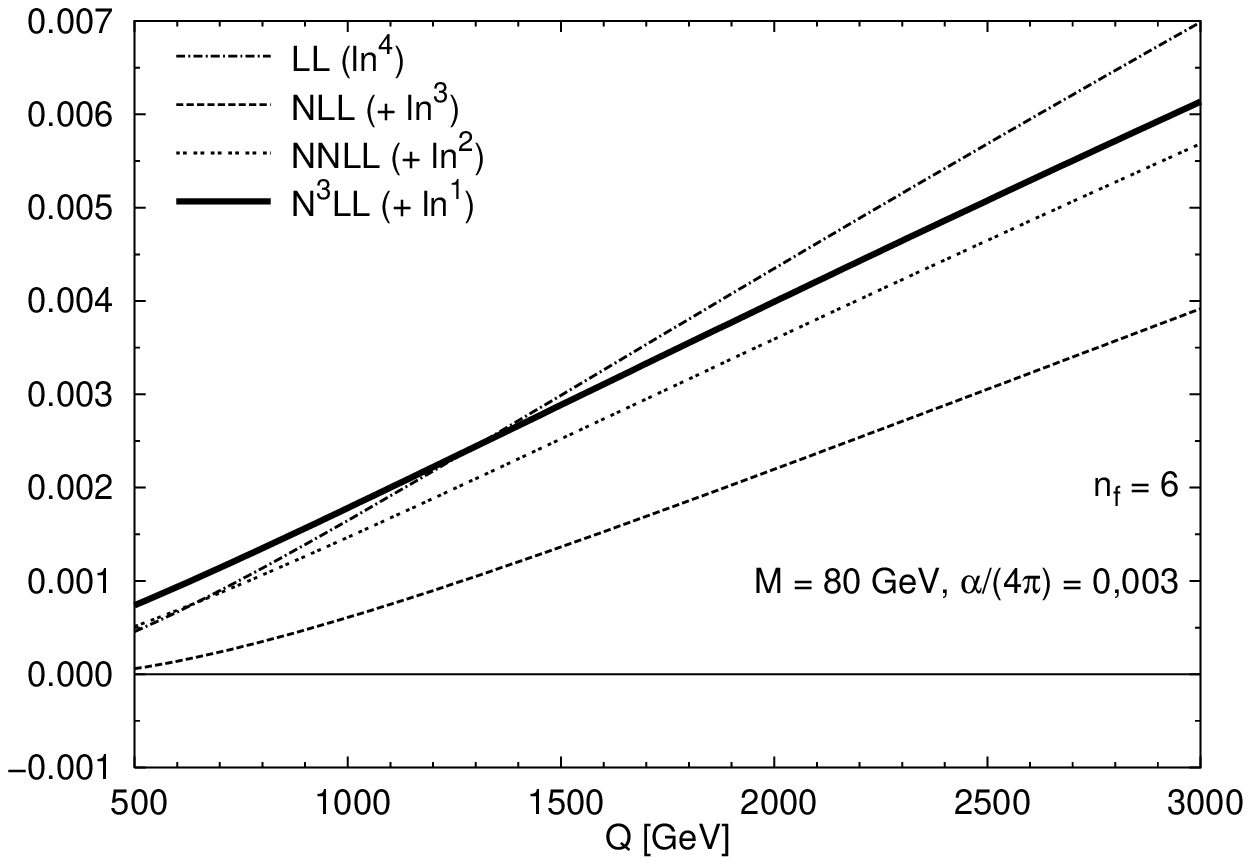}
  \caption{Kompletter $SU(2)$-Zweischleifenformfaktor in sukzessiven
    logarithmischen N"aherungen (mit $n_f=6$ Fermion-Dubletts)}
  \label{fig:plotF2}
\end{figure}
Der komplette Zweischleifenformfaktor ist in Abb.~\ref{fig:plotF2}
graphisch als Funktion des Impuls"ubertrags~$Q$ f"ur
$n_f=6$ Fermion-Dubletts (siehe Abschnitt~$\ref{eq:EWparam}$) dargestellt.
Dabei sind wieder wie in Abb.~\ref{fig:plotabelsch}
die sukzessiven logarithmischen N"aherungen gezeichnet.
Der f"uhrende $\lqm^4$ approximiert zwar um $Q\sim 1000\,\GeV$ ganz gut das
\NNNLL-Ergebnis des Hochenergielimes.
Aber f"ur h"ohere Impuls"ubertr"age entfernt sich die LL-N"aherung
wieder davon.
Durch die Hinzunahme des n"achstf"uhrenden kubischen Logarithmus~$\lqm^3$
wird die N"aherung zun"achst schlechter.
Erst mit dem quadratischen Logarithmus in \NNLL-N"aherung kommt man
dem Ergebnis wieder n"aher.
Und der lineare Logarithmus liefert einen zwar kleinen, aber durchaus noch
signifikanten Sprung der Kurve nach oben.

\begin{figure}[pt]
  \centering
  \includegraphics{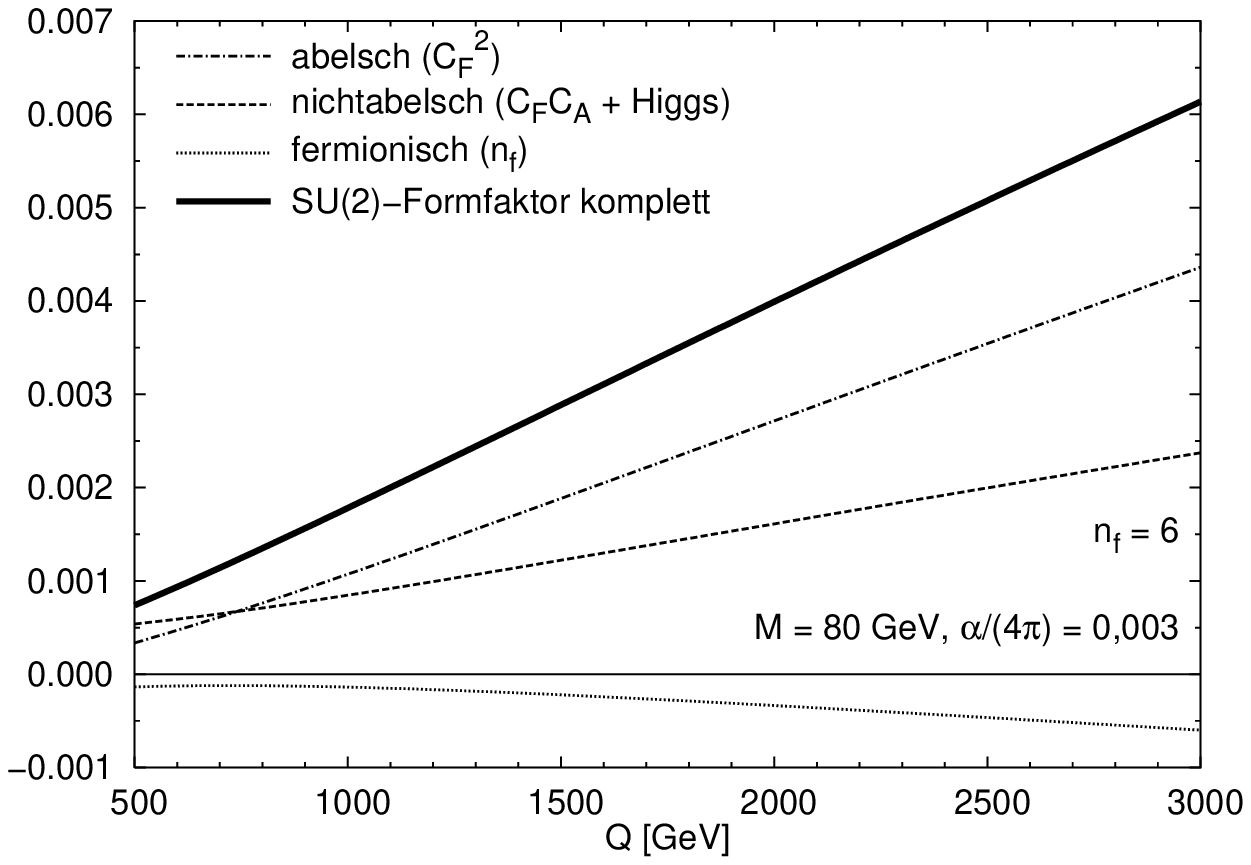}
  \caption{Beitr"age zum $SU(2)$-Zweischleifenformfaktor
    (mit $n_f=6$\protect\\ Fermion"=Dubletts), jeweils in \NNNLL-N"aherung}
  \label{fig:plotF2cont}
\end{figure}

In Abb.~\ref{fig:plotF2cont} sind die einzelnen Beitr"age zum
$SU(2)$-Zweischleifenformfaktor dargestellt.
Der gr"o"ste Anteil kommt von den abelschen und den nichtabelschen
Beitr"agen, der fermionische Beitrag spielt nur eine untergeordnete Rolle.

Der in diesem Abschnitt vorgestellte Formfaktor in Zweischleifenn"aherung
wird in Kapitel~\ref{chap:EW2loop} mit der reduzierten Amplitude aus
Abschnitt~\ref{sec:Sudakov} zur Vierfermion"-amplitude und zu Vorhersagen von
elektroschwachen Streuquerschnitten kombiniert.


\clearemptypage

\chapter{\boldmath Das $SU(2) \times U(1)$-Modell mit Massenl"ucke}
\chaptermark{Das $SU(2) \times U(1)$-Modell mit Massenl"ucke}
\label{chap:SU2U1}

Die bisherige Analyse beschr"ankt sich auf den Fall eines massiven
$SU(2)$-Modells mit spontaner Symmetriebrechung.
Dabei wird die zus"atzliche Existenz von masselosen $U(1)$-Eichbosonen
(im Standardmodell die Photonen) "ubergangen.
Das vorliegende Kapitel besch"aftigt sich mit dem Modell, das eine
schwache $SU(2)$-Wechselwirkung mit einer
$U(1)$-Hyperladungswechselwirkung kombiniert.
Dieses Modell wurde in Abschnitt~\ref{sec:SU2U1} beschrieben.
Gegen"uber dem Standardmodell besitzt das Higgs-Dublett hier die
Hyperladung~$Y_\Phi=0$, so dass lediglich die $SU(2)$-Eichbosonen massiv
werden, welche auch nicht mit den $U(1)$-Eichbosonen mischen.
Letztere bleiben masselos.

Der Einfachkeit halber werden in diesem Kapitel nur Fermionen mit gleicher
Hyperladung~$Y$ betrachtet, und $Y$ wird in die $U(1)$-Kopplung~$\alpha'$
integriert, um die Hyperladung nicht "uberall mit angeben zu m"ussen.
Anstelle von $\alpha'$ muss man also $\alpha' \, (Y/2)^2$ lesen.
Die Standardmodell-Fermionen mit unterschiedlichen Hyperladungen
entsprechend der Tabelle~\ref{tab:SMfermionen} k"onnen sp"ater leicht
wieder ber"ucksichtigt werden.

Die Ergebnisse dieses Kapitels sind in \cite{Feucht:2004rp} ver"offentlicht.

\section{Faktorisierung der IR-Singularit"aten}

\subsection[Der $SU(2)\times U(1)$-Formfaktor]
  {\boldmath Der $SU(2)\times U(1)$-Formfaktor}

Der Formfaktor der massiven $SU(2)$-Wechselwirkung,
\begin{equation}
  F_{SU(2)}(Q^2,M^2) =
  1 + \frac{\alpha}{4\pi} \, F_{SU(2)}^{(1)}
  + \left(\frac{\alpha}{4\pi}\right)^2 \, F_{SU(2)}^{(2)}
  + \Oc(\alpha^3)
  \,,
\end{equation}
ist in der logarithmischen \NNNLL-N"aherung des Hochenergielimes bekannt:
$F_{SU(2)}^{(1)}$ ist in Gl.~(\ref{eq:F1SUN}) angegeben,
$\left(\frac{\alpha}{4\pi}\right)^2 \, F_{SU(2)}^{(2)} \equiv F_2$
findet sich in Gl.~(\ref{eq:F2erg}).

Die Strahlungskorrekturen mit dem masselosen $U(1)$-Eichboson erzeugen
IR"=Divergenzen, die regularisiert werden m"ussen.
F"uhrt man dazu eine infinitesimale, aber endliche Masse~$\lambda$ f"ur das
$U(1)$-Eichboson ein, dann ist auch der Formfaktor
\begin{equation}
  F_{U(1)}(Q^2,\lambda^2) =
  1 + \frac{\alpha'}{4\pi} \, F_{U(1)}^{(1)}
  + \left(\frac{\alpha'}{4\pi}\right)^2 \, F_{U(1)}^{(2)}
  + \Oc({\alpha'}^3)  
\end{equation}
f"ur die reine $U(1)$-Wechselwirkung aus dem $SU(2)$-Ergebnis bekannt:
Der $U(1)$-Formfaktor folgt aus dem $SU(2)$-Formfaktor durch Ersetzen der
Kopplung, $\alpha \to \alpha'$, Ersetzen der Masse, $M \to \lambda$,
Einsetzen der $U(1)$-Werte f"ur $C_F=1$, $C_A=0$, $T_F=1$
entsprechend~(\ref{eq:CasimirU1}) und Weglassen der Higgs-Beitr"age, da das
$U(1)$-Eichboson nicht an das Higgs-Dublett koppelt.

Arbeitet man stattdessen mit dem physikalischen Wert $\lambda=0$ und
benutzt die dimensionale Regularisierung f"ur die IR-Divergenzen, dann ist
der masselose $U(1)$-Formfaktor in $d=4-2\eps$ Dimensionen durch
\begin{equation}
\label{eq:FU1eps}
  F_{U(1)}(Q^2,\lambda^2=0;\eps) =
  1 + \frac{\alpha'}{4\pi} \, F_{U(1),\eps}^{(1)}
  + \left(\frac{\alpha'}{4\pi}\right)^2 \, F_{U(1),\eps}^{(2)}
  + \Oc({\alpha'}^3)  
\end{equation}
gegeben mit den Ein- und Zweischleifenbeitr"agen aus (\ref{eq:F1herg}) und
(\ref{eq:F2hherg}) sowie $C_F=1$, $C_A=0$, $T_F=1$ und $n_s=0$.

Der Formfaktor des kombinierten $SU(2)\times U(1)$-Modells
hat in Zweischleifenn"aherung die Struktur
\begin{multline}
\label{eq:hatFlambda}
  \hat F(Q^2,M^2,\lambda^2) = 1
  + \frac{\alpha}{4\pi} \, F_{SU(2)}^{(1)}
  + \frac{\alpha'}{4\pi} \, F_{U(1)}^{(1)}
  + \left(\frac{\alpha}{4\pi}\right)^2 \, F_{SU(2)}^{(2)}
  + \left(\frac{\alpha'}{4\pi}\right)^2 \, F_{U(1)}^{(2)}
\\*
  + \frac{\alpha \alpha'}{(4\pi)^2} \, F_\inter^{(2)}
  + \Oc(\{\alpha,\alpha'\}^3)
  \,,
\end{multline}
wobei $\hat F(Q^2,M^2,\lambda^2)$ Korrekturen von beiden Eichgruppen
erh"alt und nur der Interferenzterm $F_\inter^{(2)}$ mit
Strahlungskorrekturen von je einem $SU(2)$- und $U(1)$-Boson bisher
unbekannt ist.

F"ur die dimensionsbehafteten Parameter des $SU(2)\times U(1)$-Modells
gilt
\begin{equation}
  \lambda \ll M \ll Q \,.
\end{equation}
Deshalb kann die Abh"angigkeit des kombinierten Formfaktors
$\hat F(Q^2,M^2,\lambda^2)$ von der infinitesimalen Masse~$\lambda$ durch eine
\emph{infrarote Evolutionsgleichung}\cite{Fadin:2000bq} beschrieben werden.
Die IR-Divergenzen des Formfaktors im Limes $\lambda\to0$ werden nach dem
Theorem von Kinoshita, Lee und
Nauenberg~\cite{Kinoshita:1962ur,Lee:1964is}
von den IR-Divergenzen in den Korrekturen reell abgestrahlter
$U(1)$-Bosonen kompensiert.
Letztere werden von masselosen $U(1)$-Bosonen verursacht,
die mit verschwindender Energie oder kollinear zu einem der externen
Fermionen abgestrahlt werden;
sie werden von der Pr"asenz der zus"atzlichen $SU(2)$-Wechselwirkung nicht
beeinflusst.
Deshalb ist auch das singul"are Verhalten des kombinierten $SU(2)\times
U(1)$-Formfaktors gleich dem des reinen $U(1)$-Formfaktors und kann als
separater Faktor in $\hat F(Q^2,M^2,\lambda^2)$ geschrieben werden.
Der Einfachkeit halber werden nicht nur die IR-Divergenzen, sondern der
ganze $U(1)$-Formfaktor $F_{U(1)}(Q^2,\lambda^2)$ vom kombinierten
Formfaktor abgespalten.
Die L"osung der infraroten Evolutionsgleichung l"asst sich in der folgenden
faktorisierten Form schreiben:
\begin{equation}
\label{eq:SU2U1fakt}
  \hat F(Q^2,M^2,\lambda^2) =
  F_{U(1)}(Q^2,\lambda^2) \cdot \tilde F(Q^2,M^2)
  + \Oc\!\left(\frac{\lambda^2}{M^2}\right)
  \,.
\end{equation}
Alle IR-Divergenzen sind im ersten Faktor $F_{U(1)}(Q^2,\lambda^2)$
enthalten.
Der zweite Faktor $\tilde F(Q^2,M^2)$ ist IR-konvergent und h"angt somit
nur noch von~$M^2$, nicht mehr von~$\lambda^2$ ab.
Die Faktorisierung~(\ref{eq:SU2U1fakt}) ist f"ur $\lambda \ll M \ll Q$
g"ultig, dabei werden Terme der Ordnung~$\lambda^2/M^2$ vernachl"assigt.

Die IR-konvergente Funktion~$\tilde F(Q^2,M^2)$ kann aus dem Quotienten der
beiden IR-singul"aren Formfaktoren bestimmt werden:
\begin{equation}
\label{eq:tildeFlambda}
  \tilde F(Q^2,M^2) = \lim_{\lambda\to0} \,
    \frac{\hat F(Q^2,M^2,\lambda^2)}{F_{U(1)}(Q^2,\lambda^2)}
  \,.
\end{equation}
Weil die Funktion~$\tilde F(Q^2,M^2)$ als IR-konvergente Gr"o"se nicht von
der Art der Regularisierung der IR-Divergenzen abh"angt, kann
Gl.~(\ref{eq:tildeFlambda}) auch in dimensionaler Regularisierung mit
$\lambda=0$ und $\eps \ne 0$ geschrieben werden:
\begin{equation}
\label{eq:tildeFeps}
  \tilde F(Q^2,M^2) = \lim_{\eps\to0} \,
    \frac{\hat F(Q^2,M^2,\lambda^2=0;\eps)}{F_{U(1)}(Q^2,\lambda^2=0;\eps)}
  \,.
\end{equation}
Dabei ist
\begin{multline}
\label{eq:hatFeps}
  \hat F(Q^2,M^2,\lambda^2=0;\eps) = 1
  + \frac{\alpha}{4\pi} \, F_{SU(2)}^{(1)}
  + \frac{\alpha'}{4\pi} \, F_{U(1),\eps}^{(1)}
  + \left(\frac{\alpha}{4\pi}\right)^2 \, F_{SU(2)}^{(2)}
\\*
  + \left(\frac{\alpha'}{4\pi}\right)^2 \, F_{U(1),\eps}^{(2)}
  + \frac{\alpha \alpha'}{(4\pi)^2} \, F_{\inter,\eps}^{(2)}
  + \Oc(\{\alpha,\alpha'\}^3)
\end{multline}
der $SU(2)\times U(1)$-Formfaktor f"ur $\lambda=0$
in dimensionaler Regularisierung.
Analog zu~(\ref{eq:hatFlambda}) ist er in Zweischleifenn"aherung bekannt,
mit Ausnahme des Interferenzterms~$F^{(2)}_{\inter,\eps}$, dessen Berechnung
im folgenden Abschnitt vorgestellt wird.
Der Vorteil der dimensionalen Regularisierung f"ur die IR-Divergenzen
besteht darin, im Folgenden nicht Zweischleifendiagramme mit Eichbosonen
zweier verschiedener, endlicher Massen $M$ und $\lambda$ ausrechnen zu
m"ussen. Stattdessen sind nur Diagramme mit \emph{einer} endlichen
Eichbosonmasse erforderlich.

\subsection{Strahlungskorrekturen mit massiven und masselosen Eich\-bosonen}
\label{sec:SU2U1int}

Zur Berechnung des Interferenzbeitrags~$F^{(2)}_{\inter,\eps}$ aus
Gl.~(\ref{eq:hatFeps}) mit dimensionaler Regularisierung der
IR-Divergenzen werden Zweischleifendiagramme mit einem massiven
$SU(2)$-Eichboson und einem masselosen $U(1)$-Eichboson ben"otigt.
Das $U(1)$-Boson koppelt weder ans Higgs-Boson oder an die
Goldstone-Bosonen noch an die $SU(2)$-Eichbosonen, da keine Mischung der
beiden Eichgruppen vorliegt.
Ein "Ubergang zwischen den Eichbosonen beider Gruppen "uber eine
Fermionschleife findet ebenfalls nicht statt, da die Spur eines einzigen
$SU(2)$-Generators verschwindet.
Deshalb m"ussen hier nur die abelschen Beitr"age aus
Kapitel~\ref{chap:abelsch} ber"ucksichtigt werden. Dabei wird jeweils eines
der beiden $SU(2)$-Eichbosonen durch ein $U(1)$-Eichboson ersetzt.

Die Berechnung der Schleifendiagramme erfolgt analog zu
Kapitel~\ref{chap:abelsch}.
Die mathematische Komplexit"at ist geringer als f"ur den rein massiven
Formfaktor, da in allen Diagrammen jetzt nur ein einziger massiver
Propagator vorhanden ist.

Im Rahmen der \emph{Expansion by Regions} fallen manche Regionen, die im
$SU(2)$-Formfaktor Beitr"age geliefert haben, ganz heraus. Andere Beitr"age
werden einfacher, z.B. durch weniger Mellin-Barnes-Integrale.
Der Beitrag der harten (h-h)-Region entspricht in allen Diagrammen dem
jeweiligen Ergebnis aus Kapitel~\ref{chap:abelsch}, da das masselose
Diagramm nicht davon abh"angt, wieviele Massen urspr"unglich vorhanden
waren.

Alle Diagramme mit einem $SU(2)$- und einem $U(1)$-Eichboson besitzen den
Farbfaktor
\[
  t^a t^a = C_F \, \unity = \frac{3}{4} \, \unity \,.
\]
Die Rechnungen werden hier nicht im Detail aufgef"uhrt, nur die
Beitr"age der einzelnen Feynman-Diagramme werden angegeben.

\subsubsection{Planares Vertexdiagramm}

Wenn im planaren Vertexdiagramm (vgl. Abschnitt~\ref{sec:LA}) der
Propagator~5 massiv und der Propagator~6 masselos ist,
sind die gleichen Regionen relevant wie im rein massiven Fall.
Allerdings sind die Regionen (1c-1c) und (2c-2c) durch ein einfaches statt
zweifaches Mellin-Barnes-Integral gegeben.
Der Beitrag des Feynman-Diagramms lautet:
\begin{align}
\label{eq:LAm5erg}
  F_{v,\LA M5} &=
    C_F
    \, \frac{\alpha \alpha'}{(4\pi)^2}
    \left(\frac{\mu^2}{M^2}\right)^{2\eps} S_\eps^2
    \, \Biggl\{
  \frac{1}{2\eps^2}
  + \frac{1}{\eps} \left[
    - \lqm^2 + 3 \lqm - \frac{2}{3}\pi^2 - \frac{11}{4} \right]
  \nonumber \\ & \qquad
  + \frac{1}{6} \lqm^4
  + \left(\frac{2}{3}\pi^2 - 1\right) \lqm^2
  + \left(-24\zeta_3 - \pi^2 + \frac{11}{2}\right) \lqm
  \nonumber \\ & \qquad
  + \frac{13}{45}\pi^4 + 46\zeta_3 + \frac{13}{12}\pi^2 - \frac{41}{8}
  \Biggr\}
  + \Oc(\eps) + \Oc\!\left(\frac{M^2}{Q^2}\right)
  ,
\end{align}
mit $\lqm = \ln(Q^2/M^2)$ und $S_\eps = (4\pi)^\eps \, e^{-\eps\gamma_E}$.

Beim planaren Vertexdiagramm mit massivem Propagator~6 und masselosem
Propagator~5 fallen die Regionen (1c-h) und (2c-h) weg.
Nur die Regionen (h-h), (1c-1c) und (2c-2c) tragen noch zum Ergebnis bei:
\begin{align}
\label{eq:LAm6erg}
  F_{v,\LA M6} &=
    C_F
    \, \frac{\alpha \alpha'}{(4\pi)^2}
    \left(\frac{\mu^2}{M^2}\right)^{2\eps} S_\eps^2
    \, \Biggl\{
  -\frac{2}{\eps^3}
  + \frac{1}{\eps^2} \left[
    2 \lqm^2 - 4 \lqm + \frac{4}{3}\pi^2 + \frac{9}{2} \right]
  \nonumber \\* & \qquad
  + \frac{1}{\eps} \left[
    -\frac{4}{3} \lqm^3 + 4 \lqm^2 + \left(\frac{2}{3}\pi^2-17\right) \lqm
    + 12\zeta_3 - \frac{7}{3}\pi^2 + \frac{85}{4} \right]
  \nonumber \\* & \qquad
  + \frac{2}{3} \lqm^4
  - \frac{8}{3} \lqm^3
  + \left(\frac{\pi^2}{3} + 17\right) \lqm^2
  + \left(-36\zeta_3 + \frac{2}{3}\pi^2 - \frac{101}{2}\right) \lqm
  \nonumber \\* & \qquad
  + \frac{107}{90}\pi^4 + \frac{184}{3}\zeta_3 - \frac{59}{12}\pi^2
  + \frac{599}{8}
  \Biggr\}
  + \Oc(\eps) + \Oc\!\left(\frac{M^2}{Q^2}\right)
  .
\end{align}
Dieses Diagramm ist IR-singul"ar, daher kommt der $\eps^{-3}$-Pol.

\subsubsection{Nichtplanares Vertexdiagramm}

Das nichtplanare Vertexdiagramm aus dem Abschnitt~\ref{sec:NP} ist
symmetrisch unter einer horizontalen Spiegelung, siehe
Gl.~(\ref{eq:NPsym}). Deshalb sind die beiden Beitr"age mit nur einer
Eichbosonmasse gleich.
Im Fall, wenn der Propagator~5 massiv und der Propagator~6 masselos ist,
fallen die Regionen (h-2c), (1c-2c), (2c'-2c), (us'"~us'), (1c"~us') und
(us'-2c) weg; nur die Regionen (h-h), (1c-h), (1c-1c), (2c-2c) und (1c-1c')
sind noch relevant.
\begin{align}
\label{eq:NPm5erg}
  F_{v,\NP M5} &= F_{v,\NP M6}
  \nonumber \\* &=
    C_F
    \, \frac{\alpha \alpha'}{(4\pi)^2}
    \left(\frac{\mu^2}{M^2}\right)^{2\eps} S_\eps^2
    \, \Biggl\{
  \frac{1}{\eps} \left[
    -\frac{2}{3} \lqm^3 + 4 \lqm^2 - 12 \lqm
    - 12\zeta_3 + \pi^2 + 14 \right]
  \nonumber \\* & \qquad
  + \frac{1}{2} \lqm^4
  - 4 \lqm^3
  + \left(-\frac{5}{3}\pi^2 + 22\right) \lqm^2
  + \left(56\zeta_3 + \frac{11}{3}\pi^2 - 68\right) \lqm
  \nonumber \\* & \qquad
  - \frac{67}{90}\pi^4 - 90\zeta_3 - 4\pi^2 + 96
  \Biggr\}
  + \Oc(\eps) + \Oc\!\left(\frac{M^2}{Q^2}\right)
  \,.
\end{align}

\subsubsection{Vertexdiagramm mit Benz-Topologie}

Wenn im abelschen Vertexdiagramm mit Benz-Topologie aus dem
Abschnitt~\ref{sec:BE} der Propagator~3 massiv und der Propagator~4
masselos ist, sind nur noch die Regionen (h-h), (1c-h) und (1c-1c)
relevant, die Regionen (h-2c), (us-2c), (2c-2c) und (1c-2c) fallen weg.
\begin{align}
\label{eq:BEm3erg}
  F_{v,\BE M3} &=
    C_F
    \, \frac{\alpha \alpha'}{(4\pi)^2}
    \left(\frac{\mu^2}{M^2}\right)^{2\eps} S_\eps^2
    \, \Biggl\{
  -\frac{2}{\eps^3}
  + \frac{1}{\eps^2} \left[
    2 \lqm - \frac{5}{2} \right]
  \nonumber \\* & \qquad
  + \frac{1}{\eps} \left[
    - 2 \lqm^2 + 7 \lqm - \frac{\pi^2}{3} - \frac{53}{4} \right]
  + \frac{4}{3} \lqm^3
  + \left(\frac{\pi^2}{3} - 9\right) \lqm^2
  \nonumber \\* & \qquad
  + \left(8\zeta_3 - \frac{7}{3}\pi^2 + \frac{73}{2}\right) \lqm
  + \frac{11}{45}\pi^4 - \frac{32}{3}\zeta_3 + \frac{17}{4}\pi^2
  - \frac{479}{8}
  \Biggr\}
  \nonumber \\* & \quad
  + \Oc(\eps) + \Oc\!\left(\frac{M^2}{Q^2}\right)
  .
\end{align}

Im anderen Fall, wenn der Propagator~4 massiv und der Propagator~3 masselos
ist, sind die Regionen (h-h), (h-2c), (1c-1c) und (2c-2c) relevant,
die Regionen (1c-h), (us-2c) und (1c-2c) fallen dann weg.
\begin{align}
\label{eq:BEm4erg}
  F_{v,\BE M4} &=
    C_F
    \, \frac{\alpha \alpha'}{(4\pi)^2}
    \left(\frac{\mu^2}{M^2}\right)^{2\eps} S_\eps^2
    \, \Biggl\{
  \nonumber \\* & \qquad
  \frac{1}{2\eps^2}
  + \frac{1}{\eps} \left[
    - \lqm^2 + \left(-\frac{2}{3}\pi^2 + 7\right) \lqm
    + 4\zeta_3 + \frac{\pi^2}{3} - \frac{53}{4} \right]
  \nonumber \\* & \qquad
  + \lqm^3
  + \left(\frac{2}{3}\pi^2 - 9\right) \lqm^2
  + \left(-4\zeta_3 - 3\pi^2 + \frac{89}{2}\right) \lqm
  \nonumber \\* & \qquad
  - \frac{13}{90}\pi^4 + 16\zeta_3 + \frac{79}{12}\pi^2 - \frac{655}{8}
  \Biggr\}
  + \Oc(\eps) + \Oc\!\left(\frac{M^2}{Q^2}\right)
  .
\end{align}

\subsubsection{Vertexdiagramm mit Fermion-Selbstenergie}

Wenn im Vertexdiagramm mit Fermion-Selbstenergie aus dem
Abschnitt~\ref{sec:fc} der Propagator~3 massiv und der Propagator~5
masselos ist, tragen die gleichen Regionen wie im rein massiven Fall
bei. Allerdings sind die Regionen (h-s) und (1c'-s) nicht nur unterdr"uckt,
sondern gar nicht vorhanden.
\begin{align}
\label{eq:fcm3erg}
  F_{v,\fc M3} &=
    C_F
    \, \frac{\alpha \alpha'}{(4\pi)^2}
    \left(\frac{\mu^2}{M^2}\right)^{2\eps} S_\eps^2
    \, \Biggl\{
  -\frac{1}{2\eps^2}
  + \frac{1}{\eps} \left[
    \lqm^2 - 3 \lqm + \frac{2}{3}\pi^2 + \frac{13}{4} \right]
  \nonumber \\* & \qquad
  - \lqm^3
  + 5 \lqm^2
  - \frac{33}{2} \lqm
  - 4\zeta_3 + \frac{\pi^2}{12} + \frac{163}{8}
  \Biggr\}
  + \Oc(\eps) + \Oc\!\left(\frac{M^2}{Q^2}\right)
  .
\end{align}

Wenn der Propagator~5 massiv und der Propagator~3 masselos ist,
sind nur die Regionen (h-h), (2c-2c) und (h-s) vorhanden, wobei letztere
unterdr"uckt ist. Die Regionen (1c'-h) und (1c'-s) fallen weg.
\begin{align}
\label{eq:fcm5erg}
  F_{v,\fc M5} &=
    C_F
    \, \frac{\alpha \alpha'}{(4\pi)^2}
    \left(\frac{\mu^2}{M^2}\right)^{2\eps} S_\eps^2
    \, \Biggl\{
  \frac{2}{\eps^3}
  + \frac{1}{\eps^2} \left[
    -2 \lqm + \frac{5}{2} \right]
  + \frac{1}{\eps} \left[
    2 \lqm^2 - 7 \lqm + \frac{\pi^2}{3} + \frac{53}{4} \right]
  \nonumber \\* & \qquad
  - \frac{4}{3} \lqm^3
  + 7 \lqm^2
  + \left(\frac{\pi^2}{3} - \frac{53}{2}\right) \lqm
  - \frac{40}{3}\zeta_3 - \frac{13}{12}\pi^2 + \frac{355}{8}
  \Biggr\}
  \nonumber \\* & \quad
  + \Oc(\eps) + \Oc\!\left(\frac{M^2}{Q^2}\right)
  .
\end{align}

\subsubsection{Selbstenergiekorrektur mit T1-Topologie}

Die beiden Selbstenergiediagramme mit T1-Topologie aus dem
Abschnitt~\ref{sec:T1}, im ersten Fall mit massivem Propagator~2 und
masselosem Propagator~3, im zweiten Fall umgekehrt, liefern aufgrund der
Symmetrie~(\ref{eq:T1sym}) den gleichen Beitrag.
\begin{align}
\label{eq:T1m2erg}
  \Sigma_{\Tone M2} &= \Sigma_{\Tone M3} =
    C_F
    \, \frac{\alpha \alpha'}{(4\pi)^2}
    \left(\frac{\mu^2}{M^2}\right)^{2\eps} S_\eps^2
    \left(
  \frac{1}{2\eps}
  - \frac{3}{4}
  \right)
  + \Oc(\eps)
  \,.
\end{align}

\subsubsection{Selbstenergiekorrektur mit T2-Topologie}

Wenn im Selbstenergiediagramm mit T2-Topologie aus dem
Abschnitt~\ref{sec:T2} der Propagator~2 massiv und der Propagator~4
masselos ist, dann gilt:
\begin{align}
\label{eq:T2m2erg}
  \Sigma_{\Ttwo M2} &=
    C_F
    \, \frac{\alpha \alpha'}{(4\pi)^2}
    \left(\frac{\mu^2}{M^2}\right)^{2\eps} S_\eps^2
    \left(
  \frac{1}{2\eps^2}
  - \frac{1}{4\eps}
  + \frac{\pi^2}{4} - \frac{1}{8}
  \right)
  + \Oc(\eps)
  \,.
\end{align}

Im anderen Fall, mit massivem Propagator~4 und masselosem Propagator~2,
gilt:
\begin{align}
\label{eq:T2m4erg}
  \Sigma_{\Ttwo M4} &=
    C_F
    \, \frac{\alpha \alpha'}{(4\pi)^2}
    \left(\frac{\mu^2}{M^2}\right)^{2\eps} S_\eps^2
    \left(
  -\frac{1}{2\eps^2}
  + \frac{3}{4\eps}
  - \frac{\pi^2}{4} - \frac{1}{8}
  \right)
  + \Oc(\eps)
  \,.
\end{align}

\subsubsection{Sonstige Beitr"age mit Selbstenergiekorrekturen}

Das Produkt aus Einschleifendiagrammen aus dem
Abschnitt~\ref{sec:ecorrsonst} (Abb.~\ref{fig:fcorr1x1}) taucht im
\mbox{$SU(2) \times U(1)$}-Interferenzterm nicht auf, da immer eine der
beiden Einschleifenkorrekturen verschwindet, wenn die entsprechende
Eichbosonmasse zu Null gesetzt wird und somit skalenlose Integrale
entstehen.

Der letzte fehlende Beitrag ist das Produkt aus dem masselosen
$U(1)$"=Einschleifenvertexdiagramm
$F_{v,1}^{M=0} = \frac{\alpha'}{4\pi} F^{(1)}_{M=0}$ in Gl.~(\ref{eq:F1herg})
mit der massiven $SU(2)$"=Einschleifenselbstenergie~$\Sigma_1$ in
Gl.~(\ref{eq:Sigma1erg}).
Da $F_{v,1}^{M=0}$ einen $\eps^{-2}$-Pol enth"alt, wird $\Sigma_1$ bis zur
Ordnung~$\eps^2$ ben"otigt.
Andererseits enth"alt $\Sigma_1$ nur einen $\eps^{-1}$-Pol, so dass bei
$F_{v,1}^{M=0}$ die lineare Ordnung in~$\eps$ gen"ugt.
Dieser Beitrag lautet:
\begin{align}
  F_{v,1}^{M=0} \cdot \Sigma_1 &=
    C_F
    \, \frac{\alpha \alpha'}{(4\pi)^2}
    \left(\frac{\mu^2}{M^2}\right)^{2\eps} S_\eps^2
    \, \Biggl\{
  \frac{2}{\eps^3}
  + \frac{1}{\eps^2} \,\Bigl[
    -2 \lqm + 2 \Bigr]
  + \frac{1}{\eps} \,\Bigl[
    \lqm^2 - 2 \lqm + 6 \Bigr]
  \nonumber \\* & \qquad
  - \frac{1}{3} \lqm^3
  + \lqm^2
  - 6 \lqm
  - \frac{16}{3}\zeta_3 + 11
  \Biggr\}
  + \Oc(\eps) + \Oc\!\left(\frac{M^2}{Q^2}\right)
  .
\end{align}

\subsubsection{Zusammenfassung der Strahlungskorrekturen}

Der in Gl.~(\ref{eq:hatFeps}) eingef"uhrte
Interferenzterm~$F^{(2)}_{\inter,\eps}$, der Strahlungskorrekturen von je einem
$SU(2)$- und $U(1)$-Eichboson beinhaltet, l"asst sich nun vollst"andig
angeben:
\begin{align}
  \frac{\alpha \alpha'}{(4\pi)^2} \, F_{\inter,\eps}^{(2)} &=
  F_{v,\LA M5} + F_{v,\LA M6}
  + F_{v,\NP M5} + F_{v,\NP M6}
  + 2\,F_{v,\BE M3} + 2\,F_{v,\BE M4}
\nonumber \\*[-1ex] & \quad
  + 2\,F_{v,\fc M3} + 2\,F_{v,\fc M5}
  + \Sigma_{\Tone M2} + \Sigma_{\Tone M3}
  + \Sigma_{\Ttwo M2} + \Sigma_{\Ttwo M4}
\nonumber \\* & \quad
  + F_{v,1}^{M=0} \cdot \Sigma_1
  \,,
\end{align}
mit
\begin{align}
\label{eq:F2intepserg}
  F^{(2)}_{\inter,\eps} &=
    C_F
    \left(\frac{\mu^2}{M^2}\right)^{2\eps} S_\eps^2
    \, \Biggl\{
  \frac{1}{\eps^2} \left[
    2 \lqm^2 - 6 \lqm + \frac{4}{3}\pi^2 + 7 \right]
  \nonumber \\* & \qquad
  + \frac{1}{\eps} \left[
    -\frac{8}{3} \lqm^3 + 12 \lqm^2
    + \left(-\frac{2}{3}\pi^2 - 32\right) \lqm
    - 4\zeta_3 + \pi^2 + 34 \right]
  \nonumber \\* & \qquad
  + \frac{11}{6} \lqm^4
  - 11 \lqm^3
  + \left(-\frac{\pi^2}{3} + 49\right) \lqm^2
  + \Bigl(60\zeta_3 - 3\pi^2 - 111\Bigr) \, \lqm
  \nonumber \\* & \qquad
  + \frac{17}{90}\pi^4 - 102\zeta_3 + \frac{47}{6}\pi^2 + 117
  \Biggr\}
  + \Oc(\eps) + \Oc\!\left(\frac{M^2}{Q^2}\right)
  .
\end{align}
Damit ist der Formfaktor $\hat F(Q^2,M^2,\lambda^2=0;\eps)$ des
kombinierten $SU(2) \times U(1)$-Modells in
dimensionaler Regularisierung bis in Zweischleifenn"aherung bekannt.

\subsection{Auswertung der Faktorisierung}

Den endlichen Faktor~$\tilde F(Q^2,M^2)$ der Faktorisierung erh"alt man
aus Gl.~(\ref{eq:tildeFeps}), indem man f"ur $\hat
F(Q^2,M^2,\lambda^2=0;\eps)$~(\ref{eq:hatFeps}) und
$F_{U(1)}(Q^2,\lambda^2=0;\eps)$~(\ref{eq:FU1eps})
die Entwicklung in der jeweiligen Kopplungskonstanten einsetzt:
\begin{align}
\label{eq:tildeFepsdev}
  \tilde F(Q^2,M^2) &=
    \lim_{\eps\to0} \,
    \frac{\hat F(Q^2,M^2,\lambda^2=0;\eps)}{F_{U(1)}(Q^2,\lambda^2=0;\eps)}
\nonumber \\ &=
    \lim_{\eps\to0}
    \,\biggl\{ 1 + \frac{\alpha}{4\pi} \, F^{(1)}_{SU(2)}
      + \left(\frac{\alpha}{4\pi}\right)^2 \, F^{(2)}_{SU(2)}
  \nonumber \\* & \qquad\qquad
      + \frac{\alpha \alpha'}{(4\pi)^2} \left[
        F^{(2)}_{\inter,\eps} - F^{(1)}_{SU(2)} \cdot F^{(1)}_{U(1),\eps}
	\right]
    \biggr\}
  + \Oc(\{\alpha,\alpha'\}^3)
\nonumber \\ &=
  F_{SU(2)}(Q^2,M^2) \cdot \left\{ 1 +
    \frac{\alpha \alpha'}{(4\pi)^2} \lim_{\eps\to0} \left[
      F^{(2)}_{\inter,\eps} - F^{(1)}_{SU(2)} \cdot F^{(1)}_{U(1),\eps}
      \right]
    \right\}
  \nonumber \\* & \qquad
  + \Oc(\{\alpha,\alpha'\}^3)
  \,.
\end{align}
Durch Einsetzen des Ergebnisses~(\ref{eq:F2intepserg}) sowie des
massiven Einschleifenformfaktors~(\ref{eq:F1erg}) bis $\Oc(\eps^2)$
und des masselosen Einschleifenformfaktors~(\ref{eq:F1herg}) ergibt sich:
\begin{align}
\label{eq:tildeFerg}
  \frac{\tilde F(Q^2,M^2)}{F_{SU(2)}(Q^2,M^2)} &=
  1 + \frac{\alpha \alpha'}{(4\pi)^2} \, C_F \,\biggl[
    \Bigl( 48\zeta_3 - 4\pi^2 + 3 \Bigr) \,\lqm
    + \frac{7}{45}\pi^4 - 84\zeta_3 + \frac{20}{3}\pi^2 - 2
    \biggr]
  \nonumber \\*[1ex] & \qquad
  + \Oc(\{\alpha,\alpha'\}^3)
  \,.
\end{align}

Dieses Ergebnis ist im Limes $\eps\to0$ endlich, dadurch ist die in
Gl.~(\ref{eq:SU2U1fakt}) behauptete Faktorisierung explizit gezeigt.
Au"serdem enth"alt die rechte Seite von Gl.~(\ref{eq:tildeFerg}) nur einen
linearen Logarithmus, keine h"oheren Logarithmenpotenzen.
In \NNLL-N"aherung gilt folglich sogar die naive Faktorisierung
\begin{align}
  \hat F(Q^2,M^2,\lambda^2) &\stackrel{\NNLL}{\sim}
  F_{U(1)}(Q^2,\lambda^2) \cdot F_{SU(2)}(Q^2,M^2)
\nonumber \\* & \qquad\qquad
  + \Oc(\alpha \alpha' \lqm^1) + \Oc(\{\alpha,\alpha'\}^3)
  + \Oc\!\left(\frac{\lambda^2}{M^2}\right)
  \,.
\end{align}
Die nichtlogarithmische Konstante auf der rechten Seite
von~(\ref{eq:tildeFerg}) wird nicht ben"otigt, da der $SU(2)$-Formfaktor in
der Zweischleifenordnung auch nur in \NNNLL-N"aherung bekannt ist.

Die Faktorisierung
\begin{equation}
\label{eq:SU2U1fakt2}
  \hat F(Q^2,M^2,\lambda^2;\eps) =
  F_{U(1)}(Q^2,\lambda^2;\eps) \cdot \tilde F(Q^2,M^2)
  + \Oc\!\left(\frac{\lambda^2}{M^2}\right)
\end{equation}
gilt unabh"angig davon, ob die IR-Divergenzen mit der infinitesimalen
Eichbosonmasse~$\lambda$ oder dimensional mit~$\eps$ regularisiert
werden.
Sie erlaubt die Wahl der IR"~Regularisierung unabh"angig
von den Korrekturen zu~$\tilde F(Q^2,M^2)$.

\section[Das $SU(2) \times U(1)$-Modell mit gleichen Massen]
  {\boldmath Das $SU(2) \times U(1)$-Modell mit gleichen\\ Massen}

F"ur den Formfaktor war es vergleichsweise einfach, die
Strahlungskorrekturen mit massiven \emph{und} masselosen Eichbosonen explizit
auszurechnen. Dies w"urde sich f"ur die Vierfermionamplitude schwieriger
gestalten, da hier die Zweischleifenergebnisse mittels der
Evolutionsgleichung~(\ref{eq:Aredevol}) berechnet werden.

Es ist jedoch leicht m"oglich, die $SU(2)$-Ergebnisse auf ein $SU(2) \times
U(1)$-Modell mit gleichen Massen $M = \lambda$ zu erweitern.
F"ur den Zweischleifenformfaktor muss dazu in allen abelschen
Beitr"agen
\begin{equation}
  \alpha^2 C_F^2 \to
  \Bigl( \alpha C_F + \alpha' \Bigr)^2
\end{equation}
ersetzt werden.
In den fermionischen Beitr"agen gilt:
\begin{equation}
  \alpha^2 C_F T_F n_f \to
  \alpha^2 C_F T_F n_f + {\alpha'}^2 n_f
  \,.
\end{equation}
An den nichtabelschen Anteilen und den Higgs-Beitr"agen des
$SU(2)$-Formfaktors "andert sich nichts durch die Erweiterung um die
massive $U(1)$-Eichgruppe.

Das Ergebnis f"ur das $SU(2)\times U(1)$-Modell mit gleichen Massen muss
mit den Gr"o"sen in der Faktorisierung~(\ref{eq:SU2U1fakt2}) in Verbindung
gebracht werden, um daraus auch Aussagen f"ur den Fall $\lambda\ne M$
treffen zu k"onnen.
Weil die mit $\lambda^2/M^2$ unterdr"uckten und vernachl"assigten
Terme in~(\ref{eq:SU2U1fakt2}) nicht bekannt sind, werden sie f"ur
$\lambda=M$ mit einer
zus"atzlichen Funktion $C(Q^2,M^2)$ parametrisiert:
\begin{align}
\label{eq:SU2U1faktMM}
  \hat F(Q^2,M^2,\lambda^2=M^2) =
  F_{U(1)}(Q^2,\lambda^2=M^2) \cdot \tilde F(Q^2,M^2) \cdot C(Q^2,M^2)
  \,.
\end{align}
In Gl.~(\ref{eq:SU2U1faktMM}) sind alle Gr"o"sen bis auf die
Funktion~$C(Q^2,M^2)$ bekannt. Aufl"osen der Gleichung nach $C(Q^2,M^2)$
und Einsetzen der Funktionen $\hat F$, $F_{U(1)}$ und $\tilde F$ ergibt:
\begin{multline}
  C(Q^2,M^2) = 
  1 + \frac{\alpha \alpha'}{(4\pi)^2} \, C_F \,\biggl[
    512\,\Li4\!\left(\frac{1}{2}\right)
    + \frac{64}{3}\,\ln^4{2} - \frac{64}{3}\pi^2\,\ln^2{2}
    - \frac{113}{15}\pi^4
  \\*
    + 244\zeta_3
    + \frac{70}{3}\pi^2 + \frac{59}{4}
    \biggr]
  + \Oc(\{\alpha,\alpha'\}^3)
  \,.
\end{multline}
In Zweischleifenordnung ist $C(Q^2,M^2)$ lediglich eine Konstante, die wie
alle nichtlogarithmischen Konstanten vernachl"assigt werden kann.
Die Faktorisierung
\begin{align}
\label{eq:SU2U1faktMMlogs}
  \hat F(Q^2,M^2,\lambda^2=M^2) \stackrel{\NNNLL}{\sim}
  F_{U(1)}(Q^2,\lambda^2=M^2) \cdot \tilde F(Q^2,M^2)
\end{align}
ohne den Faktor~$C(Q^2,M^2)$ ist also f"ur alle logarithmischen Terme in
Zweischleifenordnung g"ultig.
Diese Feststellung erm"oglicht die Berechnung der
Faktorisierungsfunktion~$\tilde F(Q^2,M^2)$ ohne explizite
Schleifenrechnungen mit massiven und masselosen Eichbosonen:
\begin{align}
\label{eq:tildeFlogs}
  \tilde F(Q^2,M^2) \stackrel{\NNNLL}{\sim}
  \frac{\hat F(Q^2,M^2,\lambda^2=M^2)}{F_{U(1)}(Q^2,\lambda^2=M^2)}
  \,.
\end{align}
Daraus k"onnen auch f"ur die gew"unschte Skalenhierarchie $\lambda \ll M \ll Q$
alle Zweischleifenlogarithmen des $SU(2)\times U(1)$-Formfaktors
folgenderma"sen bestimmt werden:
\begin{align}
\label{eq:SU2U1faktlambdalogs}
  \hat F(Q^2,M^2,\lambda^2) =
  F_{U(1)}(Q^2,\lambda^2) \,
  \frac{\hat F(Q^2,M^2,M^2)}{F_{U(1)}(Q^2,M^2)}
  + \Oc(\alpha \alpha' \Lc^0)
  \,.
\end{align}

Das "`Rezept"' f"ur die Berechnung aller logarithmischen
Zweischleifenbeitr"age im $SU(2)\times U(1)$-Modell mit der Massenl"ucke
$\lambda \ll M$ lautet also:
\begin{enumerate}
\item
  Den $SU(2)\times U(1)$-Formfaktor im Fall gleicher Massen $\lambda=M$
  durch eine Erweiterung des reinen $SU(2)$-Formfaktors bestimmen.
\item
  Durch die $U(1)$-Korrekturen bei $\lambda=M$ teilen.
\item
  Multiplizieren mit dem $U(1)$-Formfaktor im gew"unschten
  Regularisierungsschema der IR-Divergenzen (z.B. mit
  Eichbosonmasse~$\lambda$).
\end{enumerate}
Dies war auch die in
\cite{Fadin:2000bq,Kuhn:2000nn,Kuhn:2000hx,Kuhn:2001hz}
gew"ahlte Vorgehensweise, die hier durch explizite Rechnung best"atigt
wird.

Im allgemeinen Fall der Kombination zweier Eichgruppen w"urde die
Relation~(\ref{eq:SU2U1faktlambdalogs}) nur in \NNLL-N"aherung gelten,
da nur in dieser N"aherung die universellen Funktionen $\gamma$ und $\zeta$
(siehe Abschnitt~\ref{sec:formfaktor}) zur Bestimmung des Ergebnisses
gen"ugen.
Im vorliegenden $SU(2)\times U(1)$-Modell gilt
Gl.~(\ref{eq:SU2U1faktlambdalogs}) f"ur alle logarithmischen
Zweischleifenterme, weil die Eichbosonen der verschiedenen Gruppen nicht
wie im Standardmodell mischen und sich die Interferenzterme mit Korrekturen
von beiden Eichgruppen darum auf abelsche Beitr"age beschr"anken.


\clearemptypage

\chapter{Vierfermionstreuung}
\label{chap:EW2loop}

In diesem Kapitel werden die Ergebnisse aus den vorigen Kapiteln
zusammengesetzt, um elektroschwache Korrekturen zur Vierfermionstreuung zu
erhalten.
Zun"achst wird in Abschnitt~\ref{sec:sigmaSU2} beschrieben, wie
Wirkungsquerschnitte f"ur die Vierfermionstreuung in der reinen $SU(2)$-Theorie
aus der Vierfermionamplitude berechnet werden.
In Abschnitt~\ref{sec:EWcorr} geht es dann um Korrekturen im
elektroschwachen Standardmodell. Hier werden Ergebnisse f"ur
Zweischleifenkorrekturen zu verschiedenen Wirkungsquerschnitten vorgestellt
und diskutiert.
Das Kapitel schlie"st mit einer Zusammenfassung der vorliegenden Arbeit.

Die in diesem Kapitel vorgestellten Ergebnisse wurden zur Ver"offentlichung
eingereicht\cite{Jantzen:2005xi}.
Die Ver"offentlichung weiterer Details ist in Vorbereitung.

\section[$SU(2)_L$-Wirkungsquerschnitte]
  {\boldmath $SU(2)_L$-Wirkungsquerschnitte}
\label{sec:sigmaSU2}

In diesem Abschnitt geht es zun"achst um Wirkungsquerschnitte in einem
reinen $SU(2)$-Modell mit spontaner Symmetriebrechung, ohne
Ber"ucksichtigung der $U(1)$-Wechselwirkung.

Zur Berechnung von Wirkungsquerschnitten f"ur Vierfermionprozesse
$f \bar f \to f' \bar f'$ ist die Vierfermionamplitude~$\Ac$ erforderlich.
Diese wurde in Abschnitt~\ref{sec:Sudakov} analysiert. In
Gl.~(\ref{eq:Afaktor}) wurde sie als Produkt aus dem Quadrat des
Formfaktors~$F$ und der reduzierten Amplitude~$\tilde\Ac$ geschrieben.

Die Betrachtungen in Abschnitt~\ref{sec:Sudakov} gelten allgemein f"ur eine
$SU(2)$-Eichgruppe, die an links- und rechtsh"andige Fermionen koppelt.
Hier soll wie im Standardmodell eine $SU(2)_L$-Wechselwirkung nur f"ur
linksh"andige Fermionen betrachtet werden, d.h. es werden keine
rechtsh"andigen Fermionen in der Amplitude betrachtet.
Aus der Vierfermionamplitude im Raum der Chiralit"atsstruktur wird also nur
die Komponente f"ur linksh"andige Fermionen im Anfangs- und im Endzustand
ben"otigt.

Der $SU(2)$-Formfaktor~$F$ wurde in den vorangegangenen Kapiteln in
logarithmischer N"aherung berechnet,
das Einschleifenergebnis steht in Gl.~(\ref{eq:F1SUN}), das
Zweischleifenergebnis in Gl.~(\ref{eq:F2erg}).
Auch von der reduzierten Amplitude~$\tilde\Ac$ sind in
Zweischleifenn"aherung alle logarithmischen Terme bekannt,
wenn man die Ergebnisse aus Abschnitt~\ref{sec:4famplitude} in
Gl.~(\ref{eq:Atildechis}) einsetzt.

In (\ref{eq:4fbasisfarbe}) und (\ref{eq:4fbasischiral})
wurden die Basisvektoren der
Vierfermionamplitude im $SU(2)$-Farb\-raum definiert, der dem Isospinraum
entspricht.
Hier werden Prozesse $f \bar f \to f' \bar f'$ betrachtet,
die in Bornscher N"aherung "uber
den neutralen Strom vermittelt werden.
Bei diesen produzieren die $SU(2)$-Generatoren~$t^a$ im
Basisvektor~$\Ac^\lambda_{LL} = \bar v^L(-p_2) \, \gamma^\mu t^a \, u^L(p_1)
  \cdot \bar u^L(p_4) \, \gamma_\mu t^a \, v^L(-p_3)$
als Eigenwerte die Komponenten~$I^3_W$ des schwachen
Isospins des jeweiligen Fermion-Antifermion-Paars (siehe
Tabelle~\ref{tab:SMfermionen}).
Es kann also
\begin{equation}
  \Ac^\lambda_{LL} \to I^3_f I^3_{f'} \, \Ac^d_{LL}
\end{equation}
ersetzt werden, wobei $I^3_f$ und $I^3_{f'}$ f"ur den Isospin des
Fermion-Antifermion-Paars im Anfangs- respektive Endzustand stehen
und $\Ac^d_{LL}$ der Basisvektor ohne $SU(2)$-Generatoren ist.

F"ur die Korrekturen zur reduzierten Amplitude~$\tilde\Ac$ muss noch die
linksh"andige Chiralit"at der Fermionen beachtet werden.
Dies kann durch die Einf"uhrung des Projektors $(1-\gamma_5)/2$ in die
Basisamplituden $\Ac^\lambda$ und $\Ac^d$  geschehen,
bevor "uber die Spins der Fermionen im Anfangszustand gemittelt und "uber die
Spins der Fermionen im Endzustand summiert wird.
Die dabei entstehenden Spuren "uber Spinsummen und $\gamma$-Matrizen werden
mit den "ublichen Methoden ausgewertet (siehe z.B.\cite{Peskin:1995ev}).

Abschlie"send wird die Integration "uber den Winkel~$\theta$ in
Gl.~(\ref{eq:xpm}) durchgef"uhrt, also "uber den Winkel zwischen den
r"aumlichen Komponenten von $p_1$ und $p_4$ im Schwerpunktsystem.
Der totale Wirkungsquerschnitt~$\sigma$ wird auf die
Born-N"aherung~$\sigma_B$ normiert und folgenderma"sen in der
Kopplungskonstanten entwickelt:
\begin{equation}
\label{eq:sigmadev}
  \frac{\sigma}{\sigma_B} = 1 + \frac{\alpha}{4\pi} \, r^{(1)}
    + \left(\frac{\alpha}{4\pi}\right)^2 \, r^{(2)}
    + \Oc(\alpha^3)
  \,.
\end{equation}
Dabei ist die Kopplung~$\alpha$ in der Entwicklung~(\ref{eq:sigmadev}) bei
der Skala $\mu=M$ renormiert, w"ahrend die Kopplung~$g$ im
Born-Wirkungsquerschnitt~$\sigma_B$ bei $\mu=\sqrt s$ renormiert wird,
siehe Gl.~(\ref{eq:Afaktor}) und die anschlie"sende Diskussion.

Die Korrekturen h"angen davon ab, ob die Fermionen im Anfangs- und
Endzustand den gleichen Isospin ($I^3_f = I^3_{f'}$,
Bezeichnung mit $r^{(1,2)}_+$)
oder einen entgegengesetzten Isospin ($I^3_f = -I^3_{f'}$,
Bezeichnung mit $r^{(1,2)}_-$) haben.

Der Einschleifenbeitrag\cite{Kuhn:2001hz} lautet:
\begin{align}
  r^{(1)}_+ &=
    - 3\lqm^2(s)
    + \frac{80}{3} \lqm(s) 
    + \frac{35}{9} - \frac{20}{9} n_f
    - 3\pi^2
    \,,
\\
  r^{(1)}_- &=
    - 3\lqm^2(s)
    + \frac{26}{3} \lqm(s) 
    + \frac{278}{9} - \frac{20}{9} n_f
    - 3\pi^2
    \,.
\end{align}
Dabei ist $\Lc(s) = \ln(s/M^2)$ definiert, und
wegen $Q^2 = -(s+i0)$ gilt der Zusammenhang
\begin{equation}
  \ln\left(\frac{Q^2}{M^2}\right) =
  \ln\left(\frac{s}{M^2}\right) - i\pi
  \,.
\end{equation}

Der Zweischleifenbeitrag lautet:
\begin{align}
\label{eq:r2p}
  r^{(2)}_+ &=
    \frac{9}{2} \lqm^4(s)
    - \left[ \frac{437}{6} + \frac{2}{3} n_f \right] \lqm^3(s)
    + \left[
      \frac{37}{3}\pi^2 + \frac{3835}{18} + \frac{170}{9} n_f
      \right] \lqm^2(s)
  \nonumber \\* & \qquad
    + \biggl[
      26\sqrt{3}\,\Cl2\!\left(\frac{\pi}{3}\right)
      + 15\sqrt{3}\pi - 122\zeta_3
      - \left(\frac{1643}{18} + \frac{2}{3} n_f\right) \pi^2
      + \frac{96529}{216}
  \nonumber \\* & \qquad\qquad
      - \frac{1900}{27} n_f
      - \frac{40}{27} n_f^2
      \biggr]\, \lqm(s)
    + \Oc(\Lc^0)
    \,,
\end{align}
f"ur $I^3_f = I^3_{f'}$ und
\begin{align}
\label{eq:r2m}
  r^{(2)}_- &=
    \frac{9}{2} \lqm^4(s)
    - \left[ \frac{113}{6} + \frac{2}{3} n_f \right] \lqm^3(s)
    + \left[
      \frac{37}{3}\pi^2 - \frac{1147}{9} + \frac{116}{9} n_f
      \right] \lqm^2(s)
  \nonumber \\* & \qquad
    + \biggl[
      26\sqrt{3}\,\Cl2\!\left(\frac{\pi}{3}\right)
      + 15\sqrt{3}\pi - 122\zeta_3
      - \left(\frac{347}{18} + \frac{2}{3} n_f\right) \pi^2
      + \frac{35941}{216}
   \nonumber \\* & \qquad\qquad
      + \frac{206}{27} n_f
      - \frac{40}{27} n_f^2
      \biggr]\, \lqm(s)
    + \Oc(\Lc^0)
    \,,
\end{align}
f"ur $I^3_f = -I^3_{f'}$.
Die $SU(2)_L$-Wirkungsquerschnitte sind also in der Zweischleifenordnung in
\NNNLL-N"aherung bekannt,
d.h. in vollst"andiger logarithmischer N"aherung unter
Vernachl"assigung der nichtlogarithmischen Konstante.

\section{Elektroschwache Korrekturen}
\label{sec:EWcorr}

Dieser Abschnitt behandelt Strahlungskorrekturen im Rahmen des
elektroschwachen Standardmodells. Das $SU(2)$-Modell aus dem vorigen
Abschnitt wird um die $U(1)$-Eichgruppe der Hyperladung bzw. der
elektromagnetischen Wechselwirkung erweitert, die Chiralit"at der schwachen
Wechselwirkung wird beachtet und die Hyperladungen der Fermionen korrekt
eingef"uhrt.
Im Folgenden werden die Faktorisierung der
elektromagnetischen IR-Divergenzen (Abschnitt~\ref{sec:emfakt}), die
Einf"uhrung der elektroschwachen Parameter (Abschnitt~\ref{eq:EWparam}),
und die Ber"ucksichtigung von $M_W \ne M_Z$ (Abschnitt~\ref{eq:deltaM})
betrachtet, bevor in Abschnitt~\ref{sec:sigmaEW} Ergebnisse f"ur
elektroschwache Wirkungsquerschnitte pr"asentiert werden.

\subsection[Faktorisierung der elektromagnetischen IR-Divergenzen]
  {Faktorisierung der elektromagnetischen\\ IR-Divergenzen}
\label{sec:emfakt}

F"ur die exakte Berechnung von elektroschwachen Korrekturen bei
Vierfermionprozessen m"usste das Standardmodell mit der spontan gebrochenen
$SU(2)\times U(1)$-Symmetrie zugrunde gelegt werden.
Die Masseneigenzust"ande der Eichbosonen sind dort die $W^\pm$-
und $Z$-Bosonen mit "ahnlicher, aber unterschiedlicher Masse sowie das
masselose Photon.
Die beiden urspr"unglichen Eichgruppen mischen, und die ungebrochene
$U(1)$-Symmetrie hat nicht die Hyperladung, sondern die elektrische
Ladung als Generator.

In Kapitel~\ref{chap:SU2U1} wurde gezeigt, dass f"ur ein $SU(2)\times
U(1)$-Modell, in dem die beiden Eichgruppen nicht mischen, alle
logarithmischen Zweischleifenbeitr"age des Formfaktors
mit den am Ende des Kapitels
beschriebenen Schritten erhalten werden k"onnen.

Eine "ahnliche Rechnung wurde in\cite{Kuhn:2001hz} f"ur elektroschwache
Korrekturen zur Vierfermionamplitude durchgef"uhrt:
Die Strahlungskorrekturen werden f"ur den Fall berechnet, dass alle
Eichbosonen, $W$, $Z$ und Photon, die gleiche Masse~$M$ besitzen.
Dann werden die elektromagnetischen QED-Korrekturen mit einer Photonmasse,
die ebenfalls auf $M$ gesetzt wird, abfaktorisiert.
Das Ergebnis beschreibt den IR-endlichen Anteil der elektroschwachen
Korrekturen und kann bei Bedarf mit den QED-Korrekturen
multipliziert werden, wobei die IR-Divergenzen z.B. durch eine Photonmasse
$\lambda \ll M$ regularisiert werden.

Diese Rechnung ist f"ur das elektroschwache Standardmodell
in \NNLL-N"aherung g"ultig -- bis auf Terme, welche vom
Massenunterschied zwischen den $W$- und $Z$-Bosonen herr"uhren.
Das liegt daran, dass alle f"ur die \NNLL-N"aherung ben"otigten
Zweischleifenkoeffizienten $\gamma^{(2)}$, $\zeta^{(2)}$ und $\chi^{(2)}$
universell sind und weder von der Higgs-Masse noch vom Mechanismus der
spontanen Symmetriebrechung abh"angen.
Die Bestimmung des linearen Zweischleifenlogarithmus erfordert hingegen die
Verwendung der richtigen Masseneigenzust"ande f"ur die Eichbosonen.

Legt man aber eine vereinfachtes $SU(2)\times U(1)$-Modell zugrunde, in dem
die Hyperladung des Higgs-Dubletts verschwindet, so dass die Eichbosonen
der verschiedenen Eichgruppen nicht mischen, dann kann die oben
beschriebene Prozedur auch zur Berechnung des linearen
Zweischleifenlogarithmus verwendet werden.
Analog zum Kapitel~\ref{chap:SU2U1} lautet die Rechenvorschrift:
\begin{enumerate}
\item
  Berechnung der $SU(2)\times U(1)$-Korrekturen im spontan gebrochenen
  Modell mit der gleichen Masse~$M$ sowohl f"ur alle $SU(2)$-Eichbosonen
  als auch f"ur das $U(1)$-Hyperladungsboson.
\item
  Abfaktorisierung der QED-Korrekturen unter Verwendung der schweren
  Eichbosonmasse~$M$ f"ur die Photonmasse.
\end{enumerate}
Mit den QED-Korrekturen sind dabei tats"achlich die Korrekturen der
elektromagnetischen Eichsymmetrie gemeint, deren Generator proportional zur
elektrischen Ladung ist.

Durch die Annahme, dass das Higgs-Dublett keine Hyperladung hat und die
Eichgruppen folglich nicht mischen, vernachl"assigt man Terme, die den
linearen Zweischleifenlogarithmus beeinflussen.
Diese Beitr"age sind jedoch proportional zu $\sin^2\theta_W \approx 0{,}2$,
wobei $\theta_W$ der schwache Mischungswinkel ist.
Man kann also davon ausgehen, dass der Koeffizient des linearen Logarithmus
im vereinfachten Modell gegen"uber dem Standardmodellergebnis um etwa 20\%
abweichen wird.
Zus"atzlich f"uhrt die Massendifferenz $M_W\ne M_Z$ zu weiteren kubischen
und quadratischen Logarithmen, die aber in Abschnitt~\ref{eq:deltaM}
ber"ucksichtigt werden.
Da der Beitrag des linearen Logarithmus in den Zweischleifenergebnissen
insgesamt relativ klein ist (vgl. die Diskussion der Ergebnisse in
Abschnitt~\ref{sec:sigmaEW}), sagt das vereinfachte Modell ohne Mischung
die Standardmodellkorrekturen mit ausreichender Genauigkeit voraus.

\subsection{Elektroschwache Parameter}
\label{eq:EWparam}

Den bisherigen Rechnungen in den vorhergehenden Kapiteln liegt
das in Abschnitt~\ref{sec:SU2U1} beschriebene
$SU(2)\times U(1)$-Modell mit
spontaner Symmetriebrechung zugrunde.
Nicht ber"ucksichtigt wurde dabei jedoch die unterschiedliche Kopplung der
Standardmodellfermionen an die Eichgruppen in Abh"angigkeit von der
Chiralit"at und der Hyperladung.

F"ur die Vierfermionstreuung im elektroschwachen Modell m"ussen
$SU(2)$-Eich\-bosonen, die ausschlie"slich an linksh"andige Fermionen
koppeln, und $U(1)$-Bosonen, deren Kopplung an Fermionen von der
Hyperladung abh"angt, betrachtet werden.
Die Amplitude des Vierfermionprozesses $f \bar f \to f' \bar f'$ lautet in
Born-N"aherung:
\begin{equation}
  \Ac_B = \frac{i g^2}{s} \, \sum_{I,J=L,R} \left(
    I^3_f I^3_{f'} + \tan^2\theta_W \, \frac{Y_f Y_{f'}}{4} \right)
    \Ac^{f f'}_{I J}
  \,,
\end{equation}
mit
\begin{equation}
  \Ac^{f f'}_{I J} = \bar f_I \gamma^\mu f_I \cdot
    \bar f'_J \gamma_\mu f'_J
  \,,
\end{equation}
wobei $f_I$, $\bar f_I$ usw. die Spinoren der externen Fermionen mit
Chiralit"at~$I,J=L,R$ sind.
$I^3_f$ ist die Isospinkomponente des Fermions~$f$ mit
$I^3_f=0$ f"ur rechtsh"andige Singletts und $I^3_f=\pm\frac12$ f"ur die
Komponenten von linksh"andigen Dubletts.
$Y_f$ ist die Hyperladung des Fermions~$f$ entsprechend der
Tabelle~\ref{tab:SMfermionen} auf Seite~\pageref{tab:SMfermionen}.
Durch den Faktor $\tan^2\theta_W$ mit dem schwachen Mischungswinkel wurde
entsprechend $g' = g\,\tan\theta_W$~(\ref{eq:eggprel}) die
$U(1)$-Kopplung~$g'$ durch die $SU(2)$-Kopplung~$g$ ausgedr"uckt.

F"ur die Schleifenkorrekturen kann man den Isospin der Fermionen durch die
folgende Ersetzung der Basisvektoren im Raum der Isospin- und
Chiralit"atsstruktur~(\ref{eq:4fbasischiral}) ins Spiel bringen:
\begin{equation}
  \Ac^\lambda_{IJ} \to I^3_f I^3_{f'} \, \Ac^{f f'}_{IJ} \,,\quad
  \Ac^d_{IJ} \to \Ac^{f f'}_{IJ}
  \,.
\end{equation}

Im hier betrachteten Fall von Prozessen mit $f\ne f'$,
die in Bornscher N"aherung "uber den neutralen Strom vermittelt werden,
besitzen die
masselosen Fermionen $f$ und $\bar f$ im Anfangszustand und genauso die
Fermionen $f'$ und $\bar f'$ im Endzustand die gleiche Chiralit"at.
Diese kann dadurch ber"ucksichtigt werden, dass man den Anteil der
schwachen $SU(2)$-Wechselwirkung jeweils nur f"ur linksh"andige Fermionen
einberechnet.

Die Chiralit"at und die Hyperladung von Fermionen, die in den
Strahlungskorrekturen in geschlossenen Fermionschleifen auftauchen,
kann "uber den Parameter~$n_f$ ber"ucksichtigt werden.
F"ur die $SU(2)$-Wechselwirkung z"ahlt
dieser Parameter einfach die Anzahl der Dubletts von Dirac-Fermionen,
die in der Schleife laufen k"onnen.
In der ersten Fermionfamilie m"ussen hier das Leptondublett aus
$e^-$ und $\nu_e$ sowie das Quarkdublett aus $u$ und $d$ in drei
QCD-Farben gez"ahlt werden.
Allerdings laufen nur linksh"andige Fermionfreiheitsgrade in einer
Schleife, die an $SU(2)$-Bosonen koppelt. Daher kommt ein Faktor~$1/2$
hinzu.
Pro Fermionfamilie m"ussen also $(1+3)/2=2$ Fermion-Dubletts
ber"ucksichtigt werden.
Bei insgesamt $N_g=3$ Familien (oder Generationen) von Fermionen muss f"ur
die $SU(2)$-Wechselwirkung
\[
  n_f \to 2\,N_g = 6
\]
gesetzt werden.
F"ur die $U(1)$-Wechselwirkung muss $n_f$ durch eine Summe "uber die
Quadrate der Hyperladungen der Fermionen ersetzt werden.
Geladene Leptonsingletts tragen zu dieser Summe $(Y/2)^2=1$ bei,
Komponenten von Lepton\-dubletts $(Y/2)^2=1/4$,
$u$-artige Quarksingletts $(Y/2)^2=4/9$,
$d$-artige Quarksingletts $(Y/2)^2=1/9$
und Komponenten von Quarkdubletts $(Y/2)^2=1/36$.
Wieder korrigiert ein zus"atzlicher Faktor~$1/2$, dass jedes Fermion als
links- und rechtsh"andiges Teilchen doppelt gez"ahlt wird.
In der Summe ist der Beitrag jeder Fermionfamilie unter Ber"ucksichtigung
der drei QCD-Farben gerade $5/3$.
F"ur die $U(1)$-Wechselwirkung muss also
\[
  n_f \to \frac{5}{3}\,N_g = 5
\]
gesetzt werden.

\subsection[Ber"ucksichtigung des Massenunterschieds $M_W\ne M_Z$]
  {\boldmath Ber"ucksichtigung des Massenunterschieds $M_W\ne M_Z$}
\label{eq:deltaM}

Die bisherigen "Uberlegungen sind davon ausgegangen, das alle
$SU(2)$-Eichbosonen die gleiche Masse~$M$ besitzen.
Im Standardmodell hat jedoch das $Z$-Boson eine gr"o"sere Masse als das
$W^\pm$-Boson.
Dieser Massenunterschied kann durch eine Entwicklung um den Fall gleicher
Masse ber"ucksichtigt werden.
Als Entwicklungsparameter wird dazu
\begin{equation}
  \delta_M = 1 - \frac{M_W^2}{M_Z^2} \approx 0{,}2
\end{equation}
benutzt.
In Einschleifenn"aherung kann die "Anderung der Amplitude mit $M_Z$ bei
festgehaltenem $M=M_W$ in
f"uhrender Ordnung in~$\delta_M$ explizit berechnet werden.
Dies f"uhrt zu einer "Anderung der massenabh"angigen Koeffizienten
$\xi^{(1)}$, $F_0^{(1)}$ und $\Ac_0^{(1)}$ aus der L"osung der
Evolutionsgleichungen (\ref{eq:Atildechis}) und (\ref{eq:F2gzx}).
Diese Information gen"ugt, um in der Zweischleifenordnung die "Anderung der
Koeffizienten des kubischen und des quadratischen Logarithmus zu bestimmen.
Die Koeffizienten der anderen Funktionen $\gamma$, $\zeta$ und $\chi$ sind
massenunabh"angig und werden nicht vom Unterschied $M_W\ne M_Z$
beeinflusst.

Dadurch wurde der Unterschied zwischen der $W$- und der $Z$-Masse 
in \NNLL-N"aherung in f"uhrender Ordnung in~$\delta_M$ ber"ucksichtigt.
Die Gr"o"se dieses Effekts ist vergleichbar mit dem Beitrag des linearen
Logarithmus, also vergleichsweise klein.
Korrekturen h"oherer Ordnung in~$\delta_M$ k"onnen daher vernachl"assigt
werden.
Die "Anderung des linearen Logarithmus durch den Massenunterschied d"urfte
in der Gr"o"senordnung der vernachl"assigten nichtlogarithmischen Konstante
liegen und kann ebenfalls weggelassen werden.

\subsection{Elektroschwache Wirkungsquerschnitte}
\label{sec:sigmaEW}

Mit den "Uberlegungen aus den vorangegangenen Abschnitten
k"onnen nun elektroschwache Korrekturen zu Vierfermionprozessen berechnet
werden.
Beispielhaft werden hier totale Wirkungsquerschnitte angegeben, aber auch
die Vorw"arts-R"uckw"arts-Asymmetrie und die Links-Rechts-Asymmetrie
wurden berechnet.

Entsprechend der in Abschnitt~\ref{sec:emfakt} beschriebenen Faktorisierung
wird der elektroschwache Wirkungsquerschnitt~$\sigma$ auf den
Wirkungsquerschnitt~$\sigma_{\text{em}}$ mit reinen 
elektromagnetischen QED"=Korrekturen normiert.
F"ur die Berechnung von $\sigma_{\text{em}}$ wird eine endliche
Photonmasse~$\lambda$ 
benutzt, und $\sigma_{\text{em}}$ selbst wird so normiert, dass die virtuellen
QED-Korrekturen bei $s = \lambda^2$ verschwinden.

F"ur die St"orungsreihe wird die
$U(1)$-Hyperladungskopplung~$\alpha'$ mittels des schwachen Mischungswinkels
zugunsten der schwachen Kopplung~$\alpha$ eliminiert:
\begin{equation}
  \alpha' = \alpha \, \tan^2\theta_W
  \,.
\end{equation}
Der Wirkungsquerschnitt des Vierfermionprozesses $f\bar f \to f'\bar f'$
wird in der dann einzigen
Kopplung~$\alpha$ entwickelt:
\begin{equation} 
  \mathcal{R}_{ff'} = \frac{\sigma}{\sigma_{\text{em}}} =
  1 + \frac{\alpha}{4\pi} \, r^{(1)}_{ff'}
    + \left(\frac{\alpha}{4\pi}\right)^2 \, r^{(2)}_{ff'}
    + \Oc(\alpha^3)
  \,.  
\end{equation}
Die Entwicklung erfolgt in der bei $\mu=M=M_W$ renormierten Kopplung~$\alpha$,
w"ahrend die in der Born-N"aherung enthaltenen Kopplungen bei $\mu=\sqrt{s}$
renormiert werden.

Das Einschleifenergebnis ist nicht nur in logarithmischer N"aherung,
sondern exakt bekannt\cite{Beenakker:1991ca}.
F"ur die Zweischleifenbeitr"age zu verschiedenen Vierfermionprozessen
erh"alt man numerisch:
\begin{align}
\label{eq:r2llerg}
  r^{(2)}_{ll'}
  &= 1{,}42\,\Lc^4(s) -20{,}33\,\Lc^3(s) +112{,}57\,\Lc^2(s) -312{,}90\,\Lc(s)
  + \Oc(\Lc^0) \,, \\
  r^{(2)}_{lQ}
  &= 1{,}93\,\Lc^4(s) -11{,}28\,\Lc^3(s) +33{,}79\,\Lc^2(s) -60{,}87\,\Lc(s)
  + \Oc(\Lc^0) \,, \\
  r^{(2)}_{lq}
  &= 2{,}79\,\Lc^4(s) -51{,}98\,\Lc^3(s) +321{,}20\,\Lc^2(s) -757{,}35\,\Lc(s)
  + \Oc(\Lc^0) \,, \\
  r^{(2)}_{QQ'}
  &= 2{,}67\,\Lc^4(s) -46{,}64\,\Lc^3(s) +278{,}94\,\Lc^2(s) -666{,}05\,\Lc(s)
  + \Oc(\Lc^0) \,, \\
  r^{(2)}_{Qq}
  &= 3{,}53\,\Lc^4(s) -20{,}39\,\Lc^3(s) +65{,}20\,\Lc^2(s) -91{,}92\,\Lc(s)
  + \Oc(\Lc^0) \,, \\
\label{eq:r2qqerg}
  r^{(2)}_{qq'}
  &= 4{,}20\,\Lc^4(s) -71{,}87\,\Lc^3(s) +423{,}61\,\Lc^2(s) -919{,}35\,\Lc(s)
  + \Oc(\Lc^0) \,,
\end{align}
mit $\Lc(s) = \ln(s/M_W^2)$.
Der Index~$l$ bezeichnet ein geladenes Lepton, $Q$ ein Quark~$u$, $c$ oder
$t$ und $q$ ein Quark~$d$, $s$ oder $b$.
F"ur die numerische Berechnung wurde der Wert $\sin^2\theta_W=0{,}231$ f"ur
den \MSbar-renormierten Mischungswinkel an der Skala~$M_Z$ eingesetzt.

\begin{figure}[ht]
  \centering
  \includegraphics{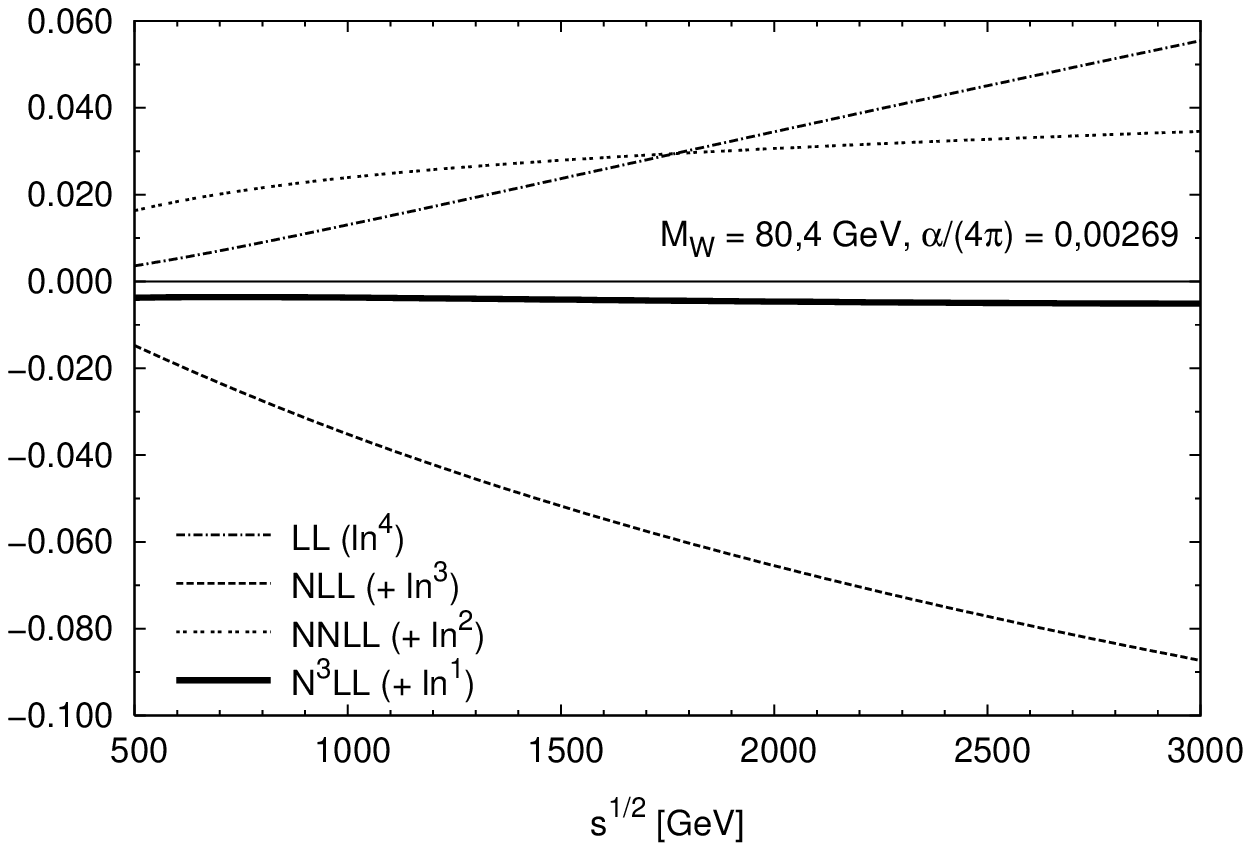}
  \caption{Zweischleifenkorrekturen~$r^{(2)}_{lq}$ zu $l\bar l \to q\bar q$
    in sukzessiven logarithmischen N"aherungen}
  \label{fig:plotr2lqlogs}
\end{figure}

F"ur die Zweischleifenkorrekturen $r^{(2)}_{lq}$ zum Prozess
$l\bar l \to q\bar q$
werden die sukzessiven logarithmischen N"aherungen in
Abb.~\ref{fig:plotr2lqlogs} inklusive des Vorfaktors~$(\alpha/4\pi)^2$
als Funktion von~$\sqrt s$ dargestellt.
Die eingesetzten Werte f"ur die $W$-Masse $M_W = 80{,}4\,\GeV$
und die schwache Kopplung~$\alpha/(4\pi) = 0{,}00269$ in \MSbar-Renormierung
bei der Skala~$M_W$ entsprechen den
Standardmodellwerten.

\begin{figure}[pt]
  \centering
  \includegraphics{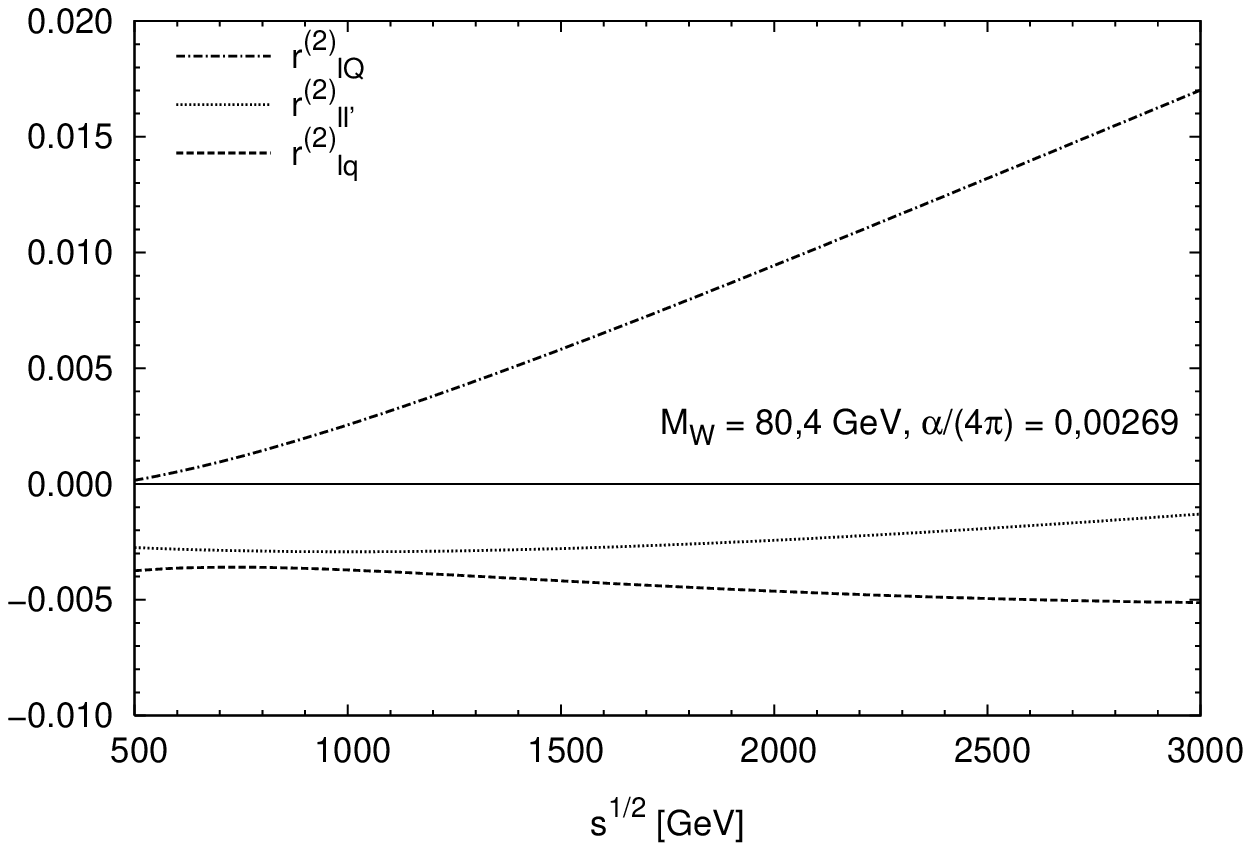}
  \caption{Zweischleifenkorrekturen zu $l\bar l \to Q\bar Q$,
      $l\bar l \to l'\bar l'$ und $l\bar l \to q\bar q$}
  \label{fig:plotr2ll}
\end{figure}
\begin{figure}[pt]
  \centering
  \includegraphics{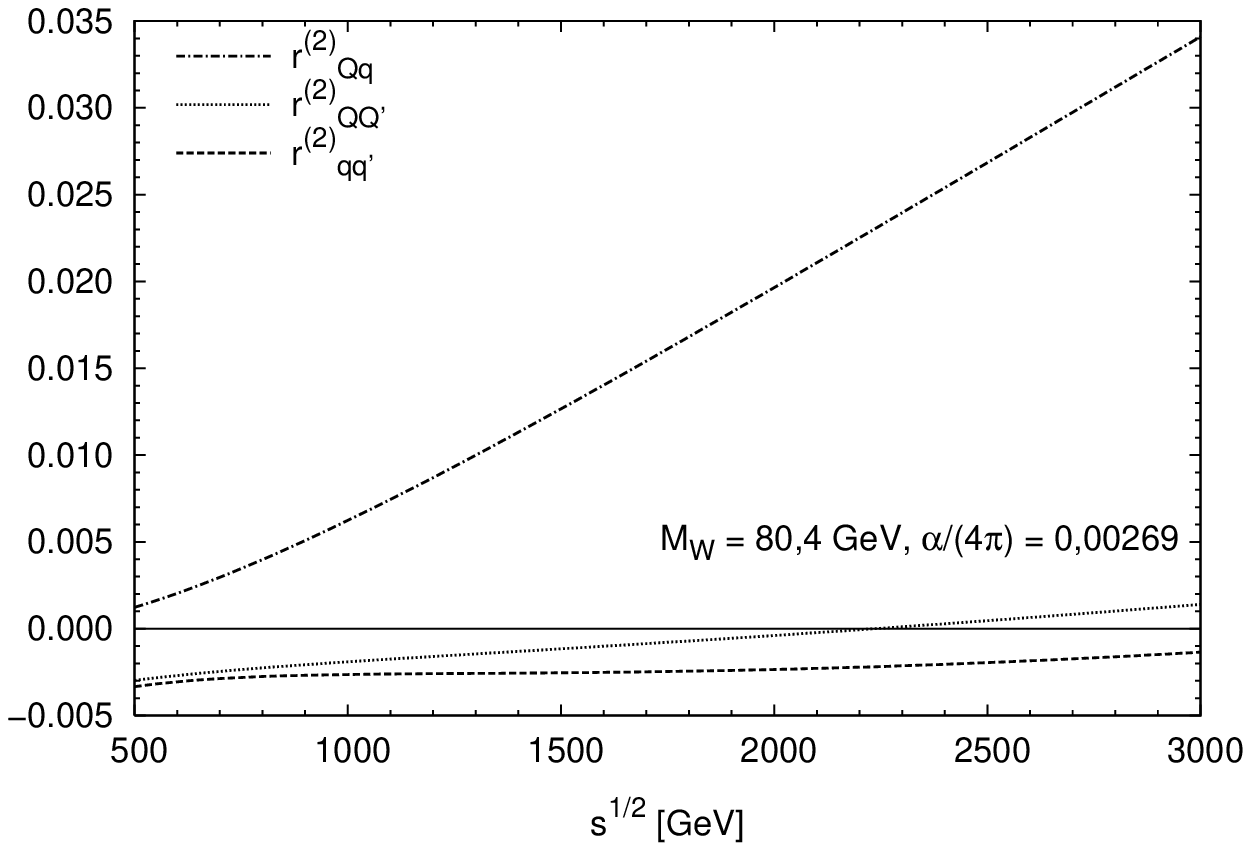}
  \caption{Zweischleifenkorrekturen zu $Q\bar Q \to q\bar q$,
    $Q\bar Q \to Q'\bar Q'$ und $q\bar q \to q'\bar q'$}
  \label{fig:plotr2qq}
\end{figure}

Die in LL-N"aherung sanft ansteigende Kurve wird durch den viel
gr"o"seren Logarithmus $\Lc^3$ stark ins Negative (\NLL)
und durch die Hinzunahme des Logarithmus $\Lc^2$ wieder ebensoweit nach
oben zur"uck verschoben (\NNLL).
Die Logarithmen $\Lc^3$ und $\Lc^2$ liefern im TeV-Bereich
den gr"o"sten Beitrag, die Verschiebung durch den linearen Logarithmus ist
bereits wieder kleiner.
Das gesamte \NNNLL-Ergebnis ist mit etwa einem halben Prozent wesentlich
kleiner als die Beitr"age der einzelnen logarithmischen Terme,
die im TeV-Bereich bis zu 10\% darstellen k"onnen.

Die Zweischleifenkorrekturen zu den Vierfermionprozessen
in den Gleichungen (\ref{eq:r2llerg}) bis (\ref{eq:r2qqerg})
sind in den Abbildungen \ref{fig:plotr2ll} und \ref{fig:plotr2qq} dargestellt.
Im Gegensatz zu den Beitr"agen der einzelnen logarithmischen Terme liegt die
Gr"o"se der gesamten Zweischleifenkorrekturen im $1\,\TeV$-Bereich
bei allen Prozessen unterhalb von 1\%.

Die Genauigkeit der elektroschwachen Zweischleifenkorrekturen
(\ref{eq:r2llerg}) bis (\ref{eq:r2qqerg}) kann wie folgt abgesch"atzt
werden:
\begin{itemize}
\item
  Auf der Basis der Ergebnisse f"ur den fermionischen und den abelschen
  Anteil des Formfaktors (Kapitel \ref{chap:nfns} und \ref{chap:abelsch})
  wird der Fehler durch die vernachl"assigte
  nichtlogarithmische Konstante in der Zweischleifenordnung auf 
  wenige Promille gesch"atzt.
\item
  Die durch $M^2/Q^2$ unterdr"uckten Beitr"age, die in dieser Arbeit nicht
  betrachtet wurden, k"onnen auf Grund des Vergleichs mit dem exakten
  Ergebnis f"ur den fermionischen Formfaktor (Kapitel~\ref{chap:nfns}) f"ur
  $\sqrt s > 500\,\GeV$ als deutlich unter einem Promille angenommen
  werden.
\item
  Die in f"uhrender Ordnung ber"ucksichtigten Effekte durch den
  Massenunterschied $M_W \ne M_Z$ betragen in Zweischleifenordnung weniger
  als f"unf Promille. Die vernachl"assigten h"oheren Ordnungen in der
  Massendifferenz und die nicht beachtete "Anderung des linearen
  Zweischleifenlogarithmus f"uhren zu einem Fehler im Promillebereich.
\item
  Die verschwindende Hyperladung des Higgs-Bosons und die dadurch
  vernachl"assigte Mischung der Eichbosonen wird als 20\%-Effekt relativ
  zum Koeffizienten des linearen Logarithmus veranschlagt, was zu einem
  Fehler von wenigen Promille f"ur die Zweischleifenkorrekturen
  f"uhrt.
\item
  Die Annahme, dass die Higgs-Masse der Eichbosonmasse entspricht,
  f"uhrt zu einer "Anderung des linearen Logarithmus
  von maximal~5\%, da die Higgs-Beitr"age insgesamt sehr klein sind.
  Die vernachl"assigte Masse von virtuellen Topquarks in Fermionschleifen
  "andert den linearen Logarithmus ebenfalls im Prozentbereich.
  Diese beiden Effekte addieren sich zu einem Fehler von einem bis wenigen
  Promille in den Zweischleifenkorrekturen.
\end{itemize}
Insgesamt d"urfte der Fehler in den Zweischleifenkorrekturen f"ur die
Produktion leichter Leptonen und Quarks, relativ zur
Born-N"aherung, im Bereich von einigen Promille bis maximal einem Prozent
liegen.
Wenn allerdings Topquarks als externe Teilchen produziert werden, liegt der
Fehler in der vernachl"assigten Topmasse aufgrund der hohen
Top-Yukawa-Kopplung deutlich h"oher.

\section{Zusammenfassung}

In dieser Arbeit wurden analytische Zweischleifenkorrekturen zum Formfaktor
eines abelschen Vektorstroms und zu Wirkungsquerschnitten von
Vierfermionprozessen berechnet, die in Born-N"aherung "uber den neutralen Strom
vermittelt werden.
Als Grundlage diente ein vereinfachtes $SU(2)\times U(1)$-Modell mit
spontaner Symmetrie\-brechung, dessen Ergebnisse unter den in der Arbeit
beschriebenen N"aherungen auf elektroschwache Korrekturen "ubertragen
wurden.
Dadurch konnten elektroschwache Zweischleifenkorrekturen mit einem
gesch"atzten Fehler im Promillebereich erzielt werden.

Diese Analyse liefert theoretische
Zweischleifenvorhersagen in ausreichender Genauigkeit f"ur die
Pr"azisionsexperimente am LHC und am International Linear Collider.

%
%
\begin{appendix}
\clearemptypage
\addtocontents{toc}{\protect{\vspace{1cm}}}
\addtocontents{toc}{\contentsline{chapter}{Anhang}{}}

\renewcommand{\chaptername}{\appendixname}


\clearemptypage

\chapter{Feynman-Regeln}
\label{chap:feynman}

In Abschnitt~\ref{sec:SU2U1} wird das in dieser Arbeit verwendete Modell
und die ihm zugrunde liegende Lagrange-Dichte beschrieben.  Daraus erh"alt
man in allgemeiner $R_\xi$"~Eichung die folgenden Feynman-Regeln des $SU(2)
\times U(1)$-Modells ohne Mischung zwischen den Eichgruppen.

\paragraph{Propagatoren:}
\begin{alignat}{4}
  \text{Fermion:} &&
    \vcentergraphics{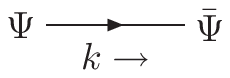}
    &= \frac{i}{\dslash k} = \frac{i\dslash k}{k^2} \,,
\\
  \text{Eichbosonen:} &&
    \vcentergraphics{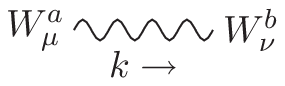}
    &= \frac{-i \delta^{ab}}{k^2-M^2} \left(
      g_{\mu\nu} - (1-\xi) \frac{k_\mu k_\nu}{k^2-\xi M^2} \right) ,
\\*
  &&
    \vcentergraphics{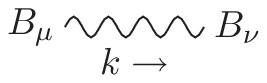}
    &= \frac{-i}{k^2} \left(
      g_{\mu\nu} - (1-\xi) \frac{k_\mu k_\nu}{k^2} \right) ,
\\
  \text{Higgs:} &&
    \vcentergraphics{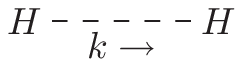}
    &= \frac{i}{k^2-M_H^2} \,,
\\
  \text{Goldstone:} &&
    \vcentergraphics{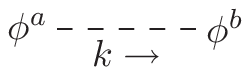}
    &= \frac{i \delta^{ab}}{k^2-\xi M^2} \,,
\\
  \text{Geist:} &&
    \vcentergraphics{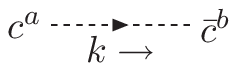}
    &= \frac{i \delta^{ab}}{k^2-\xi M^2} \,.
\end{alignat}
Auf die Angabe der infinitesimalen positiven Imagin"arteile~$+i0$ in den
Propagatornennern wurde verzichtet.
In dieser Arbeit wird ausschlie"slich die Feynman-'t~Hooft-Eichung mit
$\xi=1$ verwendet, um besonders einfache Eichboson-Propagatoren zu
erhalten.

\paragraph{Vertizes:}
\begin{gather}
  \vcentergraphics{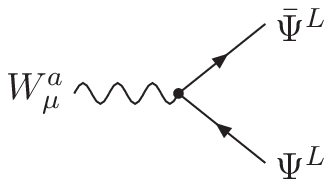} = ig \gamma_\mu t^a \,,
  \qquad
  \vcentergraphics{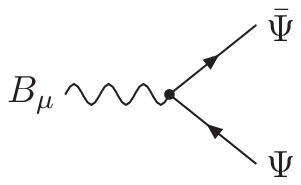} = ig' \gamma_\mu \frac{Y_\Psi}{2} \,,
\\[1ex]
  \vcentergraphics{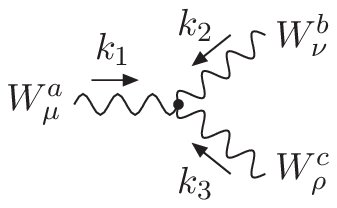} =
    g\,f^{abc} \Bigl[
      g_{\mu\nu} (k_1-k_2)_\rho
      + g_{\nu\rho} (k_2-k_3)_\mu
      + g_{\rho\mu} (k_3-k_1)_\nu
    \Bigr] \,,
\\[1ex]
  \vcentergraphics{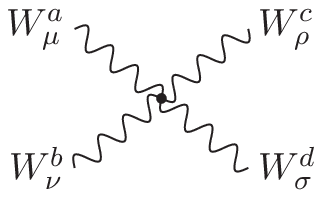} =
    \left\{ \begin{aligned}
    -i g^2 \Bigl[ &
      f^{abe}f^{cde} (g_{\mu\rho}g_{\nu\sigma} - g_{\mu\sigma}g_{\nu\rho})
    \\ + &
      f^{ace}f^{bde} (g_{\mu\nu}g_{\rho\sigma} - g_{\mu\sigma}g_{\nu\rho})
    \\ + &
      f^{ade}f^{bce} (g_{\mu\nu}g_{\rho\sigma} - g_{\mu\rho}g_{\nu\sigma})
    \Bigr] \,,
    \end{aligned} \right.
\\[1ex]
  \vcentergraphics{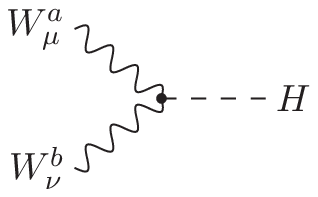} = ig M\,g_{\mu\nu} \delta^{ab} \,,
  \qquad
  \vcentergraphics{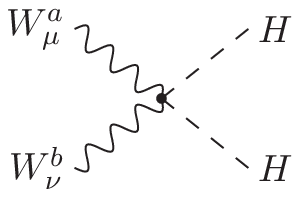} = \frac{i}{2} g^2 \,
    g_{\mu\nu} \delta^{ab} \,,
\\[1ex]
  \vcentergraphics{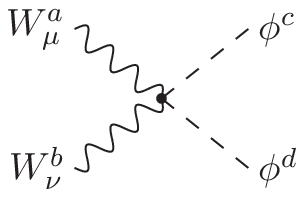} = \frac{i}{2} g^2 \,
    g_{\mu\nu} \delta^{ab} \delta^{cd} \,,
  \quad
  \vcentergraphics{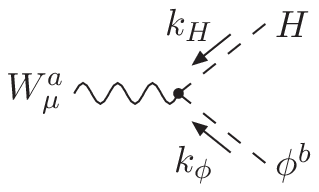} = \frac{g}{2}
    \delta^{ab} (k_H-k_\phi)_\mu
    \,,
\\[1ex]
  \vcentergraphics{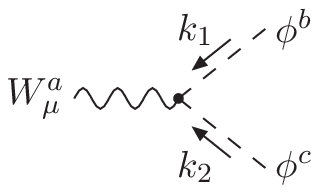} = -\frac{g}{2}
    f^{abc} (k_1-k_2)_\mu \,,
  \quad
  \vcentergraphics{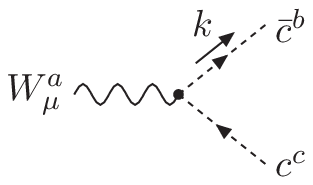} = -g\,f^{abc} k_\mu \,,
\\[1ex]
  \vcentergraphics{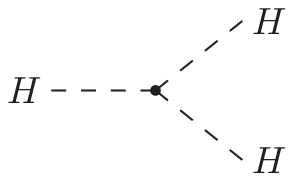} = -\frac{3}{2}i g \frac{M_H^2}{M} \,,
  \qquad
  \vcentergraphics{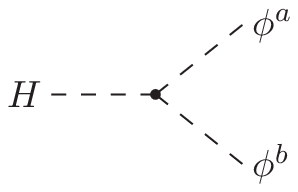} =
    -\frac{i}{2} g \frac{M_H^2}{M} \delta^{ab} \,,
\\[1ex]
  \vcentergraphics{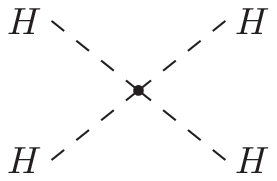} =
    -\frac{3}{4}i g^2 \frac{M_H^2}{M^2} \,,
  \qquad
  \vcentergraphics{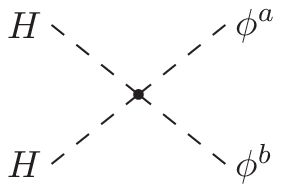} =
    -\frac{i}{4} g^2 \frac{M_H^2}{M^2} \delta^{ab} \,,
\\[1ex]
  \vcentergraphics{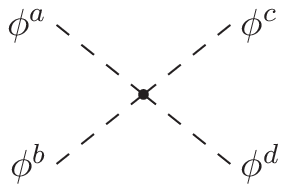} =
    -\frac{i}{4} g^2 \frac{M_H^2}{M^2} \bigl(
      \delta^{ab}\delta^{cd}
      + \delta^{ac}\delta^{bd}
      + \delta^{ad}\delta^{bc}
    \bigr) \,,
\\[1ex]
  \vcentergraphics{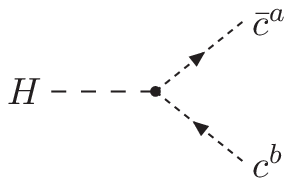} =
    -\frac{i}{2} g \, \xi M \, \delta^{ab} \,,
  \qquad
  \vcentergraphics{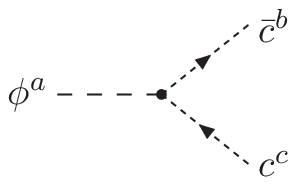} =
    \frac{i}{2} g \, \xi M \, f^{abc} \,.
\end{gather}

Au"serdem werden Spinoren f"ur externe Fermionen ben"otigt.
Ein (Anti"~)Fermion im Anfangszustand mit Impuls~$p$ und wird durch $u(p)$
($\bar v(p)$) beschrieben,
ein (Anti"~)Fermion im Endzustand durch $\bar u(p)$ ($v(p)$).
F"ur masselose Fermionen gilt:
\begin{gather}
  \dslash p \, u(p) = \dslash p \, v(p) = 0 \,,\quad
  \bar u(p) \, \dslash p = \bar v(p) \, \dslash p = 0
  \,.
\end{gather}
Der Unterschied zwischen $u(p)$ und $v(p)$ macht sich nur bei endlicher
Fermionmasse bemerkbar.
Genau genommen fehlt bei diesen Spinoren die Angabe der Spinorientierung
bzw. der Chiralit"at.
F"ur die schwache $SU(2)$-Wechselwirkung werden nur
linksh"andige Fermionen betrachtet,
f"ur die $U(1)$-Wechselwirkung der Hyperladung spielt die Chiralit"at keine
Rolle.

"Uber den Impuls jeder geschlossenen Schleife wird bei dimensionaler
Regularisierung in $d=4-2\eps$ Raum-Zeit-Dimensionen folgenderma"sen
integriert:
\begin{gather}
\label{eq:muloopint}
  \mu^{4-d} \loopint dk =
  \frac{i}{16\pi^2} \left(\frac{\mu^2}{M^2}\right)^{\eps} S_\eps
  \left[ e^{\eps\gamma_E} (M^2)^\eps \loopintf dk \right]
\end{gather}
Das Multiplizieren des Impulsintegrals mit $e^{\eps\gamma_E}$ pro
Schleife sorgt daf"ur, dass das Integral nicht mehr von der Eulerschen
Konstanten~$\gamma_E$ abh"angt.
Der dadurch entstehende Vorfaktor
\begin{equation}
  S_\eps = (4\pi)^\eps \, e^{-\eps\gamma_E}
\end{equation}
wird im Rahmen der \MSbar-Vorschrift in die Renormierungsskala~$\mu$
absorbiert.
Wenn zudem $(M^2)^{-\eps}$ pro Schleife abfaktorisiert wird (au"ser bei
masselosen Integralen), kann das Integral dimensionslos definiert
werden. Der Vorfaktor $(\mu^2/M^2)^{\eps}$ f"allt heraus, wenn sp"ater die
Skalenwahl $\mu=M$ getroffen wird.

Der Nenner $16\pi^2$ im Vorfaktor wird zusammen mit dem Quadrat der
Kopplungskonstanten zu
\[
  \frac{g^2}{16\pi^2} = \frac{\alpha}{4\pi}
  \qquad \text{bzw.} \qquad
  \frac{{g'}^2}{16\pi^2} = \frac{\alpha'}{4\pi}
\]
zusammengefasst.
Ein Faktor~$i$ wird aus dem Schleifenintegral herausgezogen, damit das
Integral selbst f"ur raumartige Impuls"ubertr"age real ist.

Geschlossene Fermion- oder Geistschleifen werden mit (-1) multipliziert.
Au"serdem wird bei Fermionschleifen die Spur im Spinorraum sowie im
Darstellungsraum der fermionischen Eichgruppe gebildet.


\clearemptypage

\chapter{Mathematische Funktionen und Konstanten}
\label{chap:math}

\subsubsection{\boldmath
  Die Eulersche $\Gamma$-Funktion und die Riemannsche $\zeta$-Funktion}

F"ur $\Rep z>0$ ist die \emph{Eulersche $\Gamma$-Funktion} "uber die folgende
Integraldarstellung definiert\cite{Gradshteyn:Table}:
\begin{equation}
  \Gamma(z) = \int_0^\infty \! \dd t \, t^{z-1} \, e^{-t}
  \,.
\end{equation}
Die $\Gamma$-Funktion l"asst sich auf die ganze komplexe Ebene analytisch
fortsetzen. Sie hat dann einfache Pole f"ur nichtpositive ganzzahlige Werte
von~$z$ mit den folgenden Residuen:
\begin{equation}
\label{eq:ResGamma}
  \Res \, \Gamma(z) \Big|_{z=-n} = \frac{(-1)^n}{n!} \,,\ n=0,1,2,\ldots
\end{equation}
Es gilt:
\begin{equation}
  \Gamma(z+1) = z \, \Gamma(z)
\end{equation}
und
\begin{equation}
  \Gamma(1) = \Gamma(2) = 1 \,,\quad
  \Gamma(\tfrac{1}{2}) = \sqrt\pi
  \,.
\end{equation}
Bei Produkten von $\Gamma$-Funktionen kann die Relation
\begin{equation}
  \Gamma(z) \Gamma(1-z) = \frac{\pi}{\sin(\pi z)}
\end{equation}
hilfreich sein.
Die sogenannte Verdopplungsformel lautet:
\begin{equation}
  \Gamma(2z) = \frac{2^{2z-1}}{\sqrt\pi} \, \Gamma(z) \Gamma(z+\tfrac12)
  \,.
\end{equation}
Die Entwicklung des Logarithmus der $\Gamma$-Funktion um das Argument~1
ist f"ur $|z|<1$ folgenderma"sen gegeben:
\begin{equation}
\label{eq:Gammadev}
  \ln \Gamma(1+z) =
  -\gamma_E \, z
  + \sum_{k=2}^\infty \, \zeta_k \, \frac{(-z)^k}{k}
  \,,
\end{equation}
mit der Eulerschen Konstanten $\gamma_E \approx 0{,}577216$
und der \emph{Riemannschen $\zeta$-Funktion}
\begin{equation}
\label{eq:zetan}
  \zeta_n = \sum_{i=1}^\infty \, \frac{1}{i^n} \,,\ n=2,3,\ldots
  \,,
\end{equation}
wobei
\begin{equation}
  \zeta_2 = \frac{\pi^2}{6} \,,\quad
  \zeta_3 \approx 1{,}20206 \,,\quad
  \zeta_4 = \frac{\pi^4}{90} \,.
\end{equation}

\subsubsection{Harmonische Summen und Polylogarithmen}

Die einfache \emph{harmonische Summe} ist durch
\begin{equation}
\label{eq:HarmonicSum}
  S_m(n) = \sum_{i=1}^n \, \frac{1}{i^m}
\end{equation}
definiert. Sie kann auf mehrere Indizes erweitert
werden\cite{Vermaseren:1999uu}:
\begin{equation}
  S_{m_1,m_2,\ldots,m_k}(n) =
  \sum_{i=1}^n \, \frac{1}{i^{m_1}} \, S_{m_2,\ldots,m_k}(i)
  \,.
\end{equation}
Produkte von harmonischen Summen lassen sich auf Linearkombinationen von
harmonischen Summen mit einer gr"o"seren Zahl von Indizes transformieren.
Eine verallgemeinerte Form von harmonischen Summen sind Z-Summen und
S-Summen\cite{Moch:2001zr}.
Diese wiederum enthalten als Speziallfall u.a. die
\emph{Polylogarithmen}\cite{Lewin:Polylog}
\begin{equation}
\label{eq:polylog}
  \Li n(x) = \sum_{i=1}^\infty \, \frac{x^i}{i^n}
  = \int_0^x \!\dd t \, \frac{\Li{n-1}(t)}{t}
  \,,\quad \text{mit } \Li1(x) = -\ln(1-x)
  \,,
\end{equation}
wobei die Summendarstellung nur f"ur $|x|<1$ definiert ist.
Es gilt:
\begin{equation}
  \Li n(1) = \zeta_n \,,\ n=2,3,\ldots
\end{equation}
In dieser Arbeit wird der Funktionswert
\begin{equation}
  \Li4\!\left(\frac12\right) \approx 0{,}517479
\end{equation}
als nicht weiter reduzierbare analytische Konstante ben"otigt.

\subsubsection{Clausen-Funktion}

Die \emph{Clausen-Funktion} ist folgenderma"sen definiert
(siehe z.B.\cite{Weinzierl:2004bn}):
\begin{equation}
\label{eq:Clausen}
  \Cl n(\theta) =
  \begin{cases}
    \tfrac{1}{2i} \Bigl[ \Li n(e^{i\theta}) - \Li n(e^{-i\theta}) \Bigr]
    \,,\ n \text{ gerade}\,, \\[1ex]
    \tfrac{1}{2} \Bigl[ \Li n(e^{i\theta}) + \Li n(e^{-i\theta}) \Bigr]
    \,,\ n \text{ ungerade}\,.
  \end{cases}
\end{equation}
In dieser Arbeit wird der Funktionswert
\begin{equation} 
   \Cl2\!\left(\frac{\pi}{3}\right) \approx 1{,}014942
\end{equation}
als analytische Konstante ben"otigt.
Die Funktion~$\Cl2$ besitzt die folgende Integraldarstellung:
\begin{equation}
  \Cl2(\theta)
  = -\int_0^\theta\!\dd\theta' \, \ln\left|2\sin\frac{\theta'}{2}\right|
  \,.
\end{equation}


\clearemptypage

\chapter{Methoden}
\label{chap:methoden}

\section{Parametrisierung von Schleifenintegralen}
\label{sec:loopparam}

\subsubsection{Feynman-Parameter}

Schleifenintegrale mit mehreren Propagatoren m"ussen parametrisiert
werden, um die Impulsintegrationen ausf"uhren zu k"onnen.
Am bekanntesten sind die \emph{Feynman-Parameter}
(siehe z.B.\cite{Peskin:1995ev}):
\begin{equation}
\label{eq:feynmanparam}
  \prod_k \frac{1}{(A_k+i0)^{n_k}} =
  \frac{\Gamma(\sum_k n_k)}{\prod_k \Gamma(n_k)}
  \left(\prod_k \int_0^1\!\dd x_k \, x_k^{n_k-1}\right)
  \frac{\delta(\sum_k x_k - 1)}{\bigl(\sum_k x_k A_k+i0\bigr)^{\sum_k n_k}}
  \,,
\end{equation}
mit den Propagatornennern~$A_k$.
Der Term~$i0$ steht f"ur einen infinitesimalen positiven Imagin"arteil.
Die Parametrisierung gilt auch f"ur nicht ganzzahlige Potenzen~$n_k$.

Damit werden alle Propagatoren $A_k$ zu einem einzigen Propagator
$(\sum_k x_k A_k)$ zusammengefasst.
Wenn alle urspr"unglichen Propagatoren von der Form $A_k = \ell_k^2 - m_k^2$
sind, dann besitzt auch der kombinierte Propagator diese Form,
ggf. nach einer
quadratischen Erg"anzung f"ur den Schleifenimpuls im neuen Propagator.
Es ist auch m"oglich, nicht gleich alle Propagatoren auf einmal mit
Gl.~(\ref{eq:feynmanparam}) zusammenzufassen, sondern z.B. wiederholt
paarweise.

Die Impulsintegration wird mit der Relation
\begin{equation}
  \loopintf d\ell \, \frac{1}{(\ell^2-\Delta+i0)^n} =
  e^{-n i\pi} \, (\Delta-i0)^{\frac d2-n} \,
  \frac{\Gamma(n-\frac d2)}{\Gamma(n)}
\end{equation}
ausgef"uhrt, ggf. mehrmals hintereinander bei Mehrschleifenintegralen.
Anschlie"send m"ussen die Integrale "uber die Feynman-Parameter~$x_k$ unter
Beachtung der $\delta$-Funktion ausgef"uhrt werden.
Eine hilfreiche Relation daf"ur ist\cite{Gradshteyn:Table}
\begin{equation}
  \int_0^1\!\dd x \, x^{\alpha-1} (1-x)^{\beta-1} =
  \frac{\Gamma(\alpha) \Gamma(\beta)}{\Gamma(\alpha+\beta)}
  \,,
\end{equation}
oder allgemeiner
\begin{equation}
\label{eq:feynmanintdeltagamma}
  \left(\prod_k \int_0^1\!\dd x_k \, x_k^{n_k-1}\right)
  \delta\!\left(\sum_k x_k - 1\right) =
  \frac{\prod_k \Gamma(n_k)}{\Gamma(\sum_k n_k)}
  \,.
\end{equation}

\subsubsection{Schwinger-Parameter}

\emph{Schwinger-Parameter}, auch $\alpha$-Parameter genannt, stellen eine
alternative Methode zur Parametrisierung von Schleifenintegralen dar
(siehe z.B.\cite{Itzykson:1980rh,Smirnov:2004ym}).
Jeder Propagator~$1/A^n$ wird entsprechend
\begin{equation}
\label{eq:schwingerparam}
  \frac{1}{(A+i0)^n} = \frac{e^{-i\pi(n/2)}}{\Gamma(n)}
  \int_0^\infty\!\dd\alpha \, \alpha^{n-1} \, e^{i\alpha(A+i0)}
\end{equation}
umgeschrieben.
Beliebig viele Propagatoren k"onnen zusammengefasst werden:
\begin{equation}
  \prod_k \frac{1}{(A_k + i0)^{n_k}} =
  \left(\prod_k \frac{e^{-i\pi(n_k/2)}}{\Gamma(n_k)}
    \int_0^\infty\!\dd\alpha_k \, \alpha_k^{n_k-1}\right)
  \exp\!\left(i \sum_k \alpha_k(A_k+i0)\right)
  .
\end{equation}
Alle Propagatornenner~$A_k$ finden sich im Argument einer Exponentialfunktion
wieder.
Auf eine "ahnliche Weise k"onnen auch Terme mit ganzzahligen Potenzen im
Z"ahler des Integrals (z.B. irreduzible Skalarprodukte) parametrisiert
werden:
\begin{equation}
  A^n = \left.\left(\frac{1}{i} \, \frac{\partial}{\partial\alpha}\right)^n
  e^{i\alpha A}\right|_{\alpha=0}
  \,.
\end{equation}

Die Schleifenintegration kann mit Hilfe der folgenden Relation ausgef"uhrt
werden:
\begin{equation}
  \loopintf d\ell \, e^{i(\alpha\ell^2 + 2p\cdot\ell)} =
  e^{-i\pi(d/4)} \, \alpha^{-d/2} \,
  \exp\!\left(-i\frac{p^2}{\alpha}\right)
  \,.
\end{equation}
Im Ergebnis dieses Integrals steht der verbleibende Impuls~$p$ wieder
quadratisch im
Argument einer Exponentialfunktion, so dass auch mehrere
Schleifenintegrationen hintereinander ausgef"uhrt werden k"onnen.

Zur Integration "uber die Schwinger-Parameter kann in einfachen F"allen die
Relation~(\ref{eq:schwingerparam}) in umgekehrter Richtung benutzt werden:
\begin{equation}
\label{eq:schwingerinvers}
  \int_0^\infty\!\dd\alpha \, \alpha^{n-1} \, e^{i\alpha(A+i0)} =
  \frac{e^{i\pi(n/2)} \, \Gamma(n)}{(A+i0)^n} =
  \frac{e^{-i\pi(n/2)} \, \Gamma(n)}{(-A-i0)^n} \,.  
\end{equation}
Wenn keine Parameterintegrale mehr mittels~(\ref{eq:schwingerinvers})
gel"ost werden k"onnen, empfiehlt sich die Reskalierung der
Schwinger-Parameter.
Dazu wird zun"achst die Zahl~1 in der folgenden Form ins Parameterintegral
eingef"ugt:
\[
  1 = \int_0^\infty\!\dd\eta \,
      \delta\biggl(\sum_{j\in S} \alpha_j - \eta\biggr) \,,
\]
wobei die Menge~$S$, "uber welche die Summe in der $\delta$-Funktion
l"auft, eine beliebige, nicht leere Teilmenge der Indizes der
Schwinger-Parameter darstellt. Das Ergebnis h"angt von der Wahl von~$S$
nicht ab.
Anschlie"send werden die Schwinger-Parameter unter dem $\eta$-Integral
mit~$\eta$ reskaliert:
$\alpha_k \to \eta \, \alpha_k \, \forall k$.
Jedes $\alpha$-Integral ergibt bei dieser Reskalierung einen Faktor~$\eta$,
und die $\delta$-Funktion wird entsprechend
\[
  \delta\biggl(\sum_{j\in S} \alpha_j - \eta\biggr) \to
  \frac{1}{\eta} \, \delta\biggl(\sum_{j\in S} \alpha_j - 1\biggr)
\]
transformiert.
Das $\eta$-Integral kann mittels~(\ref{eq:schwingerinvers}) ausgef"uhrt
werden.
"Ubrig bleiben die Integrale "uber die Schwinger-Parameter, die "ahnlich
wie bei den Feynman-Parametern durch eine $\delta$-Funktion eingeschr"ankt
sind. Tats"achlich ist so eine Transformation von Schwinger-Parametern auf
Feynman-Parameter immer m"oglich. In den Integralen "uber Parameter~$\alpha_j$,
die wegen $j\in S$ durch die $\delta$-Funktion eingeschr"ankt sind, kann
die obere Integralgrenze von $\infty$ nach~$1$ gesetzt werden.

Die Reskalierung kann auch f"ur eine Untermenge der Schwinger-Parameter
durchgef"uhrt werden, wenn dies im konkreten Fall sinnvoller ist.
Au"serdem besteht ein Vorteil dieses Zugangs darin, dass der Inhalt der
Menge~$S$ zun"achst offen gelassen werden kann und erst dann festgelegt
werden muss, wenn die jeweiligen Integrationen "uber die
Schwinger-Parameter ausgef"uhrt werden.

Zur Integration "uber einen Schwinger-Parameter, der nicht durch die
$\delta$-Funktion eingeschr"ankt ist, dient die Relation
\begin{equation}
  \int_0^\infty\!\dd\alpha \, \alpha^{\nu-1} (A+\alpha B)^{-\rho}
  = \frac{\Gamma(\nu) \Gamma(\rho-\nu)}{\Gamma(\rho)} \,
    B^{-\nu} \, A^{\nu-\rho}
  \,,
\end{equation}
und zur Integration "uber mehrere, durch die $\delta$-Funktion beschr"ankte
Parameter kann
\begin{equation}
\label{eq:schwingerintdeltagamma}
  \left(\prod_k \int_0^\infty\!\dd\alpha_k \, \alpha_k^{\nu_k-1}\right)
  \frac{\delta\bigl(\sum_{j\in S} \alpha_j - 1\bigr)}
    {\bigl(\sum_k \alpha_k\bigr)^{\sum_k \nu_k}}
  = \frac{\prod_k \Gamma(\nu_k)}{\Gamma(\sum_k \nu_k)}
\end{equation}
verwendet werden, wobei $S$ wieder eine beliebige, nicht leere Teilmenge
der Indizes ist.
Gl.~(\ref{eq:schwingerintdeltagamma}) ist eine Verallgemeinerung von
Gl.~(\ref{eq:feynmanintdeltagamma}).

\section{Tensorreduktion}
\label{sec:tensorred}

Gegeben sei ein Schleifenintegral, das den Schleifenimpuls
au"ser in den Propagatornennern auch im Z"ahler enth"alt.
Die Terme im Z"ahler k"onnen aus Schleifenimpulsen mit offenen
Lorentz-Indizes, aus irreduziblen Skalarprodukten oder aus
Schleifenimpulsen, die mit $\gamma$-Matrizen kontrahiert sind, bestehen.
Diese F"alle sind "aquivalent, denn mit $\dslash \ell = \gamma_\mu \ell^\mu$
und $p\cdot\ell = p_\mu \ell^\mu$ k"onnen alle $\ell$-abh"angigen Objekte
im Z"ahler des Integrals als Tensor
$\ell^\mu \ell^\nu \ell^\rho \cdots$ geschrieben
werden.

Die \emph{Tensorreduktion}\cite{Passarino:1979jh} transformiert
Schleifenintegrale mit einer Tensorstruktur des Schleifenimpulses im Z"ahler
auf Linearkombinationen von konstanten Tensoren, deren Koeffizienten
skalare Schleifenintegrale sind.
Als Beispiel wird der Fall einer Zweipunktfunktion diskutiert, die nur
einen unabh"angigen "au"seren Impuls enth"alt.

Das Tensorintegral 1.~Stufe,
\begin{equation}
\label{eq:2punkttensorred1ansatz}
  B^\mu(k) = \int\!\dd^d\ell \, f(k,\ell) \, \ell^\mu \,,
\end{equation}
mit einer skalaren Funktion $f(k,\ell)$,
h"angt nach der Integration "uber den Schleifenimpuls~$\ell$ nur noch vom
externen Impuls~$k$ ab. $B^\mu(k)$ ist ein Lorentz-Tensor 1.~Stufe und muss
daher proportional zu~$k^\mu$ sein.
Man macht den einzig m"oglichen Ansatz~$B^\mu(k) = k^\mu B_1(k^2)$,
setzt diesen
in~(\ref{eq:2punkttensorred1ansatz}) ein und kontrahiert beide Seiten der
Gleichung mit~$k_\mu$. Daraus erh"alt man:
\begin{align}
\label{eq:2punkttensorred1}
  \int\!\dd^d\ell \, f(k,\ell) \, \ell^\mu &=
    \frac{k^\mu}{k^2} \int\!\dd^d\ell \, f(k,\ell) \, \ell\cdot k
  \,.
\end{align}
Das Skalarprodukt $\ell\cdot k$ im Integral l"asst sich in der Regel mit
den Propagator\-nennern k"urzen.
Das Tensorintegral 2.~Stufe,
\begin{equation}
\label{eq:2punkttensorred2ansatz}
  B^{\mu\nu}(k) = \int\!\dd^d\ell \, f(k,\ell) \, \ell^\mu \ell^\nu \,,
\end{equation}
ist ein Lorentz-Tensor 2.~Stufe. Der allgemeinste Ansatz
lautet daher:
$B^{\mu\nu}(k) = g^{\mu\nu} B_{00}(k^2) + k^\mu k^\nu B_{11}(k^2)$.
Nach dem Einsetzen in~(\ref{eq:2punkttensorred2ansatz}) wird die Gleichung
einmal mit $g_{\mu\nu}$ und einmal mit $k_\mu k_\nu$ kontrahiert.
Das dabei entstehende Gleichungssystem liefert die folgende L"osung:
\begin{align}
\label{eq:2punkttensorred2}
  \int\!\dd^d\ell \, f(k,\ell) \, \ell^\mu \ell^\nu &=
    \frac{1}{d-1} \biggl[
    g^{\mu\nu} \int\!\dd^d\ell \, f(k,\ell) \left(
      \ell^2 - \frac{(\ell\cdot k)^2}{k^2} \right)
    \nonumber \\* & \qquad\qquad
    + \frac{k^\mu k^\nu}{k^2} \int\!\dd^d\ell \, f(k,\ell) \left(
      d \frac{(\ell\cdot k)^2}{k^2}  - \ell^2 \right)
    \biggr]
  \,.
\end{align}
Mit der gleichen Methode k"onnen Tensorintegrale mit mehr externen Impulsen auf
konstante Lorentz-Tensoren zur"uckgef"uhrt werden.



\section{Mellin-Barnes-Darstellung}
\label{sec:MB}

Mit Hilfe von Feynman- oder Schwinger-Parametern kann in
Schleifenintegralen die Integration "uber die Schleifenimpulse
immer ausgef"uhrt werden.
Aber h"aufig sind auch die verbliebenen Integrale "uber die Feynman- oder
Schwinger-Parameter zu aufw"andig, um sie direkt l"osen zu k"onnen.

In vielen F"allen scheitert die Parameterintegration,
weil die Abh"angigkeit des Integranden von den Parametern zu komplex ist.
Mit der \emph{Mellin-Barnes-Darstellung}\cite{Boos:1990rg}
(siehe z.B.\cite{Davydychev:1993mt,Smirnov:2004ym})
k"onnen Summen in Produkte
umgeformt werden, deren Faktoren einfacher zu integrieren sind.
Der Preis ist die zus"atzliche Einf"uhrung von Mellin-Barnes-Integralen in
der komplexen Ebene.

Die Mellin-Barnes-Darstellung lautet:
\begin{equation}
\label{eq:MBdef}
  \frac{1}{(A+B)^n} = \frac{1}{\Gamma(n)} \MBint z \,
  \Gamma(-z) \Gamma(n+z) \frac{B^z}{A^{n+z}}
  \,.
\end{equation}
$A$ und $B$ k"onnen z.B. Funktionen der Feynman- oder Schwinger-Parameter sein.
Durch die Mellin-Barnes-Darstellung erh"alt man aus der Summe~$(A+B)^{-n}$
getrennte Faktoren $A^{-n-z}$ und $B^z$, die leichter zu integrieren sind.
Oder es kann sich um einen massiven Propagator handeln mit $A=\ell^2$ und
$B=-m^2$, der durch die Mellin-Barnes-Darstellung zu einem masselosen
Propagator wird.

Der Integrationsweg der Mellin-Barnes-Integrale in~(\ref{eq:MBdef})
verl"auft so von $-i\infty$ nach $+i\infty$,
dass Pole aus $\Gamma$-Funktionen der Form
$\Gamma(\ldots+z)$ links des Integrationswegs liegen ("`IR-Pole"')
und Pole der Form $\Gamma(\ldots-z)$ rechts des Integrationswegs liegen
("`UV-Pole"').
Abh"angig vom Parameter~$n$ kann der Integrationsweg u.U. nicht gerade
sein, sondern muss durch Windungen und Kurven alle Pole von
$\Gamma$-Funktionen auf der jeweils richtigen Seite umfahren.

Wenn $|A|<|B|$ ist, dann kann das Integral "uber~$z$ auf der linken Seite
im Unendlichen (bei $\Rep z\to-\infty$) geschlossen werden. Die Summe
"uber die Residuen der Funktion $\Gamma(n+z)$~(\ref{eq:ResGamma}) liefert
genau die Taylor-Entwicklung der linken Seite von Gl.~(\ref{eq:MBdef}) f"ur
$|A|<|B|$.
Im anderen Fall $|A|>|B|$ wird das Integral auf der rechten Seite im
Unendlichen (bei $\Rep z\to+\infty$) geschlossen. Hier liefert die
Summe "uber die Residuen der Funktion~$\Gamma(-z)$ die entsprechende
Taylor-Entwicklung f"ur $|A|>|B|$.
F"ur $|A|=|B|$ kann das Mellin-Barnes-Integral wahlweise auf der linken oder
auf der rechten Seite geschlossen werden, da die $\Gamma$-Funktionen im
Unendlichen schnell genug abfallen.
Die Mellin-Barnes-Darstellung ist also vergleichbar mit einer Taylorreihe,
aber sie deckt beide F"alle $|A|<|B|$ und $|A|>|B|$ ab.

"Ublicherweise benutzt man die Mellin-Barnes-Darstellung, um eine
Schleifenintegration oder Parameterintegration zu vereinfachen. Dann wird
nach der Einf"uhrung der Mellin-Barnes-Darstellung zun"achst die Schleifen-
oder Parameterintegration ausgef"uhrt. Dabei entstehen weitere
$\Gamma$-Funktionen, die bei der Auswertung des Mellin-Barnes-Integrals
zus"atzliche Residuen liefern. Die Definition des Integrationswegs bleibt
auch bei mehr als zwei $\Gamma$-Funktionen die gleiche wie oben beschrieben.

Besonders einfache Mellin-Barnes-Integrale k"onnen durch das
erste Barnsche Lemma\cite{Barnes:1908,Bailey:1935} gel"ost werden:
\begin{multline}
\label{eq:BarnesLemma1}
  \MBint z \, \Gamma(\lambda_1-z) \Gamma(\lambda_2-z)
  \Gamma(\lambda_3+z) \Gamma(\lambda_4+z) = \\*
  \frac{\Gamma(\lambda_1+\lambda_3) \Gamma(\lambda_1+\lambda_4)
    \Gamma(\lambda_2+\lambda_3) \Gamma(\lambda_2+\lambda_4)}{
    \Gamma(\lambda_1+\lambda_2+\lambda_3+\lambda_4)}
  \,.
\end{multline}

Zur Aufsummierung der Residuen in einem Mellin-Barnes-Integral ist die
folgende Relation hilfreich:
\begin{equation}
  \Gamma(\alpha-k) = (-1)^k \,
  \frac{\Gamma(\alpha) \Gamma(1-\alpha)}{\Gamma(1-\alpha+k)}
  \,,
\end{equation}
mit ganzzahligem~$k$.
H"aufig m"ussen die $\Gamma$-Funktionen, die in den Residuen vorkommen,
im Parameter~$\eps$
der dimensionalen Regularisierung oder in anderen infinitesimalen Parametern
entwickelt werden.
Algorithmen dazu finden sich z.B.
in\cite{Moch:2001zr,Weinzierl:2004bn,Grozin:2003ak}.
In den unendlichen Reihen, die bei der Entwicklung in~$\eps$ entstehen,
tauchen beispielsweise harmonische Summen (siehe Anhang~\ref{chap:math})
und Verallgemeinerungen davon auf.

\section{Expansion by Regions}
\label{sec:ExpReg}

Die Methode der \emph{Expansion by Regions}%
\cite{Beneke:1998zp,Smirnov:1998vk,Smirnov:1999bz,Smirnov:2002pj}
erm"oglicht die asymptotische Entwicklung von Feynman-Diagrammen auch in
schwierigen kinematischen Limites.
Ein Beispiel daf"ur ist der Sudakov-Limes $M^2\ll Q^2$ des Formfaktors
(vgl. Abschnitt~\ref{sec:formfaktor}). Hier treten im Limes $M\to0$
infrarote und kollineare Divergenzen auf.

Die Methode der Expansion by Regions geht f"ur jedes Feynman-Diagramm
nach dem folgenden Schema vor:
\begin{enumerate}
\item
  Das Integrationsgebiet (also der $d$-dimensionale Impulsraum pro
  Schleifenintegration) wird in Regionen f"ur die Schleifenimpulse
  aufgeteilt.
\item
  In jeder Region wird der Integrand in eine Taylor-Reihe entwickelt
  bez"uglich der Parameter, die in dieser Region als klein betrachtet
  werden.
\item
  Anschlie"send werden jedoch die Integrale f"ur jede Region nicht nur
  "uber den Impulsbereich der jeweiligen Region ausgef"uhrt, sondern "uber
  den ganzen urspr"unglichen Impulsraum.
\end{enumerate}
Die Summe der Beitr"age aller Regionen reproduziert das exakte Ergebnis.
Durch die Entwicklungen in den Regionen liegt dieses Ergebnis jedoch als
asymptotische Entwicklung in $M^2/Q^2$ vor.

In der Praxis gen"ugt es, sich Regionen anzuschauen, in denen das Integral
im Limes $M\to0$ Singularit"aten aufweist.
Die Regionen m"ussen dann nicht exakt durch ihre Grenzen im Impulsraum
definiert werden. Die Angabe eines typischen Punktes im Impulsraum ist
ausreichend, um die Entwicklung in der Region durchf"uhren zu k"onnen.
Solche Angaben sind z.B.: "`alle Komponenten des Schleifenimpulses sind von
der Gr"o"senordnung~$M$"', oder "`der Schleifenimpuls ist kollinear zum
externen Impuls~$p_1$"'.

Das in einer bestimmten Region entwickelte Integral ist formal au"serhalb
dieser Region nicht unbedingt konvergent. Die Konvergenz wird aber durch
die dimensionale Regularisierung erreicht. Insbesondere bei kollinearen
Regionen kommt es jedoch vor, dass die dimensionale Regularisierung alleine
nicht ausreicht. Hier m"ussen zus"atzliche Parameter wie Propagatorpotenzen
von ihren ganzzahligen Werten infinitesimal in die komplexe Ebene verschoben
werden, um die Integrale der Regionen zu regularisieren. In der Summe aller
Regionen heben sich die durch eine solche analytische
Regularisierung parametrisierten Singularit"aten wieder auf.

In dieser Arbeit werden nur Terme betrachtet, die nicht mit $M^2/Q^2$
unterdr"uckt sind. F"ur die Expansion by Regions bedeutet dies, dass von
jeder Entwicklung in einer Region jeweils nur der f"uhrende Beitrag
ben"otigt wird.
Beispielsweise entspricht die harte Region, in der alle Komponenten des
Schleifenimpulses von der Gr"o"senordnung~$Q$ sind, gerade dem masselosen
Diagramm, weil die Masse in den Propagatoren gegen"uber dem harten
Schleifenimpuls vernachl"assigt wird.


\clearemptypage

\chapter{Skalare Integrale}
\label{chap:Skalar}

\section{Skalare Integrale der abelschen Beitr"age}

\subsection{Planares Vertexdiagramm}
\label{sec:SkalarLA}

Der Beitrag des planaren Vertexdiagramms zum Formfaktor
kann wie folgt in skalare Integrale~(\ref{eq:LAskalar}) zerlegt werden:
\begin{alignat}{4}
\label{eq:LAzerlegung}
  \lefteqn{F_{v,\LA} =
    C_F^2 \left(\frac{\alpha}{4\pi}\right)^2 \, i^2
    \left(\frac{\mu^2}{M^2}\right)^{2\eps} S_\eps^2
    \, \Bigl\{
  } \quad \nonumber \\
    & F_\LA(0, 0, 1, 0, 1, 1, 0) &\cdot &\bigl( -4 \bigr)
\nonumber \\
  +& F_\LA(0, 0, 1, 1, 0, 1, 0) &\cdot &\bigl( 8 - 2\*d \bigr)
\nonumber \\
  +& F_\LA(0, 0, 1, 1, 1, 1, 0) &\cdot &\bigl( -4 + (8 - 2\*d)\*\Mr \bigr)
\nonumber \\
  +& F_\LA(1, -1, 0, 1, 1, 1, 0) &\cdot &\bigl( 4 \bigr)
\nonumber \\
  +& F_\LA(1, -1, 1, 1, 0, 1, 0) &\cdot &\bigl( -4 \bigr)
\nonumber \\
  +& F_\LA(1, -1, 1, 1, 1, 1, 0) &\cdot &\bigl( -4\*\Mr \bigr)
\nonumber \\
  +& F_\LA(1, 0, -1, 1, 1, 1, 0) &\cdot &\bigl( 44 - \tfrac{16}{d-2} - 12\*d + d^2 \bigr)
\nonumber \\
  +& F_\LA(1, 0, 0, 0, 1, 1, 0) &\cdot &\bigl( -44 + \tfrac{16}{d-2} + 12\*d - d^2 \bigr)
\nonumber \\
  +& F_\LA(1, 0, 0, 1, 0, 1, 0) &\cdot &\bigl( -40 + \tfrac{16}{d-2} + 12\*d - d^2 \bigr)
\nonumber \\
  +& F_\LA(1, 0, 0, 1, 1, 0, 0) &\cdot &\bigl( -4 \bigr)
\nonumber \\
  +& F_\LA(1, 0, 0, 1, 1, 1, 0) &\cdot &\bigl( 24 - \tfrac{16}{d-2} - 4\*d + (-44 + \tfrac{16}{d-2} + 12\*d - d^2)\*\Mr \bigr)
\nonumber \\
  +& F_\LA(1, 0, 0, 1, 1, 1, 1) &\cdot &\bigl( -8 + 2\*d \bigr)
\nonumber \\
  +& F_\LA(1, 0, 1, 0, 0, 1, 0) &\cdot &\bigl( 40 - \tfrac{16}{d-2} - 12\*d + d^2 \bigr)
\nonumber \\
  +& F_\LA(1, 0, 1, 0, 1, 0, 0) &\cdot &\bigl( 4 \bigr)
\nonumber \\
  +& F_\LA(1, 0, 1, 0, 1, 1, 0) &\cdot &\bigl( -8 + (44 - \tfrac{16}{d-2} - 12\*d + d^2)\*\Mr \bigr)
\nonumber \\
  +& F_\LA(1, 0, 1, 0, 1, 1, 1) &\cdot &\bigl( 4 \bigr)
\nonumber \\
  +& F_\LA(1, 0, 1, 1, -1, 1, 0) &\cdot &\bigl( 8 - 4\*d \bigr)
\nonumber \\
  +& F_\LA(1, 0, 1, 1, 0, 0, 0) &\cdot &\bigl( -8 + 4\*d \bigr)
\nonumber \\
  +& F_\LA(1, 0, 1, 1, 0, 1, 0) &\cdot &\bigl( 24 - \tfrac{16}{d-2} - 4\*d + (8 - 4\*d)\*\Mr \bigr)
\nonumber \\
  +& F_\LA(1, 0, 1, 1, 0, 1, 1) &\cdot &\bigl( -4 + 2\*d \bigr)
\nonumber \\
  +& F_\LA(1, 0, 1, 1, 1, 0, 0) &\cdot &\bigl( 8 + (-8 + 4\*d)\*\Mr \bigr)
\nonumber \\
  +& F_\LA(1, 0, 1, 1, 1, 1, 0) &\cdot &\bigl( -8 + (32 - \tfrac{16}{d-2} - 4\*d)\*\Mr \bigr)
\nonumber \\
  +& F_\LA(1, 0, 1, 1, 1, 1, 1) &\cdot &\bigl( 4 + (-4 + 2\*d)\*\Mr \bigr)
\nonumber \\
  +& F_\LA(1, 1, 0, 0, 0, 1, 0) &\cdot &\bigl( -48 + \tfrac{16}{d-2} + 14\*d - d^2 \bigr)
\nonumber \\
  +& F_\LA(1, 1, 0, 0, 1, 0, 0) &\cdot &\bigl( 48 - \tfrac{16}{d-2} - 14\*d + d^2 \bigr)
\nonumber \\
  +& F_\LA(1, 1, 0, 0, 1, 1, 0) &\cdot &\bigl( -44 + \tfrac{16}{d-2} + 12\*d - d^2 \bigr)
\nonumber \\
  +& F_\LA(1, 1, 0, 0, 1, 1, 1) &\cdot &\bigl( 48 - \tfrac{16}{d-2} - 14\*d + d^2 \bigr)
\nonumber \\
  +& F_\LA(1, 1, 1, -1, 0, 1, 0) &\cdot &\bigl( 44 - \tfrac{16}{d-2} - 12\*d + d^2 \bigr)
\nonumber \\
  +& F_\LA(1, 1, 1, -1, 1, 0, 0) &\cdot &\bigl( -44 + \tfrac{16}{d-2} + 12\*d - d^2 \bigr)
\nonumber \\
  +& F_\LA(1, 1, 1, -1, 1, 1, 1) &\cdot &\bigl( -44 + \tfrac{16}{d-2} + 12\*d - d^2 \bigr)
\nonumber \\
  +& F_\LA(1, 1, 1, 0, -1, 1, 0) &\cdot &\bigl( 8 - 4\*d \bigr)
\nonumber \\
  +& F_\LA(1, 1, 1, 0, 0, 0, 0) &\cdot &\bigl( -8 + 4\*d \bigr)
\nonumber \\
  +& F_\LA(1, 1, 1, 0, 0, 1, 0) &\cdot &\bigl( 88 - \tfrac{48}{d-2} - 24\*d + 2\*d^2 + (8 - 4\*d)\*\Mr \bigr)
\nonumber \\
  +& F_\LA(1, 1, 1, 0, 0, 1, 1) &\cdot &\bigl( -12 + 6\*d \bigr)
\nonumber \\
  +& F_\LA(1, 1, 1, 0, 1, 0, 0) &\cdot &\bigl( \tfrac{16}{d-2} + (-8 + 4\*d)\*\Mr \bigr)
\nonumber \\
  +& F_\LA(1, 1, 1, 0, 1, 0, 1) &\cdot &\bigl( 4 - 2\*d \bigr)
\nonumber \\
  +& F_\LA(1, 1, 1, 0, 1, 1, 0) &\cdot &\bigl( -8 + (88 - \tfrac{32}{d-2} - 24\*d + 2\*d^2)\*\Mr \bigr)
\nonumber \\
  +& F_\LA(1, 1, 1, 0, 1, 1, 1) &\cdot &\bigl( -44 + \tfrac{32}{d-2} + 12\*d - d^2 + (-8 + 4\*d)\*\Mr \bigr)
\nonumber \\
  +& F_\LA(1, 1, 1, 0, 1, 1, 2) &\cdot &\bigl( 4 - 2\*d \bigr)
\nonumber \\
  +& F_\LA(1, 1, 1, 1, -1, 0, 0) &\cdot &\bigl( -4 + 2\*d \bigr)
\nonumber \\
  +& F_\LA(1, 1, 1, 1, -1, 1, 0) &\cdot &\bigl( 12 - 4\*d + (-4 + 2\*d)\*\Mr \bigr)
\nonumber \\
  +& F_\LA(1, 1, 1, 1, 0, -1, 0) &\cdot &\bigl( 4 - 2\*d \bigr)
\nonumber \\
  +& F_\LA(1, 1, 1, 1, 0, 0, 0) &\cdot &\bigl( -52 + \tfrac{16}{d-2} + 16\*d - d^2 \bigr)
\nonumber \\
  +& F_\LA(1, 1, 1, 1, 0, 0, 1) &\cdot &\bigl( 4 - 2\*d \bigr)
\nonumber \\
  +& F_\LA(1, 1, 1, 1, 0, 1, 0) &\cdot &\bigl( 16 - \tfrac{16}{d-2} - 2\*d + (-28 + \tfrac{16}{d-2} + 8\*d - d^2)\*\Mr + (-4 + 2\*d)\*\Mr^2 \bigr)
\nonumber \\
  +& F_\LA(1, 1, 1, 1, 0, 1, 1) &\cdot &\bigl( -12 + 4\*d + (4 - 2\*d)\*\Mr \bigr)
\nonumber \\
  +& F_\LA(1, 1, 1, 1, 1, -1, 0) &\cdot &\bigl( -4 + (4 - 2\*d)\*\Mr \bigr)
\nonumber \\
  +& F_\LA(1, 1, 1, 1, 1, 0, 0) &\cdot &\bigl( 16 - 2\*d + (-60 + \tfrac{16}{d-2} + 16\*d - d^2)\*\Mr + (4 - 2\*d)\*\Mr^2 \bigr)
\nonumber \\
  +& F_\LA(1, 1, 1, 1, 1, 0, 1) &\cdot &\bigl( -4 + (4 - 2\*d)\*\Mr \bigr)
\nonumber \\
  +& F_\LA(1, 1, 1, 1, 1, 1, 0) &\cdot &\bigl( -4 + (32 - \tfrac{16}{d-2} - 4\*d)\*\Mr + (-44 + \tfrac{16}{d-2} + 12\*d - d^2)\*\Mr^2 \bigr)
\nonumber \\
  +& F_\LA(1, 1, 1, 1, 1, 1, 1) &\cdot &\bigl( \tfrac{8}{d-2} + (-16 + 4\*d)\*\Mr + (4 - 2\*d)\*\Mr^2 \bigr)
\nonumber \\
  +& F_\LA(1, 1, 1, 1, 1, 1, 2) &\cdot &\bigl( 2 - d \bigr)

  \Bigr\}
  \,,
\end{alignat}
mit $S_\eps = (4\pi)^\eps \, e^{-\eps\gamma_E}$ und $z = M^2/Q^2$.
Die von Null verschiedenen Ergebnisse der skalaren Integrale lauten
in f"uhrender Ordnung in $M^2/Q^2$:
\begin{align}
\label{eq:LAergskalar}
    F_\LA(1, -1, 0, 1, 1, 1, 0) &= -\tfrac{33}{32} - \tfrac{1}{8\*\eps^2} - \tfrac{7}{16\*\eps} - \tfrac{1}{8}\*\lqm + \tfrac{1}{8}\*\lqm^2 + \tfrac{1}{48}\*\pi^2 \,,
\nonumber \\
  F_\LA(1, -1, 1, 1, 0, 1, 0) &= -\tfrac{89}{16} - \tfrac{1}{4\*\eps^2} + (-\tfrac{11}{8} + \tfrac{1}{2}\*\lqm)\tfrac{1}{\eps} + \tfrac{11}{4}\*\lqm - \tfrac{1}{2}\*\lqm^2 + \tfrac{1}{24}\*\pi^2 \,,
\nonumber \\
  F_\LA(1, -1, 1, 1, 1, 1, 0) &= \tfrac{1}{4} - \tfrac{1}{2}\*\lqm + \tfrac{3}{4}\*\lqm^2 - \tfrac{1}{4}\*\lqm^3 + \tfrac{1}{4}\*\pi^2 + (\tfrac{1}{4}\*\lqm^2 + \tfrac{1}{6}\*\pi^2)\tfrac{1}{\eps} - 2\*\zeta_3 \,,
\nonumber \\
  F_\LA(1, 0, -1, 1, 1, 1, 0) &= -\tfrac{7}{16} - \tfrac{1}{4\*\eps^2} - \tfrac{5}{8\*\eps} - \tfrac{3}{4}\*\lqm + \tfrac{1}{4}\*\lqm^2 + \tfrac{1}{24}\*\pi^2 \,,
\nonumber \\
  F_\LA(1, 0, 0, 1, 0, 1, 0) &= \tfrac{13}{8} + \tfrac{1}{4\*\eps} - \tfrac{1}{2}\*\lqm \,,
\nonumber \\
  F_\LA(1, 0, 0, 1, 1, 1, 0) &= \tfrac{5}{2} + \tfrac{1}{2\*\eps^2} + \tfrac{3}{2\*\eps} + \lqm - \tfrac{1}{2}\*\lqm^2 - \tfrac{1}{12}\*\pi^2 \,,
\nonumber \\
  F_\LA(1, 0, 0, 1, 1, 1, 1) &= \tfrac{85}{32} + \tfrac{1}{8\*\eps^2} + \tfrac{11}{16\*\eps} - \tfrac{3}{8}\*\lqm - \tfrac{1}{8}\*\lqm^2 - \tfrac{1}{48}\*\pi^2 \,,
\nonumber \\
  F_\LA(1, 0, 1, 0, 1, 1, 0) &= \tfrac{7}{2} + \tfrac{1}{2\*\eps^2} + \tfrac{3}{2\*\eps} - \tfrac{1}{12}\*\pi^2 \,,
\nonumber \\
  F_\LA(1, 0, 1, 1, -1, 1, 0) &= \tfrac{89}{16} + \tfrac{1}{4\*\eps^2} + (\tfrac{11}{8} - \tfrac{1}{2}\*\lqm)\tfrac{1}{\eps} - \tfrac{11}{4}\*\lqm + \tfrac{1}{2}\*\lqm^2 - \tfrac{1}{24}\*\pi^2 \,,
\nonumber \\
  F_\LA(1, 0, 1, 1, 0, 1, 0) &= \tfrac{19}{2} + \tfrac{1}{2\*\eps^2} + (\tfrac{5}{2} - \lqm)\tfrac{1}{\eps} - 5\*\lqm + \lqm^2 - \tfrac{1}{12}\*\pi^2 \,,
\nonumber \\
  F_\LA(1, 0, 1, 1, 0, 1, 1) &= \tfrac{241}{32} + \tfrac{3}{8\*\eps^2} + (\tfrac{31}{16} - \tfrac{3}{4}\*\lqm)\tfrac{1}{\eps} - \tfrac{31}{8}\*\lqm + \tfrac{3}{4}\*\lqm^2 - \tfrac{1}{16}\*\pi^2 \,,
\nonumber \\
  F_\LA(1, 0, 1, 1, 1, 1, 0) &= -\lqm^2 + \tfrac{1}{2}\*\lqm^3 - \tfrac{1}{3}\*\pi^2 + (-\tfrac{1}{2}\*\lqm^2 - \tfrac{1}{3}\*\pi^2)\tfrac{1}{\eps} + 4\*\zeta_3 \,,
\nonumber \\
  F_\LA(1, 0, 1, 1, 1, 1, 1) &= 13 + \tfrac{1}{2\*\eps^2} + (3 - \tfrac{3}{2}\*\lqm)\tfrac{1}{\eps} - 8\*\lqm + \tfrac{7}{4}\*\lqm^2 - \tfrac{1}{12}\*\pi^2 \,,
\nonumber \\
  F_\LA(1, 1, 0, 0, 1, 1, 0) &= \tfrac{19}{2} + \tfrac{1}{2\*\eps^2} + (\tfrac{5}{2} - \lqm)\tfrac{1}{\eps} - 5\*\lqm + \lqm^2 + \tfrac{1}{12}\*\pi^2 \,,
\nonumber \\
  F_\LA(1, 1, 0, 0, 1, 1, 1) &= \tfrac{63}{16} + \tfrac{1}{4\*\eps^2} + (\tfrac{9}{8} - \tfrac{1}{2}\*\lqm)\tfrac{1}{\eps} - \tfrac{9}{4}\*\lqm + \tfrac{1}{2}\*\lqm^2 + \tfrac{1}{24}\*\pi^2 \,,
\nonumber \\
  F_\LA(1, 1, 1, -1, 0, 1, 0) &= -\tfrac{89}{16} - \tfrac{1}{4\*\eps^2} + (-\tfrac{11}{8} + \tfrac{1}{2}\*\lqm)\tfrac{1}{\eps} + \tfrac{11}{4}\*\lqm - \tfrac{1}{2}\*\lqm^2 + \tfrac{1}{24}\*\pi^2 \,,
\nonumber \\
  F_\LA(1, 1, 1, -1, 1, 0, 0) &= -\tfrac{35}{8} - \tfrac{1}{2\*\eps^2} + (-\tfrac{7}{4} + \tfrac{1}{2}\*\lqm)\tfrac{1}{\eps} + \tfrac{7}{4}\*\lqm - \tfrac{1}{4}\*\lqm^2 \,,
\nonumber \\
  F_\LA(1, 1, 1, -1, 1, 1, 1) &= \tfrac{43}{16} + \tfrac{1}{4\*\eps^2} + (\tfrac{5}{8} - \tfrac{1}{2}\*\lqm)\tfrac{1}{\eps} + \tfrac{1}{2}\*\lqm^2 - \tfrac{5}{24}\*\pi^2 + \lqm\*(-\tfrac{5}{2} + \tfrac{1}{6}\*\pi^2) + \zeta_3 \,,
\nonumber \\
  F_\LA(1, 1, 1, 0, -1, 1, 0) &= \tfrac{63}{32} + \tfrac{1}{8\*\eps^2} + (\tfrac{9}{16} - \tfrac{1}{4}\*\lqm)\tfrac{1}{\eps} - \tfrac{9}{8}\*\lqm + \tfrac{1}{4}\*\lqm^2 - \tfrac{1}{48}\*\pi^2 \,,
\nonumber \\
  F_\LA(1, 1, 1, 0, 0, 1, 0) &= \tfrac{19}{2} + \tfrac{1}{2\*\eps^2} + (\tfrac{5}{2} - \lqm)\tfrac{1}{\eps} - 5\*\lqm + \lqm^2 - \tfrac{1}{12}\*\pi^2 \,,
\nonumber \\
  F_\LA(1, 1, 1, 0, 0, 1, 1) &= \tfrac{241}{32} + \tfrac{3}{8\*\eps^2} + (\tfrac{31}{16} - \tfrac{3}{4}\*\lqm)\tfrac{1}{\eps} - \tfrac{31}{8}\*\lqm + \tfrac{3}{4}\*\lqm^2 - \tfrac{1}{16}\*\pi^2 \,,
\nonumber \\
  F_\LA(1, 1, 1, 0, 1, 0, 0) &= 7 + \tfrac{1}{\eps^2} + (3 - \lqm)\tfrac{1}{\eps} - 3\*\lqm + \tfrac{1}{2}\*\lqm^2 \,,
\nonumber \\
  F_\LA(1, 1, 1, 0, 1, 0, 1) &= \tfrac{35}{16} + \tfrac{1}{4\*\eps^2} + (\tfrac{7}{8} - \tfrac{1}{4}\*\lqm)\tfrac{1}{\eps} - \tfrac{7}{8}\*\lqm + \tfrac{1}{8}\*\lqm^2 \,,
\nonumber \\
  F_\LA(1, 1, 1, 0, 1, 1, 0) &= -\tfrac{1}{12}\*\pi^2\*\lqm^2 - \tfrac{7}{360}\*\pi^4 - \zeta_3\*\lqm \,,
\nonumber \\
  F_\LA(1, 1, 1, 0, 1, 1, 1) &= 5 + \tfrac{1}{\eps} + \lqm\*(-1 - \tfrac{1}{6}\*\pi^2) - \zeta_3 \,,
\nonumber \\
  F_\LA(1, 1, 1, 0, 1, 1, 2) &= \tfrac{21}{2} + \tfrac{1}{4\*\eps^2} + (\tfrac{9}{4} - \tfrac{1}{2}\*\lqm)\tfrac{1}{\eps} + \tfrac{1}{2}\*\lqm^2 - \tfrac{1}{8}\*\pi^2 + \lqm\*(-\tfrac{31}{8} - \tfrac{1}{12}\*\pi^2) - \tfrac{1}{2}\*\zeta_3 \,,
\nonumber \\
  F_\LA(1, 1, 1, 1, -1, 0, 0) &= 6 + \tfrac{1}{2\*\eps^2} + (2 - \lqm)\tfrac{1}{\eps} - 4\*\lqm + \lqm^2 - \tfrac{1}{12}\*\pi^2 \,,
\nonumber \\
  F_\LA(1, 1, 1, 1, -1, 1, 0) &= \tfrac{19}{2} + \tfrac{1}{2\*\eps^2} + (\tfrac{5}{2} - \lqm)\tfrac{1}{\eps} - 5\*\lqm + \lqm^2 - \tfrac{1}{12}\*\pi^2 - 3\*\zeta_3 \,,
\nonumber \\
  F_\LA(1, 1, 1, 1, 0, -1, 0) &= 6 + \tfrac{1}{2\*\eps^2} + (2 - \lqm)\tfrac{1}{\eps} - 4\*\lqm + \lqm^2 - \tfrac{1}{12}\*\pi^2 \,,
\nonumber \\
  F_\LA(1, 1, 1, 1, 0, 0, 0) &= 12 + \tfrac{1}{\eps^2} + (4 - 2\*\lqm)\tfrac{1}{\eps} - 8\*\lqm + 2\*\lqm^2 - \tfrac{1}{6}\*\pi^2 \,,
\nonumber \\
  F_\LA(1, 1, 1, 1, 0, 0, 1) &= 6 + \tfrac{1}{2\*\eps^2} + (2 - \lqm)\tfrac{1}{\eps} - 4\*\lqm + \lqm^2 - \tfrac{1}{12}\*\pi^2 \,,
\nonumber \\
  F_\LA(1, 1, 1, 1, 0, 1, 0) &= -6\*\zeta_3 \,,
\nonumber \\
  F_\LA(1, 1, 1, 1, 0, 1, 1) &= 7 + \tfrac{1}{\eps} - 2\*\lqm - 6\*\zeta_3 \,,
\nonumber \\
  F_\LA(1, 1, 1, 1, 1, -1, 0) &= 7 + \tfrac{1}{\eps^2} + \tfrac{1}{3}\*\lqm^3 - \tfrac{1}{3}\*\pi^2 + (3 - \lqm - \tfrac{1}{4}\*\lqm^2 - \tfrac{1}{6}\*\pi^2)\tfrac{1}{\eps} \nonumber\\*&\qquad + \lqm\*(-3 + \tfrac{1}{12}\*\pi^2) + \tfrac{1}{2}\*\zeta_3 \,,
\nonumber \\
  F_\LA(1, 1, 1, 1, 1, 0, 0) &= -\lqm^2 + \tfrac{2}{3}\*\lqm^3 - \tfrac{2}{3}\*\pi^2 + \tfrac{1}{6}\*\pi^2\*\lqm + (-\tfrac{1}{2}\*\lqm^2 - \tfrac{1}{3}\*\pi^2)\tfrac{1}{\eps} + \zeta_3 \,,
\nonumber \\
  F_\LA(1, 1, 1, 1, 1, 0, 1) &= 5 + (1 - \lqm)\tfrac{1}{\eps} - 5\*\lqm + \tfrac{3}{2}\*\lqm^2 - \tfrac{1}{6}\*\pi^2 \,,
\nonumber \\
  F_\LA(1, 1, 1, 1, 1, 1, 0) &= \tfrac{1}{24}\*\lqm^4 + \tfrac{1}{3}\*\pi^2\*\lqm^2 + \tfrac{31}{180}\*\pi^4 - 6\*\zeta_3\*\lqm \,,
\nonumber \\
  F_\LA(1, 1, 1, 1, 1, 1, 1) &= \tfrac{1}{3}\*\pi^2\*\lqm - 10\*\zeta_3 \,,
\nonumber \\
  F_\LA(1, 1, 1, 1, 1, 1, 2) &= \tfrac{37}{2} + \tfrac{1}{2\*\eps^2} + (\tfrac{7}{2} - \lqm)\tfrac{1}{\eps} + \lqm^2 + \tfrac{1}{12}\*\pi^2 + \lqm\*(-8 + \tfrac{1}{6}\*\pi^2) - 11\*\zeta_3

  \,,
\end{align}
mit $\lqm = \ln(Q^2/M^2)$.

\subsection{Nichtplanares Vertexdiagramm}
\label{sec:SkalarNP}

Die Zerlegung in skalare Integrale~(\ref{eq:NPskalar}) lautet:
\begin{alignat}{4}
\label{eq:NPzerlegung}
  \lefteqn{F_{v,\NP} =
    \left( C_F^2 - \frac{1}{2} C_F C_A \right)
    \left(\frac{\alpha}{4\pi}\right)^2 \, i^2
    \left(\frac{\mu^2}{M^2}\right)^{2\eps} S_\eps^2
    \, \Bigl\{
  } \quad \nonumber \\
    & F_\NP(0, -1, 1, 1, 1, 1, 0) &\cdot &\bigl( 8 - \tfrac{16}{d-2} - 2\*d \bigr)
\nonumber \\
  +& F_\NP(0, 0, 1, 0, 1, 1, 0) &\cdot &\bigl( -56 + \tfrac{32}{d-2} + 16\*d - d^2 \bigr)
\nonumber \\
  +& F_\NP(0, 0, 1, 1, 1, 0, 0) &\cdot &\bigl( \tfrac{32}{d-2} + 2\*d \bigr)
\nonumber \\
  +& F_\NP(0, 0, 1, 1, 1, 1, 0) &\cdot &\bigl( -8 - \tfrac{8}{d-2} + (\tfrac{32}{d-2} + 2\*d)\*\Mr \bigr)
\nonumber \\
  +& F_\NP(0, 0, 1, 1, 1, 1, 1) &\cdot &\bigl( -36 + \tfrac{32}{d-2} + 10\*d - \tfrac{1}{2}\*d^2 \bigr)
\nonumber \\
  +& F_\NP(1, -1, 0, 1, 1, 1, 0) &\cdot &\bigl( -16 + \tfrac{16}{d-2} + 2\*d \bigr)
\nonumber \\
  +& F_\NP(1, -1, 1, 1, 0, 1, 0) &\cdot &\bigl( 8 \bigr)
\nonumber \\
  +& F_\NP(1, -1, 1, 1, 1, 0, 0) &\cdot &\bigl( -8 + \tfrac{16}{d-2} + 2\*d \bigr)
\nonumber \\
  +& F_\NP(1, -1, 1, 1, 1, 1, 0) &\cdot &\bigl( -8 + (\tfrac{16}{d-2} + 2\*d)\*\Mr \bigr)
\nonumber \\
  +& F_\NP(1, -1, 1, 1, 1, 1, 1) &\cdot &\bigl( -8 + \tfrac{16}{d-2} + 2\*d \bigr)
\nonumber \\
  +& F_\NP(1, 0, -1, 1, 1, 1, 0) &\cdot &\bigl( 48 - \tfrac{16}{d-2} - 14\*d + d^2 \bigr)
\nonumber \\
  +& F_\NP(1, 0, 0, 0, 1, 1, 0) &\cdot &\bigl( 48 - \tfrac{16}{d-2} - 14\*d + d^2 \bigr)
\nonumber \\
  +& F_\NP(1, 0, 0, 1, 0, 1, 0) &\cdot &\bigl( 16 - \tfrac{32}{d-2} - 2\*d \bigr)
\nonumber \\
  +& F_\NP(1, 0, 0, 1, 1, 0, 0) &\cdot &\bigl( 80 - \tfrac{48}{d-2} - 20\*d + d^2 \bigr)
\nonumber \\
  +& F_\NP(1, 0, 0, 1, 1, 1, 0) &\cdot &\bigl( -16 + \tfrac{16}{d-2} + 4\*d + (96 - \tfrac{80}{d-2} - 22\*d + d^2)\*\Mr \bigr)
\nonumber \\
  +& F_\NP(1, 0, 0, 1, 1, 1, 1) &\cdot &\bigl( 120 - \tfrac{64}{d-2} - 32\*d + 2\*d^2 \bigr)
\nonumber \\
  +& F_\NP(1, 0, 1, 0, 0, 1, 0) &\cdot &\bigl( 8 - \tfrac{16}{d-2} - 2\*d \bigr)
\nonumber \\
  +& F_\NP(1, 0, 1, 0, 1, 0, 0) &\cdot &\bigl( 56 - \tfrac{32}{d-2} - 16\*d + d^2 \bigr)
\nonumber \\
  +& F_\NP(1, 0, 1, 0, 1, 1, 0) &\cdot &\bigl( 8 + (64 - \tfrac{48}{d-2} - 18\*d + d^2)\*\Mr \bigr)
\nonumber \\
  +& F_\NP(1, 0, 1, 0, 1, 1, 1) &\cdot &\bigl( 56 - \tfrac{32}{d-2} - 16\*d + d^2 \bigr)
\nonumber \\
  +& F_\NP(1, 0, 1, 1, -1, 1, 0) &\cdot &\bigl( -16 \bigr)
\nonumber \\
  +& F_\NP(1, 0, 1, 1, 0, 0, 0) &\cdot &\bigl( -40 - \tfrac{16}{d-2} + 4\*d \bigr)
\nonumber \\
  +& F_\NP(1, 0, 1, 1, 0, 1, 0) &\cdot &\bigl( 48 - 4\*d + (-72 - \tfrac{16}{d-2} + 4\*d)\*\Mr \bigr)
\nonumber \\
  +& F_\NP(1, 0, 1, 1, 0, 1, 1) &\cdot &\bigl( -8 - \tfrac{32}{d-2} - 2\*d \bigr)
\nonumber \\
  +& F_\NP(1, 0, 1, 1, 1, -1, 0) &\cdot &\bigl( -8 - \tfrac{16}{d-2} \bigr)
\nonumber \\
  +& F_\NP(1, 0, 1, 1, 1, 0, 0) &\cdot &\bigl( 40 + \tfrac{16}{d-2} - 4\*d + (-56 - \tfrac{48}{d-2} + 4\*d)\*\Mr \bigr)
\nonumber \\
  +& F_\NP(1, 0, 1, 1, 1, 0, 1) &\cdot &\bigl( 56 - \tfrac{64}{d-2} - 18\*d + d^2 \bigr)
\nonumber \\
  +& F_\NP(1, 0, 1, 1, 1, 1, 0) &\cdot &\bigl( -16 + (88 + \tfrac{16}{d-2} - 8\*d)\*\Mr + (-64 - \tfrac{32}{d-2} + 4\*d)\*\Mr^2 \bigr)
\nonumber \\
  +& F_\NP(1, 0, 1, 1, 1, 1, 1) &\cdot &\bigl( 8 + \tfrac{32}{d-2} + 2\*d + (48 - \tfrac{96}{d-2} - 20\*d + d^2)\*\Mr \bigr)
\nonumber \\
  +& F_\NP(1, 0, 1, 1, 1, 1, 2) &\cdot &\bigl( 64 - \tfrac{48}{d-2} - 18\*d + d^2 \bigr)
\nonumber \\
  +& F_\NP(1, 1, 0, 0, 1, 0, 0) &\cdot &\bigl( -112 + \tfrac{32}{d-2} + 32\*d - 2\*d^2 \bigr)
\nonumber \\
  +& F_\NP(1, 1, 0, 0, 1, 1, 0) &\cdot &\bigl( 48 - \tfrac{16}{d-2} - 14\*d + d^2 + (-112 + \tfrac{32}{d-2} + 32\*d - 2\*d^2)\*\Mr \bigr)
\nonumber \\
  +& F_\NP(1, 1, 0, 0, 1, 1, 1) &\cdot &\bigl( -56 + \tfrac{16}{d-2} + 16\*d - d^2 \bigr)
\nonumber \\
  +& F_\NP(1, 1, 1, -1, 0, 1, 0) &\cdot &\bigl( -48 + \tfrac{16}{d-2} + 14\*d - d^2 \bigr)
\nonumber \\
  +& F_\NP(1, 1, 1, -1, 1, 0, 0) &\cdot &\bigl( -48 + \tfrac{16}{d-2} + 14\*d - d^2 \bigr)
\nonumber \\
  +& F_\NP(1, 1, 1, -1, 1, 1, 0) &\cdot &\bigl( (-96 + \tfrac{32}{d-2} + 28\*d - 2\*d^2)\*\Mr \bigr)
\nonumber \\
  +& F_\NP(1, 1, 1, -1, 1, 1, 1) &\cdot &\bigl( -48 + \tfrac{16}{d-2} + 14\*d - d^2 \bigr)
\nonumber \\
  +& F_\NP(1, 1, 1, 0, -1, 1, 0) &\cdot &\bigl( \tfrac{16}{d-2} \bigr)
\nonumber \\
  +& F_\NP(1, 1, 1, 0, 0, 0, 0) &\cdot &\bigl( \tfrac{32}{d-2} \bigr)
\nonumber \\
  +& F_\NP(1, 1, 1, 0, 0, 1, 0) &\cdot &\bigl( -8 - \tfrac{16}{d-2} + \tfrac{64}{d-2}\*\Mr \bigr)
\nonumber \\
  +& F_\NP(1, 1, 1, 0, 0, 1, 1) &\cdot &\bigl( -104 + \tfrac{64}{d-2} + 30\*d - 2\*d^2 \bigr)
\nonumber \\
  +& F_\NP(1, 1, 1, 0, 1, -1, 0) &\cdot &\bigl( \tfrac{16}{d-2} \bigr)
\nonumber \\
  +& F_\NP(1, 1, 1, 0, 1, 0, 0) &\cdot &\bigl( -8 - \tfrac{16}{d-2} + \tfrac{64}{d-2}\*\Mr \bigr)
\nonumber \\
  +& F_\NP(1, 1, 1, 0, 1, 0, 1) &\cdot &\bigl( -104 + \tfrac{64}{d-2} + 30\*d - 2\*d^2 \bigr)
\nonumber \\
  +& F_\NP(1, 1, 1, 0, 1, 1, 0) &\cdot &\bigl( 8 + (-16 - \tfrac{32}{d-2})\*\Mr + \tfrac{64}{d-2}\*\Mr^2 \bigr)
\nonumber \\
  +& F_\NP(1, 1, 1, 0, 1, 1, 1) &\cdot &\bigl( 40 - \tfrac{32}{d-2} - 14\*d + d^2 + (-208 + \tfrac{128}{d-2} + 60\*d - 4\*d^2)\*\Mr \bigr)
\nonumber \\
  +& F_\NP(1, 1, 1, 0, 1, 1, 2) &\cdot &\bigl( -104 + \tfrac{48}{d-2} + 30\*d - 2\*d^2 \bigr)
\nonumber \\
  +& F_\NP(1, 1, 1, 1, 0, -1, 0) &\cdot &\bigl( 40 - 4\*d \bigr)
\nonumber \\
  +& F_\NP(1, 1, 1, 1, 0, 0, 0) &\cdot &\bigl( -104 + \tfrac{16}{d-2} + 22\*d - d^2 + (80 - 8\*d)\*\Mr \bigr)
\nonumber \\
  +& F_\NP(1, 1, 1, 1, 0, 0, 1) &\cdot &\bigl( 24 + \tfrac{16}{d-2} - 2\*d \bigr)
\nonumber \\
  +& F_\NP(1, 1, 1, 1, 1, -2, 0) &\cdot &\bigl( 8 \bigr)
\nonumber \\
  +& F_\NP(1, 1, 1, 1, 1, -1, 0) &\cdot &\bigl( -40 + 4\*d + (64 - 4\*d)\*\Mr \bigr)
\nonumber \\
  +& F_\NP(1, 1, 1, 1, 1, -1, 1) &\cdot &\bigl( 16 + \tfrac{16}{d-2} \bigr)
\nonumber \\
  +& F_\NP(1, 1, 1, 1, 1, 0, 0) &\cdot &\bigl( 40 - 4\*d + (-288 + \tfrac{32}{d-2} + 52\*d - 2\*d^2)\*\Mr + (144 - 12\*d)\*\Mr^2 \bigr)
\nonumber \\
  +& F_\NP(1, 1, 1, 1, 1, 0, 1) &\cdot &\bigl( -48 - \tfrac{32}{d-2} + 4\*d + (80 + \tfrac{64}{d-2} - 4\*d)\*\Mr \bigr)
\nonumber \\
  +& F_\NP(1, 1, 1, 1, 1, 0, 2) &\cdot &\bigl( -48 + \tfrac{48}{d-2} + 16\*d - d^2 \bigr)
\nonumber \\
  +& F_\NP(1, 1, 1, 1, 1, 1, 0) &\cdot &\bigl( -4 + (40 - 4\*d)\*\Mr + (-144 + \tfrac{16}{d-2} + 26\*d - d^2)\*\Mr^2 \nonumber\\*&&&\quad + (48 - 4\*d)\*\Mr^3 \bigr)
\nonumber \\
  +& F_\NP(1, 1, 1, 1, 1, 1, 1) &\cdot &\bigl( 8 + \tfrac{8}{d-2} + (-48 - \tfrac{32}{d-2} + 4\*d)\*\Mr + (40 + \tfrac{32}{d-2} - 2\*d)\*\Mr^2 \bigr)
\nonumber \\
  +& F_\NP(1, 1, 1, 1, 1, 1, 2) &\cdot &\bigl( 24 - \tfrac{24}{d-2} - 8\*d + \tfrac{1}{2}\*d^2 + (-48 + \tfrac{48}{d-2} + 16\*d - d^2)\*\Mr \bigr)
\nonumber \\
  +& F_\NP(1, 1, 1, 1, 1, 1, 3) &\cdot &\bigl( -28 + \tfrac{16}{d-2} + 8\*d - \tfrac{1}{2}\*d^2 \bigr)

  \Bigr\}
  \,.
\end{alignat}
Ergebnisse der skalaren Integrale:
\begin{align}
\label{eq:NPergskalar}
    F_\NP(0, -1, 1, 1, 1, 1, 0) &= -\tfrac{11}{16} - \tfrac{1}{4\*\eps^2} - \tfrac{1}{2\*\eps} - \tfrac{1}{24}\*\pi^2 \,,
\nonumber \\
  F_\NP(0, 0, 1, 1, 1, 1, 0) &= 3 + \tfrac{1}{\eps^2} + \tfrac{2}{\eps} + \tfrac{1}{6}\*\pi^2 \,,
\nonumber \\
  F_\NP(0, 0, 1, 1, 1, 1, 1) &= \tfrac{23}{16} + \tfrac{1}{4\*\eps^2} + \tfrac{3}{4\*\eps} + \tfrac{1}{24}\*\pi^2 \,,
\nonumber \\
  F_\NP(1, -1, 0, 1, 1, 1, 0) &= -\tfrac{33}{32} - \tfrac{1}{8\*\eps^2} - \tfrac{7}{16\*\eps} - \tfrac{1}{8}\*\lqm + \tfrac{1}{8}\*\lqm^2 + \tfrac{1}{48}\*\pi^2 \,,
\nonumber \\
  F_\NP(1, -1, 1, 1, 0, 1, 0) &= -\tfrac{77}{32} - \tfrac{3}{8\*\eps^2} - \tfrac{17}{16\*\eps} - \tfrac{3}{16}\*\pi^2 \,,
\nonumber \\
  F_\NP(1, -1, 1, 1, 1, 0, 0) &= -\tfrac{73}{32} - \tfrac{1}{8\*\eps^2} - \tfrac{11}{16\*\eps} + \tfrac{1}{8}\*\lqm + \tfrac{1}{8}\*\lqm^2 - \tfrac{1}{48}\*\pi^2 \,,
\nonumber \\
  F_\NP(1, -1, 1, 1, 1, 1, 0) &= 10 + \tfrac{1}{\eps^2} + \tfrac{2}{\eps} - \tfrac{5}{6}\*\lqm^3 + \tfrac{1}{8}\*\lqm^4 + \pi^2 + \tfrac{1}{30}\*\pi^4 + \lqm^2\*(\tfrac{7}{2} + \tfrac{1}{4}\*\pi^2) \nonumber\\*&\qquad - 4\*\zeta_3 + \lqm\*(-7 - \tfrac{2}{3}\*\pi^2 + \zeta_3) \,,
\nonumber \\
  F_\NP(1, -1, 1, 1, 1, 1, 1) &= -\tfrac{27}{8} + \tfrac{1}{4\*\eps^2} + \tfrac{5}{4\*\eps} - \tfrac{15}{8}\*\lqm^2 + \tfrac{1}{4}\*\lqm^3 - \tfrac{3}{8}\*\pi^2 + \lqm\*(\tfrac{45}{8} + \tfrac{1}{4}\*\pi^2) + \tfrac{1}{2}\*\zeta_3 \,,
\nonumber \\
  F_\NP(1, 0, -1, 1, 1, 1, 0) &= \tfrac{33}{32} + \tfrac{1}{8\*\eps^2} + \tfrac{7}{16\*\eps} + \tfrac{1}{8}\*\lqm - \tfrac{1}{8}\*\lqm^2 - \tfrac{1}{48}\*\pi^2 \,,
\nonumber \\
  F_\NP(1, 0, 0, 1, 1, 0, 0) &= \tfrac{13}{8} + \tfrac{1}{4\*\eps} - \tfrac{1}{2}\*\lqm \,,
\nonumber \\
  F_\NP(1, 0, 0, 1, 1, 1, 0) &= \tfrac{5}{2} + \tfrac{1}{2\*\eps^2} + \tfrac{3}{2\*\eps} + \lqm - \tfrac{1}{2}\*\lqm^2 - \tfrac{1}{12}\*\pi^2 \,,
\nonumber \\
  F_\NP(1, 0, 0, 1, 1, 1, 1) &= -\tfrac{19}{32} + \tfrac{1}{8\*\eps^2} + \tfrac{3}{16\*\eps} + \tfrac{5}{8}\*\lqm - \tfrac{1}{8}\*\lqm^2 - \tfrac{1}{48}\*\pi^2 \,,
\nonumber \\
  F_\NP(1, 0, 1, 0, 1, 1, 0) &= \tfrac{7}{2} + \tfrac{1}{2\*\eps^2} + \tfrac{3}{2\*\eps} - \tfrac{1}{12}\*\pi^2 \,,
\nonumber \\
  F_\NP(1, 0, 1, 1, -1, 1, 0) &= -\tfrac{35}{32} - \tfrac{1}{8\*\eps^2} - \tfrac{7}{16\*\eps} - \tfrac{1}{16}\*\pi^2 \,,
\nonumber \\
  F_\NP(1, 0, 1, 1, 0, 1, 0) &= \tfrac{7}{2} + \tfrac{1}{2\*\eps^2} + \tfrac{3}{2\*\eps} + \tfrac{1}{4}\*\pi^2 \,,
\nonumber \\
  F_\NP(1, 0, 1, 1, 0, 1, 1) &= \tfrac{35}{16} + \tfrac{1}{4\*\eps^2} + \tfrac{7}{8\*\eps} + \tfrac{1}{8}\*\pi^2 \,,
\nonumber \\
  F_\NP(1, 0, 1, 1, 1, -1, 0) &= -\tfrac{7}{32} + \tfrac{1}{8\*\eps^2} + \tfrac{3}{16\*\eps} + \tfrac{3}{8}\*\lqm - \tfrac{1}{8}\*\lqm^2 + \tfrac{1}{48}\*\pi^2 \,,
\nonumber \\
  F_\NP(1, 0, 1, 1, 1, 0, 0) &= \tfrac{5}{2} + \tfrac{1}{2\*\eps^2} + \tfrac{3}{2\*\eps} + \lqm - \tfrac{1}{2}\*\lqm^2 + \tfrac{1}{12}\*\pi^2 \,,
\nonumber \\
  F_\NP(1, 0, 1, 1, 1, 0, 1) &= \tfrac{21}{32} + \tfrac{1}{8\*\eps^2} + \tfrac{7}{16\*\eps} + \tfrac{3}{8}\*\lqm - \tfrac{1}{8}\*\lqm^2 + \tfrac{1}{48}\*\pi^2 \,,
\nonumber \\
  F_\NP(1, 0, 1, 1, 1, 1, 0) &= -\tfrac{1}{8}\*\lqm^4 - \tfrac{1}{4}\*\pi^2\*\lqm^2 - \tfrac{1}{30}\*\pi^4 - \zeta_3\*\lqm \,,
\nonumber \\
  F_\NP(1, 0, 1, 1, 1, 1, 1) &= 1 - \tfrac{1}{\eps} + \tfrac{5}{2}\*\lqm^2 - \tfrac{1}{2}\*\lqm^3 + \tfrac{2}{3}\*\pi^2 + \lqm\*(-5 - \tfrac{1}{2}\*\pi^2) - \zeta_3 \,,
\nonumber \\
  F_\NP(1, 0, 1, 1, 1, 1, 2) &= -\tfrac{7}{16} - \tfrac{3}{8\*\eps} + \tfrac{9}{8}\*\lqm^2 - \tfrac{1}{4}\*\lqm^3 + \tfrac{1}{4}\*\pi^2 + \lqm\*(-\tfrac{15}{8} - \tfrac{1}{4}\*\pi^2) - \tfrac{1}{2}\*\zeta_3 \,,
\nonumber \\
  F_\NP(1, 1, 0, 0, 1, 1, 0) &= \tfrac{19}{2} + \tfrac{1}{2\*\eps^2} + (\tfrac{5}{2} - \lqm)\tfrac{1}{\eps} - 5\*\lqm + \lqm^2 + \tfrac{1}{12}\*\pi^2 \,,
\nonumber \\
  F_\NP(1, 1, 0, 0, 1, 1, 1) &= \tfrac{63}{16} + \tfrac{1}{4\*\eps^2} + (\tfrac{9}{8} - \tfrac{1}{2}\*\lqm)\tfrac{1}{\eps} - \tfrac{9}{4}\*\lqm + \tfrac{1}{2}\*\lqm^2 + \tfrac{1}{24}\*\pi^2 \,,
\nonumber \\
  F_\NP(1, 1, 1, -1, 0, 1, 0) &= \tfrac{63}{32} + \tfrac{1}{8\*\eps^2} + (\tfrac{9}{16} - \tfrac{1}{4}\*\lqm)\tfrac{1}{\eps} - \tfrac{9}{8}\*\lqm + \tfrac{1}{4}\*\lqm^2 - \tfrac{1}{48}\*\pi^2 \,,
\nonumber \\
  F_\NP(1, 1, 1, -1, 1, 0, 0) &= \tfrac{35}{16} + \tfrac{1}{4\*\eps^2} + (\tfrac{7}{8} - \tfrac{1}{4}\*\lqm)\tfrac{1}{\eps} - \tfrac{7}{8}\*\lqm + \tfrac{1}{8}\*\lqm^2 \,,
\nonumber \\
  F_\NP(1, 1, 1, -1, 1, 1, 0) &= 11 + \tfrac{1}{\eps^2} + (3 - \lqm)\tfrac{1}{\eps} - \tfrac{1}{3}\*\pi^2 - \tfrac{7}{360}\*\pi^4 + \lqm^2\*(\tfrac{3}{2} - \tfrac{1}{12}\*\pi^2) \nonumber\\*&\qquad + \lqm\*(-6 + \tfrac{1}{3}\*\pi^2 - \zeta_3) + 2\*\zeta_3 \,,
\nonumber \\
  F_\NP(1, 1, 1, -1, 1, 1, 1) &= -\tfrac{33}{16} - \tfrac{1}{8\*\eps} + \tfrac{1}{4}\*\pi^2 + \lqm\*(\tfrac{7}{8} - \tfrac{1}{12}\*\pi^2) - \tfrac{1}{2}\*\zeta_3 \,,
\nonumber \\
  F_\NP(1, 1, 1, 0, -1, 1, 0) &= \tfrac{63}{32} + \tfrac{1}{8\*\eps^2} + (\tfrac{9}{16} - \tfrac{1}{4}\*\lqm)\tfrac{1}{\eps} - \tfrac{9}{8}\*\lqm + \tfrac{1}{4}\*\lqm^2 - \tfrac{1}{48}\*\pi^2 \,,
\nonumber \\
  F_\NP(1, 1, 1, 0, 0, 1, 0) &= \tfrac{19}{2} + \tfrac{1}{2\*\eps^2} + (\tfrac{5}{2} - \lqm)\tfrac{1}{\eps} - 5\*\lqm + \lqm^2 - \tfrac{1}{12}\*\pi^2 \,,
\nonumber \\
  F_\NP(1, 1, 1, 0, 0, 1, 1) &= \tfrac{115}{32} + \tfrac{1}{8\*\eps^2} + (\tfrac{13}{16} - \tfrac{1}{4}\*\lqm)\tfrac{1}{\eps} - \tfrac{13}{8}\*\lqm + \tfrac{1}{4}\*\lqm^2 - \tfrac{1}{48}\*\pi^2 \,,
\nonumber \\
  F_\NP(1, 1, 1, 0, 1, -1, 0) &= \tfrac{21}{16} + \tfrac{1}{4\*\eps^2} + (\tfrac{5}{8} - \tfrac{1}{4}\*\lqm)\tfrac{1}{\eps} - \tfrac{5}{8}\*\lqm + \tfrac{1}{8}\*\lqm^2 \,,
\nonumber \\
  F_\NP(1, 1, 1, 0, 1, 0, 0) &= 7 + \tfrac{1}{\eps^2} + (3 - \lqm)\tfrac{1}{\eps} - 3\*\lqm + \tfrac{1}{2}\*\lqm^2 \,,
\nonumber \\
  F_\NP(1, 1, 1, 0, 1, 0, 1) &= \tfrac{35}{16} + \tfrac{1}{4\*\eps^2} + (\tfrac{7}{8} - \tfrac{1}{4}\*\lqm)\tfrac{1}{\eps} - \tfrac{7}{8}\*\lqm + \tfrac{1}{8}\*\lqm^2 \,,
\nonumber \\
  F_\NP(1, 1, 1, 0, 1, 1, 0) &= -\tfrac{1}{12}\*\pi^2\*\lqm^2 - \tfrac{7}{360}\*\pi^4 - \zeta_3\*\lqm \,,
\nonumber \\
  F_\NP(1, 1, 1, 0, 1, 1, 1) &= -14 - \tfrac{1}{\eps^2} - 2\*\lqm^2 + (-4 + 2\*\lqm)\tfrac{1}{\eps} + \tfrac{1}{6}\*\pi^2 + \lqm\*(9 - \tfrac{1}{6}\*\pi^2) - \zeta_3 \,,
\nonumber \\
  F_\NP(1, 1, 1, 0, 1, 1, 2) &= -\tfrac{47}{4} - \tfrac{3}{4\*\eps^2} - \tfrac{3}{2}\*\lqm^2 + (-\tfrac{13}{4} + \tfrac{3}{2}\*\lqm)\tfrac{1}{\eps} + \tfrac{1}{24}\*\pi^2 \nonumber\\*&\qquad + \lqm\*(\tfrac{57}{8} - \tfrac{1}{12}\*\pi^2) - \tfrac{1}{2}\*\zeta_3 \,,
\nonumber \\
  F_\NP(1, 1, 1, 1, 0, -1, 0) &= -\tfrac{63}{32} - \tfrac{1}{8\*\eps^2} + (-\tfrac{9}{16} + \tfrac{1}{4}\*\lqm)\tfrac{1}{\eps} + \tfrac{9}{8}\*\lqm - \tfrac{1}{4}\*\lqm^2 - \tfrac{1}{48}\*\pi^2 \,,
\nonumber \\
  F_\NP(1, 1, 1, 1, 0, 0, 0) &= \tfrac{19}{2} + \tfrac{1}{2\*\eps^2} + (\tfrac{5}{2} - \lqm)\tfrac{1}{\eps} - 5\*\lqm + \lqm^2 + \tfrac{1}{12}\*\pi^2 \,,
\nonumber \\
  F_\NP(1, 1, 1, 1, 0, 0, 1) &= \tfrac{19}{2} + \tfrac{1}{2\*\eps^2} + (\tfrac{5}{2} - \lqm)\tfrac{1}{\eps} - 5\*\lqm + \lqm^2 + \tfrac{1}{12}\*\pi^2 \,,
\nonumber \\
  F_\NP(1, 1, 1, 1, 1, -2, 0) &= \tfrac{157}{16} + \tfrac{1}{4\*\eps^2} + (\tfrac{3}{8} - \tfrac{1}{4}\*\lqm)\tfrac{1}{\eps} - \tfrac{3}{2}\*\pi^2 - \tfrac{11}{180}\*\pi^4 + \lqm^2\*(\tfrac{7}{8} - \tfrac{1}{12}\*\pi^2) \nonumber\\*&\qquad + \lqm\*(-\tfrac{11}{2} + \tfrac{3}{4}\*\pi^2 - 2\*\zeta_3) + 9\*\zeta_3 \,,
\nonumber \\
  F_\NP(1, 1, 1, 1, 1, -1, 0) &= 11 + \tfrac{1}{\eps^2} + (3 - \lqm)\tfrac{1}{\eps} - \tfrac{1}{3}\*\pi^2 - \tfrac{11}{180}\*\pi^4 + \lqm^2\*(\tfrac{3}{2} - \tfrac{1}{12}\*\pi^2) \nonumber\\*&\qquad + \lqm\*(-6 + \tfrac{1}{2}\*\pi^2 - 2\*\zeta_3) + 6\*\zeta_3 \,,
\nonumber \\
  F_\NP(1, 1, 1, 1, 1, -1, 1) &= 1 + \tfrac{1}{2\*\eps^2} + (\tfrac{15}{8} - \tfrac{1}{2}\*\lqm)\tfrac{1}{\eps} + \tfrac{3}{8}\*\lqm^2 + \tfrac{19}{24}\*\pi^2 + \lqm\*(-\tfrac{3}{8} - \tfrac{1}{6}\*\pi^2) - 2\*\zeta_3 \,,
\nonumber \\
  F_\NP(1, 1, 1, 1, 1, 0, 0) &= -\tfrac{1}{12}\*\pi^2\*\lqm^2 - \tfrac{11}{180}\*\pi^4 - 2\*\zeta_3\*\lqm \,,
\nonumber \\
  F_\NP(1, 1, 1, 1, 1, 0, 1) &= -17 - \tfrac{1}{\eps^2} - \tfrac{5}{2}\*\lqm^2 + (-4 + 2\*\lqm)\tfrac{1}{\eps} + \tfrac{1}{3}\*\pi^2 + \lqm\*(11 - \tfrac{1}{3}\*\pi^2) \!-\! 4\*\zeta_3 \,,
\nonumber \\
  F_\NP(1, 1, 1, 1, 1, 0, 2) &= -\tfrac{325}{16} - \tfrac{5}{4\*\eps^2} - \tfrac{23}{8}\*\lqm^2 + (-\tfrac{11}{2} + \tfrac{5}{2}\*\lqm)\tfrac{1}{\eps} - \tfrac{1}{6}\*\pi^2 \nonumber\\*&\qquad + \lqm\*(\tfrac{103}{8} - \tfrac{1}{6}\*\pi^2) - 2\*\zeta_3 \,,
\nonumber \\
  F_\NP(1, 1, 1, 1, 1, 1, 0) &= \tfrac{7}{12}\*\lqm^4 - \tfrac{1}{6}\*\pi^2\*\lqm^2 - \tfrac{31}{180}\*\pi^4 + 20\*\zeta_3\*\lqm \,,
\nonumber \\
  F_\NP(1, 1, 1, 1, 1, 1, 1) &= \tfrac{1}{4}\*\lqm^4 - \tfrac{1}{6}\*\pi^2\*\lqm^2 - \tfrac{1}{90}\*\pi^4 + 14\*\zeta_3\*\lqm \,,
\nonumber \\
  F_\NP(1, 1, 1, 1, 1, 1, 2) &= 38 + \tfrac{2}{\eps^2} + (7 - 4\*\lqm)\tfrac{1}{\eps} - \lqm^3 + \tfrac{1}{4}\*\lqm^4 + \tfrac{1}{3}\*\pi^2 - \tfrac{1}{90}\*\pi^4 \nonumber\\*&\qquad + \lqm^2\*(9 - \tfrac{1}{6}\*\pi^2) - 4\*\zeta_3 + \lqm\*(-30 + 14\*\zeta_3) \,,
\nonumber \\
  F_\NP(1, 1, 1, 1, 1, 1, 3) &= \tfrac{571}{8} + \tfrac{7}{2\*\eps^2} + (\tfrac{111}{8} - 7\*\lqm)\tfrac{1}{\eps} - \tfrac{3}{2}\*\lqm^3 + \tfrac{1}{4}\*\lqm^4 + \tfrac{3}{4}\*\pi^2 - \tfrac{1}{90}\*\pi^4 \nonumber\\*&\qquad + \lqm^2\*(\tfrac{59}{4} - \tfrac{1}{6}\*\pi^2) - 6\*\zeta_3 + \lqm\*(-\tfrac{211}{4} + 14\*\zeta_3) 

  \,.
\end{align}

\subsection{Vertexdiagramm mit Benz-Topologie}
\label{sec:SkalarBE}

Zerlegung des Benz-Vertexdiagramms in skalare Integrale~(\ref{eq:BEskalar}):
\begin{alignat}{4}
\label{eq:BEzerlegung}
  \lefteqn{F_{v,\BE} =
    \left( C_F^2 - \frac{1}{2} C_F C_A \right)
    \left(\frac{\alpha}{4\pi}\right)^2 \, i^2
    \left(\frac{\mu^2}{M^2}\right)^{2\eps} S_\eps^2
    \, \Bigl\{
  } \quad \nonumber \\
    & F_\BE(-1, 1, 1, 1, 1, 1, 1) &\cdot &\bigl( 22 - \tfrac{8}{d-2} - 6\*d + \tfrac{1}{2}\*d^2 \bigr)
\nonumber \\
  +& F_\BE(0, 0, 0, 1, 1, 1, 0) &\cdot &\bigl( 22 - \tfrac{8}{d-2} - 6\*d + \tfrac{1}{2}\*d^2 \bigr)
\nonumber \\
  +& F_\BE(0, 0, 1, 0, 1, 1, 0) &\cdot &\bigl( 4 \bigr)
\nonumber \\
  +& F_\BE(0, 0, 1, 1, 0, 1, 0) &\cdot &\bigl( -10 + \tfrac{8}{d-2} + 4\*d - \tfrac{1}{2}\*d^2 \bigr)
\nonumber \\
  +& F_\BE(0, 0, 1, 1, 1, 0, 0) &\cdot &\bigl( -4 \bigr)
\nonumber \\
  +& F_\BE(0, 0, 1, 1, 1, 1, 0) &\cdot &\bigl( (26 - \tfrac{8}{d-2} - 6\*d + \tfrac{1}{2}\*d^2)\*\Mr \bigr)
\nonumber \\
  +& F_\BE(0, 0, 1, 1, 1, 1, 1) &\cdot &\bigl( -2 \bigr)
\nonumber \\
  +& F_\BE(0, 1, 0, 0, 1, 1, 0) &\cdot &\bigl( -22 + \tfrac{8}{d-2} + 6\*d - \tfrac{1}{2}\*d^2 \bigr)
\nonumber \\
  +& F_\BE(0, 1, 0, 1, 1, 1, 0) &\cdot &\bigl( (-22 + \tfrac{8}{d-2} + 6\*d - \tfrac{1}{2}\*d^2)\*\Mr \bigr)
\nonumber \\
  +& F_\BE(0, 1, 0, 1, 1, 1, 1) &\cdot &\bigl( 2 - d \bigr)
\nonumber \\
  +& F_\BE(0, 1, 1, -1, 1, 1, 0) &\cdot &\bigl( -4 \bigr)
\nonumber \\
  +& F_\BE(0, 1, 1, 0, 0, 1, 0) &\cdot &\bigl( 10 - \tfrac{8}{d-2} - 4\*d + \tfrac{1}{2}\*d^2 \bigr)
\nonumber \\
  +& F_\BE(0, 1, 1, 0, 1, 0, 0) &\cdot &\bigl( 4 \bigr)
\nonumber \\
  +& F_\BE(0, 1, 1, 0, 1, 1, 0) &\cdot &\bigl( 22 - \tfrac{8}{d-2} - 6\*d + \tfrac{1}{2}\*d^2 + (-30 + \tfrac{8}{d-2} + 6\*d - \tfrac{1}{2}\*d^2)\*\Mr \bigr)
\nonumber \\
  +& F_\BE(0, 1, 1, 0, 1, 1, 1) &\cdot &\bigl( -22 + \tfrac{16}{d-2} + 6\*d - \tfrac{1}{2}\*d^2 \bigr)
\nonumber \\
  +& F_\BE(0, 1, 1, 1, 0, 1, 0) &\cdot &\bigl( 26 - \tfrac{8}{d-2} - 8\*d + \tfrac{1}{2}\*d^2 + (10 - \tfrac{8}{d-2} - 4\*d + \tfrac{1}{2}\*d^2)\*\Mr \bigr)
\nonumber \\
  +& F_\BE(0, 1, 1, 1, 0, 1, 1) &\cdot &\bigl( d \bigr)
\nonumber \\
  +& F_\BE(0, 1, 1, 1, 1, 0, 0) &\cdot &\bigl( -22 + \tfrac{8}{d-2} + 6\*d - \tfrac{1}{2}\*d^2 + 4\*\Mr \bigr)
\nonumber \\
  +& F_\BE(0, 1, 1, 1, 1, 0, 1) &\cdot &\bigl( -2 \bigr)
\nonumber \\
  +& F_\BE(0, 1, 1, 1, 1, 1, 0) &\cdot &\bigl( (22 - \tfrac{8}{d-2} - 6\*d + \tfrac{1}{2}\*d^2)\*\Mr + (-26 + \tfrac{8}{d-2} + 6\*d - \tfrac{1}{2}\*d^2)\*\Mr^2 \bigr)
\nonumber \\
  +& F_\BE(0, 1, 1, 1, 1, 1, 1) &\cdot &\bigl( -2 + d + (-20 + \tfrac{16}{d-2} + 5\*d - \tfrac{1}{2}\*d^2)\*\Mr \bigr)
\nonumber \\
  +& F_\BE(0, 1, 1, 1, 1, 1, 2) &\cdot &\bigl( 2 - d \bigr)
\nonumber \\
  +& F_\BE(1, -1, 0, 1, 1, 1, 0) &\cdot &\bigl( -2 \bigr)
\nonumber \\
  +& F_\BE(1, -1, 1, 1, 0, 1, 0) &\cdot &\bigl( 2 \bigr)
\nonumber \\
  +& F_\BE(1, -1, 1, 1, 1, 1, 0) &\cdot &\bigl( -2\*\Mr \bigr)
\nonumber \\
  +& F_\BE(1, 0, -1, 1, 1, 1, 0) &\cdot &\bigl( 2 - d \bigr)
\nonumber \\
  +& F_\BE(1, 0, 0, 0, 1, 1, 0) &\cdot &\bigl( -14 + \tfrac{8}{d-2} + 2\*d \bigr)
\nonumber \\
  +& F_\BE(1, 0, 0, 1, 0, 1, 0) &\cdot &\bigl( -6 + 2\*d \bigr)
\nonumber \\
  +& F_\BE(1, 0, 0, 1, 1, 0, 0) &\cdot &\bigl( 2 \bigr)
\nonumber \\
  +& F_\BE(1, 0, 0, 1, 1, 1, 0) &\cdot &\bigl( 2 - d + (-10 + \tfrac{8}{d-2})\*\Mr \bigr)
\nonumber \\
  +& F_\BE(1, 0, 0, 1, 1, 1, 1) &\cdot &\bigl( 2 - d \bigr)
\nonumber \\
  +& F_\BE(1, 0, 1, -1, 1, 1, 0) &\cdot &\bigl( -4 \bigr)
\nonumber \\
  +& F_\BE(1, 0, 1, 0, 0, 1, 0) &\cdot &\bigl( 2 - \tfrac{8}{d-2} \bigr)
\nonumber \\
  +& F_\BE(1, 0, 1, 0, 1, 0, 0) &\cdot &\bigl( 4 \bigr)
\nonumber \\
  +& F_\BE(1, 0, 1, 0, 1, 1, 0) &\cdot &\bigl( 2 + (-22 + \tfrac{8}{d-2} + 2\*d)\*\Mr \bigr)
\nonumber \\
  +& F_\BE(1, 0, 1, 0, 1, 1, 1) &\cdot &\bigl( 2 \bigr)
\nonumber \\
  +& F_\BE(1, 0, 1, 1, -1, 1, 0) &\cdot &\bigl( 4 - d \bigr)
\nonumber \\
  +& F_\BE(1, 0, 1, 1, 0, 0, 0) &\cdot &\bigl( -2 \bigr)
\nonumber \\
  +& F_\BE(1, 0, 1, 1, 0, 1, 0) &\cdot &\bigl( 4 - d + (-4 - \tfrac{8}{d-2} + 2\*d)\*\Mr \bigr)
\nonumber \\
  +& F_\BE(1, 0, 1, 1, 0, 1, 1) &\cdot &\bigl( -2 + d \bigr)
\nonumber \\
  +& F_\BE(1, 0, 1, 1, 1, 0, 0) &\cdot &\bigl( -2 + 6\*\Mr \bigr)
\nonumber \\
  +& F_\BE(1, 0, 1, 1, 1, 1, 0) &\cdot &\bigl( (4 - d)\*\Mr + (-16 + \tfrac{8}{d-2} + d)\*\Mr^2 \bigr)
\nonumber \\
  +& F_\BE(1, 0, 1, 1, 1, 1, 1) &\cdot &\bigl( (4 - d)\*\Mr \bigr)
\nonumber \\
  +& F_\BE(1, 1, -1, 0, 1, 1, 0) &\cdot &\bigl( -2 + d \bigr)
\nonumber \\
  +& F_\BE(1, 1, -1, 1, 1, 1, 0) &\cdot &\bigl( (-2 + d)\*\Mr \bigr)
\nonumber \\
  +& F_\BE(1, 1, 0, -1, 1, 1, 0) &\cdot &\bigl( 16 - \tfrac{8}{d-2} - 2\*d \bigr)
\nonumber \\
  +& F_\BE(1, 1, 0, 0, 0, 1, 0) &\cdot &\bigl( 6 - 2\*d \bigr)
\nonumber \\
  +& F_\BE(1, 1, 0, 0, 1, 0, 0) &\cdot &\bigl( -2 \bigr)
\nonumber \\
  +& F_\BE(1, 1, 0, 0, 1, 1, 0) &\cdot &\bigl( -48 + \tfrac{16}{d-2} + 14\*d - d^2 + (28 - \tfrac{16}{d-2} - 2\*d)\*\Mr \bigr)
\nonumber \\
  +& F_\BE(1, 1, 0, 0, 1, 1, 1) &\cdot &\bigl( -8 + 2\*d \bigr)
\nonumber \\
  +& F_\BE(1, 1, 0, 1, 0, 1, 0) &\cdot &\bigl( -2 + d + (6 - 2\*d)\*\Mr \bigr)
\nonumber \\
  +& F_\BE(1, 1, 0, 1, 1, 0, 0) &\cdot &\bigl( 2 - d - 2\*\Mr \bigr)
\nonumber \\
  +& F_\BE(1, 1, 0, 1, 1, 1, 0) &\cdot &\bigl( (-48 + \tfrac{16}{d-2} + 14\*d - d^2)\*\Mr + (12 - \tfrac{8}{d-2})\*\Mr^2 \bigr)
\nonumber \\
  +& F_\BE(1, 1, 0, 1, 1, 1, 1) &\cdot &\bigl( 4 - 2\*d + (-8 + 2\*d)\*\Mr \bigr)
\nonumber \\
  +& F_\BE(1, 1, 1, -2, 1, 1, 0) &\cdot &\bigl( 4 \bigr)
\nonumber \\
  +& F_\BE(1, 1, 1, -1, 0, 1, 0) &\cdot &\bigl( -4 + \tfrac{8}{d-2} \bigr)
\nonumber \\
  +& F_\BE(1, 1, 1, -1, 1, 0, 0) &\cdot &\bigl( -4 \bigr)
\nonumber \\
  +& F_\BE(1, 1, 1, -1, 1, 1, 0) &\cdot &\bigl( -16 + \tfrac{8}{d-2} + 2\*d + (28 - \tfrac{8}{d-2} - 2\*d)\*\Mr \bigr)
\nonumber \\
  +& F_\BE(1, 1, 1, -1, 1, 1, 1) &\cdot &\bigl( -\tfrac{8}{d-2} \bigr)
\nonumber \\
  +& F_\BE(1, 1, 1, 0, -1, 1, 0) &\cdot &\bigl( -4 + d \bigr)
\nonumber \\
  +& F_\BE(1, 1, 1, 0, 0, 0, 0) &\cdot &\bigl( 2 \bigr)
\nonumber \\
  +& F_\BE(1, 1, 1, 0, 0, 1, 0) &\cdot &\bigl( 6 - \tfrac{8}{d-2} + (-2 + \tfrac{16}{d-2} - 2\*d)\*\Mr \bigr)
\nonumber \\
  +& F_\BE(1, 1, 1, 0, 0, 1, 1) &\cdot &\bigl( 6 - 2\*d \bigr)
\nonumber \\
  +& F_\BE(1, 1, 1, 0, 1, 0, 0) &\cdot &\bigl( 14 - \tfrac{8}{d-2} - 2\*d - 10\*\Mr \bigr)
\nonumber \\
  +& F_\BE(1, 1, 1, 0, 1, 0, 1) &\cdot &\bigl( 2 \bigr)
\nonumber \\
  +& F_\BE(1, 1, 1, 0, 1, 1, 0) &\cdot &\bigl( -2 + d + (-80 + \tfrac{32}{d-2} + 18\*d - d^2)\*\Mr + (42 - \tfrac{16}{d-2} - 3\*d)\*\Mr^2 \bigr)
\nonumber \\
  +& F_\BE(1, 1, 1, 0, 1, 1, 1) &\cdot &\bigl( 8 - 2\*d + (-8 - \tfrac{16}{d-2} + 2\*d)\*\Mr \bigr)
\nonumber \\
  +& F_\BE(1, 1, 1, 0, 1, 1, 2) &\cdot &\bigl( -2 + d \bigr)
\nonumber \\
  +& F_\BE(1, 1, 1, 1, -1, 1, 0) &\cdot &\bigl( 4 - d + (-4 + d)\*\Mr \bigr)
\nonumber \\
  +& F_\BE(1, 1, 1, 1, 0, 0, 0) &\cdot &\bigl( -6 + d + 2\*\Mr \bigr)
\nonumber \\
  +& F_\BE(1, 1, 1, 1, 0, 1, 0) &\cdot &\bigl( 2 - d + (4 - \tfrac{8}{d-2} + d)\*\Mr + (2 + \tfrac{8}{d-2} - 2\*d)\*\Mr^2 \bigr)
\nonumber \\
  +& F_\BE(1, 1, 1, 1, 0, 1, 1) &\cdot &\bigl( -2 + d + (6 - 2\*d)\*\Mr \bigr)
\nonumber \\
  +& F_\BE(1, 1, 1, 1, 1, -1, 0) &\cdot &\bigl( 2 \bigr)
\nonumber \\
  +& F_\BE(1, 1, 1, 1, 1, 0, 0) &\cdot &\bigl( 2 - d + (16 - \tfrac{8}{d-2} - 3\*d)\*\Mr - 6\*\Mr^2 \bigr)
\nonumber \\
  +& F_\BE(1, 1, 1, 1, 1, 0, 1) &\cdot &\bigl( -2 + d + 2\*\Mr \bigr)
\nonumber \\
  +& F_\BE(1, 1, 1, 1, 1, 1, 0) &\cdot &\bigl( (-2 + d)\*\Mr + (-64 + \tfrac{24}{d-2} + 16\*d - d^2)\*\Mr^2 \nonumber\\*&&&\quad + (18 - \tfrac{8}{d-2} - d)\*\Mr^3 \bigr)
\nonumber \\
  +& F_\BE(1, 1, 1, 1, 1, 1, 1) &\cdot &\bigl( (12 - 4\*d)\*\Mr + (-8 - \tfrac{8}{d-2} + 2\*d)\*\Mr^2 \bigr)
\nonumber \\
  +& F_\BE(1, 1, 1, 1, 1, 1, 2) &\cdot &\bigl( (-2 + d)\*\Mr \bigr)

  \Bigr\}
  \,.
\end{alignat}
Ergebnisse der skalaren Integrale
in f"uhrender Ordnung in $z = M^2/Q^2$:
\begin{align}
\label{eq:BEergskalar}
    F_\BE(-1, 1, 1, 1, 1, 1, 1) &= \tfrac{47}{16} + \tfrac{1}{4\*\eps^2} + \tfrac{1}{8\*\eps} - \tfrac{11}{4}\*\lqm + \tfrac{3}{4}\*\lqm^2 + \tfrac{5}{24}\*\pi^2 \,,
\nonumber \\
  F_\BE(0, 0, 1, 1, 1, 1, 0) &= \tfrac{7}{2} + \tfrac{1}{2\*\eps^2} + \tfrac{3}{2\*\eps} - \tfrac{1}{12}\*\pi^2 \,,
\nonumber \\
  F_\BE(0, 0, 1, 1, 1, 1, 1) &= \tfrac{7}{16} + \tfrac{1}{4\*\eps^2} + \tfrac{3}{8\*\eps} + \tfrac{1}{24}\*\pi^2 \,,
\nonumber \\
  F_\BE(0, 1, 0, 1, 1, 1, 0) &= \tfrac{7}{2} + \tfrac{1}{2\*\eps^2} + \tfrac{3}{2\*\eps} + \tfrac{1}{4}\*\pi^2 \,,
\nonumber \\
  F_\BE(0, 1, 1, -1, 1, 1, 0) &= -\tfrac{7}{32} + \tfrac{1}{8\*\eps^2} + \tfrac{3}{16\*\eps} + \tfrac{3}{8}\*\lqm - \tfrac{1}{8}\*\lqm^2 + \tfrac{1}{48}\*\pi^2 \,,
\nonumber \\
  F_\BE(0, 1, 1, 0, 0, 1, 0) &= \tfrac{13}{8} + \tfrac{1}{4\*\eps} - \tfrac{1}{2}\*\lqm \,,
\nonumber \\
  F_\BE(0, 1, 1, 0, 1, 1, 0) &= \tfrac{5}{2} + \tfrac{1}{2\*\eps^2} + \tfrac{3}{2\*\eps} + \lqm - \tfrac{1}{2}\*\lqm^2 + \tfrac{1}{12}\*\pi^2 \,,
\nonumber \\
  F_\BE(0, 1, 1, 0, 1, 1, 1) &= -\tfrac{7}{16} + \tfrac{1}{4\*\eps^2} + \tfrac{3}{8\*\eps} + \tfrac{3}{4}\*\lqm - \tfrac{1}{4}\*\lqm^2 + \tfrac{1}{24}\*\pi^2 \,,
\nonumber \\
  F_\BE(0, 1, 1, 1, 0, 1, 0) &= \tfrac{5}{2} + \tfrac{1}{2\*\eps^2} + \tfrac{3}{2\*\eps} + \lqm - \tfrac{1}{2}\*\lqm^2 - \tfrac{1}{12}\*\pi^2 \,,
\nonumber \\
  F_\BE(0, 1, 1, 1, 0, 1, 1) &= \tfrac{33}{16} + \tfrac{1}{4\*\eps^2} + \tfrac{7}{8\*\eps} + \tfrac{1}{4}\*\lqm - \tfrac{1}{4}\*\lqm^2 - \tfrac{1}{24}\*\pi^2 \,,
\nonumber \\
  F_\BE(0, 1, 1, 1, 1, 0, 0) &= 3 + \tfrac{1}{\eps^2} + \tfrac{2}{\eps} + \tfrac{1}{6}\*\pi^2 \,,
\nonumber \\
  F_\BE(0, 1, 1, 1, 1, 0, 1) &= \tfrac{7}{8} + \tfrac{1}{2\*\eps^2} + \tfrac{3}{4\*\eps} + \tfrac{1}{12}\*\pi^2 \,,
\nonumber \\
  F_\BE(0, 1, 1, 1, 1, 1, 0) &= -\tfrac{1}{8}\*\lqm^4 - \tfrac{1}{4}\*\pi^2\*\lqm^2 - \tfrac{1}{30}\*\pi^4 - \zeta_3\*\lqm \,,
\nonumber \\
  F_\BE(0, 1, 1, 1, 1, 1, 1) &= 2 + \lqm^2 - \tfrac{1}{3}\*\lqm^3 + \tfrac{1}{6}\*\pi^2 + \lqm\*(-2 - \tfrac{1}{6}\*\pi^2) - 3\*\zeta_3 \,,
\nonumber \\
  F_\BE(0, 1, 1, 1, 1, 1, 2) &= 1 + \tfrac{1}{2}\*\lqm^2 - \tfrac{1}{6}\*\lqm^3 + \tfrac{1}{8}\*\pi^2 + \lqm\*(-1 - \tfrac{1}{12}\*\pi^2) - \tfrac{3}{2}\*\zeta_3 \,,
\nonumber \\
  F_\BE(1, -1, 0, 1, 1, 1, 0) &= -\tfrac{35}{16} - \tfrac{1}{4\*\eps^2} - \tfrac{7}{8\*\eps} - \tfrac{1}{8}\*\pi^2 \,,
\nonumber \\
  F_\BE(1, -1, 1, 1, 0, 1, 0) &= -\tfrac{49}{16} - \tfrac{1}{4\*\eps^2} - \tfrac{9}{8\*\eps} + \tfrac{1}{8}\*\pi^2 \,,
\nonumber \\
  F_\BE(1, -1, 1, 1, 1, 1, 0) &= \bigl(\tfrac{1}{4} + \tfrac{1}{6}\*\pi^2\bigr)/\Mr \,,
\nonumber \\
  F_\BE(1, 0, 0, 1, 1, 1, 0) &= \tfrac{7}{2} + \tfrac{1}{2\*\eps^2} + \tfrac{3}{2\*\eps} + \tfrac{1}{4}\*\pi^2 \,,
\nonumber \\
  F_\BE(1, 0, 0, 1, 1, 1, 1) &= \tfrac{35}{32} + \tfrac{1}{8\*\eps^2} + \tfrac{7}{16\*\eps} + \tfrac{1}{16}\*\pi^2 \,,
\nonumber \\
  F_\BE(1, 0, 1, 0, 1, 1, 0) &= \tfrac{7}{2} + \tfrac{1}{2\*\eps^2} + \tfrac{3}{2\*\eps} + \tfrac{1}{4}\*\pi^2 \,,
\nonumber \\
  F_\BE(1, 0, 1, 0, 1, 1, 1) &= \tfrac{21}{16} + \tfrac{1}{4\*\eps^2} + \tfrac{5}{8\*\eps} + \tfrac{1}{8}\*\pi^2 \,,
\nonumber \\
  F_\BE(1, 0, 1, 1, 0, 1, 0) &= \tfrac{9}{2} + \tfrac{1}{2\*\eps^2} + \tfrac{3}{2\*\eps} + \tfrac{1}{4}\*\pi^2 \,,
\nonumber \\
  F_\BE(1, 0, 1, 1, 0, 1, 1) &= \tfrac{113}{32} + \tfrac{3}{8\*\eps^2} + \tfrac{21}{16\*\eps} - \tfrac{7}{48}\*\pi^2 \,,
\nonumber \\
  F_\BE(1, 0, 1, 1, 1, 0, 0) &= 3 + \tfrac{1}{\eps^2} + \tfrac{2}{\eps} + \tfrac{1}{6}\*\pi^2 \,,
\nonumber \\
  F_\BE(1, 0, 1, 1, 1, 1, 0) &= -\tfrac{1}{3}\*\pi^2/\Mr \,,
\nonumber \\
  F_\BE(1, 0, 1, 1, 1, 1, 1) &= \bigl(\tfrac{1}{4} - \tfrac{1}{6}\*\pi^2\bigr)/\Mr \,,
\nonumber \\
  F_\BE(1, 1, -1, 0, 1, 1, 0) &= -\tfrac{63}{32} - \tfrac{1}{8\*\eps^2} + (-\tfrac{9}{16} + \tfrac{1}{4}\*\lqm)\tfrac{1}{\eps} + \tfrac{9}{8}\*\lqm - \tfrac{1}{4}\*\lqm^2 - \tfrac{1}{48}\*\pi^2 \,,
\nonumber \\
  F_\BE(1, 1, -1, 1, 1, 1, 0) &= -3 + (-\tfrac{1}{2} + \tfrac{1}{2}\*\lqm)\tfrac{1}{\eps} + \tfrac{5}{2}\*\lqm - \tfrac{1}{2}\*\lqm^2 + \tfrac{1}{12}\*\pi^2 \,,
\nonumber \\
  F_\BE(1, 1, 0, -1, 1, 1, 0) &= \tfrac{63}{16} + \tfrac{1}{4\*\eps^2} + (\tfrac{9}{8} - \tfrac{1}{2}\*\lqm)\tfrac{1}{\eps} - \tfrac{9}{4}\*\lqm + \tfrac{1}{2}\*\lqm^2 + \tfrac{1}{24}\*\pi^2 \,,
\nonumber \\
  F_\BE(1, 1, 0, 0, 1, 1, 0) &= \tfrac{19}{2} + \tfrac{1}{2\*\eps^2} + (\tfrac{5}{2} - \lqm)\tfrac{1}{\eps} - 5\*\lqm + \lqm^2 + \tfrac{1}{12}\*\pi^2 \,,
\nonumber \\
  F_\BE(1, 1, 0, 0, 1, 1, 1) &= \tfrac{63}{32} + \tfrac{1}{8\*\eps^2} + (\tfrac{9}{16} - \tfrac{1}{4}\*\lqm)\tfrac{1}{\eps} - \tfrac{9}{8}\*\lqm + \tfrac{1}{4}\*\lqm^2 + \tfrac{1}{48}\*\pi^2 \,,
\nonumber \\
  F_\BE(1, 1, 0, 1, 1, 1, 0) &= -\lqm^2 + \tfrac{1}{3}\*\lqm^3 - \tfrac{2}{3}\*\pi^2 - \tfrac{1}{6}\*\pi^2\*\lqm + (-\tfrac{1}{2}\*\lqm^2 - \tfrac{1}{3}\*\pi^2)\tfrac{1}{\eps} + \zeta_3 \,,
\nonumber \\
  F_\BE(1, 1, 0, 1, 1, 1, 1) &= 3 + (\tfrac{1}{2} - \tfrac{1}{2}\*\lqm)\tfrac{1}{\eps} - \tfrac{5}{2}\*\lqm + \tfrac{1}{2}\*\lqm^2 - \tfrac{1}{12}\*\pi^2 \,,
\nonumber \\
  F_\BE(1, 1, 1, -2, 1, 1, 0) &= \tfrac{59}{32} + \tfrac{3}{8\*\eps^2} + \tfrac{17}{16\*\eps} - \tfrac{3}{8}\*\lqm^2 + \tfrac{5}{48}\*\pi^2 + \lqm\*(\tfrac{5}{8} - \tfrac{1}{12}\*\pi^2) - \zeta_3 \,,
\nonumber \\
  F_\BE(1, 1, 1, -1, 0, 1, 0) &= \tfrac{89}{16} + \tfrac{1}{4\*\eps^2} + (\tfrac{11}{8} - \tfrac{1}{2}\*\lqm)\tfrac{1}{\eps} - \tfrac{11}{4}\*\lqm + \tfrac{1}{2}\*\lqm^2 - \tfrac{1}{24}\*\pi^2 \,,
\nonumber \\
  F_\BE(1, 1, 1, -1, 1, 0, 0) &= \tfrac{7}{2} + \tfrac{1}{2\*\eps^2} + (\tfrac{3}{2} - \tfrac{1}{2}\*\lqm)\tfrac{1}{\eps} - \tfrac{3}{2}\*\lqm + \tfrac{1}{4}\*\lqm^2 \,,
\nonumber \\
  F_\BE(1, 1, 1, -1, 1, 1, 0) &= \tfrac{5}{2} + \tfrac{1}{2\*\eps^2} + \tfrac{3}{2\*\eps} - \tfrac{1}{2}\*\lqm^2 + \tfrac{1}{4}\*\pi^2 + \lqm\*(1 - \tfrac{1}{6}\*\pi^2) - 2\*\zeta_3 \,,
\nonumber \\
  F_\BE(1, 1, 1, -1, 1, 1, 1) &= -\tfrac{13}{16} + \tfrac{1}{4\*\eps^2} + \tfrac{5}{8\*\eps} - \tfrac{1}{4}\*\lqm^2 + \tfrac{11}{24}\*\pi^2 + \lqm\*(\tfrac{5}{4} - \tfrac{1}{6}\*\pi^2) - 2\*\zeta_3 \,,
\nonumber \\
  F_\BE(1, 1, 1, 0, -1, 1, 0) &= \tfrac{63}{32} + \tfrac{1}{8\*\eps^2} + (\tfrac{9}{16} - \tfrac{1}{4}\*\lqm)\tfrac{1}{\eps} - \tfrac{9}{8}\*\lqm + \tfrac{1}{4}\*\lqm^2 - \tfrac{1}{48}\*\pi^2 \,,
\nonumber \\
  F_\BE(1, 1, 1, 0, 0, 1, 0) &= \tfrac{19}{2} + \tfrac{1}{2\*\eps^2} + (\tfrac{5}{2} - \lqm)\tfrac{1}{\eps} - 5\*\lqm + \lqm^2 - \tfrac{1}{12}\*\pi^2 \,,
\nonumber \\
  F_\BE(1, 1, 1, 0, 0, 1, 1) &= \tfrac{241}{32} + \tfrac{3}{8\*\eps^2} + (\tfrac{31}{16} - \tfrac{3}{4}\*\lqm)\tfrac{1}{\eps} - \tfrac{31}{8}\*\lqm + \tfrac{3}{4}\*\lqm^2 - \tfrac{1}{16}\*\pi^2 \,,
\nonumber \\
  F_\BE(1, 1, 1, 0, 1, 0, 0) &= 7 + \tfrac{1}{\eps^2} + (3 - \lqm)\tfrac{1}{\eps} - 3\*\lqm + \tfrac{1}{2}\*\lqm^2 \,,
\nonumber \\
  F_\BE(1, 1, 1, 0, 1, 0, 1) &= \tfrac{21}{8} + \tfrac{1}{2\*\eps^2} + (\tfrac{5}{4} - \tfrac{1}{2}\*\lqm)\tfrac{1}{\eps} - \tfrac{5}{4}\*\lqm + \tfrac{1}{4}\*\lqm^2 \,,
\nonumber \\
  F_\BE(1, 1, 1, 0, 1, 1, 0) &= -\tfrac{1}{12}\*\pi^2\*\lqm^2 - \tfrac{11}{180}\*\pi^4 - 2\*\zeta_3\*\lqm \,,
\nonumber \\
  F_\BE(1, 1, 1, 0, 1, 1, 1) &= 3 - \tfrac{1}{3}\*\pi^2 - \tfrac{11}{180}\*\pi^4 + \lqm^2\*(\tfrac{1}{2} - \tfrac{1}{12}\*\pi^2) + \lqm\*(-2 + \tfrac{1}{3}\*\pi^2 - 2\*\zeta_3) \nonumber\\*&\qquad + 4\*\zeta_3 \,,
\nonumber \\
  F_\BE(1, 1, 1, 0, 1, 1, 2) &= \tfrac{91}{16} - \tfrac{3}{4}\*\pi^2 - \tfrac{11}{180}\*\pi^4 + \lqm^2\*(\tfrac{5}{8} - \tfrac{1}{12}\*\pi^2) + \lqm\*(-\tfrac{13}{4} + \tfrac{1}{2}\*\pi^2 - 2\*\zeta_3) \nonumber\\*&\qquad + 6\*\zeta_3 \,,
\nonumber \\
  F_\BE(1, 1, 1, 1, -1, 1, 0) &= \tfrac{7}{2} + (\tfrac{1}{2} - \tfrac{1}{2}\*\lqm)\tfrac{1}{\eps} - 3\*\lqm + \tfrac{3}{4}\*\lqm^2 \,,
\nonumber \\
  F_\BE(1, 1, 1, 1, 0, 1, 0) &= -\lqm^2 + \tfrac{1}{2}\*\lqm^3 - \tfrac{1}{3}\*\pi^2 + (-\tfrac{1}{2}\*\lqm^2 - \tfrac{1}{3}\*\pi^2)\tfrac{1}{\eps} + 4\*\zeta_3 \,,
\nonumber \\
  F_\BE(1, 1, 1, 1, 0, 1, 1) &= \tfrac{11}{4} - \tfrac{5}{2}\*\lqm + \tfrac{1}{4}\*\lqm^3 - \tfrac{1}{12}\*\pi^2 + (\tfrac{1}{2} - \tfrac{1}{2}\*\lqm - \tfrac{1}{4}\*\lqm^2 - \tfrac{1}{6}\*\pi^2)\tfrac{1}{\eps} + 2\*\zeta_3 \,,
\nonumber \\
  F_\BE(1, 1, 1, 1, 1, -1, 0) &= \tfrac{21}{4} + \tfrac{1}{\eps^2} + (\tfrac{5}{2} - \tfrac{1}{2}\*\lqm)\tfrac{1}{\eps} - \tfrac{7}{4}\*\lqm + \tfrac{1}{4}\*\lqm^2 + \tfrac{1}{12}\*\pi^2 \,,
\nonumber \\
  F_\BE(1, 1, 1, 1, 1, 0, 0) &= -\tfrac{1}{2}\*\lqm^2 + \tfrac{1}{6}\*\lqm^3 - \tfrac{1}{3}\*\pi^2 - \tfrac{1}{6}\*\pi^2\*\lqm + (-\tfrac{1}{2}\*\lqm^2 - \tfrac{1}{3}\*\pi^2)\tfrac{1}{\eps} + \zeta_3 \,,
\nonumber \\
  F_\BE(1, 1, 1, 1, 1, 0, 1) &= -\tfrac{1}{8}\*\lqm^2 + \tfrac{1}{12}\*\lqm^3 - \tfrac{1}{12}\*\pi^2 - \tfrac{1}{12}\*\pi^2\*\lqm + (-\tfrac{1}{4}\*\lqm^2 - \tfrac{1}{6}\*\pi^2)\tfrac{1}{\eps} + \tfrac{1}{2}\*\zeta_3 \,,
\nonumber \\
  F_\BE(1, 1, 1, 1, 1, 1, 0) &= \bigl(\tfrac{19}{144}\*\pi^4 + \tfrac{1}{3}\*\pi^2\*\ln^2{2} - \tfrac{1}{3}\*\ln^4{2} - 8\*\,\Li4(\tfrac12)\bigr)/\Mr \,,
\nonumber \\
  F_\BE(1, 1, 1, 1, 1, 1, 1) &= \bigl(-1 - \pi^2 + \tfrac{19}{48}\*\pi^4 + \pi^2\*\ln^2{2} - \ln^4{2} - 24\*\,\Li4(\tfrac12) - 14\*\zeta_3\bigr)/\Mr \,,
\nonumber \\
  F_\BE(1, 1, 1, 1, 1, 1, 2) &= \bigl(-\tfrac{73}{16} - \tfrac{31}{6}\*\pi^2 + \tfrac{247}{144}\*\pi^4 + \tfrac{13}{3}\*\pi^2\*\ln^2{2} - \tfrac{13}{3}\*\ln^4{2} - 104\*\,\Li4(\tfrac12) \nonumber\\*&\qquad - 63\*\zeta_3\bigr)/\Mr 

  \,.
\end{align}
Alle sechs skalaren Integrale der Form $F_\BE(1,n_2,1,1,1,1,n_7)$
sind proportional zu $1/z$, d.h. ihr f"uhrender Beitrag ist von der
Ordnung~$(M^2/Q^2)^{-1}$.
Diese Singularit"at in $M^2$ wird allerdings durch den jeweiligen Vorfaktor
in der Zerlegung kompensiert.

\subsection{Vertexdiagramm mit Fermion-Selbstenergie}
\label{sec:Skalarfc}

Zerlegung in skalare Integrale~(\ref{eq:fcskalar}):
\begin{alignat}{4}
\label{eq:fczerlegung}
  \lefteqn{F_{v,\fc} =
    C_F^2 \left(\frac{\alpha}{4\pi}\right)^2 \, i^2
    \left(\frac{\mu^2}{M^2}\right)^{2\eps} S_\eps^2
    \, \Bigl\{
  } \quad \nonumber \\
    & F_\fc(0, 0, 1, 1, 1) &\cdot &\bigl( -2 + d \bigr)
\nonumber \\
  +& F_\fc(0, 1, 0, 1, 1) &\cdot &\bigl( 2 - d \bigr)
\nonumber \\
  +& F_\fc(0, 1, 1, 0, 1) &\cdot &\bigl( -2 + d \bigr)
\nonumber \\
  +& F_\fc(0, 1, 1, 1, 0) &\cdot &\bigl( 2 - d \bigr)
\nonumber \\
  +& F_\fc(0, 1, 1, 1, 1) &\cdot &\bigl( -2 + d + (4 - 2\*d)\*\Mr \bigr)
\nonumber \\
  +& F_\fc(0, 2, 0, 0, 1) &\cdot &\bigl( 2 - d \bigr)
\nonumber \\
  +& F_\fc(0, 2, 0, 1, 0) &\cdot &\bigl( -2 + d \bigr)
\nonumber \\
  +& F_\fc(0, 2, 0, 1, 1) &\cdot &\bigl( (-2 + d)\*\Mr \bigr)
\nonumber \\
  +& F_\fc(0, 2, 1, 0, 1) &\cdot &\bigl( -2 + d + (2 - d)\*\Mr \bigr)
\nonumber \\
  +& F_\fc(0, 2, 1, 1, 0) &\cdot &\bigl( 2 - d + (-2 + d)\*\Mr \bigr)
\nonumber \\
  +& F_\fc(0, 2, 1, 1, 1) &\cdot &\bigl( (2 - d)\*\Mr + (-2 + d)\*\Mr^2 \bigr)
\nonumber \\
  +& F_\fc(1, 0, 0, 1, 1) &\cdot &\bigl( 2 - d \bigr)
\nonumber \\
  +& F_\fc(1, 0, 1, 1, 1) &\cdot &\bigl( -2 + d + (2 - d)\*\Mr \bigr)
\nonumber \\
  +& F_\fc(1, 1, -1, 1, 1) &\cdot &\bigl( -2 + d \bigr)
\nonumber \\
  +& F_\fc(1, 1, 0, 0, 1) &\cdot &\bigl( 2 - d \bigr)
\nonumber \\
  +& F_\fc(1, 1, 0, 1, 0) &\cdot &\bigl( -2 + d \bigr)
\nonumber \\
  +& F_\fc(1, 1, 0, 1, 1) &\cdot &\bigl( 8 - 5\*d + \tfrac{1}{2}\*d^2 + (-6 + 3\*d)\*\Mr \bigr)
\nonumber \\
  +& F_\fc(1, 1, 1, 0, 1) &\cdot &\bigl( -2 + d + (2 - d)\*\Mr \bigr)
\nonumber \\
  +& F_\fc(1, 1, 1, 1, 0) &\cdot &\bigl( 2 - d + (-2 + d)\*\Mr \bigr)
\nonumber \\
  +& F_\fc(1, 1, 1, 1, 1) &\cdot &\bigl( -2 + d + (10 - 6\*d + \tfrac{1}{2}\*d^2)\*\Mr + (-4 + 2\*d)\*\Mr^2 \bigr)
\nonumber \\
  +& F_\fc(1, 2, -1, 0, 1) &\cdot &\bigl( -2 + d \bigr)
\nonumber \\
  +& F_\fc(1, 2, -1, 1, 0) &\cdot &\bigl( 2 - d \bigr)
\nonumber \\
  +& F_\fc(1, 2, -1, 1, 1) &\cdot &\bigl( (2 - d)\*\Mr \bigr)
\nonumber \\
  +& F_\fc(1, 2, 0, 0, 1) &\cdot &\bigl( 8 - 5\*d + \tfrac{1}{2}\*d^2 + (-4 + 2\*d)\*\Mr \bigr)
\nonumber \\
  +& F_\fc(1, 2, 0, 1, 0) &\cdot &\bigl( -8 + 5\*d - \tfrac{1}{2}\*d^2 + (4 - 2\*d)\*\Mr \bigr)
\nonumber \\
  +& F_\fc(1, 2, 0, 1, 1) &\cdot &\bigl( (-8 + 5\*d - \tfrac{1}{2}\*d^2)\*\Mr + (4 - 2\*d)\*\Mr^2 \bigr)
\nonumber \\
  +& F_\fc(1, 2, 1, 0, 1) &\cdot &\bigl( -2 + d + (8 - 5\*d + \tfrac{1}{2}\*d^2)\*\Mr + (-2 + d)\*\Mr^2 \bigr)
\nonumber \\
  +& F_\fc(1, 2, 1, 1, 0) &\cdot &\bigl( 2 - d + (-8 + 5\*d - \tfrac{1}{2}\*d^2)\*\Mr + (2 - d)\*\Mr^2 \bigr)
\nonumber \\
  +& F_\fc(1, 2, 1, 1, 1) &\cdot &\bigl( (2 - d)\*\Mr + (-8 + 5\*d - \tfrac{1}{2}\*d^2)\*\Mr^2 + (2 - d)\*\Mr^3 \bigr)

  \Bigr\}
  \,.
\end{alignat}
Ergebnisse der skalaren Integrale
in f"uhrender Ordnung in $z = M^2/Q^2$:
\begin{align}
\label{eq:fcergskalar}
    F_\fc(0, 1, 1, 1, 1) &= \tfrac{7}{2} + \tfrac{1}{2\*\eps^2} + \tfrac{3}{2\*\eps} - \tfrac{1}{12}\*\pi^2 \,,
\nonumber \\
  F_\fc(0, 2, 0, 1, 1) &= -\tfrac{1}{2} - \tfrac{1}{2\*\eps^2} - \tfrac{1}{2\*\eps} - \tfrac{1}{4}\*\pi^2 \,,
\nonumber \\
  F_\fc(0, 2, 1, 0, 1) &= 3 + \tfrac{1}{\eps^2} + \tfrac{2}{\eps} + \tfrac{1}{6}\*\pi^2 \,,
\nonumber \\
  F_\fc(0, 2, 1, 1, 1) &= \bigl(4 + \tfrac{1}{\eps^2} + \tfrac{2}{\eps} + \tfrac{1}{6}\*\pi^2\bigr)/\Mr \,,
\nonumber \\
  F_\fc(1, 0, 0, 1, 1) &= \tfrac{13}{8} + \tfrac{1}{4\*\eps} - \tfrac{1}{2}\*\lqm \,,
\nonumber \\
  F_\fc(1, 0, 1, 1, 1) &= \tfrac{5}{2} + \tfrac{1}{2\*\eps^2} + \tfrac{3}{2\*\eps} + \lqm - \tfrac{1}{2}\*\lqm^2 - \tfrac{1}{12}\*\pi^2 \,,
\nonumber \\
  F_\fc(1, 1, -1, 1, 1) &= \tfrac{89}{16} + \tfrac{1}{4\*\eps^2} + (\tfrac{11}{8} - \tfrac{1}{2}\*\lqm)\tfrac{1}{\eps} - \tfrac{11}{4}\*\lqm + \tfrac{1}{2}\*\lqm^2 - \tfrac{1}{24}\*\pi^2 \,,
\nonumber \\
  F_\fc(1, 1, 0, 1, 1) &= \tfrac{19}{2} + \tfrac{1}{2\*\eps^2} + (\tfrac{5}{2} - \lqm)\tfrac{1}{\eps} - 5\*\lqm + \lqm^2 - \tfrac{1}{12}\*\pi^2 \,,
\nonumber \\
  F_\fc(1, 1, 1, 1, 1) &= -\lqm^2 + \tfrac{1}{2}\*\lqm^3 - \tfrac{1}{3}\*\pi^2 + (-\tfrac{1}{2}\*\lqm^2 - \tfrac{1}{3}\*\pi^2)\tfrac{1}{\eps} + 4\*\zeta_3 \,,
\nonumber \\
  F_\fc(1, 2, -1, 1, 1) &= \tfrac{11}{2} + \tfrac{1}{2\*\eps^2} + (\tfrac{3}{2} - \lqm)\tfrac{1}{\eps} - 3\*\lqm + \lqm^2 - \tfrac{1}{12}\*\pi^2 \,,
\nonumber \\
  F_\fc(1, 2, 0, 1, 1) &= 2 + \tfrac{1}{\eps^2} + (1 - \lqm)\tfrac{1}{\eps} - \lqm + \lqm^2 + \tfrac{1}{6}\*\pi^2 \,,
\nonumber \\
  F_\fc(1, 2, 1, 0, 1) &= -1 - \tfrac{1}{\eps^2} - \tfrac{1}{\eps} - \tfrac{1}{6}\*\pi^2 \,,
\nonumber \\
  F_\fc(1, 2, 1, 1, 1) &= \bigl(-\tfrac{3}{2} - \tfrac{1}{\eps^2} - \tfrac{1}{\eps} - \tfrac{1}{3}\*\pi^2\bigr)/\Mr 

  \,.
\end{align}
Die beiden skalaren Integrale der Form $F_\fc(n_1,2,1,1,1)$
sind proportional zu $1/z$, d.h. ihr f"uhrender Beitrag ist von der
Ordnung~$(M^2/Q^2)^{-1}$.
Diese Singularit"at in $M^2$ wird jeweils durch den Vorfaktor
in der Zerlegung kompensiert.

\subsection{Selbstenergiekorrektur mit T1-Topologie}
\label{sec:SkalarT1}

Zerlegung der T1-Selbstenergiekorrektur in skalare
Integrale (\ref{eq:T1skalar0}) und (\ref{eq:T1skalarD0}):
\begin{alignat}{4}
\label{eq:T1zerlegung}
  \lefteqn{\Sigma_\Tone =
    \left( C_F^2 - \frac{1}{2} C_F C_A \right)
    \left(\frac{\alpha}{4\pi}\right)^2 \, i^2
    \left(\frac{\mu^2}{M^2}\right)^{2\eps} S_\eps^2
    \, \Bigl\{
  } \quad \nonumber \\
    & \Bm_\Tone(0, 1, 1, 0, 1) &\cdot &\bigl( (4 - 3\*d + \tfrac{1}{2}\*d^2)/\pMr \bigr)
\nonumber \\
  +& \Bm_\Tone(1, 0, 0, 1, 1) &\cdot &\bigl( (-8 + 5\*d - \tfrac{1}{2}\*d^2)/\pMr \bigr)
\nonumber \\
  +& \Bm_\Tone(1, 0, 1, 0, 1) &\cdot &\bigl( (4 - 2\*d)/\pMr \bigr)
\nonumber \\
  +& \Bm_\Tone(1, 1, -1, 1, 1) &\cdot &\bigl( (-4 + 2\*d)/\pMr \bigr)
\nonumber \\
  +& \Bm_\Tone(1, 1, 0, 0, 1) &\cdot &\bigl( (4 - 2\*d)/\pMr \bigr)
\nonumber \\
  +& \Bm_\Tone(1, 1, 0, 1, 0) &\cdot &\bigl( (4 - 2\*d)/\pMr \bigr)
\nonumber \\
  +& \Bm_\Tone(1, 1, 0, 1, 1) &\cdot &\bigl( 4 - 2\*d + (-24 + 14\*d - d^2)/\pMr \bigr)
\nonumber \\
  +& \Bm_\Tone(1, 1, 1, 0, 0) &\cdot &\bigl( (-4 + 2\*d)/\pMr \bigr)
\nonumber \\
  +& \Bm_\Tone(1, 1, 1, 0, 1) &\cdot &\bigl( -4 + 2\*d + (8 - 4\*d)/\pMr \bigr)
\nonumber \\
  +& \Bm_\Tone(1, 1, 1, 1, 0) &\cdot &\bigl( 4 - 3\*d + \tfrac{1}{2}\*d^2 + (4 - 2\*d)/\pMr \bigr)
\nonumber \\
  +& \Bm_\Tone(1, 1, 1, 1, 1) &\cdot &\bigl( 4 - 2\*d + (-12 + 7\*d - \tfrac{1}{2}\*d^2)/\pMr \bigr)
\nonumber \\
  +& \DBm_\Tone(0, 1, 1, 0, 1) &\cdot &\bigl( 4 - 3\*d + \tfrac{1}{2}\*d^2 \bigr)
\nonumber \\
  +& \DBm_\Tone(1, 0, 0, 1, 1) &\cdot &\bigl( -8 + 5\*d - \tfrac{1}{2}\*d^2 \bigr)
\nonumber \\
  +& \DBm_\Tone(1, 0, 1, 0, 1) &\cdot &\bigl( 4 - 2\*d \bigr)
\nonumber \\
  +& \DBm_\Tone(1, 1, -1, 1, 1) &\cdot &\bigl( -4 + 2\*d \bigr)
\nonumber \\
  +& \DBm_\Tone(1, 1, 0, 0, 1) &\cdot &\bigl( 4 - 2\*d \bigr)
\nonumber \\
  +& \DBm_\Tone(1, 1, 0, 1, 0) &\cdot &\bigl( 4 - 2\*d \bigr)
\nonumber \\
  +& \DBm_\Tone(1, 1, 0, 1, 1) &\cdot &\bigl( -24 + 14\*d - d^2 \bigr)
\nonumber \\
  +& \DBm_\Tone(1, 1, 1, 0, 0) &\cdot &\bigl( -4 + 2\*d \bigr)
\nonumber \\
  +& \DBm_\Tone(1, 1, 1, 0, 1) &\cdot &\bigl( 8 - 4\*d \bigr)
\nonumber \\
  +& \DBm_\Tone(1, 1, 1, 1, 0) &\cdot &\bigl( 4 - 2\*d \bigr)
\nonumber \\
  +& \DBm_\Tone(1, 1, 1, 1, 1) &\cdot &\bigl( -12 + 7\*d - \tfrac{1}{2}\*d^2 \bigr)

  \Bigr\}
  \,,
\end{alignat}
wobei $y = p^2/M^2$ der infinitesimale Parameter ist, der bei der
Projektion~(\ref{eq:fcorrproj}) auftritt.
Im Gesamtergebnis fallen die Pole in $y$ heraus.

Ergebnisse der skalaren Integrale:
\begin{align}
\label{eq:T1ergskalar}
    \Bm_\Tone(0, 1, 1, 0, 1) &= 7 + \tfrac{1}{\eps^2} + \tfrac{3}{\eps} + \tfrac{1}{6}\*\pi^2 \,,
\nonumber \\
  \Bm_\Tone(1, 0, 1, 0, 1) &= \tfrac{7}{2} + \tfrac{1}{2\*\eps^2} + \tfrac{3}{2\*\eps} + \tfrac{1}{4}\*\pi^2 \,,
\nonumber \\
  \Bm_\Tone(1, 1, -1, 1, 1) &= -\tfrac{7}{2} - \tfrac{1}{2\*\eps^2} - \tfrac{3}{2\*\eps} - \tfrac{1}{4}\*\pi^2 \,,
\nonumber \\
  \Bm_\Tone(1, 1, 0, 1, 1) &= \tfrac{7}{2} + \tfrac{1}{2\*\eps^2} + \tfrac{3}{2\*\eps} + \tfrac{1}{4}\*\pi^2 \,,
\nonumber \\
  \Bm_\Tone(1, 1, 1, 0, 0) &= 3 + \tfrac{1}{\eps^2} + \tfrac{2}{\eps} + \tfrac{1}{6}\*\pi^2 \,,
\nonumber \\
  \Bm_\Tone(1, 1, 1, 0, 1) &= \tfrac{7}{2} + \tfrac{1}{2\*\eps^2} + \tfrac{3}{2\*\eps} - \tfrac{1}{12}\*\pi^2 \,,
\nonumber \\
  \Bm_\Tone(1, 1, 1, 1, 0) &= 3 + \tfrac{1}{\eps^2} + \tfrac{2}{\eps} + \tfrac{1}{6}\*\pi^2 \,,
\nonumber \\
  \Bm_\Tone(1, 1, 1, 1, 1) &= -\tfrac{1}{3}\*\pi^2 \,,
\nonumber \\
  \DBm_\Tone(0, 1, 1, 0, 1) &= \tfrac{1}{8} - \tfrac{1}{4\*\eps} \,,
\nonumber \\
  \DBm_\Tone(1, 1, -1, 1, 1) &= \tfrac{7}{16} + \tfrac{1}{4\*\eps^2} + \tfrac{3}{8\*\eps} + \tfrac{1}{8}\*\pi^2 \,,
\nonumber \\
  \DBm_\Tone(1, 1, 0, 1, 1) &= \tfrac{7}{8} + \tfrac{1}{4\*\eps} \,,
\nonumber \\
  \DBm_\Tone(1, 1, 1, 0, 0) &= \tfrac{5}{4} + \tfrac{1}{2\*\eps} \,,
\nonumber \\
  \DBm_\Tone(1, 1, 1, 0, 1) &= \tfrac{11}{4} + \tfrac{1}{2\*\eps} - \tfrac{1}{6}\*\pi^2 \,,
\nonumber \\
  \DBm_\Tone(1, 1, 1, 1, 0) &= \tfrac{5}{2} + \tfrac{1}{\eps} \,,
\nonumber \\
  \DBm_\Tone(1, 1, 1, 1, 1) &= \tfrac{11}{8} + \tfrac{1}{4\*\eps} - \tfrac{1}{6}\*\pi^2 

  \,.
\end{align}

\subsection{Selbstenergiekorrektur mit T2-Topologie}
\label{sec:SkalarT2}

Zerlegung der T2-Selbstenergiekorrektur in skalare
Integrale $\Bm_\Ttwo$ und $\DBm_\Ttwo$ entsprechend (\ref{eq:T2skalar}):
\begin{alignat}{4}
\label{eq:T2zerlegung}
  \lefteqn{\Sigma_\Ttwo =
    C_F^2 \left(\frac{\alpha}{4\pi}\right)^2 \, i^2
    \left(\frac{\mu^2}{M^2}\right)^{2\eps} S_\eps^2
    \, \Bigl\{
  } \quad \nonumber \\
    & \Bm_\Ttwo(0, 1, 1, 1) &\cdot &\bigl( (-1 + d - \tfrac{1}{4}\*d^2)/\pMr \bigr)
\nonumber \\
  +& \Bm_\Ttwo(1, 0, 1, 1) &\cdot &\bigl( (1 - d + \tfrac{1}{4}\*d^2)/\pMr \bigr)
\nonumber \\
  +& \Bm_\Ttwo(1, 1, 0, 1) &\cdot &\bigl( (-1 + d - \tfrac{1}{4}\*d^2)/\pMr \bigr)
\nonumber \\
  +& \Bm_\Ttwo(1, 1, 1, 0) &\cdot &\bigl( (1 - d + \tfrac{1}{4}\*d^2)/\pMr \bigr)
\nonumber \\
  +& \Bm_\Ttwo(1, 1, 1, 1) &\cdot &\bigl( -1 + d - \tfrac{1}{4}\*d^2 + (2 - 2\*d + \tfrac{1}{2}\*d^2)/\pMr \bigr)
\nonumber \\
  +& \Bm_\Ttwo(2, 0, 0, 1) &\cdot &\bigl( (1 - d + \tfrac{1}{4}\*d^2)/\pMr \bigr)
\nonumber \\
  +& \Bm_\Ttwo(2, 0, 1, 0) &\cdot &\bigl( (-1 + d - \tfrac{1}{4}\*d^2)/\pMr \bigr)
\nonumber \\
  +& \Bm_\Ttwo(2, 0, 1, 1) &\cdot &\bigl( (-1 + d - \tfrac{1}{4}\*d^2)/\pMr \bigr)
\nonumber \\
  +& \Bm_\Ttwo(2, 1, 0, 1) &\cdot &\bigl( -1 + d - \tfrac{1}{4}\*d^2 + (1 - d + \tfrac{1}{4}\*d^2)/\pMr \bigr)
\nonumber \\
  +& \Bm_\Ttwo(2, 1, 1, 0) &\cdot &\bigl( 1 - d + \tfrac{1}{4}\*d^2 + (-1 + d - \tfrac{1}{4}\*d^2)/\pMr \bigr)
\nonumber \\
  +& \Bm_\Ttwo(2, 1, 1, 1) &\cdot &\bigl( 1 - d + \tfrac{1}{4}\*d^2 + (-1 + d - \tfrac{1}{4}\*d^2)/\pMr \bigr)
\nonumber \\
  +& \DBm_\Ttwo(0, 1, 1, 1) &\cdot &\bigl( -1 + d - \tfrac{1}{4}\*d^2 \bigr)
\nonumber \\
  +& \DBm_\Ttwo(1, 0, 1, 1) &\cdot &\bigl( 1 - d + \tfrac{1}{4}\*d^2 \bigr)
\nonumber \\
  +& \DBm_\Ttwo(1, 1, 0, 1) &\cdot &\bigl( -1 + d - \tfrac{1}{4}\*d^2 \bigr)
\nonumber \\
  +& \DBm_\Ttwo(1, 1, 1, 0) &\cdot &\bigl( 1 - d + \tfrac{1}{4}\*d^2 \bigr)
\nonumber \\
  +& \DBm_\Ttwo(1, 1, 1, 1) &\cdot &\bigl( 2 - 2\*d + \tfrac{1}{2}\*d^2 \bigr)
\nonumber \\
  +& \DBm_\Ttwo(2, 0, 0, 1) &\cdot &\bigl( 1 - d + \tfrac{1}{4}\*d^2 \bigr)
\nonumber \\
  +& \DBm_\Ttwo(2, 0, 1, 0) &\cdot &\bigl( -1 + d - \tfrac{1}{4}\*d^2 \bigr)
\nonumber \\
  +& \DBm_\Ttwo(2, 0, 1, 1) &\cdot &\bigl( -1 + d - \tfrac{1}{4}\*d^2 \bigr)
\nonumber \\
  +& \DBm_\Ttwo(2, 1, 0, 1) &\cdot &\bigl( 1 - d + \tfrac{1}{4}\*d^2 \bigr)
\nonumber \\
  +& \DBm_\Ttwo(2, 1, 1, 0) &\cdot &\bigl( -1 + d - \tfrac{1}{4}\*d^2 \bigr)
\nonumber \\
  +& \DBm_\Ttwo(2, 1, 1, 1) &\cdot &\bigl( -1 + d - \tfrac{1}{4}\*d^2 \bigr)

  \Bigr\}
  \,,
\end{alignat}
mit $y = p^2/M^2$.
Ergebnisse der skalaren Integrale:
\begin{align}
\label{eq:T2ergskalar}
    \Bm_\Ttwo(0, 1, 1, 1) &= 7 + \tfrac{1}{\eps^2} + \tfrac{3}{\eps} + \tfrac{1}{6}\*\pi^2 \,,
\nonumber \\
  \Bm_\Ttwo(1, 0, 1, 1) &= \tfrac{7}{2} + \tfrac{1}{2\*\eps^2} + \tfrac{3}{2\*\eps} + \tfrac{1}{4}\*\pi^2 \,,
\nonumber \\
  \Bm_\Ttwo(1, 1, 0, 1) &= 3 + \tfrac{1}{\eps^2} + \tfrac{2}{\eps} + \tfrac{1}{6}\*\pi^2 \,,
\nonumber \\
  \Bm_\Ttwo(1, 1, 1, 1) &= \tfrac{7}{2} + \tfrac{1}{2\*\eps^2} + \tfrac{3}{2\*\eps} - \tfrac{1}{12}\*\pi^2 \,,
\nonumber \\
  \Bm_\Ttwo(2, 0, 1, 1) &= -\tfrac{1}{2} - \tfrac{1}{2\*\eps^2} - \tfrac{1}{2\*\eps} - \tfrac{1}{4}\*\pi^2 \,,
\nonumber \\
  \Bm_\Ttwo(2, 1, 0, 1) &= 3 + \tfrac{1}{\eps^2} + \tfrac{2}{\eps} + \tfrac{1}{6}\*\pi^2 \,,
\nonumber \\
  \Bm_\Ttwo(2, 1, 1, 1) &= 4 + \tfrac{1}{\eps^2} + \tfrac{2}{\eps} + \tfrac{1}{6}\*\pi^2 \,,
\nonumber \\
  \DBm_\Ttwo(0, 1, 1, 1) &= \tfrac{1}{8} - \tfrac{1}{4\*\eps} \,,
\nonumber \\
  \DBm_\Ttwo(1, 1, 0, 1) &= \tfrac{5}{4} + \tfrac{1}{2\*\eps} \,,
\nonumber \\
  \DBm_\Ttwo(1, 1, 1, 1) &= \tfrac{11}{4} + \tfrac{1}{2\*\eps} - \tfrac{1}{6}\*\pi^2 \,,
\nonumber \\
  \DBm_\Ttwo(2, 1, 0, 1) &= \tfrac{27}{4} + \tfrac{1}{\eps^2} + \tfrac{7}{2\*\eps} + \tfrac{1}{6}\*\pi^2 \,,
\nonumber \\
  \DBm_\Ttwo(2, 1, 1, 1) &= \tfrac{41}{4} + \tfrac{1}{\eps^2} + \tfrac{7}{2\*\eps} - \tfrac{1}{6}\*\pi^2 

  \,.
\end{align}

%
%
\section{Skalare Integrale der nichtabelschen Beitr"age}

\subsection{Nichtabelsches Vertexdiagramm mit Benz-Topologie}
\label{sec:SkalarBECA}

Zerlegung in skalare Integrale~(\ref{eq:BECAskalar}):
\begin{alignat}{4}
\label{eq:BECAzerlegung}
  \lefteqn{F_{v,\BECA} =
    C_F C_A \left(\frac{\alpha}{4\pi}\right)^2 \, i^2
    \left(\frac{\mu^2}{M^2}\right)^{2\eps} S_\eps^2
    \, \Bigl\{
  } \quad \nonumber \\
    & F_\BECA(-1, 0, 1, 1, 1, 1, 0) &\cdot &\bigl( 2 \bigr)
\nonumber \\
  +& F_\BECA(-1, 1, 1, 0, 1, 1, 0) &\cdot &\bigl( -2 \bigr)
\nonumber \\
  +& F_\BECA(-1, 1, 1, 1, 1, 1, 0) &\cdot &\bigl( 1 - 2\*\Mr \bigr)
\nonumber \\
  +& F_\BECA(-1, 1, 1, 1, 1, 1, 1) &\cdot &\bigl( 5 - \tfrac{4}{d-2} - \tfrac{3}{4}\*d \bigr)
\nonumber \\
  +& F_\BECA(0, 0, 0, 1, 1, 1, 0) &\cdot &\bigl( 4 - \tfrac{4}{d-2} - \tfrac{3}{4}\*d \bigr)
\nonumber \\
  +& F_\BECA(0, 0, 1, 0, 1, 1, 0) &\cdot &\bigl( -2 \bigr)
\nonumber \\
  +& F_\BECA(0, 0, 1, 1, 0, 1, 0) &\cdot &\bigl( -6 + \tfrac{4}{d-2} + \tfrac{3}{4}\*d \bigr)
\nonumber \\
  +& F_\BECA(0, 0, 1, 1, 1, 0, 0) &\cdot &\bigl( -2 \bigr)
\nonumber \\
  +& F_\BECA(0, 0, 1, 1, 1, 1, 0) &\cdot &\bigl( 2 + (-10 + \tfrac{4}{d-2} + \tfrac{3}{4}\*d)\*\Mr \bigr)
\nonumber \\
  +& F_\BECA(0, 0, 1, 1, 1, 1, 1) &\cdot &\bigl( -\tfrac{3}{2} + \tfrac{1}{4}\*d \bigr)
\nonumber \\
  +& F_\BECA(0, 1, 0, 0, 1, 1, 0) &\cdot &\bigl( -4 + \tfrac{4}{d-2} + \tfrac{3}{4}\*d \bigr)
\nonumber \\
  +& F_\BECA(0, 1, 0, 1, 1, 1, 0) &\cdot &\bigl( -2 + (-4 + \tfrac{4}{d-2} + \tfrac{3}{4}\*d)\*\Mr \bigr)
\nonumber \\
  +& F_\BECA(0, 1, 0, 1, 1, 1, 1) &\cdot &\bigl( -1 + \tfrac{1}{2}\*d \bigr)
\nonumber \\
  +& F_\BECA(0, 1, 1, -1, 1, 1, 0) &\cdot &\bigl( 2 \bigr)
\nonumber \\
  +& F_\BECA(0, 1, 1, 0, 0, 1, 0) &\cdot &\bigl( 6 - \tfrac{4}{d-2} - \tfrac{3}{4}\*d \bigr)
\nonumber \\
  +& F_\BECA(0, 1, 1, 0, 1, 0, 0) &\cdot &\bigl( 2 \bigr)
\nonumber \\
  +& F_\BECA(0, 1, 1, 0, 1, 1, 0) &\cdot &\bigl( -3 - \tfrac{4}{d-2} + \tfrac{1}{4}\*d + (12 - \tfrac{4}{d-2} - \tfrac{3}{4}\*d)\*\Mr \bigr)
\nonumber \\
  +& F_\BECA(0, 1, 1, 0, 1, 1, 1) &\cdot &\bigl( -6 + \tfrac{8}{d-2} + \tfrac{3}{4}\*d \bigr)
\nonumber \\
  +& F_\BECA(0, 1, 1, 1, 0, 1, 0) &\cdot &\bigl( 6 - \tfrac{4}{d-2} - \tfrac{3}{4}\*d + (6 - \tfrac{4}{d-2} - \tfrac{3}{4}\*d)\*\Mr \bigr)
\nonumber \\
  +& F_\BECA(0, 1, 1, 1, 0, 1, 1) &\cdot &\bigl( -\tfrac{1}{2}\*d \bigr)
\nonumber \\
  +& F_\BECA(0, 1, 1, 1, 1, 0, 0) &\cdot &\bigl( -6 + \tfrac{4}{d-2} + \tfrac{3}{4}\*d + 2\*\Mr \bigr)
\nonumber \\
  +& F_\BECA(0, 1, 1, 1, 1, 0, 1) &\cdot &\bigl( 1 \bigr)
\nonumber \\
  +& F_\BECA(0, 1, 1, 1, 1, 1, 0) &\cdot &\bigl( 1 + (-3 - \tfrac{4}{d-2} + \tfrac{1}{4}\*d)\*\Mr + (10 - \tfrac{4}{d-2} - \tfrac{3}{4}\*d)\*\Mr^2 \bigr)
\nonumber \\
  +& F_\BECA(0, 1, 1, 1, 1, 1, 1) &\cdot &\bigl( -1 + (-5 + \tfrac{8}{d-2} + \tfrac{1}{4}\*d)\*\Mr \bigr)
\nonumber \\
  +& F_\BECA(0, 1, 1, 1, 1, 1, 2) &\cdot &\bigl( -1 + \tfrac{1}{2}\*d \bigr)
\nonumber \\
  +& F_\BECA(1, -1, 0, 1, 1, 1, 0) &\cdot &\bigl( -\tfrac{3}{2} + \tfrac{1}{4}\*d \bigr)
\nonumber \\
  +& F_\BECA(1, -1, 1, 1, 0, 1, 0) &\cdot &\bigl( \tfrac{3}{2} - \tfrac{1}{4}\*d \bigr)
\nonumber \\
  +& F_\BECA(1, -1, 1, 1, 1, 1, 0) &\cdot &\bigl( (\tfrac{3}{2} - \tfrac{1}{4}\*d)\*\Mr \bigr)
\nonumber \\
  +& F_\BECA(1, 0, -1, 1, 1, 1, 0) &\cdot &\bigl( -1 + \tfrac{1}{2}\*d \bigr)
\nonumber \\
  +& F_\BECA(1, 0, 0, 0, 1, 1, 0) &\cdot &\bigl( \tfrac{1}{2} + \tfrac{4}{d-2} - \tfrac{1}{4}\*d \bigr)
\nonumber \\
  +& F_\BECA(1, 0, 0, 1, 0, 1, 0) &\cdot &\bigl( 3 - d \bigr)
\nonumber \\
  +& F_\BECA(1, 0, 0, 1, 1, 0, 0) &\cdot &\bigl( -1 \bigr)
\nonumber \\
  +& F_\BECA(1, 0, 0, 1, 1, 1, 0) &\cdot &\bigl( -3 + \tfrac{1}{2}\*d + (\tfrac{5}{2} + \tfrac{4}{d-2} - \tfrac{5}{4}\*d)\*\Mr \bigr)
\nonumber \\
  +& F_\BECA(1, 0, 0, 1, 1, 1, 1) &\cdot &\bigl( -1 + \tfrac{1}{2}\*d \bigr)
\nonumber \\
  +& F_\BECA(1, 0, 1, 0, 0, 1, 0) &\cdot &\bigl( \tfrac{3}{2} - \tfrac{4}{d-2} + \tfrac{1}{4}\*d \bigr)
\nonumber \\
  +& F_\BECA(1, 0, 1, 0, 1, 0, 0) &\cdot &\bigl( 2 \bigr)
\nonumber \\
  +& F_\BECA(1, 0, 1, 0, 1, 1, 0) &\cdot &\bigl( -\tfrac{3}{2} + \tfrac{1}{4}\*d + (\tfrac{7}{2} - \tfrac{4}{d-2} + \tfrac{1}{4}\*d)\*\Mr \bigr)
\nonumber \\
  +& F_\BECA(1, 0, 1, 0, 1, 1, 1) &\cdot &\bigl( \tfrac{3}{2} - \tfrac{1}{4}\*d \bigr)
\nonumber \\
  +& F_\BECA(1, 0, 1, 1, -1, 1, 0) &\cdot &\bigl( -2 + \tfrac{1}{2}\*d \bigr)
\nonumber \\
  +& F_\BECA(1, 0, 1, 1, 0, 0, 0) &\cdot &\bigl( 1 \bigr)
\nonumber \\
  +& F_\BECA(1, 0, 1, 1, 0, 1, 0) &\cdot &\bigl( \tfrac{3}{2} - \tfrac{1}{4}\*d + (-\tfrac{3}{2} - \tfrac{4}{d-2} + \tfrac{5}{4}\*d)\*\Mr \bigr)
\nonumber \\
  +& F_\BECA(1, 0, 1, 1, 0, 1, 1) &\cdot &\bigl( 1 - \tfrac{1}{2}\*d \bigr)
\nonumber \\
  +& F_\BECA(1, 0, 1, 1, 1, 0, 0) &\cdot &\bigl( -\tfrac{1}{2} - \tfrac{1}{4}\*d + 3\*\Mr \bigr)
\nonumber \\
  +& F_\BECA(1, 0, 1, 1, 1, 1, 0) &\cdot &\bigl( (-\tfrac{1}{2} - \tfrac{1}{4}\*d)\*\Mr + (\tfrac{5}{2} - \tfrac{4}{d-2} + \tfrac{3}{4}\*d)\*\Mr^2 \bigr)
\nonumber \\
  +& F_\BECA(1, 0, 1, 1, 1, 1, 1) &\cdot &\bigl( (\tfrac{5}{2} - \tfrac{3}{4}\*d)\*\Mr \bigr)
\nonumber \\
  +& F_\BECA(1, 1, -1, 0, 1, 1, 0) &\cdot &\bigl( 1 - \tfrac{1}{2}\*d \bigr)
\nonumber \\
  +& F_\BECA(1, 1, -1, 1, 1, 1, 0) &\cdot &\bigl( (1 - \tfrac{1}{2}\*d)\*\Mr \bigr)
\nonumber \\
  +& F_\BECA(1, 1, 0, -1, 1, 1, 0) &\cdot &\bigl( 1 - \tfrac{4}{d-2} \bigr)
\nonumber \\
  +& F_\BECA(1, 1, 0, 0, 0, 1, 0) &\cdot &\bigl( -3 + d \bigr)
\nonumber \\
  +& F_\BECA(1, 1, 0, 0, 1, 0, 0) &\cdot &\bigl( 1 \bigr)
\nonumber \\
  +& F_\BECA(1, 1, 0, 0, 1, 1, 0) &\cdot &\bigl( -5 + \tfrac{8}{d-2} + \tfrac{1}{2}\*d + (-\tfrac{8}{d-2} + d)\*\Mr \bigr)
\nonumber \\
  +& F_\BECA(1, 1, 0, 0, 1, 1, 1) &\cdot &\bigl( 4 - d \bigr)
\nonumber \\
  +& F_\BECA(1, 1, 0, 1, 0, 1, 0) &\cdot &\bigl( 1 - \tfrac{1}{2}\*d + (-3 + d)\*\Mr \bigr)
\nonumber \\
  +& F_\BECA(1, 1, 0, 1, 1, 0, 0) &\cdot &\bigl( -1 + \tfrac{1}{2}\*d + \Mr \bigr)
\nonumber \\
  +& F_\BECA(1, 1, 0, 1, 1, 1, 0) &\cdot &\bigl( -2 + (-5 + \tfrac{8}{d-2} + \tfrac{1}{2}\*d)\*\Mr + (-1 - \tfrac{4}{d-2} + d)\*\Mr^2 \bigr)
\nonumber \\
  +& F_\BECA(1, 1, 0, 1, 1, 1, 1) &\cdot &\bigl( -2 + d + (4 - d)\*\Mr \bigr)
\nonumber \\
  +& F_\BECA(1, 1, 1, -1, 0, 1, 0) &\cdot &\bigl( -3 + \tfrac{4}{d-2} \bigr)
\nonumber \\
  +& F_\BECA(1, 1, 1, -1, 1, 0, 0) &\cdot &\bigl( -2 \bigr)
\nonumber \\
  +& F_\BECA(1, 1, 1, -1, 1, 1, 0) &\cdot &\bigl( -1 + \tfrac{4}{d-2} + (-5 + \tfrac{4}{d-2})\*\Mr \bigr)
\nonumber \\
  +& F_\BECA(1, 1, 1, -1, 1, 1, 1) &\cdot &\bigl( 1 - \tfrac{4}{d-2} \bigr)
\nonumber \\
  +& F_\BECA(1, 1, 1, 0, -1, 1, 0) &\cdot &\bigl( 2 - \tfrac{1}{2}\*d \bigr)
\nonumber \\
  +& F_\BECA(1, 1, 1, 0, 0, 0, 0) &\cdot &\bigl( -1 \bigr)
\nonumber \\
  +& F_\BECA(1, 1, 1, 0, 0, 1, 0) &\cdot &\bigl( 9 - \tfrac{4}{d-2} - \tfrac{3}{2}\*d + (-3 + \tfrac{8}{d-2} - d)\*\Mr \bigr)
\nonumber \\
  +& F_\BECA(1, 1, 1, 0, 0, 1, 1) &\cdot &\bigl( -3 + d \bigr)
\nonumber \\
  +& F_\BECA(1, 1, 1, 0, 1, 0, 0) &\cdot &\bigl( 10 - \tfrac{4}{d-2} - d - 5\*\Mr \bigr)
\nonumber \\
  +& F_\BECA(1, 1, 1, 0, 1, 0, 1) &\cdot &\bigl( -1 \bigr)
\nonumber \\
  +& F_\BECA(1, 1, 1, 0, 1, 1, 0) &\cdot &\bigl( -2 + (17 - \tfrac{5}{2}\*d)\*\Mr + (-9 + \tfrac{8}{d-2} - \tfrac{1}{2}\*d)\*\Mr^2 \bigr)
\nonumber \\
  +& F_\BECA(1, 1, 1, 0, 1, 1, 1) &\cdot &\bigl( 1 + \tfrac{1}{2}\*d + (-2 - \tfrac{8}{d-2} + d)\*\Mr \bigr)
\nonumber \\
  +& F_\BECA(1, 1, 1, 0, 1, 1, 2) &\cdot &\bigl( 1 - \tfrac{1}{2}\*d \bigr)
\nonumber \\
  +& F_\BECA(1, 1, 1, 1, -1, 1, 0) &\cdot &\bigl( -2 + \tfrac{1}{2}\*d + (2 - \tfrac{1}{2}\*d)\*\Mr \bigr)
\nonumber \\
  +& F_\BECA(1, 1, 1, 1, 0, 0, 0) &\cdot &\bigl( 3 - \tfrac{1}{2}\*d - \Mr \bigr)
\nonumber \\
  +& F_\BECA(1, 1, 1, 1, 0, 1, 0) &\cdot &\bigl( (8 - \tfrac{4}{d-2} - d)\*\Mr + (\tfrac{4}{d-2} - d)\*\Mr^2 \bigr)
\nonumber \\
  +& F_\BECA(1, 1, 1, 1, 0, 1, 1) &\cdot &\bigl( 1 - \tfrac{1}{2}\*d + (-3 + d)\*\Mr \bigr)
\nonumber \\
  +& F_\BECA(1, 1, 1, 1, 1, -1, 0) &\cdot &\bigl( -1 \bigr)
\nonumber \\
  +& F_\BECA(1, 1, 1, 1, 1, 0, 0) &\cdot &\bigl( (11 - \tfrac{4}{d-2} - \tfrac{3}{2}\*d)\*\Mr - 3\*\Mr^2 \bigr)
\nonumber \\
  +& F_\BECA(1, 1, 1, 1, 1, 0, 1) &\cdot &\bigl( 1 - \tfrac{1}{2}\*d - \Mr \bigr)
\nonumber \\
  +& F_\BECA(1, 1, 1, 1, 1, 1, 0) &\cdot &\bigl( -2\*\Mr + (18 - \tfrac{4}{d-2} - \tfrac{5}{2}\*d)\*\Mr^2 + (-4 + \tfrac{4}{d-2} - \tfrac{1}{2}\*d)\*\Mr^3 \bigr)
\nonumber \\
  +& F_\BECA(1, 1, 1, 1, 1, 1, 1) &\cdot &\bigl( (3 - \tfrac{1}{2}\*d)\*\Mr + (-3 - \tfrac{4}{d-2} + d)\*\Mr^2 \bigr)
\nonumber \\
  +& F_\BECA(1, 1, 1, 1, 1, 1, 2) &\cdot &\bigl( (1 - \tfrac{1}{2}\*d)\*\Mr \bigr)

  \Bigr\}
  \,,
\end{alignat}
mit $z = M^2/Q^2$.
In den Ergebnissen der skalaren Integrale sind alle $\eps$"~Pole
und alle Logarithmen angegeben, es fehlt die nichtlogarithmische, in $d=4$
Dimensionen endliche Konstante.
Nur skalare Integrale, die einen solchen Beitrag liefern, sind aufgelistet:
\begin{align}
\label{eq:BECAergskalar}
    F_\BECA(-1, 1, 1, 0, 1, 1, 0) &= -\tfrac{1}{8\*\eps^2} - \tfrac{7}{16\*\eps} - \tfrac{1}{8}\*\lqm + \tfrac{1}{8}\*\lqm^2 \,,
\nonumber \\
  F_\BECA(-1, 1, 1, 1, 1, 1, 0) &= \tfrac{1}{2\*\eps^2} + \tfrac{3}{2\*\eps} - \tfrac{3}{2}\*\lqm^2 + \tfrac{1}{3}\*\lqm^3 + \lqm\*(3 + \tfrac{1}{3}\*\pi^2) \,,
\nonumber \\
  F_\BECA(-1, 1, 1, 1, 1, 1, 1) &= \tfrac{1}{4\*\eps^2} + \tfrac{5}{8\*\eps} - \tfrac{13}{4}\*\lqm + \tfrac{3}{4}\*\lqm^2 \,,
\nonumber \\
  F_\BECA(0, 0, 1, 1, 1, 1, 0) &= \tfrac{1}{2\*\eps^2} + \tfrac{3}{2\*\eps} \,,
\nonumber \\
  F_\BECA(0, 0, 1, 1, 1, 1, 1) &= \tfrac{1}{4\*\eps^2} + \tfrac{7}{8\*\eps} \,,
\nonumber \\
  F_\BECA(0, 1, 0, 1, 1, 1, 0) &= \tfrac{1}{2\*\eps^2} + \tfrac{3}{2\*\eps} \,,
\nonumber \\
  F_\BECA(0, 1, 1, -1, 1, 1, 0) &= \tfrac{1}{8\*\eps^2} + \tfrac{7}{16\*\eps} + \tfrac{1}{8}\*\lqm - \tfrac{1}{8}\*\lqm^2 \,,
\nonumber \\
  F_\BECA(0, 1, 1, 0, 0, 1, 0) &= \tfrac{1}{4\*\eps} - \tfrac{1}{2}\*\lqm \,,
\nonumber \\
  F_\BECA(0, 1, 1, 0, 1, 1, 0) &= \tfrac{1}{2\*\eps^2} + \tfrac{3}{2\*\eps} + \lqm - \tfrac{1}{2}\*\lqm^2 \,,
\nonumber \\
  F_\BECA(0, 1, 1, 0, 1, 1, 1) &= \tfrac{1}{4\*\eps^2} + \tfrac{7}{8\*\eps} + \tfrac{1}{4}\*\lqm - \tfrac{1}{4}\*\lqm^2 \,,
\nonumber \\
  F_\BECA(0, 1, 1, 1, 0, 1, 0) &= \tfrac{1}{2\*\eps^2} + \tfrac{3}{2\*\eps} + \lqm - \tfrac{1}{2}\*\lqm^2 \,,
\nonumber \\
  F_\BECA(0, 1, 1, 1, 0, 1, 1) &= \tfrac{1}{4\*\eps^2} + \tfrac{7}{8\*\eps} + \tfrac{1}{4}\*\lqm - \tfrac{1}{4}\*\lqm^2 \,,
\nonumber \\
  F_\BECA(0, 1, 1, 1, 1, 0, 0) &= \tfrac{1}{\eps^2} + \tfrac{2}{\eps} \,,
\nonumber \\
  F_\BECA(0, 1, 1, 1, 1, 0, 1) &= \tfrac{1}{2\*\eps^2} + \tfrac{5}{4\*\eps} \,,
\nonumber \\
  F_\BECA(0, 1, 1, 1, 1, 1, 0) &= -\tfrac{1}{12}\*\lqm^4 - \tfrac{1}{6}\*\pi^2\*\lqm^2 + \tfrac{2}{3}\*\zeta_3\*\lqm \,,
\nonumber \\
  F_\BECA(0, 1, 1, 1, 1, 1, 1) &= \lqm^2 - \tfrac{1}{3}\*\lqm^3 + \lqm\*(-2 - \tfrac{1}{3}\*\pi^2) \,,
\nonumber \\
  F_\BECA(0, 1, 1, 1, 1, 1, 2) &= \tfrac{1}{4}\*\lqm^2 - \tfrac{1}{6}\*\lqm^3 + \lqm\*(-\tfrac{1}{4} - \tfrac{1}{6}\*\pi^2) \,,
\nonumber \\
  F_\BECA(1, -1, 0, 1, 1, 1, 0) &= -\tfrac{1}{4\*\eps^2} - \tfrac{7}{8\*\eps} \,,
\nonumber \\
  F_\BECA(1, -1, 1, 1, 0, 1, 0) &= -\tfrac{1}{4\*\eps^2} - \tfrac{9}{8\*\eps} \,,
\nonumber \\
  F_\BECA(1, 0, 0, 1, 1, 1, 0) &= \tfrac{1}{2\*\eps^2} + \tfrac{3}{2\*\eps} \,,
\nonumber \\
  F_\BECA(1, 0, 0, 1, 1, 1, 1) &= \tfrac{1}{8\*\eps^2} + \tfrac{7}{16\*\eps} \,,
\nonumber \\
  F_\BECA(1, 0, 1, 0, 1, 1, 0) &= \tfrac{1}{2\*\eps^2} + \tfrac{3}{2\*\eps} \,,
\nonumber \\
  F_\BECA(1, 0, 1, 0, 1, 1, 1) &= \tfrac{1}{4\*\eps^2} + \tfrac{9}{8\*\eps} \,,
\nonumber \\
  F_\BECA(1, 0, 1, 1, 0, 1, 0) &= \tfrac{1}{2\*\eps^2} + \tfrac{3}{2\*\eps} \,,
\nonumber \\
  F_\BECA(1, 0, 1, 1, 0, 1, 1) &= \tfrac{3}{8\*\eps^2} + \tfrac{21}{16\*\eps} \,,
\nonumber \\
  F_\BECA(1, 0, 1, 1, 1, 0, 0) &= \tfrac{1}{\eps^2} + \tfrac{2}{\eps} \,,
\nonumber \\
  F_\BECA(1, 1, -1, 0, 1, 1, 0) &= -\tfrac{1}{8\*\eps^2} + (-\tfrac{9}{16} + \tfrac{1}{4}\*\lqm)\tfrac{1}{\eps} + \tfrac{9}{8}\*\lqm - \tfrac{1}{4}\*\lqm^2 \,,
\nonumber \\
  F_\BECA(1, 1, -1, 1, 1, 1, 0) &= (-\tfrac{1}{2} + \tfrac{1}{2}\*\lqm)\tfrac{1}{\eps} + \tfrac{5}{2}\*\lqm - \tfrac{1}{2}\*\lqm^2 \,,
\nonumber \\
  F_\BECA(1, 1, 0, -1, 1, 1, 0) &= \tfrac{1}{4\*\eps^2} + (\tfrac{9}{8} - \tfrac{1}{2}\*\lqm)\tfrac{1}{\eps} - \tfrac{9}{4}\*\lqm + \tfrac{1}{2}\*\lqm^2 \,,
\nonumber \\
  F_\BECA(1, 1, 0, 0, 1, 1, 0) &= \tfrac{1}{2\*\eps^2} + (\tfrac{5}{2} - \lqm)\tfrac{1}{\eps} - 5\*\lqm + \lqm^2 \,,
\nonumber \\
  F_\BECA(1, 1, 0, 0, 1, 1, 1) &= \tfrac{1}{8\*\eps^2} + (\tfrac{9}{16} - \tfrac{1}{4}\*\lqm)\tfrac{1}{\eps} - \tfrac{9}{8}\*\lqm + \tfrac{1}{4}\*\lqm^2 \,,
\nonumber \\
  F_\BECA(1, 1, 0, 1, 1, 1, 0) &= -\lqm^2 + \tfrac{1}{3}\*\lqm^3 + (-\tfrac{1}{2}\*\lqm^2 - \tfrac{1}{3}\*\pi^2)\tfrac{1}{\eps} + 2\*\sqrt{3}\*\Cl2(\tfrac{\pi}{3})\*\lqm \,,
\nonumber \\
  F_\BECA(1, 1, 0, 1, 1, 1, 1) &= (\tfrac{1}{2} - \tfrac{1}{2}\*\lqm)\tfrac{1}{\eps} - \tfrac{5}{2}\*\lqm + \tfrac{1}{2}\*\lqm^2 \,,
\nonumber \\
  F_\BECA(1, 1, 1, -1, 0, 1, 0) &= \tfrac{1}{4\*\eps^2} + (\tfrac{11}{8} - \tfrac{1}{2}\*\lqm)\tfrac{1}{\eps} - \tfrac{11}{4}\*\lqm + \tfrac{1}{2}\*\lqm^2 \,,
\nonumber \\
  F_\BECA(1, 1, 1, -1, 1, 0, 0) &= \tfrac{1}{2\*\eps^2} + (\tfrac{3}{2} - \tfrac{1}{2}\*\lqm)\tfrac{1}{\eps} - \tfrac{3}{2}\*\lqm + \tfrac{1}{4}\*\lqm^2 \,,
\nonumber \\
  F_\BECA(1, 1, 1, -1, 1, 1, 0) &= \tfrac{1}{2\*\eps^2} + \tfrac{3}{2\*\eps} - \tfrac{1}{2}\*\lqm^2 + \lqm\*(1 - \tfrac{1}{6}\*\pi^2) \,,
\nonumber \\
  F_\BECA(1, 1, 1, -1, 1, 1, 1) &= \tfrac{1}{4\*\eps^2} + \tfrac{9}{8\*\eps} - \tfrac{1}{4}\*\lqm^2 + \lqm\*(\tfrac{3}{4} - \tfrac{1}{6}\*\pi^2) \,,
\nonumber \\
  F_\BECA(1, 1, 1, 0, -1, 1, 0) &= \tfrac{1}{8\*\eps^2} + (\tfrac{9}{16} - \tfrac{1}{4}\*\lqm)\tfrac{1}{\eps} - \tfrac{9}{8}\*\lqm + \tfrac{1}{4}\*\lqm^2 \,,
\nonumber \\
  F_\BECA(1, 1, 1, 0, 0, 1, 0) &= \tfrac{1}{2\*\eps^2} + (\tfrac{5}{2} - \lqm)\tfrac{1}{\eps} - 5\*\lqm + \lqm^2 \,,
\nonumber \\
  F_\BECA(1, 1, 1, 0, 0, 1, 1) &= \tfrac{3}{8\*\eps^2} + (\tfrac{31}{16} - \tfrac{3}{4}\*\lqm)\tfrac{1}{\eps} - \tfrac{31}{8}\*\lqm + \tfrac{3}{4}\*\lqm^2 \,,
\nonumber \\
  F_\BECA(1, 1, 1, 0, 1, 0, 0) &= \tfrac{1}{\eps^2} + (3 - \lqm)\tfrac{1}{\eps} - 3\*\lqm + \tfrac{1}{2}\*\lqm^2 \,,
\nonumber \\
  F_\BECA(1, 1, 1, 0, 1, 0, 1) &= \tfrac{1}{2\*\eps^2} + (\tfrac{7}{4} - \tfrac{1}{2}\*\lqm)\tfrac{1}{\eps} - \tfrac{7}{4}\*\lqm + \tfrac{1}{4}\*\lqm^2 \,,
\nonumber \\
  F_\BECA(1, 1, 1, 0, 1, 1, 0) &= -\tfrac{1}{12}\*\pi^2\*\lqm^2 - \zeta_3\*\lqm \,,
\nonumber \\
  F_\BECA(1, 1, 1, 0, 1, 1, 1) &= \lqm^2\*(\tfrac{1}{2} - \tfrac{1}{12}\*\pi^2) + \lqm\*(-2 + \tfrac{1}{6}\*\pi^2 - \zeta_3) \,,
\nonumber \\
  F_\BECA(1, 1, 1, 0, 1, 1, 2) &= \lqm^2\*(\tfrac{5}{8} - \tfrac{1}{12}\*\pi^2) + \lqm\*(-\tfrac{5}{2} + \tfrac{1}{4}\*\pi^2 - \zeta_3) \,,
\nonumber \\
  F_\BECA(1, 1, 1, 1, -1, 1, 0) &= (\tfrac{1}{2} - \tfrac{1}{2}\*\lqm)\tfrac{1}{\eps} - 3\*\lqm + \tfrac{3}{4}\*\lqm^2 \,,
\nonumber \\
  F_\BECA(1, 1, 1, 1, 0, 1, 0) &= -\lqm^2 + \tfrac{1}{2}\*\lqm^3 + (-\tfrac{1}{2}\*\lqm^2 - \tfrac{1}{3}\*\pi^2)\tfrac{1}{\eps} \,,
\nonumber \\
  F_\BECA(1, 1, 1, 1, 0, 1, 1) &= -\tfrac{5}{2}\*\lqm + \tfrac{1}{4}\*\lqm^3 + (\tfrac{1}{2} - \tfrac{1}{2}\*\lqm - \tfrac{1}{4}\*\lqm^2 - \tfrac{1}{6}\*\pi^2)\tfrac{1}{\eps} \,,
\nonumber \\
  F_\BECA(1, 1, 1, 1, 1, -1, 0) &= \tfrac{1}{\eps^2} + (\tfrac{5}{2} - \tfrac{1}{2}\*\lqm)\tfrac{1}{\eps} - \tfrac{5}{4}\*\lqm + \tfrac{1}{4}\*\lqm^2 \,,
\nonumber \\
  F_\BECA(1, 1, 1, 1, 1, 0, 0) &= -\tfrac{1}{2}\*\lqm^2 + \tfrac{1}{6}\*\lqm^3 - \tfrac{1}{6}\*\pi^2\*\lqm + (-\tfrac{1}{2}\*\lqm^2 - \tfrac{1}{3}\*\pi^2)\tfrac{1}{\eps} \,,
\nonumber \\
  F_\BECA(1, 1, 1, 1, 1, 0, 1) &= -\tfrac{3}{8}\*\lqm^2 + \tfrac{1}{12}\*\lqm^3 - \tfrac{1}{12}\*\pi^2\*\lqm + (-\tfrac{1}{4}\*\lqm^2 - \tfrac{1}{6}\*\pi^2)\tfrac{1}{\eps} 

  \,.
\end{align}

\subsection[Nichtabelsche Selbstenergiekorrektur mit T1-Topologie]
  {Nichtabelsche Selbstenergiekorrektur mit\\ T1-Topologie}
\label{sec:SkalarT1CA}

Zerlegung in skalare 
Integrale $B_\ToneCA^0$ und $B_\ToneCA'$ entsprechend (\ref{eq:T1CAskalar}):
\begin{alignat}{4}
\label{eq:T1CAzerlegung}
  \lefteqn{\Sigma_\Tone =
    C_F C_A \left(\frac{\alpha}{4\pi}\right)^2 \, i^2
    \left(\frac{\mu^2}{M^2}\right)^{2\eps} S_\eps^2
    \, \Bigl\{
  } \quad \nonumber \\
    & \Bm_\ToneCA(0, 1, 0, 1, 1) &\cdot &\bigl( (-1 + \tfrac{1}{2}\*d)/\pMr \bigr)
\nonumber \\
  +& \Bm_\ToneCA(1, 0, 1, 0, 1) &\cdot &\bigl( (1 - \tfrac{1}{2}\*d)/\pMr \bigr)
\nonumber \\
  +& \Bm_\ToneCA(1, 1, -1, 1, 1) &\cdot &\bigl( (2 - d)/\pMr \bigr)
\nonumber \\
  +& \Bm_\ToneCA(1, 1, 0, 0, 1) &\cdot &\bigl( (-2 + d)/\pMr \bigr)
\nonumber \\
  +& \Bm_\ToneCA(1, 1, 0, 1, 0) &\cdot &\bigl( (-2 + d)/\pMr \bigr)
\nonumber \\
  +& \Bm_\ToneCA(1, 1, 0, 1, 1) &\cdot &\bigl( -2 + d + (-4 + 2\*d)/\pMr \bigr)
\nonumber \\
  +& \Bm_\ToneCA(1, 1, 1, 0, 0) &\cdot &\bigl( (2 - d)/\pMr \bigr)
\nonumber \\
  +& \Bm_\ToneCA(1, 1, 1, 0, 1) &\cdot &\bigl( -2 + d + (4 - 2\*d)/\pMr \bigr)
\nonumber \\
  +& \Bm_\ToneCA(1, 1, 1, 1, 0) &\cdot &\bigl( (2 - d)/\pMr \bigr)
\nonumber \\
  +& \Bm_\ToneCA(1, 1, 1, 1, 1) &\cdot &\bigl( -2 + d + (3 - \tfrac{3}{2}\*d)/\pMr \bigr)
\nonumber \\
  +& \DBm_\ToneCA(0, 1, 0, 1, 1) &\cdot &\bigl( -1 + \tfrac{1}{2}\*d \bigr)
\nonumber \\
  +& \DBm_\ToneCA(1, 0, 1, 0, 1) &\cdot &\bigl( 1 - \tfrac{1}{2}\*d \bigr)
\nonumber \\
  +& \DBm_\ToneCA(1, 1, -1, 1, 1) &\cdot &\bigl( 2 - d \bigr)
\nonumber \\
  +& \DBm_\ToneCA(1, 1, 0, 0, 1) &\cdot &\bigl( -2 + d \bigr)
\nonumber \\
  +& \DBm_\ToneCA(1, 1, 0, 1, 0) &\cdot &\bigl( -2 + d \bigr)
\nonumber \\
  +& \DBm_\ToneCA(1, 1, 0, 1, 1) &\cdot &\bigl( -4 + 2\*d \bigr)
\nonumber \\
  +& \DBm_\ToneCA(1, 1, 1, 0, 0) &\cdot &\bigl( 2 - d \bigr)
\nonumber \\
  +& \DBm_\ToneCA(1, 1, 1, 0, 1) &\cdot &\bigl( 4 - 2\*d \bigr)
\nonumber \\
  +& \DBm_\ToneCA(1, 1, 1, 1, 0) &\cdot &\bigl( 2 - d \bigr)
\nonumber \\
  +& \DBm_\ToneCA(1, 1, 1, 1, 1) &\cdot &\bigl( 3 - \tfrac{3}{2}\*d \bigr)

  \Bigr\}
  \,,
\end{alignat}
mit $y = p^2/M^2$.
Im Gesamtergebnis fallen die Pole in $y$ heraus.
Ergebnisse der skalaren Integrale bis $\Oc(\eps^{-1})$ (nur
nichtverschwindende Beitr"age):
\begin{align}
\label{eq:T1CAergskalar}
    \Bm_\ToneCA(0, 1, 0, 1, 1) &= \tfrac{3}{2\*\eps^2} + \tfrac{9}{2\*\eps} \,,
\nonumber \\
  \Bm_\ToneCA(1, 0, 1, 0, 1) &= \tfrac{1}{2\*\eps^2} + \tfrac{3}{2\*\eps} \,,
\nonumber \\
  \Bm_\ToneCA(1, 1, -1, 1, 1) &= \tfrac{3}{2\*\eps^2} + \tfrac{7}{2\*\eps} \,,
\nonumber \\
  \Bm_\ToneCA(1, 1, 0, 0, 1) &= \tfrac{1}{\eps^2} + \tfrac{2}{\eps} \,,
\nonumber \\
  \Bm_\ToneCA(1, 1, 0, 1, 0) &= \tfrac{1}{\eps^2} + \tfrac{2}{\eps} \,,
\nonumber \\
  \Bm_\ToneCA(1, 1, 0, 1, 1) &= \tfrac{1}{2\*\eps^2} + \tfrac{3}{2\*\eps} \,,
\nonumber \\
  \Bm_\ToneCA(1, 1, 1, 0, 1) &= \tfrac{1}{2\*\eps^2} + \tfrac{3}{2\*\eps} \,,
\nonumber \\
  \Bm_\ToneCA(1, 1, 1, 1, 0) &= \tfrac{1}{\eps^2} + \tfrac{2}{\eps} \,,
\nonumber \\
  \DBm_\ToneCA(1, 1, -1, 1, 1) &= \tfrac{1}{4\*\eps^2} + \tfrac{13}{8\*\eps} \,,
\nonumber \\
  \DBm_\ToneCA(1, 1, 0, 0, 1) &= \tfrac{1}{2\*\eps} \,,
\nonumber \\
  \DBm_\ToneCA(1, 1, 0, 1, 0) &= \tfrac{1}{2\*\eps} \,,
\nonumber \\
  \DBm_\ToneCA(1, 1, 0, 1, 1) &= \tfrac{1}{2\*\eps} \,,
\nonumber \\
  \DBm_\ToneCA(1, 1, 1, 0, 1) &= \tfrac{1}{2\*\eps} \,,
\nonumber \\
  \DBm_\ToneCA(1, 1, 1, 1, 0) &= \tfrac{1}{\eps} 

  \,.
\end{align}

\subsection{Vertexdiagramme mit Eichboson- oder Geistschleife}
\label{sec:SkalarWWcc}

Zerlegung in skalare Integrale~(\ref{eq:Wcorrskalar}):
\begin{alignat}{4}
\label{eq:WWcczerlegung}
  \lefteqn{F_{v,\WWcc} =
    C_F C_A \left(\frac{\alpha}{4\pi}\right)^2 \, i^2
    \left(\frac{\mu^2}{M^2}\right)^{2\eps} S_\eps^2
    \, \Bigl\{
  } \quad \nonumber \\
    & F_\Wc(0, 0, 2, 0, 0, 1) &\cdot &\bigl( -2 + \tfrac{2}{d-1} \bigr)
\nonumber \\
  +& F_\Wc(0, 0, 2, 0, 0, 2) &\cdot &\bigl( -1 - \tfrac{1}{d-1} + d \bigr)
\nonumber \\
  +& F_\Wc(0, 0, 2, 1, -1, 1) &\cdot &\bigl( 2 - \tfrac{2}{d-1} \bigr)
\nonumber \\
  +& F_\Wc(0, 0, 2, 1, -1, 2) &\cdot &\bigl( 1 + \tfrac{1}{d-1} - d \bigr)
\nonumber \\
  +& F_\Wc(0, 0, 2, 1, 0, 0) &\cdot &\bigl( -6 + \tfrac{4}{d-1} \bigr)
\nonumber \\
  +& F_\Wc(0, 0, 2, 1, 0, 1) &\cdot &\bigl( 2 - \tfrac{2}{d-1} \bigr)
\nonumber \\
  +& F_\Wc(0, 0, 2, 1, 1, -1) &\cdot &\bigl( -3 - \tfrac{1}{d-1} \bigr)
\nonumber \\
  +& F_\Wc(0, 0, 2, 1, 1, 0) &\cdot &\bigl( \tfrac{3}{2} + \tfrac{1}{2\*(d-1)} + (-6 + \tfrac{4}{d-1})\*\Mr \bigr)
\nonumber \\
  +& F_\Wc(0, 0, 2, 1, 1, 1) &\cdot &\bigl( (2 - \tfrac{2}{d-1})\*\Mr \bigr)
\nonumber \\
  +& F_\Wc(1, 0, 2, 0, 0, 0) &\cdot &\bigl( 4 - \tfrac{4}{d-1} \bigr)
\nonumber \\
  +& F_\Wc(1, 0, 2, 0, 0, 1) &\cdot &\bigl( -4 + \tfrac{4}{d-1} \bigr)
\nonumber \\
  +& F_\Wc(1, 0, 2, 1, -1, 0) &\cdot &\bigl( -4 + \tfrac{4}{d-1} \bigr)
\nonumber \\
  +& F_\Wc(1, 0, 2, 1, -1, 1) &\cdot &\bigl( 4 - \tfrac{4}{d-1} \bigr)
\nonumber \\
  +& F_\Wc(1, 0, 2, 1, 0, -1) &\cdot &\bigl( 12 - \tfrac{8}{d-1} \bigr)
\nonumber \\
  +& F_\Wc(1, 0, 2, 1, 0, 0) &\cdot &\bigl( -12 + \tfrac{8}{d-1} \bigr)
\nonumber \\
  +& F_\Wc(1, 0, 2, 1, 1, -2) &\cdot &\bigl( 6 + \tfrac{2}{d-1} \bigr)
\nonumber \\
  +& F_\Wc(1, 0, 2, 1, 1, -1) &\cdot &\bigl( -6 - \tfrac{2}{d-1} + (12 - \tfrac{8}{d-1})\*\Mr \bigr)
\nonumber \\
  +& F_\Wc(1, 0, 2, 1, 1, 0) &\cdot &\bigl( (-12 + \tfrac{8}{d-1})\*\Mr \bigr)
\nonumber \\
  +& F_\Wc(1, 1, 2, 0, 0, -1) &\cdot &\bigl( -2 + \tfrac{2}{d-1} \bigr)
\nonumber \\
  +& F_\Wc(1, 1, 2, 0, 0, 0) &\cdot &\bigl( 9 - \tfrac{7}{d-1} - d \bigr)
\nonumber \\
  +& F_\Wc(1, 1, 2, 0, 0, 1) &\cdot &\bigl( -2 + \tfrac{2}{d-1} \bigr)
\nonumber \\
  +& F_\Wc(1, 1, 2, 1, -1, -1) &\cdot &\bigl( 2 - \tfrac{2}{d-1} \bigr)
\nonumber \\
  +& F_\Wc(1, 1, 2, 1, -1, 0) &\cdot &\bigl( -9 + \tfrac{7}{d-1} + d \bigr)
\nonumber \\
  +& F_\Wc(1, 1, 2, 1, -1, 1) &\cdot &\bigl( 2 - \tfrac{2}{d-1} \bigr)
\nonumber \\
  +& F_\Wc(1, 1, 2, 1, 0, -2) &\cdot &\bigl( -6 + \tfrac{4}{d-1} \bigr)
\nonumber \\
  +& F_\Wc(1, 1, 2, 1, 0, -1) &\cdot &\bigl( 26 - \tfrac{14}{d-1} - 3\*d \bigr)
\nonumber \\
  +& F_\Wc(1, 1, 2, 1, 0, 0) &\cdot &\bigl( -6 + \tfrac{4}{d-1} \bigr)
\nonumber \\
  +& F_\Wc(1, 1, 2, 1, 1, -3) &\cdot &\bigl( -3 - \tfrac{1}{d-1} \bigr)
\nonumber \\
  +& F_\Wc(1, 1, 2, 1, 1, -2) &\cdot &\bigl( \tfrac{23}{2} + \tfrac{7}{2\*(d-1)} - \tfrac{3}{2}\*d + (-6 + \tfrac{4}{d-1})\*\Mr \bigr)
\nonumber \\
  +& F_\Wc(1, 1, 2, 1, 1, -1) &\cdot &\bigl( -3 - \tfrac{1}{d-1} + (26 - \tfrac{14}{d-1} - 3\*d)\*\Mr \bigr)
\nonumber \\
  +& F_\Wc(1, 1, 2, 1, 1, 0) &\cdot &\bigl( (-6 + \tfrac{4}{d-1})\*\Mr \bigr)

  \Bigr\}
  \,.
\end{alignat}

Die Ergebnisse der skalaren Integrale sind bis auf die nichtlogarithmische
Konstante in $\Oc(\eps^0)$ angegeben (nur nichtverschwindende Beitr"age):
\begin{align}
\label{eq:WWccergskalar}
    F_\Wc(0, 0, 2, 1, -1, 2) &= -\tfrac{1}{\eps} \,,
\nonumber \\
  F_\Wc(0, 0, 2, 1, 0, 1) &= -\tfrac{1}{\eps} \,,
\nonumber \\
  F_\Wc(0, 0, 2, 1, 1, 0) &= \tfrac{1}{2\*\eps^2} + \tfrac{1}{2\*\eps} \,,
\nonumber \\
  F_\Wc(0, 0, 2, 1, 1, 1) &= -\tfrac{1}{\eps}/\Mr \,,
\nonumber \\
  F_\Wc(1, 0, 2, 1, -1, 1) &= -\tfrac{1}{\eps} \,,
\nonumber \\
  F_\Wc(1, 0, 2, 1, 0, 0) &= -\tfrac{1}{\eps} \,,
\nonumber \\
  F_\Wc(1, 0, 2, 1, 1, -1) &= \tfrac{1}{2\*\eps^2} + \tfrac{1}{2\*\eps} \,,
\nonumber \\
  F_\Wc(1, 0, 2, 1, 1, 0) &= -\tfrac{1}{\eps}/\Mr \,,
\nonumber \\
  F_\Wc(1, 1, 2, 1, -1, 1) &= \lqm + \lqm\tfrac{1}{\eps} \,,
\nonumber \\
  F_\Wc(1, 1, 2, 1, 0, 0) &= \lqm + \lqm\tfrac{1}{\eps} \,,
\nonumber \\
  F_\Wc(1, 1, 2, 1, 1, -3) &= \tfrac{1}{4\*\eps^2} + (\tfrac{9}{8} - \tfrac{1}{2}\*\lqm)\tfrac{1}{\eps} - \tfrac{9}{4}\*\lqm + \tfrac{1}{2}\*\lqm^2 \,,
\nonumber \\
  F_\Wc(1, 1, 2, 1, 1, -2) &= \tfrac{1}{2\*\eps^2} + (\tfrac{5}{2} - \lqm)\tfrac{1}{\eps} - 5\*\lqm + \lqm^2 \,,
\nonumber \\
  F_\Wc(1, 1, 2, 1, 1, -1) &= -\lqm^2 + \tfrac{1}{3}\*\lqm^3 + (\lqm - \tfrac{1}{2}\*\lqm^2 - \tfrac{1}{3}\*\pi^2)\tfrac{1}{\eps} + \lqm\*(2 + \tfrac{2}{3}\*\sqrt{3}\*\Cl2(\tfrac{\pi}{3})) \,,
\nonumber \\
  F_\Wc(1, 1, 2, 1, 1, 0) &= \bigl(\lqm\tfrac{1}{\eps} + \lqm\*(2 - \tfrac{4}{3}\*\sqrt{3}\*\Cl2(\tfrac{\pi}{3}))\bigr)/\Mr 

  \,.
\end{align}
Bei den drei Integralen, die proportional zu $1/\Mr = Q^2/M^2$ sind, wird
dieser Pol in~$M^2$ durch den jeweiligen Vorfaktor in der
Zerlegung~(\ref{eq:WWcczerlegung}) gek"urzt.

\subsection[Selbstenergiekorrekturen mit Eichboson- oder Geistschleife]
  {Selbstenergiekorrekturen mit Eichboson- oder\\ Geistschleife}
\label{sec:SkalarfcorrWWcc}

Zerlegung in skalare Integrale $\Bm_\Wc$ und $\DBm_\Wc$
entsprechend~(\ref{eq:fcorrWcorrskalar}):
\begin{alignat}{4}
\label{eq:fcorrWWcczerlegung}
  \lefteqn{\Sigma_\WWcc =
    C_F C_A \left(\frac{\alpha}{4\pi}\right)^2 \, i^2
    \left(\frac{\mu^2}{M^2}\right)^{2\eps} S_\eps^2
    \, \Bigl\{
  } \quad \nonumber \\
    & \Bm_\Wc(-1, 2, 0, 0, 2) &\cdot &\bigl( (-\tfrac{1}{2} - \tfrac{1}{2\*(d-1)} + \tfrac{1}{2}\*d)/\pMr \bigr)
\nonumber \\
  +& \Bm_\Wc(-1, 2, 1, -1, 2) &\cdot &\bigl( (\tfrac{1}{2} + \tfrac{1}{2\*(d-1)} - \tfrac{1}{2}\*d)/\pMr \bigr)
\nonumber \\
  +& \Bm_\Wc(-1, 2, 1, 0, 1) &\cdot &\bigl( (1 - \tfrac{1}{d-1})/\pMr \bigr)
\nonumber \\
  +& \Bm_\Wc(-1, 2, 1, 1, 0) &\cdot &\bigl( (\tfrac{3}{4} + \tfrac{1}{4\*(d-1)})/\pMr \bigr)
\nonumber \\
  +& \Bm_\Wc(-1, 2, 1, 1, 1) &\cdot &\bigl( (1 - \tfrac{1}{d-1})/\pMr \bigr)
\nonumber \\
  +& \Bm_\Wc(0, 2, 0, 0, 1) &\cdot &\bigl( (-1 + \tfrac{1}{d-1})/\pMr \bigr)
\nonumber \\
  +& \Bm_\Wc(0, 2, 0, 0, 2) &\cdot &\bigl( 1 + \tfrac{1}{d-1} - d \bigr)
\nonumber \\
  +& \Bm_\Wc(0, 2, 1, -1, 1) &\cdot &\bigl( (1 - \tfrac{1}{d-1})/\pMr \bigr)
\nonumber \\
  +& \Bm_\Wc(0, 2, 1, -1, 2) &\cdot &\bigl( -1 - \tfrac{1}{d-1} + d \bigr)
\nonumber \\
  +& \Bm_\Wc(0, 2, 1, 0, 0) &\cdot &\bigl( (-5 + \tfrac{2}{d-1} + \tfrac{3}{2}\*d)/\pMr \bigr)
\nonumber \\
  +& \Bm_\Wc(0, 2, 1, 0, 1) &\cdot &\bigl( -2 + \tfrac{2}{d-1} \bigr)
\nonumber \\
  +& \Bm_\Wc(0, 2, 1, 1, -1) &\cdot &\bigl( (-2 - \tfrac{1}{2\*(d-1)} + \tfrac{3}{4}\*d)/\pMr \bigr)
\nonumber \\
  +& \Bm_\Wc(0, 2, 1, 1, 0) &\cdot &\bigl( -\tfrac{3}{2} - \tfrac{1}{2\*(d-1)} + (-5 + \tfrac{2}{d-1} + \tfrac{3}{2}\*d)/\pMr \bigr)
\nonumber \\
  +& \Bm_\Wc(0, 2, 1, 1, 1) &\cdot &\bigl( -2 + \tfrac{2}{d-1} \bigr)
\nonumber \\
  +& \Bm_\Wc(1, 2, 0, 0, 0) &\cdot &\bigl( (\tfrac{3}{2} - \tfrac{1}{2\*(d-1)} - \tfrac{1}{2}\*d)/\pMr \bigr)
\nonumber \\
  +& \Bm_\Wc(1, 2, 0, 0, 1) &\cdot &\bigl( -1 + \tfrac{1}{d-1} \bigr)
\nonumber \\
  +& \Bm_\Wc(1, 2, 1, -1, 0) &\cdot &\bigl( (-\tfrac{3}{2} + \tfrac{1}{2\*(d-1)} + \tfrac{1}{2}\*d)/\pMr \bigr)
\nonumber \\
  +& \Bm_\Wc(1, 2, 1, -1, 1) &\cdot &\bigl( 1 - \tfrac{1}{d-1} \bigr)
\nonumber \\
  +& \Bm_\Wc(1, 2, 1, 0, -1) &\cdot &\bigl( (4 - \tfrac{1}{d-1} - \tfrac{3}{2}\*d)/\pMr \bigr)
\nonumber \\
  +& \Bm_\Wc(1, 2, 1, 0, 0) &\cdot &\bigl( -5 + \tfrac{2}{d-1} + \tfrac{3}{2}\*d \bigr)
\nonumber \\
  +& \Bm_\Wc(1, 2, 1, 1, -2) &\cdot &\bigl( (\tfrac{5}{4} + \tfrac{1}{4\*(d-1)} - \tfrac{3}{4}\*d)/\pMr \bigr)
\nonumber \\
  +& \Bm_\Wc(1, 2, 1, 1, -1) &\cdot &\bigl( -2 - \tfrac{1}{2\*(d-1)} + \tfrac{3}{4}\*d + (4 - \tfrac{1}{d-1} - \tfrac{3}{2}\*d)/\pMr \bigr)
\nonumber \\
  +& \Bm_\Wc(1, 2, 1, 1, 0) &\cdot &\bigl( -5 + \tfrac{2}{d-1} + \tfrac{3}{2}\*d \bigr)
\nonumber \\
  +& \DBm_\Wc(-1, 2, 0, 0, 2) &\cdot &\bigl( -\tfrac{1}{2} - \tfrac{1}{2\*(d-1)} + \tfrac{1}{2}\*d \bigr)
\nonumber \\
  +& \DBm_\Wc(-1, 2, 1, -1, 2) &\cdot &\bigl( \tfrac{1}{2} + \tfrac{1}{2\*(d-1)} - \tfrac{1}{2}\*d \bigr)
\nonumber \\
  +& \DBm_\Wc(-1, 2, 1, 0, 1) &\cdot &\bigl( 1 - \tfrac{1}{d-1} \bigr)
\nonumber \\
  +& \DBm_\Wc(-1, 2, 1, 1, 0) &\cdot &\bigl( \tfrac{3}{4} + \tfrac{1}{4\*(d-1)} \bigr)
\nonumber \\
  +& \DBm_\Wc(-1, 2, 1, 1, 1) &\cdot &\bigl( 1 - \tfrac{1}{d-1} \bigr)
\nonumber \\
  +& \DBm_\Wc(0, 2, 0, 0, 1) &\cdot &\bigl( -1 + \tfrac{1}{d-1} \bigr)
\nonumber \\
  +& \DBm_\Wc(0, 2, 1, -1, 1) &\cdot &\bigl( 1 - \tfrac{1}{d-1} \bigr)
\nonumber \\
  +& \DBm_\Wc(0, 2, 1, 0, 0) &\cdot &\bigl( -5 + \tfrac{2}{d-1} + \tfrac{3}{2}\*d \bigr)
\nonumber \\
  +& \DBm_\Wc(0, 2, 1, 1, -1) &\cdot &\bigl( -2 - \tfrac{1}{2\*(d-1)} + \tfrac{3}{4}\*d \bigr)
\nonumber \\
  +& \DBm_\Wc(0, 2, 1, 1, 0) &\cdot &\bigl( -5 + \tfrac{2}{d-1} + \tfrac{3}{2}\*d \bigr)
\nonumber \\
  +& \DBm_\Wc(1, 2, 0, 0, 0) &\cdot &\bigl( \tfrac{3}{2} - \tfrac{1}{2\*(d-1)} - \tfrac{1}{2}\*d \bigr)
\nonumber \\
  +& \DBm_\Wc(1, 2, 1, -1, 0) &\cdot &\bigl( -\tfrac{3}{2} + \tfrac{1}{2\*(d-1)} + \tfrac{1}{2}\*d \bigr)
\nonumber \\
  +& \DBm_\Wc(1, 2, 1, 0, -1) &\cdot &\bigl( 4 - \tfrac{1}{d-1} - \tfrac{3}{2}\*d \bigr)
\nonumber \\
  +& \DBm_\Wc(1, 2, 1, 1, -2) &\cdot &\bigl( \tfrac{5}{4} + \tfrac{1}{4\*(d-1)} - \tfrac{3}{4}\*d \bigr)
\nonumber \\
  +& \DBm_\Wc(1, 2, 1, 1, -1) &\cdot &\bigl( 4 - \tfrac{1}{d-1} - \tfrac{3}{2}\*d \bigr)

  \Bigr\}
  \,,
\end{alignat}
mit $y = p^2/M^2$.
Im Gesamtergebnis fallen die Pole in $y$ heraus.
Ergebnisse der skalaren Integrale bis $\Oc(\eps^{-1})$ (nur
nichtverschwindende Beitr"age):
\begin{align}
\label{eq:fcorrWWccergskalar}
    \Bm_\Wc(-1, 2, 1, -1, 2) &= \tfrac{1}{\eps^2} + \tfrac{1}{\eps} \,,
\nonumber \\
  \Bm_\Wc(-1, 2, 1, 0, 1) &= \tfrac{1}{\eps^2} + \tfrac{1}{\eps} \,,
\nonumber \\
  \Bm_\Wc(-1, 2, 1, 1, 0) &= \tfrac{2}{\eps^2} + \tfrac{5}{\eps} \,,
\nonumber \\
  \Bm_\Wc(-1, 2, 1, 1, 1) &= \tfrac{1}{2\*\eps^2} + \tfrac{1}{2\*\eps} \,,
\nonumber \\
  \Bm_\Wc(0, 2, 1, -1, 1) &= \tfrac{1}{\eps^2} + \tfrac{1}{\eps} \,,
\nonumber \\
  \Bm_\Wc(0, 2, 1, -1, 2) &= -\tfrac{1}{\eps} \,,
\nonumber \\
  \Bm_\Wc(0, 2, 1, 0, 0) &= \tfrac{1}{\eps^2} + \tfrac{1}{\eps} \,,
\nonumber \\
  \Bm_\Wc(0, 2, 1, 0, 1) &= -\tfrac{1}{\eps} \,,
\nonumber \\
  \Bm_\Wc(0, 2, 1, 1, -1) &= \tfrac{2}{\eps^2} + \tfrac{5}{\eps} \,,
\nonumber \\
  \Bm_\Wc(0, 2, 1, 1, 0) &= \tfrac{1}{2\*\eps^2} + \tfrac{1}{2\*\eps} \,,
\nonumber \\
  \Bm_\Wc(0, 2, 1, 1, 1) &= -\tfrac{1}{\eps} \,,
\nonumber \\
  \Bm_\Wc(1, 2, 1, -1, 0) &= \tfrac{1}{\eps^2} + \tfrac{1}{\eps} \,,
\nonumber \\
  \Bm_\Wc(1, 2, 1, -1, 1) &= -\tfrac{1}{\eps} \,,
\nonumber \\
  \Bm_\Wc(1, 2, 1, 0, -1) &= \tfrac{1}{\eps^2} + \tfrac{1}{\eps} \,,
\nonumber \\
  \Bm_\Wc(1, 2, 1, 0, 0) &= -\tfrac{1}{\eps} \,,
\nonumber \\
  \Bm_\Wc(1, 2, 1, 1, -2) &= \tfrac{2}{\eps^2} + \tfrac{5}{\eps} \,,
\nonumber \\
  \Bm_\Wc(1, 2, 1, 1, -1) &= \tfrac{1}{2\*\eps^2} + \tfrac{1}{2\*\eps} \,,
\nonumber \\
  \Bm_\Wc(1, 2, 1, 1, 0) &= -\tfrac{1}{\eps} \,,
\nonumber \\
  \DBm_\Wc(-1, 2, 1, -1, 2) &= -\tfrac{1}{\eps} \,,
\nonumber \\
  \DBm_\Wc(-1, 2, 1, 0, 1) &= -\tfrac{1}{\eps} \,,
\nonumber \\
  \DBm_\Wc(-1, 2, 1, 1, 0) &= \tfrac{1}{2\*\eps^2} + \tfrac{1}{2\*\eps} \,,
\nonumber \\
  \DBm_\Wc(-1, 2, 1, 1, 1) &= -\tfrac{1}{\eps} \,,
\nonumber \\
  \DBm_\Wc(1, 2, 1, 1, -2) &= \tfrac{1}{4\*\eps} 

  \,.
\end{align}

\subsection{Vertexdiagramme mit Higgs- und Goldstone-Bosonen}
\label{sec:SkalarHiggsvertex}

Zerlegung der Vertexdiagramme mit Higgs- und Goldstone-Bosonen
in skalare Integrale~(\ref{eq:Wcorrskalar})
f"ur ein $SU(2)$-Higgs-Modell:
\begin{alignat}{4}
\label{eq:WHvertexzerlegung}
  \lefteqn{F_{v,\WH} =
    \left(\frac{\alpha}{4\pi}\right)^2 \, i^2
    \left(\frac{\mu^2}{M^2}\right)^{2\eps} S_\eps^2
    \, \Bigl\{
  } \quad \nonumber \\
    & F_\Wc(0, 0, 2, 1, 1, 0) &\cdot &\bigl( \tfrac{3}{2}\*\Mr \bigr)
\nonumber \\
  +& F_\Wc(1, 0, 2, 1, 1, -1) &\cdot &\bigl( -3\*\Mr \bigr)
\nonumber \\
  +& F_\Wc(1, 0, 2, 1, 1, 0) &\cdot &\bigl( 3\*\Mr \bigr)
\nonumber \\
  +& F_\Wc(1, 1, 2, 1, 1, -2) &\cdot &\bigl( \tfrac{3}{2}\*\Mr \bigr)
\nonumber \\
  +& F_\Wc(1, 1, 2, 1, 1, -1) &\cdot &\bigl( (-6 + \tfrac{3}{4}\*d)\*\Mr \bigr)
\nonumber \\
  +& F_\Wc(1, 1, 2, 1, 1, 0) &\cdot &\bigl( \tfrac{3}{2}\*\Mr \bigr)

  \Bigr\}
  \,,
\end{alignat}
\begin{alignat}{4}
\label{eq:Hphiphivertexzerlegung}
  \lefteqn{F_{v,\Hphi} = F_{v,\phiphi} =
    \left(\frac{\alpha}{4\pi}\right)^2 \, i^2
    \left(\frac{\mu^2}{M^2}\right)^{2\eps} S_\eps^2
    \, \Bigl\{
  } \quad \nonumber \\
    & F_\Wc(0, 0, 2, 0, 0, 1) &\cdot &\bigl( -\tfrac{3}{4\*(d-1)} \bigr)
\nonumber \\
  +& F_\Wc(0, 0, 2, 0, 0, 2) &\cdot &\bigl( \tfrac{3}{8} + \tfrac{3}{8\*(d-1)} \bigr)
\nonumber \\
  +& F_\Wc(0, 0, 2, 1, -1, 1) &\cdot &\bigl( \tfrac{3}{4\*(d-1)} \bigr)
\nonumber \\
  +& F_\Wc(0, 0, 2, 1, -1, 2) &\cdot &\bigl( -\tfrac{3}{8} - \tfrac{3}{8\*(d-1)} \bigr)
\nonumber \\
  +& F_\Wc(0, 0, 2, 1, 0, 0) &\cdot &\bigl( -\tfrac{3}{2\*(d-1)} \bigr)
\nonumber \\
  +& F_\Wc(0, 0, 2, 1, 0, 1) &\cdot &\bigl( \tfrac{3}{4\*(d-1)} \bigr)
\nonumber \\
  +& F_\Wc(0, 0, 2, 1, 1, -1) &\cdot &\bigl( \tfrac{3}{8\*(d-1)} \bigr)
\nonumber \\
  +& F_\Wc(0, 0, 2, 1, 1, 0) &\cdot &\bigl( -\tfrac{3}{16\*(d-1)} - \tfrac{3}{2\*(d-1)}\*\Mr \bigr)
\nonumber \\
  +& F_\Wc(0, 0, 2, 1, 1, 1) &\cdot &\bigl( \tfrac{3}{4\*(d-1)}\*\Mr \bigr)
\nonumber \\
  +& F_\Wc(1, 0, 2, 0, 0, 0) &\cdot &\bigl( \tfrac{3}{2\*(d-1)} \bigr)
\nonumber \\
  +& F_\Wc(1, 0, 2, 0, 0, 1) &\cdot &\bigl( -\tfrac{3}{2\*(d-1)} \bigr)
\nonumber \\
  +& F_\Wc(1, 0, 2, 1, -1, 0) &\cdot &\bigl( -\tfrac{3}{2\*(d-1)} \bigr)
\nonumber \\
  +& F_\Wc(1, 0, 2, 1, -1, 1) &\cdot &\bigl( \tfrac{3}{2\*(d-1)} \bigr)
\nonumber \\
  +& F_\Wc(1, 0, 2, 1, 0, -1) &\cdot &\bigl( \tfrac{3}{d-1} \bigr)
\nonumber \\
  +& F_\Wc(1, 0, 2, 1, 0, 0) &\cdot &\bigl( -\tfrac{3}{d-1} \bigr)
\nonumber \\
  +& F_\Wc(1, 0, 2, 1, 1, -2) &\cdot &\bigl( -\tfrac{3}{4\*(d-1)} \bigr)
\nonumber \\
  +& F_\Wc(1, 0, 2, 1, 1, -1) &\cdot &\bigl( \tfrac{3}{4\*(d-1)} + \tfrac{3}{d-1}\*\Mr \bigr)
\nonumber \\
  +& F_\Wc(1, 0, 2, 1, 1, 0) &\cdot &\bigl( \tfrac{-3}{d-1}\*\Mr \bigr)
\nonumber \\
  +& F_\Wc(1, 1, 2, 0, 0, -1) &\cdot &\bigl( -\tfrac{3}{4\*(d-1)} \bigr)
\nonumber \\
  +& F_\Wc(1, 1, 2, 0, 0, 0) &\cdot &\bigl( -\tfrac{3}{8} + \tfrac{21}{8\*(d-1)} \bigr)
\nonumber \\
  +& F_\Wc(1, 1, 2, 0, 0, 1) &\cdot &\bigl( -\tfrac{3}{4\*(d-1)} \bigr)
\nonumber \\
  +& F_\Wc(1, 1, 2, 1, -1, -1) &\cdot &\bigl( \tfrac{3}{4\*(d-1)} \bigr)
\nonumber \\
  +& F_\Wc(1, 1, 2, 1, -1, 0) &\cdot &\bigl( \tfrac{3}{8} - \tfrac{21}{8\*(d-1)} \bigr)
\nonumber \\
  +& F_\Wc(1, 1, 2, 1, -1, 1) &\cdot &\bigl( \tfrac{3}{4\*(d-1)} \bigr)
\nonumber \\
  +& F_\Wc(1, 1, 2, 1, 0, -2) &\cdot &\bigl( -\tfrac{3}{2\*(d-1)} \bigr)
\nonumber \\
  +& F_\Wc(1, 1, 2, 1, 0, -1) &\cdot &\bigl( -\tfrac{3}{4} + \tfrac{21}{4\*(d-1)} \bigr)
\nonumber \\
  +& F_\Wc(1, 1, 2, 1, 0, 0) &\cdot &\bigl( -\tfrac{3}{2\*(d-1)} \bigr)
\nonumber \\
  +& F_\Wc(1, 1, 2, 1, 1, -3) &\cdot &\bigl( \tfrac{3}{8\*(d-1)} \bigr)
\nonumber \\
  +& F_\Wc(1, 1, 2, 1, 1, -2) &\cdot &\bigl( \tfrac{3}{16} - \tfrac{21}{16\*(d-1)} - \tfrac{3}{2\*(d-1)}\*\Mr \bigr)
\nonumber \\
  +& F_\Wc(1, 1, 2, 1, 1, -1) &\cdot &\bigl( \tfrac{3}{8\*(d-1)} + (-\tfrac{3}{4} + \tfrac{21}{4\*(d-1)})\*\Mr \bigr)
\nonumber \\
  +& F_\Wc(1, 1, 2, 1, 1, 0) &\cdot &\bigl( \tfrac{-3}{2\*(d-1)}\*\Mr \bigr)

  \Bigr\}
  \,.
\end{alignat}
Die Ergebnisse der skalaren Integrale sind aus
Gl.~(\ref{eq:WWccergskalar}) bekannt.

\subsection[Selbstenergiekorrekturen mit Higgs- und Goldstone-Bosonen]
  {Selbstenergiekorrekturen mit Higgs- und\\ Goldstone-Bosonen}
\label{sec:SkalarHiggsfcorr}

Zerlegung der Selbstenergiediagramme mit Higgs- und Goldstone-Bosonen
in skalare Integrale $\Bm_\Wc$ und $\DBm_\Wc$
entsprechend~(\ref{eq:fcorrWcorrskalar})
f"ur ein $SU(2)$-Higgs-Modell:
\begin{alignat}{4}
\label{eq:WHfcorrzerlegung}
  \lefteqn{\Sigma_\WH =
    \left(\frac{\alpha}{4\pi}\right)^2 \, i^2
    \left(\frac{\mu^2}{M^2}\right)^{2\eps} S_\eps^2
    \, \Bigl\{
  } \quad \nonumber \\
    & \Bm_\Wc(0, 2, 1, 1, 0) &\cdot &\bigl( (\tfrac{3}{4} - \tfrac{3}{8}\*d)/\pMr \bigr)
\nonumber \\
  +& \Bm_\Wc(1, 2, 1, 1, -1) &\cdot &\bigl( (-\tfrac{3}{4} + \tfrac{3}{8}\*d)/\pMr \bigr)
\nonumber \\
  +& \Bm_\Wc(1, 2, 1, 1, 0) &\cdot &\bigl( \tfrac{3}{4} - \tfrac{3}{8}\*d \bigr)
\nonumber \\
  +& \DBm_\Wc(0, 2, 1, 1, 0) &\cdot &\bigl( \tfrac{3}{4} - \tfrac{3}{8}\*d \bigr)
\nonumber \\
  +& \DBm_\Wc(1, 2, 1, 1, -1) &\cdot &\bigl( -\tfrac{3}{4} + \tfrac{3}{8}\*d \bigr)

  \Bigr\}
  \,,
\end{alignat}
\begin{alignat}{4}
\label{eq:Hphiphifcorrzerlegung}
  \lefteqn{\Sigma_\Hphi = \Sigma_\phiphi =
    \left(\frac{\alpha}{4\pi}\right)^2 \, i^2
    \left(\frac{\mu^2}{M^2}\right)^{2\eps} S_\eps^2
    \, \Bigl\{
  } \quad \nonumber \\
    & \Bm_\Wc(-1, 2, 0, 0, 2) &\cdot &\bigl( (\tfrac{3}{16} + \tfrac{3}{16\*(d-1)})/\pMr \bigr)
\nonumber \\
  +& \Bm_\Wc(-1, 2, 1, -1, 2) &\cdot &\bigl( (-\tfrac{3}{16} - \tfrac{3}{16\*(d-1)})/\pMr \bigr)
\nonumber \\
  +& \Bm_\Wc(-1, 2, 1, 0, 1) &\cdot &\bigl( (\tfrac{3}{8\*(d-1)})/\pMr \bigr)
\nonumber \\
  +& \Bm_\Wc(-1, 2, 1, 1, 0) &\cdot &\bigl( -(\tfrac{3}{32\*(d-1)})/\pMr \bigr)
\nonumber \\
  +& \Bm_\Wc(-1, 2, 1, 1, 1) &\cdot &\bigl( (\tfrac{3}{8\*(d-1)})/\pMr \bigr)
\nonumber \\
  +& \Bm_\Wc(0, 2, 0, 0, 1) &\cdot &\bigl( -(\tfrac{3}{8\*(d-1)})/\pMr \bigr)
\nonumber \\
  +& \Bm_\Wc(0, 2, 0, 0, 2) &\cdot &\bigl( -\tfrac{3}{8} - \tfrac{3}{8\*(d-1)} \bigr)
\nonumber \\
  +& \Bm_\Wc(0, 2, 1, -1, 1) &\cdot &\bigl( (\tfrac{3}{8\*(d-1)})/\pMr \bigr)
\nonumber \\
  +& \Bm_\Wc(0, 2, 1, -1, 2) &\cdot &\bigl( \tfrac{3}{8} + \tfrac{3}{8\*(d-1)} \bigr)
\nonumber \\
  +& \Bm_\Wc(0, 2, 1, 0, 0) &\cdot &\bigl( (\tfrac{3}{8} - \tfrac{3}{4\*(d-1)})/\pMr \bigr)
\nonumber \\
  +& \Bm_\Wc(0, 2, 1, 0, 1) &\cdot &\bigl( -\tfrac{3}{4\*(d-1)} \bigr)
\nonumber \\
  +& \Bm_\Wc(0, 2, 1, 1, -1) &\cdot &\bigl( (-\tfrac{3}{32} + \tfrac{3}{16\*(d-1)})/\pMr \bigr)
\nonumber \\
  +& \Bm_\Wc(0, 2, 1, 1, 0) &\cdot &\bigl( \tfrac{3}{16\*(d-1)} + (\tfrac{3}{8} - \tfrac{3}{4\*(d-1)})/\pMr \bigr)
\nonumber \\
  +& \Bm_\Wc(0, 2, 1, 1, 1) &\cdot &\bigl( -\tfrac{3}{4\*(d-1)} \bigr)
\nonumber \\
  +& \Bm_\Wc(1, 2, 0, 0, 0) &\cdot &\bigl( (-\tfrac{3}{16} + \tfrac{3}{16\*(d-1)})/\pMr \bigr)
\nonumber \\
  +& \Bm_\Wc(1, 2, 0, 0, 1) &\cdot &\bigl( -\tfrac{3}{8\*(d-1)} \bigr)
\nonumber \\
  +& \Bm_\Wc(1, 2, 1, -1, 0) &\cdot &\bigl( (\tfrac{3}{16} - \tfrac{3}{16\*(d-1)})/\pMr \bigr)
\nonumber \\
  +& \Bm_\Wc(1, 2, 1, -1, 1) &\cdot &\bigl( \tfrac{3}{8\*(d-1)} \bigr)
\nonumber \\
  +& \Bm_\Wc(1, 2, 1, 0, -1) &\cdot &\bigl( (-\tfrac{3}{8} + \tfrac{3}{8\*(d-1)})/\pMr \bigr)
\nonumber \\
  +& \Bm_\Wc(1, 2, 1, 0, 0) &\cdot &\bigl( \tfrac{3}{8} - \tfrac{3}{4\*(d-1)} \bigr)
\nonumber \\
  +& \Bm_\Wc(1, 2, 1, 1, -2) &\cdot &\bigl( (\tfrac{3}{32} - \tfrac{3}{32\*(d-1)})/\pMr \bigr)
\nonumber \\
  +& \Bm_\Wc(1, 2, 1, 1, -1) &\cdot &\bigl( -\tfrac{3}{32} + \tfrac{3}{16\*(d-1)} + (-\tfrac{3}{8} + \tfrac{3}{8\*(d-1)})/\pMr \bigr)
\nonumber \\
  +& \Bm_\Wc(1, 2, 1, 1, 0) &\cdot &\bigl( \tfrac{3}{8} - \tfrac{3}{4\*(d-1)} \bigr)
\nonumber \\
  +& \DBm_\Wc(-1, 2, 0, 0, 2) &\cdot &\bigl( \tfrac{3}{16} + \tfrac{3}{16\*(d-1)} \bigr)
\nonumber \\
  +& \DBm_\Wc(-1, 2, 1, -1, 2) &\cdot &\bigl( -\tfrac{3}{16} - \tfrac{3}{16\*(d-1)} \bigr)
\nonumber \\
  +& \DBm_\Wc(-1, 2, 1, 0, 1) &\cdot &\bigl( \tfrac{3}{8\*(d-1)} \bigr)
\nonumber \\
  +& \DBm_\Wc(-1, 2, 1, 1, 0) &\cdot &\bigl( -\tfrac{3}{32\*(d-1)} \bigr)
\nonumber \\
  +& \DBm_\Wc(-1, 2, 1, 1, 1) &\cdot &\bigl( \tfrac{3}{8\*(d-1)} \bigr)
\nonumber \\
  +& \DBm_\Wc(0, 2, 0, 0, 1) &\cdot &\bigl( -\tfrac{3}{8\*(d-1)} \bigr)
\nonumber \\
  +& \DBm_\Wc(0, 2, 1, -1, 1) &\cdot &\bigl( \tfrac{3}{8\*(d-1)} \bigr)
\nonumber \\
  +& \DBm_\Wc(0, 2, 1, 0, 0) &\cdot &\bigl( \tfrac{3}{8} - \tfrac{3}{4\*(d-1)} \bigr)
\nonumber \\
  +& \DBm_\Wc(0, 2, 1, 1, -1) &\cdot &\bigl( -\tfrac{3}{32} + \tfrac{3}{16\*(d-1)} \bigr)
\nonumber \\
  +& \DBm_\Wc(0, 2, 1, 1, 0) &\cdot &\bigl( \tfrac{3}{8} - \tfrac{3}{4\*(d-1)} \bigr)
\nonumber \\
  +& \DBm_\Wc(1, 2, 0, 0, 0) &\cdot &\bigl( -\tfrac{3}{16} + \tfrac{3}{16\*(d-1)} \bigr)
\nonumber \\
  +& \DBm_\Wc(1, 2, 1, -1, 0) &\cdot &\bigl( \tfrac{3}{16} - \tfrac{3}{16\*(d-1)} \bigr)
\nonumber \\
  +& \DBm_\Wc(1, 2, 1, 0, -1) &\cdot &\bigl( -\tfrac{3}{8} + \tfrac{3}{8\*(d-1)} \bigr)
\nonumber \\
  +& \DBm_\Wc(1, 2, 1, 1, -2) &\cdot &\bigl( \tfrac{3}{32} - \tfrac{3}{32\*(d-1)} \bigr)
\nonumber \\
  +& \DBm_\Wc(1, 2, 1, 1, -1) &\cdot &\bigl( -\tfrac{3}{8} + \tfrac{3}{8\*(d-1)} \bigr)

  \Bigr\}
  \,.
\end{alignat}
Die Ergebnisse der skalaren Integrale sind aus
Gl.~(\ref{eq:fcorrWWccergskalar}) bekannt.


\clearemptypage



\bibliographystyle{diss-unsrt}

\bibliography{literatur}

\clearemptypage


\chapter*{Danksagung}
\addcontentsline{toc}{chapter}{Danksagung}
\pagestyle{plain}

\vspace{-0.25cm}
Mein erster Dank gilt Herrn Prof. Dr. Johann H. K"uhn, der mir mit der
Promotion eine interessante wissenschaftliche Arbeit in einer sehr
angenehmen Atmosph"are erm"oglicht hat.
Herrn Prof. Dr. Dieter Zeppenfeld danke ich f"ur die "Ubernahme des
Korreferats.
Beiden Referenten danke ich f"ur zahlreiche wichtige Hinweise und kl"arende
Fragen zur Doktorarbeit.

F"ur die fruchtbare und lehrreiche wissenschaftliche Zusammenarbeit bedanke
ich mich bei Prof. Dr. Johann H. K"uhn, Dr.~Sven Moch, Dr.~Alexander Penin
und Prof. Dr. Vladimir Smirnov. Sie alle haben sich immer viel Zeit
genommen, meine physikalischen und rechentechnischen Fragen zu beantworten,
so dass ich bei diesem Projekt eine Menge dazugelernt habe.

Besonders herzlich danke ich meiner Frau Annette Jantzen, die mich immer sehr
unterst"utzt hat, und ohne deren R"uckhalt die stressige Zeit am Ende der
Promotion kaum zu bew"altigen gewesen w"are.

Dem Graduiertenkolleg "`Hochenergiephysik und Teilchenastrophysik"' an der
Universit"at Karlsruhe danke ich f"ur die finanzielle Unterst"utzung
und f"ur bereichernde Workshops und Seminare.
Ebenso danke ich der Landesgraduiertenf"orderung Baden-W"urttemberg f"ur
das Promotionsstipendium.
Der gr"o"ste Dank gilt dem Cusanuswerk, nicht nur f"ur die l"angste
Stipendienf"orderung, sondern auch f"ur die vielen interdisziplin"aren
Begegnungen auf den Graduiertentagungen und auf dem Jahrestreffen, die
letztendlich meine eigene Biographie entscheidend gepr"agt haben.

Ich danke allen Mitarbeiten des Instituts f"ur Theoretische Teilchenphysik
f"ur die gute Atmosph"are und die Hilfe, die ich von ihnen erfahren habe.
Besonders erw"ahnen m"ochte ich Christian Sturm, mit dem ich mehrere Jahre
das B"uro geteilt und unz"ahlige interessante Diskussionen gef"uhrt habe,
sowie Michael Faisst, Michael Kraetz, Martin Melcher, Dr. Pierpaolo
Mastrolia, Stefan Bekavac, Andreas Scharf, Dr. Stefano Pozzorini,
Dr. Marcus Weber, Prof. Dr. Matthias Steinhauser, Dr. Robert Harlander
und Dr. Luminita Mihaila.
Ein weiterer Dank gilt Dr. Oleg Veretin f"ur die "Uberlassung seines
Fortran-Programms zum PSLQ-Algorithmus.

Abschlie"send bedanke ich mich bei meiner Familie und bei meinen Freunden
f"ur die Unterst"utzung und die Verbundenheit in den Jahren der Promotion.

\clearemptypage


\thispagestyle{empty}
\let\em=\slshape
\renewcommand{\arraystretch}{1.2}

\centerline{\bfseries\Large Lebenslauf}

\vspace{0.75cm}

\textbf{Pers"onliche Daten:}

\begin{tabular}{@{}p{0.3\textwidth}@{}p{0.7\textwidth}@{}}
  \em Name: & Bernd Joachim Jantzen ({\em geb.} Feucht) \\
  \em Geburtsdatum: & 3. August 1975 \\
  \em Geburtsort: & Backnang \\
  \em Staatsangeh"origkeit: & deutsch \\
  \em Familienstand: & verheiratet
\end{tabular}

\vspace{0.75cm}

\textbf{Ausbildung:}

\begin{tabular}{@{}p{0.25\textwidth}@{}lp{0.55\textwidth}@{}}
  \em Schulbesuch: &
    1982 -- 1986 & Grundschule Schwaigern \\ &
    1986 -- 1995 & Gymnasium Eppingen \\[2ex]
  \em Zivildienst: & 1995 -- 1996 &
    Caritas-Altenheim St. Elisabeth, Heilbronn \\[2ex]
  \em Studium: &
    1996 -- 2002 & Physik an der Universit"at Karlsruhe \\ &
    1999 -- 2000 & Ma\^itrise Physique an der
      Universit\'e Joseph Fourier, Grenoble (Frankreich) \\[1ex]
  \em Diplomarbeit: & \multicolumn{2}{@{}p{0.7\textwidth}@{}}{
    "`Sudakov-Logarithmen in einer massiven $U(1)$-Theorie
    in Zweischleifen-N"aherung"'} \\[1ex]
  \em Referent: & \multicolumn{2}{@{}p{0.7\textwidth}@{}}{%
     Prof.~Dr.~J.H.~K"uhn} \\[1ex]
  \em Diplom: & \multicolumn{2}{@{}p{0.7\textwidth}@{}}{%
    4. Februar 2002 mit der Note 1,0 (mit Auszeichnung)} \\[2ex]
  \em Promotion: & \multicolumn{2}{@{}p{0.7\textwidth}@{}}{%
    Physik an der Universit"at Karlsruhe seit Februar 2002 \newline
    Februar -- M"arz 2002: Stipendiat des Graduiertenkollegs
      "`Hochenergiephysik und Teilchenastrophysik"'
      an der \newline Universit"at Karlsruhe \newline
    April -- August 2002: Stipendiat der Landesgraduierten\-f"orderung
      Baden-W"urttemberg \newline
    seit September 2002: Stipendiat des Cusanuswerks} \\[1ex]
  \em Tag der Pr"ufung: & \multicolumn{2}{@{}p{0.7\textwidth}@{}}{%
    17. Juni 2005} \\[1ex]
  \em Titel der Arbeit: & \multicolumn{2}{@{}p{0.7\textwidth}@{}}{%
    "`Sudakov-Logarithmen in der elektroschwachen \newline Wechselwirkung"'}
    \\[3ex]
  \em Referent: & \multicolumn{2}{@{}p{0.7\textwidth}@{}}{%
     Prof.~Dr.~J.H.~K"uhn} \\[1ex]
  \em Korreferent: & \multicolumn{2}{@{}p{0.7\textwidth}@{}}{%
     Prof.~Dr.~D.~Zeppenfeld}
\end{tabular}

\end{appendix}

\end{document}